%% file: main.tex
\documentclass{article}
\usepackage[utf8]{inputenc}
\usepackage{graphicx}
\usepackage[portrait, margin=0.8in]{geometry}
\usepackage{hyperref}
\usepackage{authblk}
\usepackage{todonotes}
\usepackage{comment}
\usepackage{algpseudocode}
\usepackage{algorithm,algorithmicx,algcompatible}
\usepackage{braket}
\usepackage{amsmath,amsthm,amssymb}
\usepackage[sorting = none, backend = bibtex, style=numeric-comp]{biblatex}
\usepackage{adjustbox}
\usepackage{booktabs,multirow}
\usepackage{tikz}
\usetikzlibrary{quantikz}
\usepackage{enumitem}
\usetikzlibrary{external}
\newcommand{\dicke}[2]{\ket{D_{#2}^{#1}}}
\newcommand{\dsu}[1]{\textit{DSU}(#1)}
\newcommand{\scs}[1]{\textit{SCS}(#1)}
\newcommand{\ryangle}[2]{2\arccos(\sqrt{#1/#2})}

\title{Probing Quantum Telecloning on Superconducting Quantum Processors}

\author[1]{Elijah Pelofske\thanks{Email: epelofske@lanl.gov}}
\author[1]{Andreas Bärtschi\thanks{Email: baertschi@lanl.gov}}
\author[1]{Stephan Eidenbenz}
\author[2]{Bryan Garcia}
\author[2]{Boris Kiefer}
\affil[1]{Los Alamos National Laboratory, CCS-3 Information Sciences}
\affil[2]{New Mexico State University, Department of Physics}

\date{\vspace{-6ex}}

\begin{document}
\maketitle

\begin{abstract}
\input{abstract}
\end{abstract}

\input{text}

\end{document}

%% file: abstract.tex
Quantum information can not be perfectly cloned, but approximate copies of quantum information can be generated. Quantum telecloning combines approximate quantum cloning, more typically referred as quantum cloning, and quantum teleportation. Quantum telecloning allows approximate copies of quantum information to be constructed by separate parties, using the classical results of a Bell measurement made on a prepared quantum telecloning state. Quantum telecloning can be implemented as a circuit on quantum computers using a classical co-processor to compute classical feed forward instructions using if statements based on the results of a mid-circuit Bell measurement in real time. We present universal, symmetric, optimal $1 \rightarrow M$ telecloning circuits, and experimentally demonstrate these quantum telecloning circuits for $M=2$ up to $M=10$, natively executed with real time classical control systems on IBM Quantum superconducting processors, known as dynamic circuits. We perform the cloning procedure on many different message states across the Bloch sphere, on $7$ IBM Quantum processors, optionally using the error suppression technique X-X sequence digital dynamical decoupling. Two circuit optimizations are utilized, one which removes ancilla qubits for $M=2, 3$, and one which reduces the total number of gates in the circuit but still uses ancilla qubits. Parallel single qubit tomography with MLE density matrix reconstruction is used in order to compute the mixed state density matrices of the clone qubits, and clone quality is measured using quantum fidelity. These results present one of the largest and most comprehensive NISQ computer experimental analyses on (single qubit) quantum telecloning to date. The clone fidelity sharply decreases to $0.5$ for $M > 5$, but for $M=2$ we are able to achieve a mean clone fidelity of up to $0.79$ using dynamical decoupling. 

%% file: text.tex
\section{Introduction}
\label{section:introduction}

One of the fundamental properties of quantum mechanics is that unknown arbitrary quantum information can not be cloned~\cite{wootters1982single, dieks1982communication}, more specifically it can not be \emph{perfectly} cloned. This result is highly consequential, both as a tool and a hurdle that causes quantum information processing to be handled very differently from classical information processing in many domains of quantum information, including quantum networking communication~\cite{Cacciapuoti_2020, cozzolino2019high, lu2019experimental}, quantum cryptography~\cite{Bennett_2014, PhysRevA.72.032301, PhysRevLett.67.661, PhysRevA.73.012337, bouchard2017high, PhysRevLett.110.173601, PhysRevLett.126.060503, PhysRevA.69.032313, PhysRevA.72.032320, srikara2020continuous} and quantum error correction~\cite{Cerf_1997, gottesman2000introduction, roffe2019quantum}. However, it was subsequently shown that in fact \emph{approximate quantum cloning} is possible~\cite{Bu_ek_1996}. 

Since then, there have been a large number of variants of approximate quantum cloning (copying) protocols \footnote{We will typically refer to any approximate quantum cloning process simply as \emph{quantum cloning}}~\cite{fan2014quantum, RevModPhys.77.1225, cerf2006optical, PhysRevA.62.062302, Maruyama_2003}. Optimal quantum cloning refers to any process that approaches the theoretical upper limit of the cloning process with respect to clone quality~\cite{PhysRevLett.79.2153, PhysRevA.58.4377, PhysRevLett.81.2598, gisin1998quantum}, usually measured as the fidelity~\cite{jozsa1994fidelity}, of the generated clones for an input quantum state allowed by the laws of quantum mechanics. How close a cloning protocol is to the theoretical limit can be established by computing the fidelity of the prepared cloned quantum state and the original pure quantum state. Eq.\eqref{eq:theoretical-fidelity} defines the optimal \emph{universal} (e.g. state independent) quantum cloning fidelities when copying $N$ quantum states to $M$ clones. Universal quantum cloning refers to a cloning process where the quality of the clone, e.g. how closely it represents the state which is copied, is independent of the quantum state that is being cloned and all generated clones would have the same ideal state overlap with the original input state~\cite{PhysRevA.56.3446, PhysRevLett.81.5003, PhysRevA.57.2368, PhysRevA.84.034302, adhikari2007hybrid}. State depending quantum cloning~\cite{PhysRevA.57.2368, cerf2000asymmetric, PhysRevA.72.042328, PhysRevA.66.042304, adhikari2007hybrid}, in contrast to universal quantum cloning, performs a copying process that results in clones whose quality is dependent in the state that is cloned. For example, phase-covariant quantum cloning generates perfect quantum copies of equatorial single qubit states on the Bloch sphere single qubit states~\cite{Fan_2001, Bru__2000}. Symmetric quantum cloning refers to a cloning process which generates indistinguishable clones, and asymmetric quantum cloning generates clones of varying quality (e.g. not identical clones). Probabilistic quantum cloning refers to a cloning process where the expected clone quality is not deterministic, although on average it will provide the same clone quality as a deterministic quantum cloning process~\cite{PhysRevLett.106.180404, hardy1999no, PhysRevA.62.042302}. Quantum cloning can be applied to discrete variable quantum information, as well as continuous variable quantum information~\cite{PhysRevLett.86.4942, PhysRevLett.126.060503}. Quantum cloning, of all variants enumerated above, can be applied to any type of quantum information including qudits~\cite{PhysRevA.67.022317, cerf2002cloning, PhysRevA.69.032313, PhysRevA.71.042327, 10.5555/2535680.2535689, PhysRevA.79.064306, PhysRevA.58.3484}, where the quantum information unit is a discrete $d$-dimensional system. There have been a large number of experimental realizations of different quantum cloning variants using a number of different quantum technology platforms, see refs.~\cite{yang2021experimental, bouchard2017high, PhysRevLett.88.187901, PhysRevA.105.042604, PhysRevA.75.012317, gupta10achieving, PhysRevLett.92.047901, haw2016surpassing, dan2010universal} for more examples of physical quantum cloning experiments. Quantum telecloning, which is the focus of this paper, is a combination of a quantum cloning process and quantum teleportation~\cite{PhysRevA.59.156, PhysRevLett.87.247901, PhysRevA.72.032331, PhysRevA.67.012323, zhang2006cavity}. Studying quantum cloning is of fundamental interest for the field of quantum information processing, and has relevance in quantum information in physical phenomena~\cite{Adami_2015}. 

\begin{equation}
    F_{N \rightarrow M} = \frac{ MN + M + N }{ M(N+2) }
    \label{eq:theoretical-fidelity}
\end{equation}

For this article, we use an existing quantum telecloning circuit method~\cite{telecloning_circuits_1} along with a telecloning circuit ancilla qubit optimization technique~\cite{telecloning_circuits_2} to construct and execute quantum telecloning circuits using the \emph{dynamic circuit} functionality on IBMQ systems~\cite{Cross_2022, gupta2023encoding, bäumer2023efficient, C_rcoles_2021} in order to directly execute the telecloning protocol using mid-circuit measurements and real time classical co-processing. These quantum telecloning circuits are \emph{optimal}, \emph{universal}, \emph{symmetric}, and act on qubit based quantum information systems (e.g. $d=2$ quantum systems, as opposed to general $d$ dimensional systems). Importantly, these quantum telecloning circuits~\cite{telecloning_circuits_1, telecloning_circuits_2} are scalable to any number of clones $M$ for $1 \rightarrow M$; the limitation is that as $M$ increases, the resulting quantum telecloning machine (e.g. circuit) has increased gate depth, total gate count, and number of qubits used, making experimental demonstrations of very large quantum telecloning circuits difficult because of errors and noise present in the computation~\cite{Preskill2018quantumcomputingin}. Expanding on the capabilities of OpenQASM 2~\cite{https://doi.org/10.48550/arxiv.1707.03429}, OpenQASM 3~\cite{Cross_2022} provides the capability to specify these \emph{dynamic circuits}; in particular, we use mid-circuit measurements and classical conditional feed-forward operations, which subsequently cause different quantum gate instructions to be applied to the circuit while it is executing (e.g. while the state of the circuit is not measured). During the classical processing, the quantum processor could be performing additional operations, or could be idling and waiting for the result of the classical processor in order to proceed. In this case, the quantum processor is idle for a period of approximately several hundred nanoseconds~\cite{gupta2023encoding}, which could introduce errors into the computation. Because quantum telecloning circuits require native classical if-statements, conditioned on mid-circuit measurements, running quantum telecloning circuits using the \emph{dynamic circuit} capability of OpenQASM3 on IBMQ devices provide a reasonable analysis of the current achievable device performance when classical conditional operations and mid-circuit measurements are used. Note that Quantinuum quantum computers also provide mid-circuit measurement and classical conditioning capability, which was experimentally demonstrated for $1 \rightarrow 9$ on Quantinuum H1-1~\cite{telecloning_circuits_2}. 

More precisely, quantum telecloning circuits with ancilla qubits, using the ancilla optimization from ref.~\cite{telecloning_circuits_2}, for $N=1 \rightarrow M=2, 3, 4$ are executed on seven IBM Quantum computers. For $M=2$ and $M=3$, gate model circuit constructions are known which do not use ancilla qubits~\cite{telecloning_circuits_1} and for $M=2,3$ we implement quantum telecloning circuits without ancilla qubits. Importantly, these quantum telecloning circuits~\cite{telecloning_circuits_1, telecloning_circuits_2} rely on preparing Dicke state unitaries, and recent advancements in optimized Dicke state preparation circuits~\cite{B_rtschi_2019, 9951196, 9774323} allows for the quantum telecloning circuits which utilize fewer gate operations. Specifically for the experiments we present, we utilize the optimized Linear Nearest Neighbors (LNN) connectivity Dicke state preparation circuits. Dicke states are uniform superpositions of $n$ qubit states with hamming weight $k$, which interestingly turn out to be a critical building block for quantum telecloning states.

These quantum telecloning results are the most comprehensive experiments reported to date of quantum cloning, executed on quantum computers. For instance, there have been a number of experimental demonstrations of $1 \rightarrow 2$ quantum cloning~\cite{yang2021experimental}, but here we present circuits which can implement arbitrary $1 \rightarrow M$ quantum telecloning, and present experimental results for up to $M=10$ on IBM Quantum fixed-frequency superconducting qubit processors~\cite{PhysRevLett.107.080502}. Importantly, we provide explicit circuit instructions for building quantum telecloning machines on quantum computers, which has not generally been presented before. We believe that these quantum telecloning circuits, and future constructions of quantum cloning circuits, can be used in quantum computations for simulating quantum information networking protocols. These experiments also demonstrate the capability of mid-circuit measurements with feed-forward classical co-processor control (also known as \emph{dynamic circuits}) on IBMQ superconducting quantum processors. The use of dynamic circuits, or real time classical co-processor feedback, means that direct quantum telecloning can be executed, as opposed to processors where this classical feedback is missing and alternative methods such as deferred measurement and post selection must be utilized~\cite{telecloning_circuits_1}.

Figures are generated using a combination of Qiskit~\cite{Qiskit}, Matplotlib~\cite{thomas_a_caswell_2021_5194481, Hunter:2007}, QuTiP~\cite{JOHANSSON20121760, JOHANSSON20131234} and mayavi~\cite{ramachandran2011mayavi}. Data, code, and extra figures are available on a public Github repository\footnote{\url{https://github.com/lanl/Quantum-Telecloning}}.

\section{Methods}
\label{section:methods}

\subsection{Quantum Telecloning circuits}
\label{section:methods_telecloning_circuits}
Algorithm~\ref{algorithm:telecloning} describes the Quantum Telecloning protocol for distributing $M$ single qubit clones of an arbitrary $N=1$ qubit state to $M$ separate parties. For general $1\rightarrow M$ quantum telecloning, a \emph{telecloning state} $A^{(M-1)}PC^M$ is prepared of the form~\cite{PhysRevA.59.156}
\begin{align}
    \ket{A^{M-1}PC^M} = \frac{1}{\sqrt{M+1}} \sum\nolimits_{i=0}^M \dicke{M}{i}_{A^{M-1}P} \dicke{M}{i}_{C^M},
\end{align}
where $\dicke{M}{i}$ denotes the uniform superposition over all $M$-qubit states of Hamming weight $i$ with real amplitudes. Dicke state unitaries $\dsu{M}$ and Split \& Cyclic Shift unitaries $\scs{m}$~\cite{B_rtschi_2019} are defined as:
\begin{align*}
    \dsu{M}\colon & \ket{1^i 0^{M-i}} \mapsto \dicke{M}{i},   \\
    \scs{m}\colon & \ket{1^i 0^{m-i}} \mapsto \surd{\tfrac{m-i}{m}} \ket{1^i 0^{m-i}} + \surd{\tfrac{i}{m}} \ket{1^{i-1} 0^{m-i}1}.
\end{align*}
Dicke state unitaries are constructed recursively, $\dsu{M} = \dsu{M-1} \cdot \scs{M} = \scs{2} \cdot \scs{3} \cdot \ldots \cdot \scs{M}$. The observation from ref.~\cite{telecloning_circuits_2} is that previously, quantum telecloning circuits with ancilla applied a Dicke state unitary $\dsu{M}$ on the port and ancilla qubits~\cite{telecloning_circuits_1}; however, only the first SCS unitary $\scs{M}$ acts on the port qubit, the remaining $\scs{m}$ unitaries comprising a $\dsu{M-1}$ unitary act only on the to-be discarded ancilla qubits. Therefore, the remaining $\scs{m}$ unitaries can be removed from the circuit without affecting the quantum telecloning state, thus reducing the number of two qubit gates significantly (especially for large $M$).  

Figure~\ref{fig:main-circuits} details the quantum telecloning circuit construction for $1 \rightarrow 3$ quantum telecloning, using the ancilla optimization provided in ref.~\cite{telecloning_circuits_2}. Importantly, the circuits with ancilla qubits, including the the ancilla optimization provided in ref.~\cite{telecloning_circuits_2}, can be fully generalized to $1 \rightarrow M$ telecloning. Figure~\ref{fig:main-circuits} also describes the quantum telecloning circuit for $1 \rightarrow 3$ quantum telecloning circuits with no ancilla qubits from ref.~\cite{telecloning_circuits_1}. A telecloning circuit without ancilla qubits is also known for $1 \rightarrow 2$~\cite{telecloning_circuits_1}. More detailed circuit descriptions, including compiled circuit examples, are given in Appendix~\ref{section:appendix_compiled_circuits}. 

Note that the clones produced by the quantum telecloning circuits are weakly entangled, which we will describe using the entanglement measures of negativity~\cite{Vidal_2002, PhysRevA.58.883, eisert2006entanglement} and concurrence~\cite{Hill_1997, PhysRevLett.80.2245}. Importantly, the will-be clone qubits are \emph{not entangled} before the Bell measurement is performed~\cite{PhysRevA.59.156}, but after the Bell measurement they become weakly entangled. Negativity and concurrence are two quantum information measures of entanglement that are defined on a density matrix $\rho$ where the quantum state is not separable (e.g. the qubits are entangled) if the negativity or concurrence of $\rho$ greater than $0$ - if the measures are $0$ then the states are not entangled. Concurrence is defined to be $\in [0, 1]$ and negativity is defined to be $\in [0, 0.5]$. The operations are defined as $\mathit{N}(\rho)$ for negativity and $\mathit{C}(\rho)$ for concurrence. 

For a symmetric universal quantum telecloning machine the $1 \rightarrow 2$ quantum cloning process produces the subsystem of $2$ clones which have a concurrence measure of $\mathit{C}(\rho_{M=2}) = \frac{1}{3}$, and a negativity measure of $\mathit{N}(\rho_{M=2}) = \frac{\sqrt{5}-2}{6}$. The computed entanglement measures for the quantum clones are independent of the single qubit state that is cloned. The entanglement measures were computed using Qiskit to numerically simulate the circuits classically and compute the density matrices of the $2$ qubit clone sub-system. These entanglement measures are independent of whether the quantum telecloning circuit is the variant with or without ancilla qubits. The Python 3 packages Qiskit~\cite{Qiskit} and Toqito~\cite{russo2021toqito} helped with the numerical computation of the negativity and concurrence measures from the density matrices. Concurrence is defined only for two qubit states, whereas negativity is defined for any multi qubit state. 

\begin{algorithm}[t!]
\caption{Quantum $1\rightarrow M$ Telecloning Protocol}
\begin{algorithmic}[1]
\Statex \hspace*{-2.5ex}\textbf{State Preparation:}
\State A message qubit $q_m$ is prepared by a sender (e.g. $N=1$)
\State A quantum telecloning state $TC$ is constructed with
\Statex - (up to) ~$(M-1)$ ancilla qubits $A$, $1$ Port qubit $P$, and
\Statex -~$M$ clone qubits $C$ (symmetrized with Dicke state unitaries; sent to the receivers).
\Statex \hspace*{-2.5ex}\textbf{Teleportation:}
\State A Bell measurement is made between $q_m$ and $P$, and the results are communicated over a classical channel (assumed to be noiseless) to the clone holders.
\State The clone holders use the two classical bits from the bell measurement to decide whether to apply $X$- and/or $Z$-gates to the clone qubits so as to construct the approximate clones:
\Statex - $\Phi^+$: apply no gate
\hspace*{20pt} - $\Phi^-$: apply $Z$-gate
\Statex - $\Psi^+$: apply $X$-gate
\hspace*{21pt} - $\Psi^-$: apply $X$- then $Z$-gate
\Statex \hspace*{-2.5ex}\textbf{Result:}
\State $M$ approximate clones of $q_m$ have been generated, by the $M$ clone holders, with theoretical maximal single qubit fidelity given by eq.~\eqref{eq:theoretical-fidelity}.
\end{algorithmic}
\label{algorithm:telecloning}
\end{algorithm}

\newcommand{\rynewg}[2]{\gate{\sqrt{#1/#2}}}
\begin{figure*}[h!]
      \centering
      \begin{adjustbox}{width=\linewidth}
            \begin{quantikz}[row sep={24pt,between origins},execute at end picture={
                              \node[fit=(\tikzcdmatrixname-2-8)(\tikzcdmatrixname-4-9),draw,dashed,thick,rounded corners,inner xsep=14pt,inner ysep=7pt,xshift=10.5pt,yshift=-4.5pt,label={[yshift=-2pt]$\dsu{3}$}] {};                         
                              \node[fit=(\tikzcdmatrixname-5-8)(\tikzcdmatrixname-7-9),draw,thick,fill=white,inner xsep=19pt,inner ysep=10pt,xshift=0pt,yshift=0pt,label={[yshift=-43pt]$\dsu{3}$}] {}; 
                              \draw[red, thick, dotted] ($(\tikzcdmatrixname-2-9)+(-22pt,10pt)$) -- ($(\tikzcdmatrixname-3-9)+(22pt,-10pt)$) ;
                              \draw[red, thick, dotted] ($(\tikzcdmatrixname-2-9)+(22pt,10pt)$) -- ($(\tikzcdmatrixname-3-9)+(-22pt,-10pt)$) ;
                  }]
                  \lstick{$q_m\colon\ket{0}$}   & \gate{R_y(\psi)}      & \gate{R_z(\phi)}      & \qw       & \qw       & \qw       & \qw\slice{}     & \qw             & \qw             & \qw\slice{}           & \ctrl{3}  & \gate{H}        & \meter{}        & \cw       & \cwbend{6}      &     &     &     &     &     & \lstick{$q_m\colon\ket{0}$}       & \gate{R_y(\psi)}      & \gate{R_z(\phi)}      & \qw\slice{}           & \qw\slice{}           & \ctrl{3}  & \gate{H}  & \meter{}  & \cw       & \cwbend{6}      &     \\
                  \lstick{$A\colon \ket{0}$}    & \rynewg{1}{4}         & \ctrl{1}        & \qw       & \ctrl{3}  & \qw       & \qw       & \gate[3]{\scs{3}}     & \gate[2]{\dsu{2}}     & \trash{\text{discard}}&           &                 &                 &           &           &     &     &     &     &     &                       &                 &                 &                 &                 &           &           &           &           &           &     \\
                  \lstick{$A\colon \ket{0}$}    & \qw             & \rynewg{1}{3}         & \ctrl{1}  & \qw       & \ctrl{3}  & \qw       & \qw             & \qw             & \trash{}        &           &                 &                 &           &           &     &     &     &     &     &                       &                 &                 &                 &                 &           &           &           &           &           &     \\
                  \lstick{$P\colon \ket{0}$}    & \qw             & \qw             & \rynewg{1}{2}   & \qw       & \qw       & \ctrl{3}  & \qw             & \qw             & \qw             & \targ{}   & \qw             & \meter{}        & \cwbend{3}      &           &     &     &     &     &     & \lstick{$P\colon\ket{0}$}   & \rynewg{2}{3}         & \ctrl{2}        & \octrl{1}       & \qw             & \targ{}   & \qw       & \meter{}  & \cwbend{3}      &           &     \\
                  \lstick{$C\colon \ket{0}$}    & \qw             & \qw             & \qw       & \targ{}   & \qw       & \qw       & \qw             & \qw             & \qw             & \qw       & \qw             & \qw             & \gate{X}  & \gate{Z}  & \qw &     &     &     &     & \lstick{$C\colon\ket{0}$}   & \qw             & \targ{}         & \rynewg{3}{4}         & \gate[3]{\dsu{3}}     & \qw       & \qw       & \qw       & \gate{X}  & \gate{Z}  & \qw \\
                  \lstick{$C\colon \ket{0}$}    & \qw             & \qw             & \qw       & \qw       & \targ{}   & \qw       & \qw            & \qw             & \qw             & \qw       & \qw             & \qw             & \gate{X}  & \gate{Z}  & \qw &     &     &     &     & \lstick{$C\colon\ket{0}$}   & \qw             & \rynewg{1}{2}         & \qw             & \qw                   & \qw       & \qw       & \qw       & \gate{X}  & \gate{Z}  & \qw \\
                  \lstick{$C\colon \ket{0}$}    & \qw             & \qw             & \qw       & \qw       & \qw       & \targ{}   & \qw             & \qw             & \qw             & \qw       & \qw             & \qw             & \gate{X}  & \gate{Z}  & \qw &     &     &     &     & \lstick{$C\colon\ket{0}$}   & \qw             & \qw             & \qw             & \qw             & \qw       & \qw       & \qw       & \gate{X}  & \gate{Z}  & \qw             
            \end{quantikz}
      \end{adjustbox}
      \caption{Quantum Telecloning circuits for $1 \rightarrow 3$, using ancilla qubits (left) and no ancilla qubits (right). The Quantum Telecloning circuit with ancilla (left) use the gate reduction optimization introduced in ref.~\cite{telecloning_circuits_2} and can be scaled to any value of $M$ (number of clones), whereas the circuits with no ancilla are only known for $M=2$ and $M=3$. The circuit on the left shows where previously a $DSU(3)$ unitary would be acting on the port and ancilla qubits, we only apply an $SCS(3)$ unitary -- discarding the following $DSU(2)$ unitary -- and thereby saving gate operations (especially when going to $DSU(M)$ unitaries for larger $M$). 
      The top wire $q_m$ defines the single qubit which we want to clone, in this case parameterized by two single qubit gates (in general these two single qubit gates are not required for the protocol to work).
      $\sqrt{a/b}$ stands for $R_y(\ryangle{a}{b})$. Qubit wires labeled A are ancilla qubits, P is the single port qubit, and C are clone qubits. 
      The end of the circuit shows where the mid-circuit Bell measurement is made between the Port qubit and the message qubit, followed by classical conditional operations on the remaining Pauli gate rotations on the clone qubits. }
      \label{fig:main-circuits}
\end{figure*}
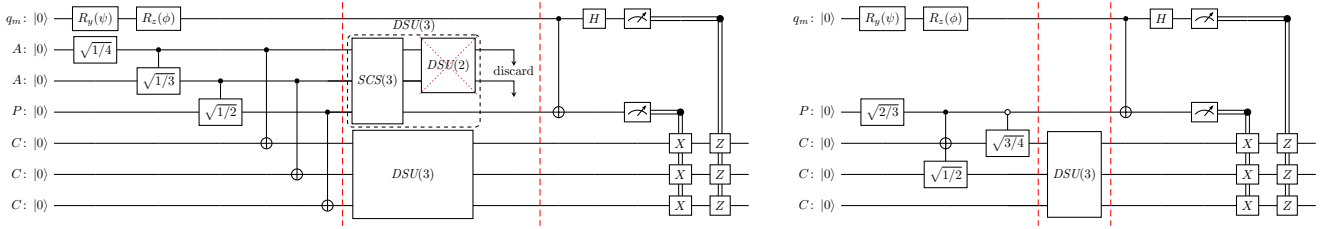

\subsection{IBM Quantum computer implementation and result characterization details}
\label{section:methods_implementation_details}
All circuits are optimized and adapted to the IBM Quantum hardware gateset using the Qiskit~\cite{Qiskit} transpiler with \texttt{optimization\_level=1} (which is the highest optimization setting that can be applied to dynamic circuits for the version of Qiskit used in these experiments, which was \texttt{qiskit-terra==0.24.1}). All circuit executions used all default settings (except the number of shots), meaning that all experiments used \texttt{meas\_level=2}. The IBM Quantum processor native gateset for all of the devices used in these experiments is \texttt{rz, cx, sx, x}. Note that on the IBM Quantum devices, the \texttt{rz} gate is \emph{virtual}~\cite{McKay_2017} meaning that it can be implemented with an error rate of $0$, so it does not contribute to the error encountered when these compiled circuits are executed. The dynamic circuit control instruction blocks are all if statements. These if statements are conditioned on the results of the Bell measurement being $1$ (e.g. if the bit is $1$, then the corresponding Pauli X or Z gate is applied to the clone, and if not nothing is applied), for each of the $M$ clones (this is step 4 of Algorithm~\ref{algorithm:telecloning}). This means that there are $2$ classical if statements for each clone (conditioned on the two bits from the Bell measurement), and therefore for every $1 \rightarrow M$ quantum telecloning circuit that is programmed and executed on an IBM Quantum computer, a total of $2M$ if statements are applied. Figure~\ref{fig:main-circuits} shows these conditional instructions, and Figures~\ref{fig:M2_detailed_circuits},~\ref{fig:M3_detailed_circuits},~\ref{fig:M4_detailed_circuit} in Appendix~\ref{section:appendix_compiled_circuits} show the complete quantum telecloning circuits with the if-else control blocks. Note that although we do not use this feature (all classical control mechanisms are programmed using if / else statements), Qiskit~\cite{Qiskit} currently supports switch statements \footnote{\url{https://qiskit.org/documentation/stubs/qiskit.circuit.SwitchCaseOp.html}}, which once available on IBM Quantum hardware could also be used to more efficiently execute the quantum telecloning protocol without as many separate classical conditional instructions. 

For each of the initial pure quantum states that we prepare, the goal is to characterize how well the quantum telecloning process performs optimal universal symmetric approximate cloning. To this end, we use the standard quantum measure of \emph{fidelity}, defined in eq.~\eqref{eq:fidelity-general}, which computes the state overlap between two density matrices $\rho_1$ and $\rho_2$. The optimal universal symmetric quantum cloning bounds in terms of fidelity is given in eq.~\eqref{eq:theoretical-fidelity} for cloning $1$ pure quantum state into $M$ approximate quantum clones. A fidelity measure of $1$ means that the states are exactly overlapping. A fidelity of $0.5$ means that although there is state overlap, the overlap is no better than choosing pairs of random density matrices and measuring the state overlap -- meaning that the single qubit clones do not provide a meaningful representation of the original pure quantum state.

\begin{equation}
    F(\rho_1, \rho_2) = Tr[ \sqrt{ \sqrt{\rho_1}\rho_2 \sqrt{\rho_1} } ]^2
    \label{eq:fidelity-general}
\end{equation}

\begin{figure*}[th!]
    \centering
    \includegraphics[width=0.24\textwidth]{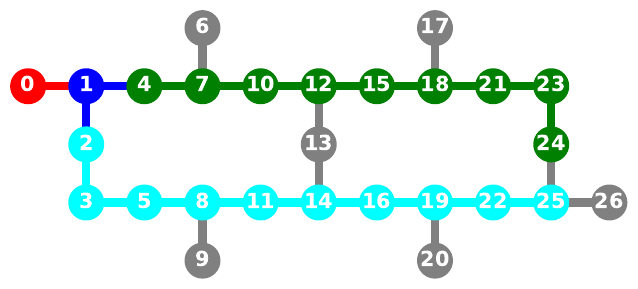}
    \includegraphics[width=0.24\textwidth]{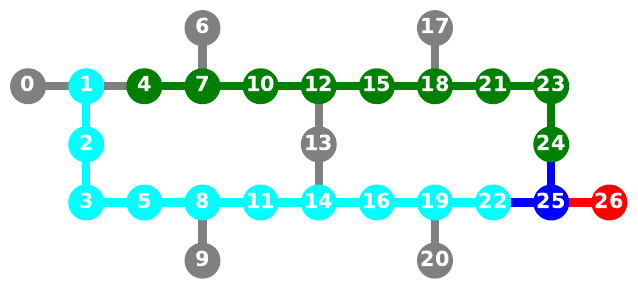}
    \includegraphics[width=0.24\textwidth]{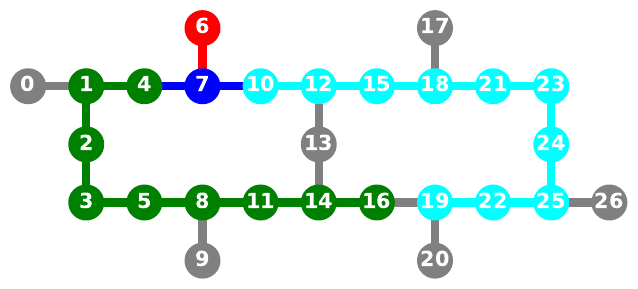}
    \includegraphics[width=0.24\textwidth]{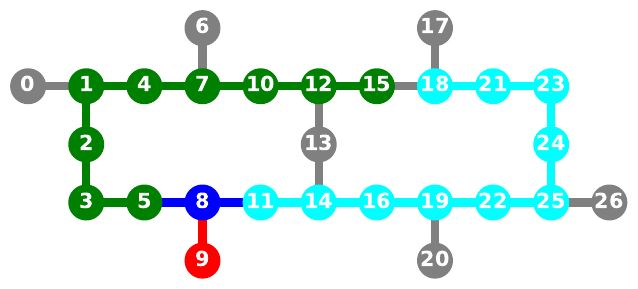}
    \includegraphics[width=0.24\textwidth]{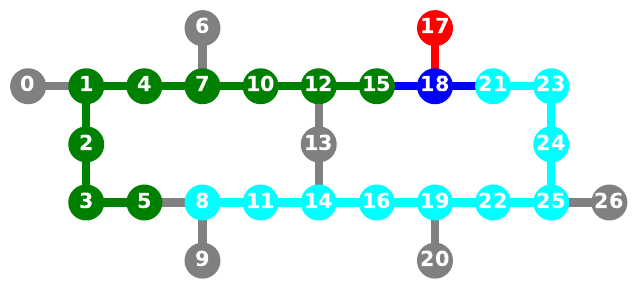}
    \includegraphics[width=0.24\textwidth]{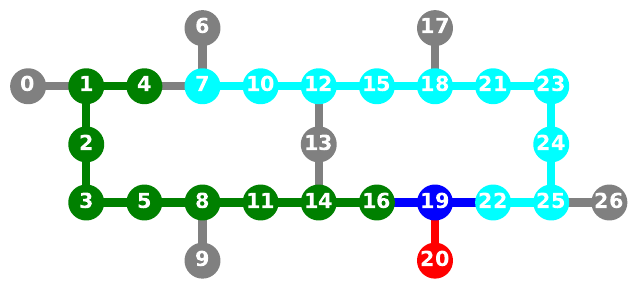}
    \includegraphics[width=0.24\textwidth]{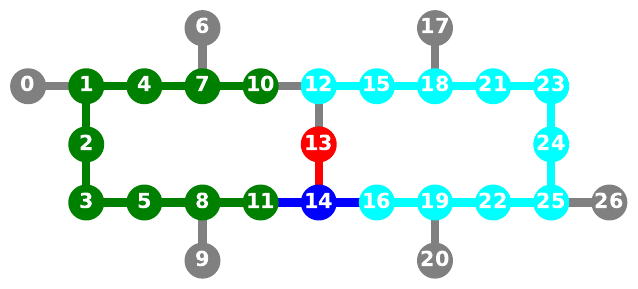}
    \caption{$7$ different qubit subgraph isomorphic layouts for compiling the LNN quantum telecloning circuits to the relatively sparse heavy-hex connectivity, with $27$ qubits. These layouts are mapped to a telecloning circuit with $N=1$ and $M=10$, which is the largest telecloning circuit that can be fit onto this architecture, without the introduction of significant overhead due to qubit swapping. The message qubit ($N=1$) is colored red, the port qubit is dark blue, the $10$ clone qubits are colored cyan, and the $9$ ancilla qubits are colored green, using a total of $21$ qubits. The unused qubits and CNOT connections are grey. This connectivity graph is identical between the IBMQ processors \texttt{ibmq\_kolkata}, \texttt{ibmq\_mumbai}, \texttt{ibm\_geneva}, \texttt{ibm\_hanoi}, \texttt{ibm\_algiers}, \texttt{ibm\_cairo} and \texttt{ibmq\_auckland}, meaning that the same compiled circuits can be used across all of the $27$ qubit backends we used in the experiments. The quantum telecloning circuits without ancilla can be compiled to these connectivity graphs by simply not using the LNN qubit line that would have been used for the Dicke state preparation of the ancilla qubit state. }
    \label{fig:qubit_layouts}
\end{figure*}

To compute the single qubit clone quality the quantum telecloning circuits generated, we need to perform quantum state tomography (QST). We use Pauli basis state tomography, although there are other types of state tomography that can be used. This procedure generally has a cost of $3^n$ for an $n$ qubit system, but in this case we only are interested in single qubit tomography of the clone qubits. The procedure we follow is to prepare the quantum telecloning circuit for a specific state we want to clone, then at the end of the circuit we insert the basis change gates to put all of $M$ clones into the Pauli X, Y, or Z basis, and then we measure the state of all $M$ of the clones. We refer to this procedure as \emph{parallel single qubit state tomography}~\cite{yang2021experimental, telecloning_circuits_1, telecloning_circuits_2}. Detailed circuit examples of the Pauli basis parallel single qubit state tomography circuits are shown in Appendix~\ref{section:appendix_compiled_circuits}. We then repeat this protocol using $10,000$ shots, and then this is repeated so that each of the three basis states have been measured. Therefore, in total, for the purpose of characterizing the clone quality of a single pure quantum state, a total of $30000$ samples are taken in order to reliably compute the true density matrix representing the physical state that was constructed (for each clone qubit), specifically to mitigate the finite sampling effect. Using these three Pauli basis measurements, we can then compute the density matrix of the mixed state for each of the $M$ quantum clones, using maximum likelihood estimation~\cite{PhysRevLett.108.070502} in Qiskit~\cite{Qiskit} Ignis (with slight modifications) with sequential least squares programming optimizer fitting. With the single qubit density matrices having been computed, the fidelity of the quantum clones can be computed using eq.~\eqref{eq:fidelity-general}.

Because the error rates of the different operations on the quantum hardware can vary significantly, the other method we will utilize is to map the circuits identically to different subgraph isomorphisms of the chip hardware, thus getting a range of the possible device performance since performance can vary significantly both over time and across a fixed superconducting quantum processor, see for example ref.~\cite{Pelofske_2022}. Figure~\ref{fig:qubit_layouts} shows the $7$ layouts that we use on the heavy-hex architecture~\cite{PhysRevX.10.011022} in order to go up to $M=10$ quantum telecloning circuits (with ancilla qubits). This construction makes use of the LNN Dicke state preparation circuits to prepare the two Dicke states on two different linear lines of qubits on the heavy hex architecture. There are other possible layouts besides the $7$ shown in Figure~\ref{fig:qubit_layouts} on the $27$ qubit heavy-hex lattice, however we used these because they are the easiest to implement -- in particular they do not require use of SWAP networks to move ancilla qubits around the lattice. Without a reduction in the number of ancilla qubits required for the telecloning circuits, and without the use of SWAP gates to move ancilla qubits around the lattice, $M=10$ is the largest number of clones that can be generated on these $27$ qubit heavy hex lattices. The $7$ circuit layouts shown in Figure~\ref{fig:qubit_layouts} can be easily adapted to any quantum telecloning circuit, with or without ancilla qubits, for $M < 10$ by removing unused qubits down the nearest neighbors line for both the clone qubits (cyan nodes) and ancilla qubits (green nodes), and fixing the location of the Port and message qubit. These are the fixed layouts that are used for executing these circuits on the various IBM Quantum computers. We note in passing that while there are larger IBM Quantum devices with heavy-hex lattices available, which could implement larger circuits, our results suggest -- as we will see -- that at $M=10$ clones most of the signal is lost as we measure fidelities very close to $0.5$, therefore larger telecloning circuits on current hardware will likely not produce high quality clones.

\begin{figure*}[h!]
    \centering
    \includegraphics[width=0.32\textwidth]{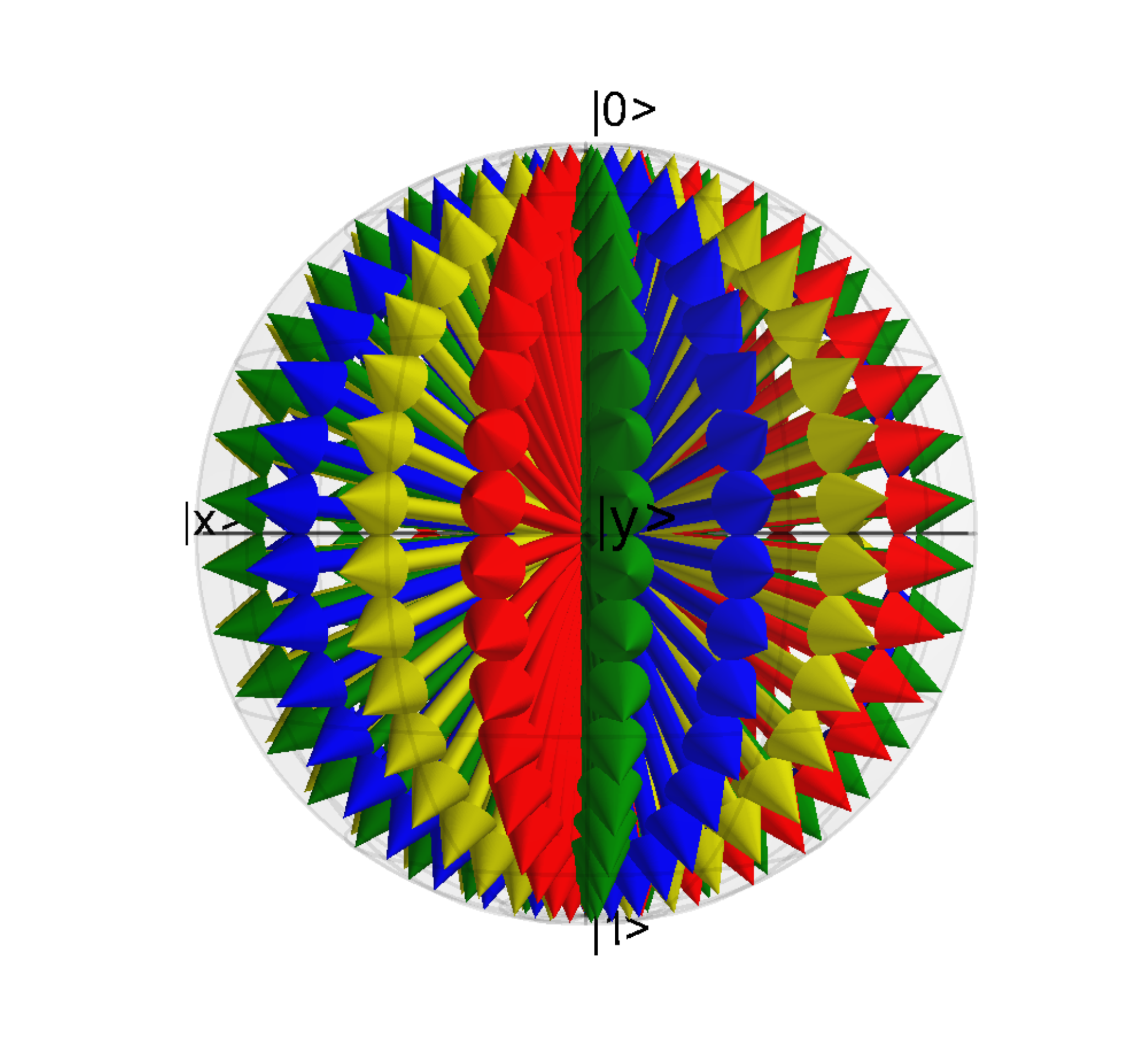}
    \includegraphics[width=0.32\textwidth]{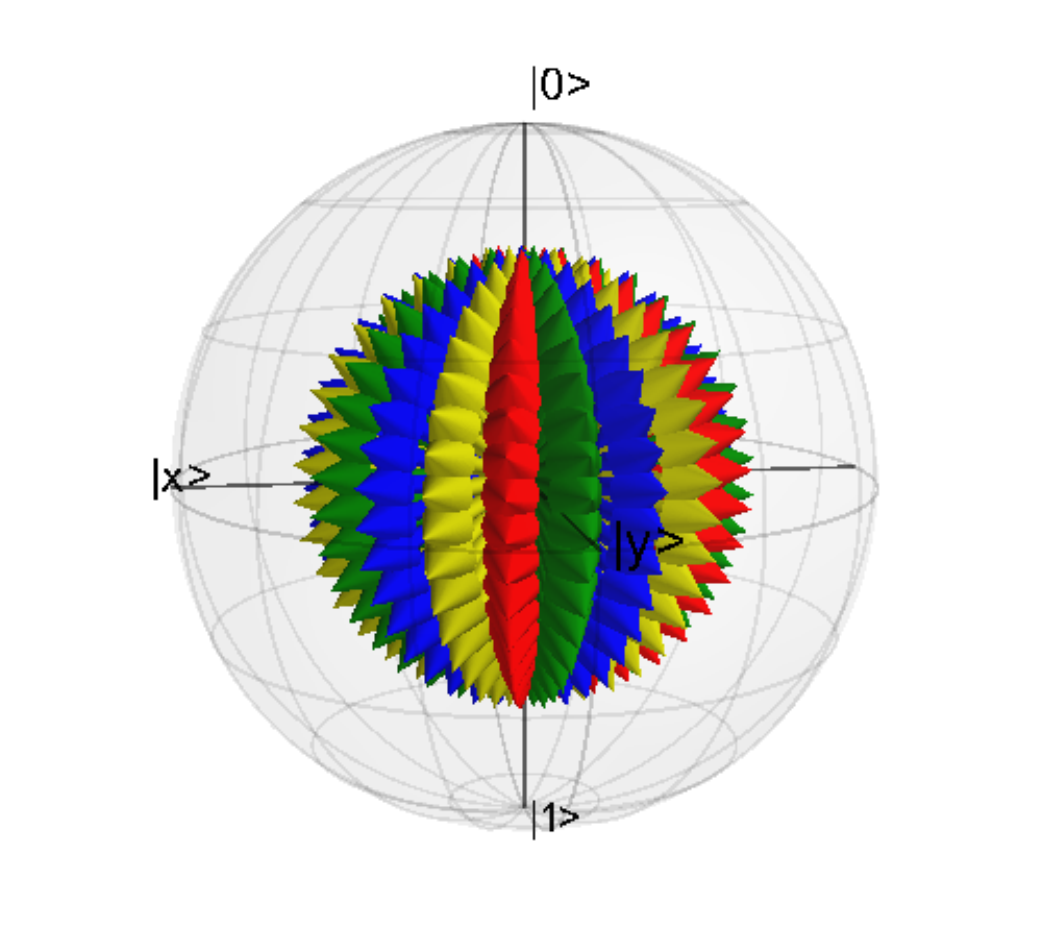}
    \includegraphics[width=0.32\textwidth]{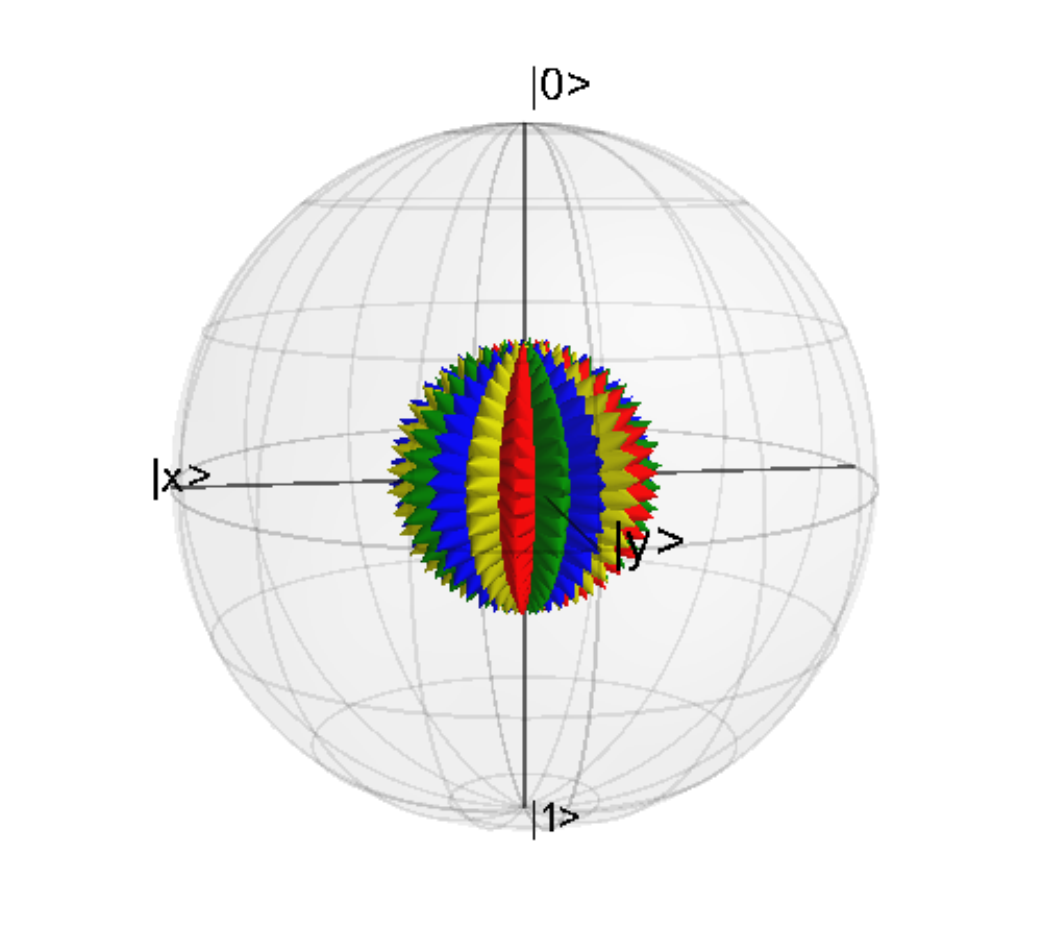}
    \caption{The left hand plot shows a Bloch sphere representation of all vectors of the pure quantum states that are cloned in the quantum telecloning circuits on the quantum hardware. The middle plot shows the corresponding Bloch sphere for those same quantum states (for a single qubit), after they have been copied by a universal symmetric $1 \rightarrow 2$ cloning machine, and the right hand plot shows a Bloch sphere representation of these same states after having been cloned by a $1 \rightarrow 10$ universal symmetric quantum cloning machine. Because the cloning is symmetric, there would be $2$ and $10$ identical copies of the qubits corresponding to the middle and right hand side plots. }
    \label{fig:bloch_sphere_vectors_initial_states}
\end{figure*}

Figure~\ref{fig:bloch_sphere_vectors_initial_states} (left) shows all pure quantum states which we aim to clone in the following experiments, represented by combining all of the vectors onto a single Bloch sphere. These states are computed by generating $20$ linearly spaced angles in the range of $R_y \in [0, \pi]$ and $R_z \in [0, 2\pi]$, where $R_y$ parameterizes the \texttt{ry} single qubit rotation on qubit $0$ in Figure~\ref{fig:main-circuits}, and $R_z$ parameterizes the subsequent \texttt{rz} on qubit $0$. Using these simple qubit rotations we can reach any point on the Bloch sphere. And in particular, this range of angles allows the experiments to cover a range of states across the entire Bloch sphere, although not uniformly.

\subsubsection{Dynamical Decoupling}
\label{section:methods_dynamical_decoupling}

With the goal of improving the computation of the quantum telecloning circuits, we test digital dynamical decoupling sequences on IBM Quantum hardware. Dynamical decoupling is an error suppression technique that can mitigate certain types of noise on idle qubits by keeping the qubits isolated from environmental interactions using sequences of gates (or pulse sequences) that are equivalent to identity gate operations~\cite{PhysRevLett.82.2417, Pokharel_2023, PhysRevLett.121.220502, ezzell2023dynamical, 9872062, Souza_2012, Zanardi1999}. For our experiments, we use the gate sequence of X-X; pairs of Pauli X gates, since the Pauli X gate is a native gate to the IBM Quantum devices we use. The sequences are scheduled using a Qiskit passmanager \footnote{\url{https://qiskit.org/documentation/locale/bn_BN/stubs/qiskit.transpiler.passes.PadDynamicalDecoupling.html}} with the As Late As Possible (ALAP) algorithm. The X-X sequences are not added during the classical control if / else statement blocks - this is a capability that is not yet available on the IBM quantum hardware as of when these experiments were executed. As a point of comparison, for the $M=10$ quantum telecloning circuits (scheduled using the ALAP algorithm), approximately $800$ Pauli X gates (approximately $400$ dynamical decoupling sequences) can be expected to be scheduled, dependent on the backend timing properties and the exact properties of the circuit (e.g. including what Pauli basis rotations are being applied for instance). For the $M=2$ quantum telecloning circuits with no ancilla qubits, we can expect on average $14$ Pauli X gates to be inserted by the dynamical decoupling pass.

\subsection{Bloch sphere representation of single qubit clones}
\label{section:methods_Bloch_sphere}

An optimal universal symmetric quantum cloning machine has the property that the clones that are generated will retain the same vector angle, e.g. for a geometric Bloch sphere representation, but at a smaller magnitude than the original pure quantum state. Specifically, the generated clones are each a \emph{mixed quantum state}. The \emph{shrinking factor} of the generated universal, symmetric, optimal clones, is given in refs.~\cite{PhysRevLett.81.2598, PhysRevA.57.2368, RevModPhys.77.1225}, and is shown in eq.~\eqref{eq:shrinking_factor} for a $d=2$ quantum system, i.e., a qubit. 

\begin{equation}
    \eta (N, M) = \frac{ N }{ M } \frac{M+2}{N+2}
    \label{eq:shrinking_factor}
\end{equation}

\begin{table*}[t!]
	\centering
	\renewcommand{\arraystretch}{1.1}
	\newcommand{\mdd}[1]{\bf{#1}}
	\begin{tabular*}{\textwidth}{@{\extracolsep{\fill}}lllrrrrrrrr@{}}
		\toprule
		\	 & \	& \		& \			& \small{Falcon r8}	& \small{Falcon r5.10}	& \multicolumn{5}{r@{}}{\small{Falcon r5.11}} \\[-3pt]
		\cmidrule(lr){5-5}
		\cmidrule(lr){6-6}
		\cmidrule(ll){7-11}
		DD	&	 Ancilla		& Clones	& \texttt{Theory}	& \texttt{geneva}	& \texttt{mumbai}	& \texttt{kolkata}	& \texttt{auckland}	& \texttt{hanoi}	& \texttt{cairo}	& \texttt{algiers}		\\
		\midrule[\heavyrulewidth]
		No   &  No	& $M=2$		& $0.833\overline{3}$	& 0.6317		& 0.6280		& 		0.7598	& 0.7641		&		-	&		0.7801	&			-	\\[-3pt]
		Yes   &   No 		&	$M=2$	&		$0.833\overline{3}$	& \mdd{0.6811}		& \mdd{0.7435}		& 		-	& \mdd{0.7756}		& \mdd{0.7900}		& \mdd{0.7291}		& \mdd{0.7274}			\\
		No	&	No & $M=3$		& $0.777\overline{7}$	& 0.5949		& 0.6211		& 		-	& 		-	& 		-	& 		-	& 		-		\\[-3pt]
		Yes	&	No &	$M=3$	&		$0.777\overline{7}$	& 		-	& \mdd{0.6550}		& 		-	& \mdd{0.6803}		& \mdd{0.6874}		& 	\mdd{0.6849}		& \mdd{0.6594}			\\
		\midrule
		No   &  Yes	& $M=2$		& $0.833\overline{3}$	& 0.6251		& 0.6280		& 0.6916		& 0.6691		&		-	&		-	&		-		\\
		No   &  Yes	& $M=3$		& $0.777\overline{7}$	& 0.5266		& 0.5437		& 		-	& 0.6029		& 		-	& 		-	& 		-		\\
		No   &  Yes	& $M=4$		& $0.7500$		& 0.5082		& 0.5475		& 		-	& 0.5620		& 		-	& 		-	& 		-		\\[-3pt]
		Yes   &  Yes	& $M=4$	&	$0.7500$		& 		-	& \mdd{0.5656}		& 		-	& \mdd{0.5847}		& \mdd{0.5721}		&	\mdd{0.5669}		& \mdd{0.5377}			\\
		No   &  Yes	& $M=5$		& $0.733\overline{3}$	& 0.5168		& 0.5334		& 0.5356		& 0.5382		& 		-	& 		-	& 			-	\\
		No   &  Yes	& $M=6$		& $0.722\overline{2}$	& 0.5058		& 		0.5282	& 		-	& 		-	& 		-	& 		-	& 		-		\\
		No   &  Yes	& $M=7$		& $0.714..$		& 0.5074		& 		-	& 		-	& 		-	& 		-	& 		-	& 			-	\\
		No   &  Yes	& $M=8$		& $0.708\overline{3}$	& 0.5002		& 		-	& 		-	& 		-	& 		-	& 		-	& 				\\
		No   &  Yes	& $M=10$	& $0.7000$		& 0.5011		& 		-	& 		-	& 		-	& 		-	& 		-	& 		-		\\[-3pt]
		Yes   &  Yes	& $M=10$	&		$0.7000$	&		-	& \mdd{0.5053}		& 		-	& \mdd{0.5111}		& \mdd{0.5093}		& \mdd{0.5039}		& \mdd{0.4996}			\\
		\bottomrule
	\end{tabular*}
	\caption{Cloning fidelity computed on $7$ IBM Quantum devices, averaged over the 400 different message qubit states and the $M$ clones.
		For each device (\texttt{ibm\_geneva}, \texttt{ibmq\_mumbai}, \texttt{ibmq\_kolkata}, \texttt{ibm\_auckland}, \texttt{ibm\_hanoi}, \texttt{ibm\_cairo}, \texttt{ibm\_algiers}), the telecloning circuits were compiled to 7 different compiled hardware layouts; only the best mean fidelity out of these is reported in these table entries.
	The results using X-X dynamical decoupling sequences are marked in bold. The top four rows show the clone fidelity results for the optimized telecloning circuits with no ancilla qubits, and the bottom $10$ rows show results for the telecloning circuits with ancilla (for which there exists a general $1 \rightarrow M$ circuits). The fourth column shows the theoretical fidelity that can be achieved by universal quantum cloning circuits (given by eq.~\eqref{eq:fidelity-general}). }
	\label{table:summary_clone_fidelity}
\end{table*}

Visually, this means that the computed density matrices of the clones can be plotted on a 3-d Bloch sphere representation. The input pure quantum states are an arbitrarily selected distribution on the Bloch sphere, and shown visually being aggregated together onto a single plot, in Figure~\ref{fig:bloch_sphere_vectors_initial_states} (left). The different vector colors in Figure~\ref{fig:bloch_sphere_vectors_initial_states} do not correspond to anything; they are intended to only make the vectors visually distinct. We expect that if the quantum cloning operation is ideal, then plotting the experimentally computed density matrices, represented as vectors on the Bloch sphere, would appear as Figure~\ref{fig:bloch_sphere_vectors_initial_states} but with every vector having been shrunk, towards the origin of the Bloch sphere by the factor $\eta$ in eq.~\eqref{eq:shrinking_factor}, but with the same angle. The middle and right and Bloch sphere plots in Figure~\ref{fig:bloch_sphere_vectors_initial_states} show what a single qubit $M=2$ and $M=10$ would look like with ideal universal quantum cloning. This representation of cloned quantum states is, as far as we are aware, a novel way to show an aggregated distribution of the $400$ quantum states. This method of visualizing experimentally measured density matrices allows for a concise summary of any asymmetry or bias of the quantum states compared to the ideal case. Visually seeing the Bloch vectors gives some additional information on the density matrix - namely it gives the angle and magnitude - compared to purely looking at state overlap. Combined, state overlap (fidelity) and the Bloch vector plots, give a good summary of the measured cloned quantum states on the quantum computers.

\section{Results}
\label{section:results}

\begin{figure*}[th!]
    \centering
    \includegraphics[width=0.13\textwidth]{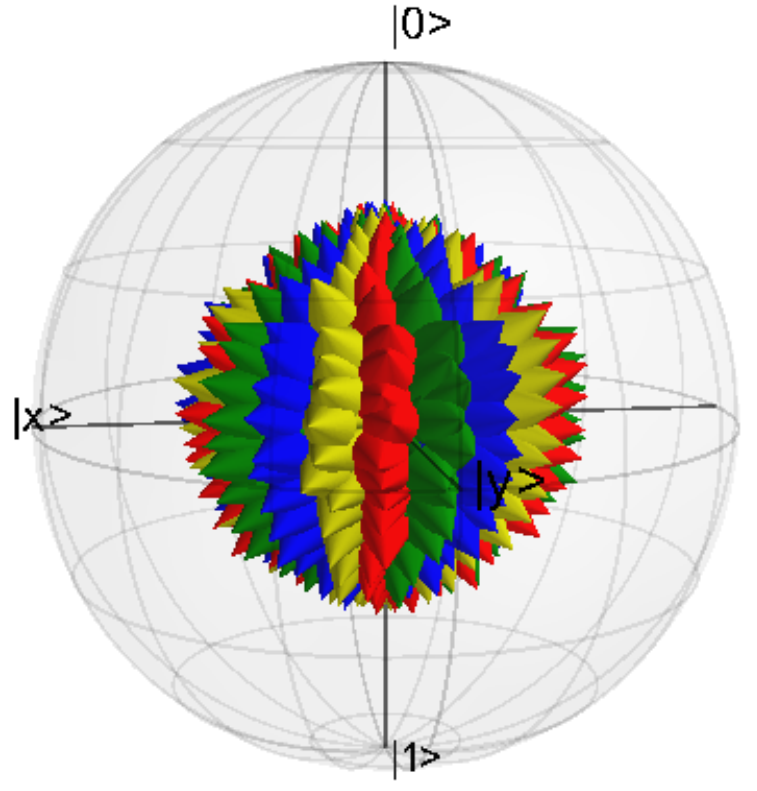}
    \includegraphics[width=0.13\textwidth]{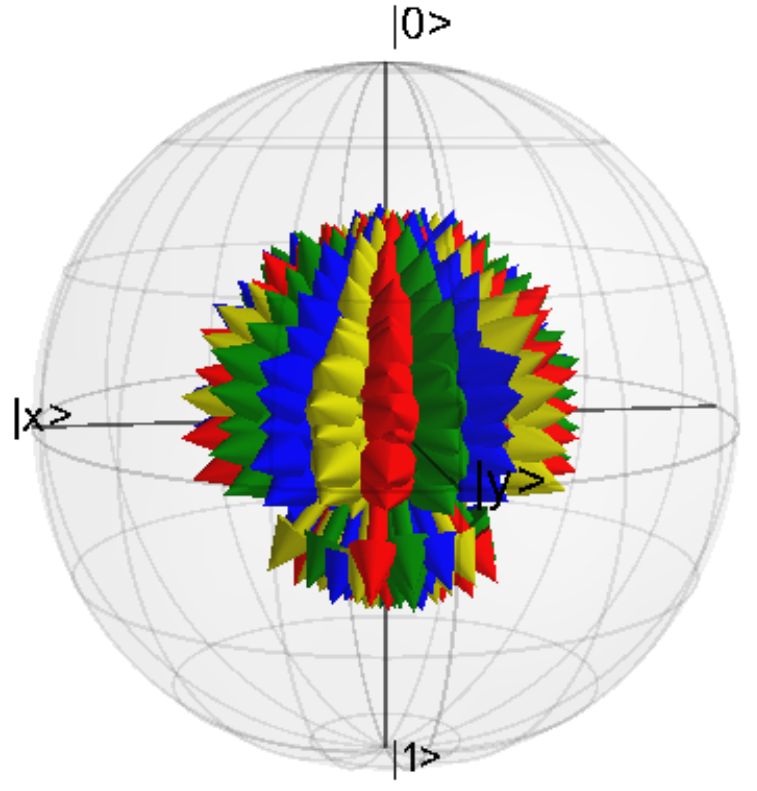}
    \includegraphics[width=0.13\textwidth]{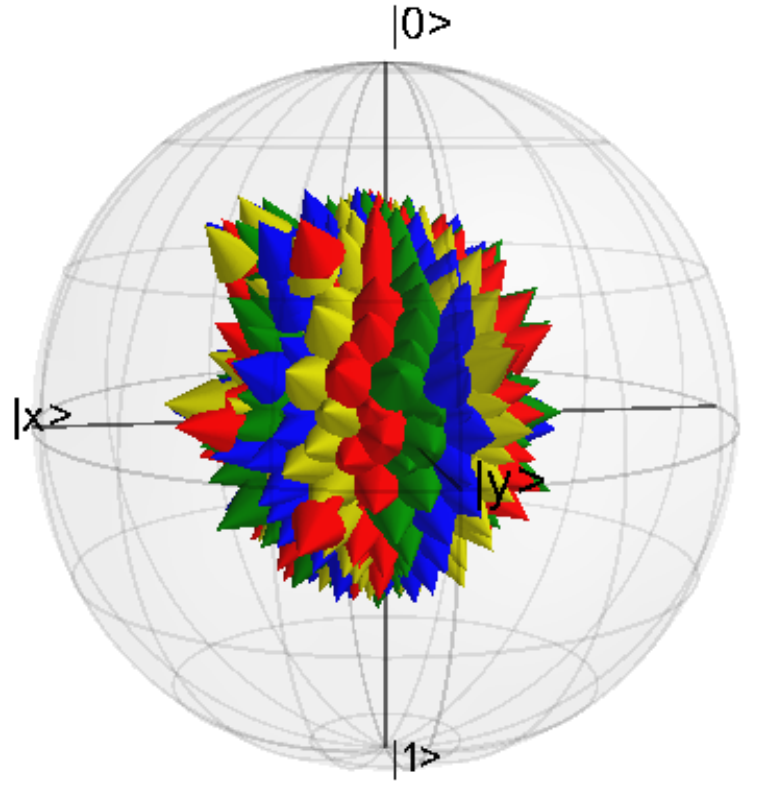}
    \includegraphics[width=0.13\textwidth]{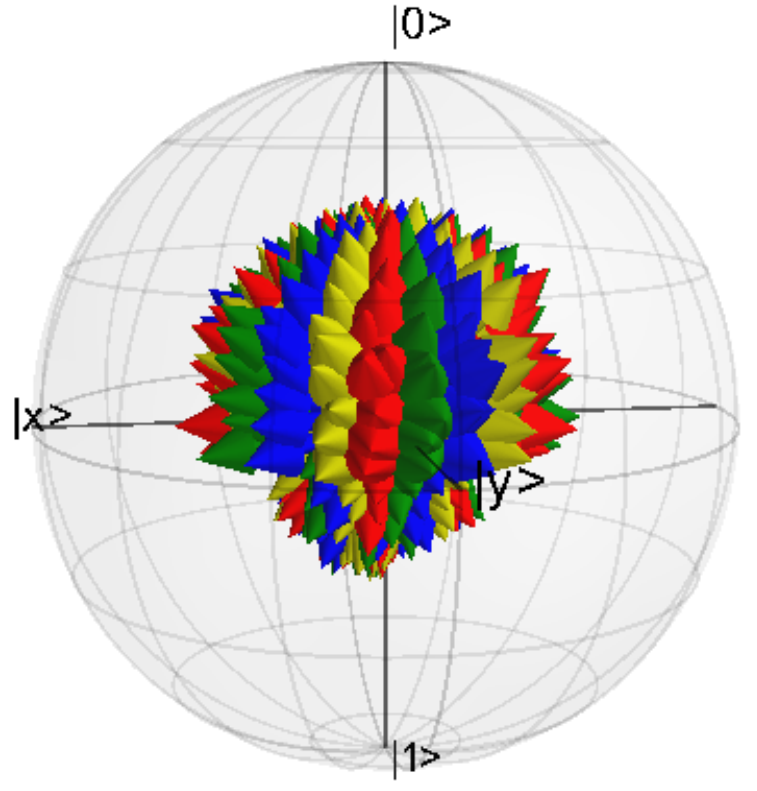}
    \includegraphics[width=0.13\textwidth]{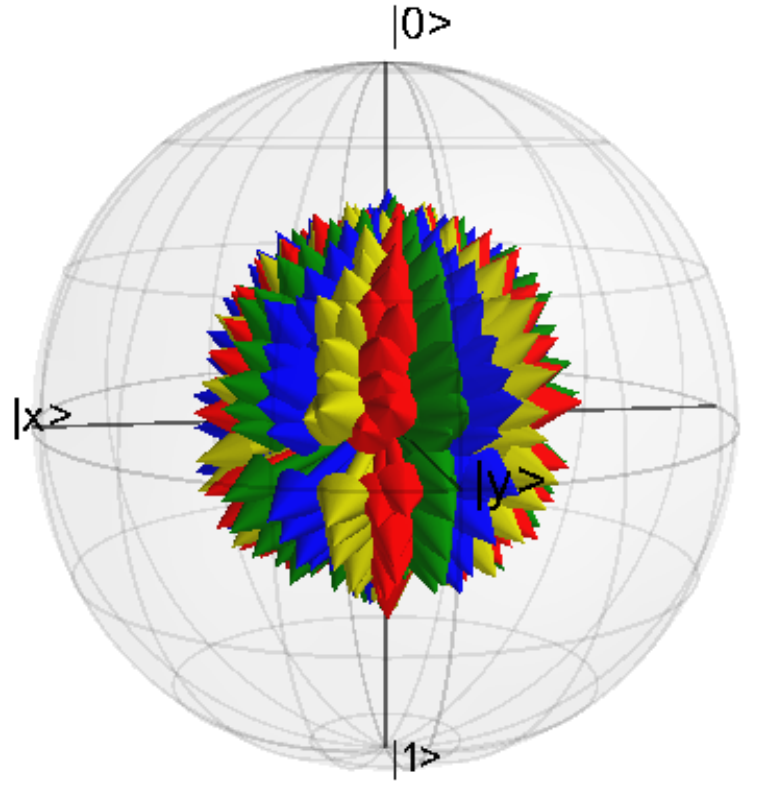}
    \includegraphics[width=0.13\textwidth]{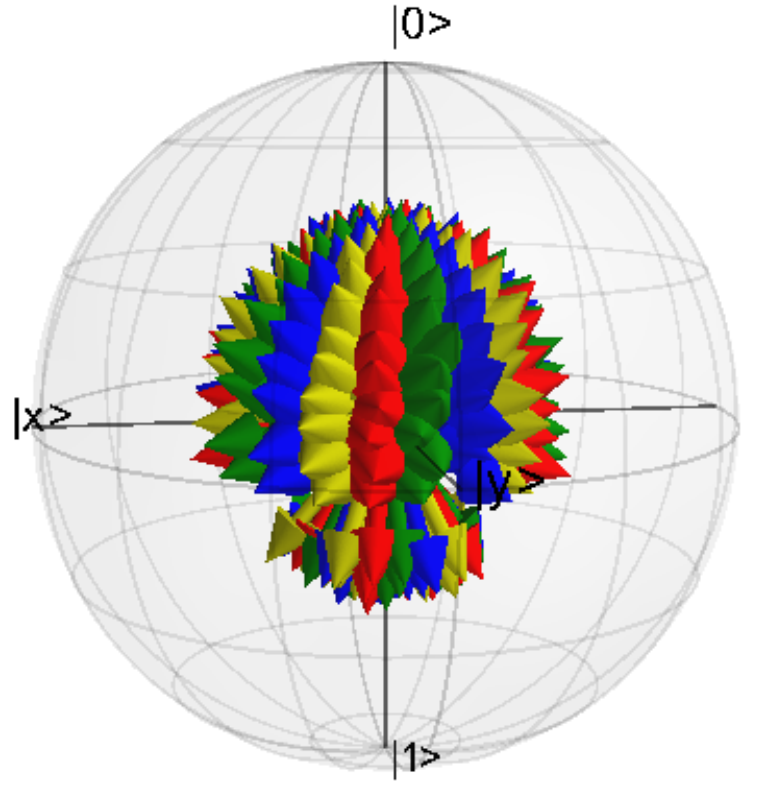}
    \includegraphics[width=0.13\textwidth]{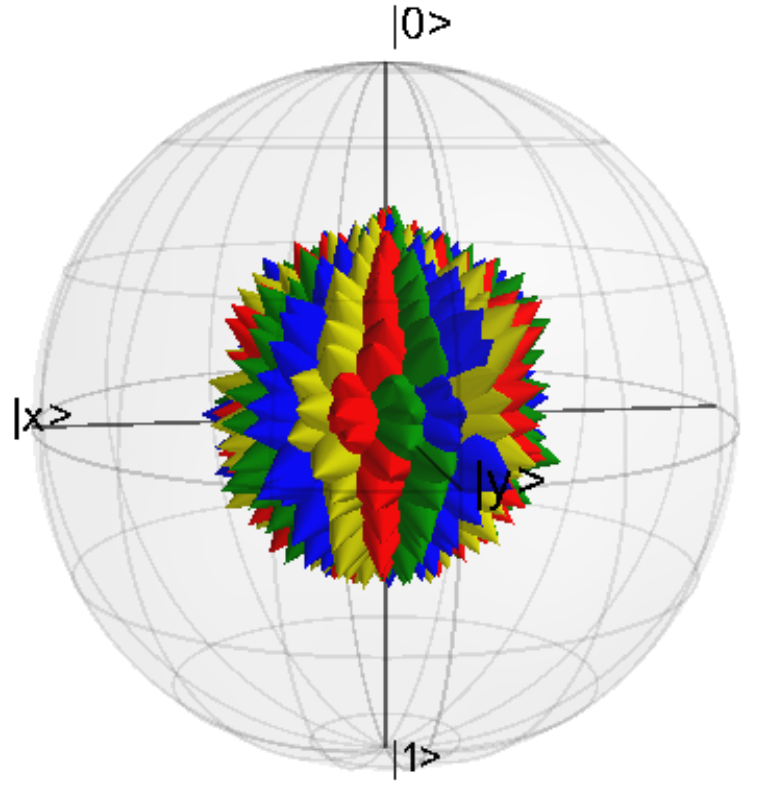}
    \includegraphics[width=0.13\textwidth]{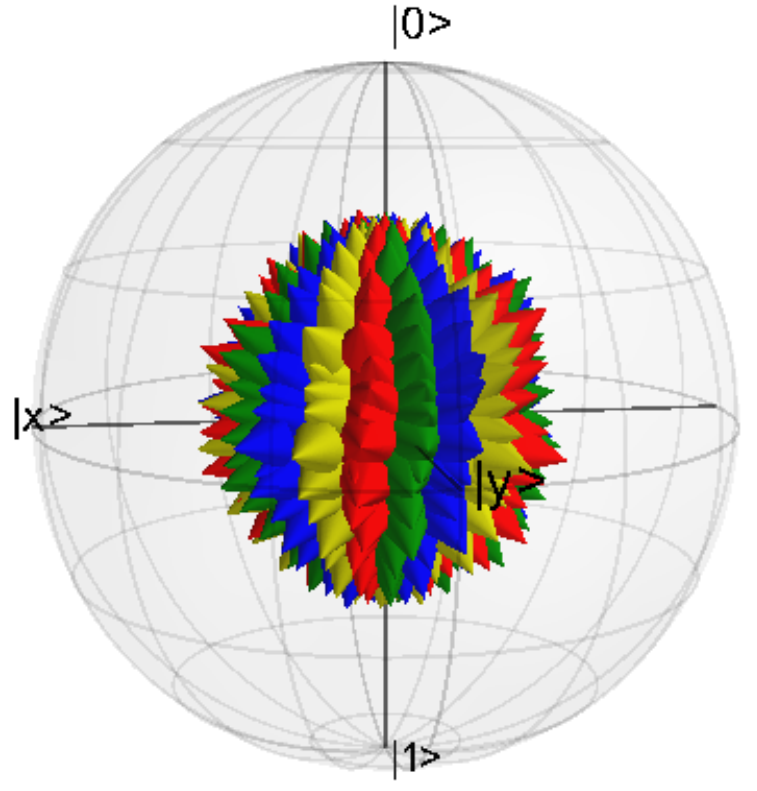}
    \includegraphics[width=0.13\textwidth]{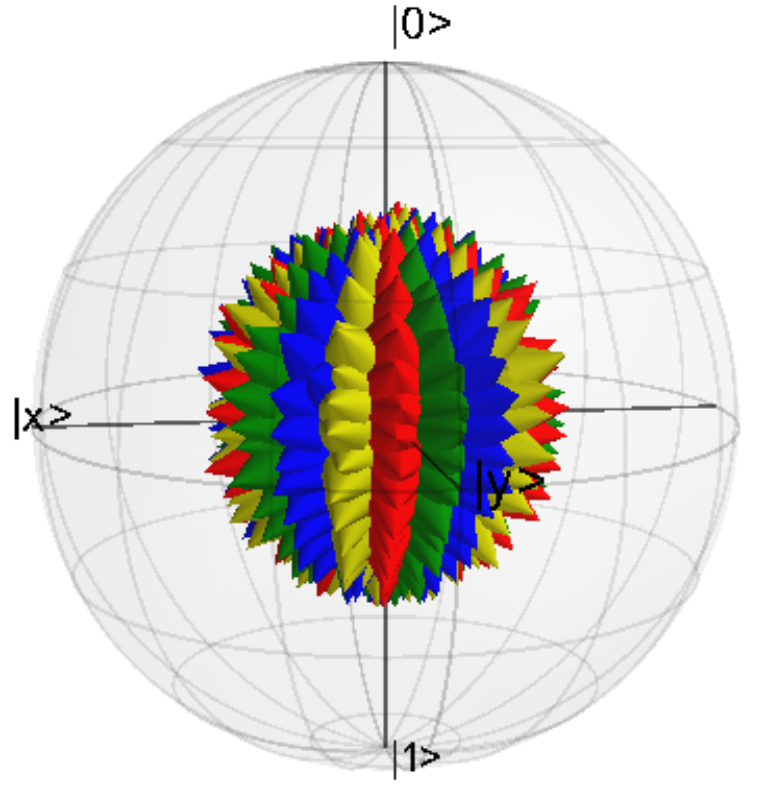}
    \includegraphics[width=0.13\textwidth]{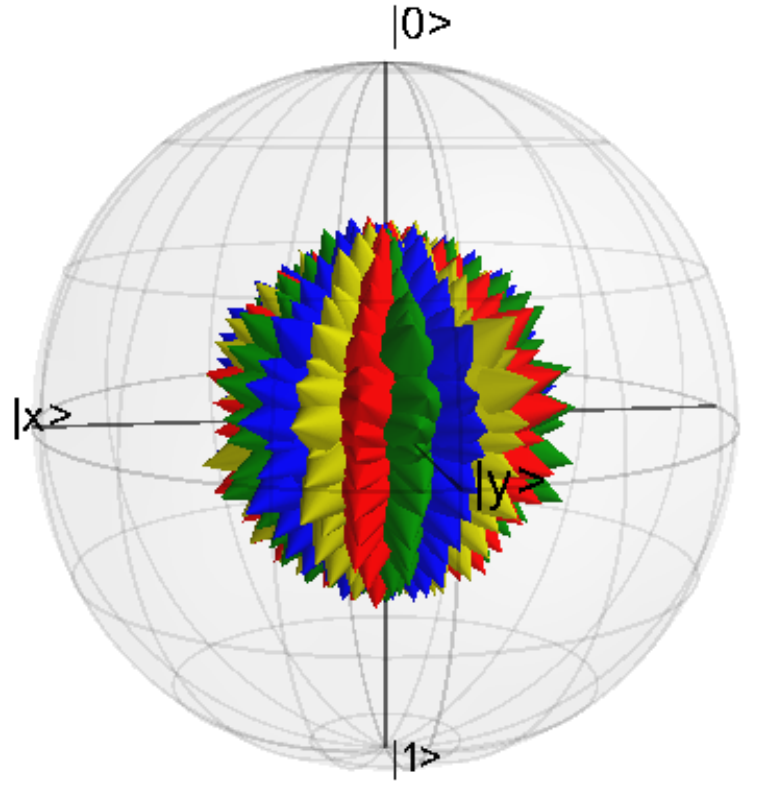}
    \includegraphics[width=0.13\textwidth]{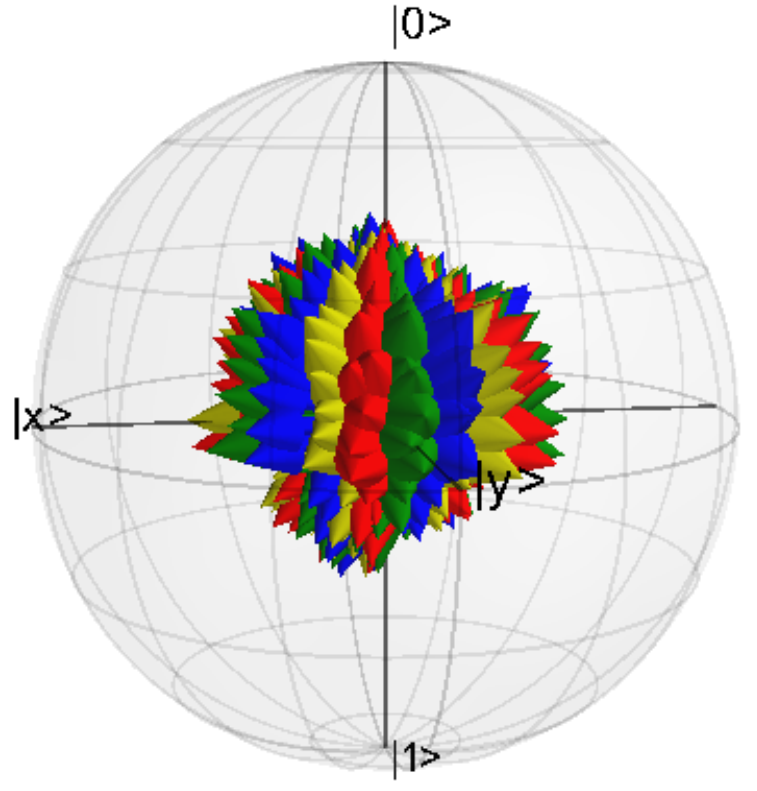}
    \includegraphics[width=0.13\textwidth]{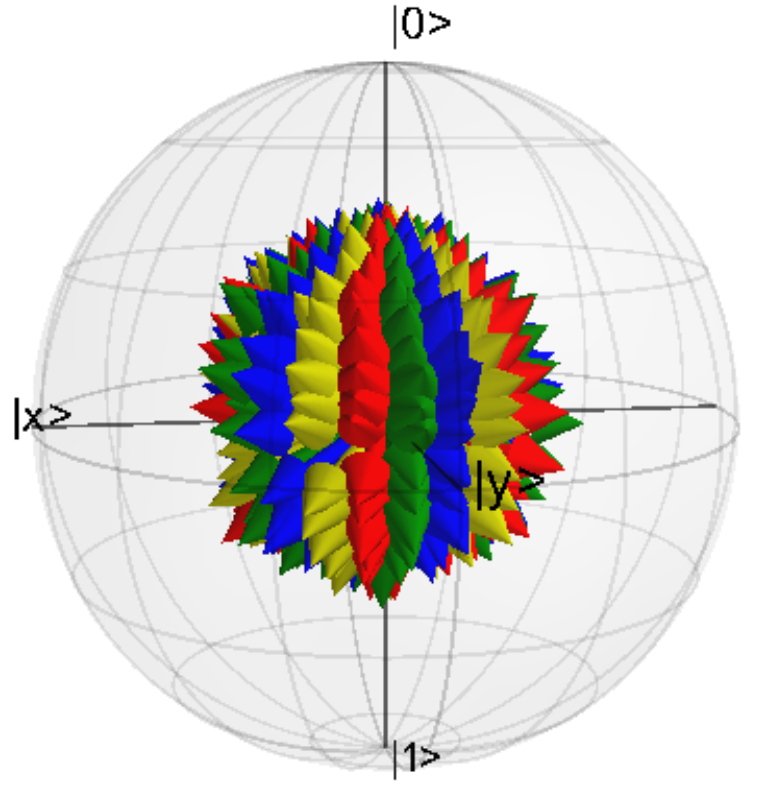}
    \includegraphics[width=0.13\textwidth]{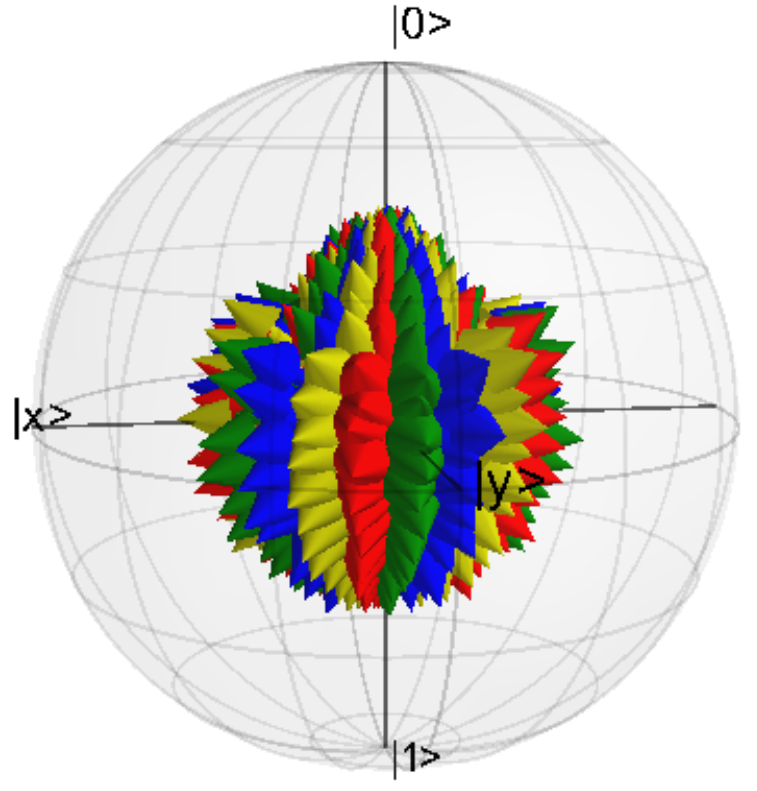}
    \includegraphics[width=0.13\textwidth]{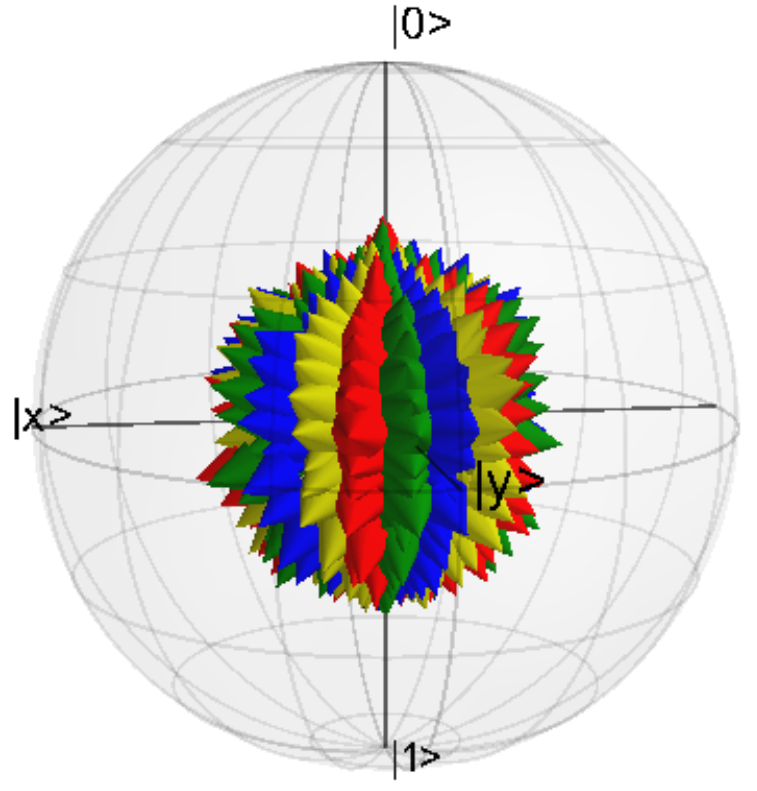}
    \includegraphics[width=0.13\textwidth]{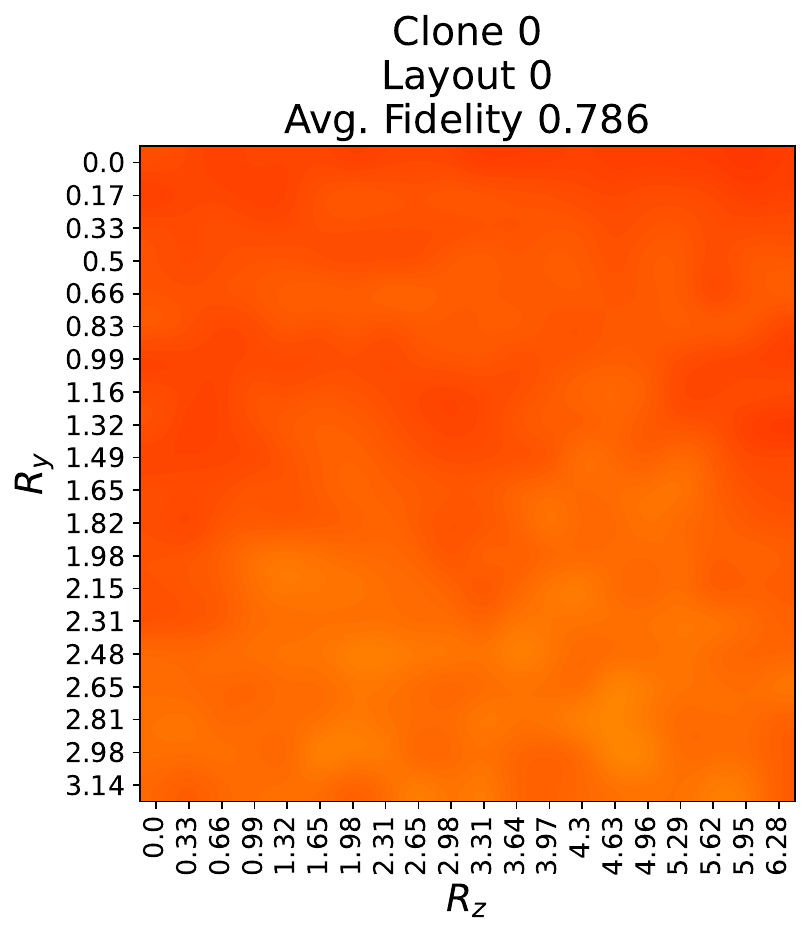}
    \includegraphics[width=0.13\textwidth]{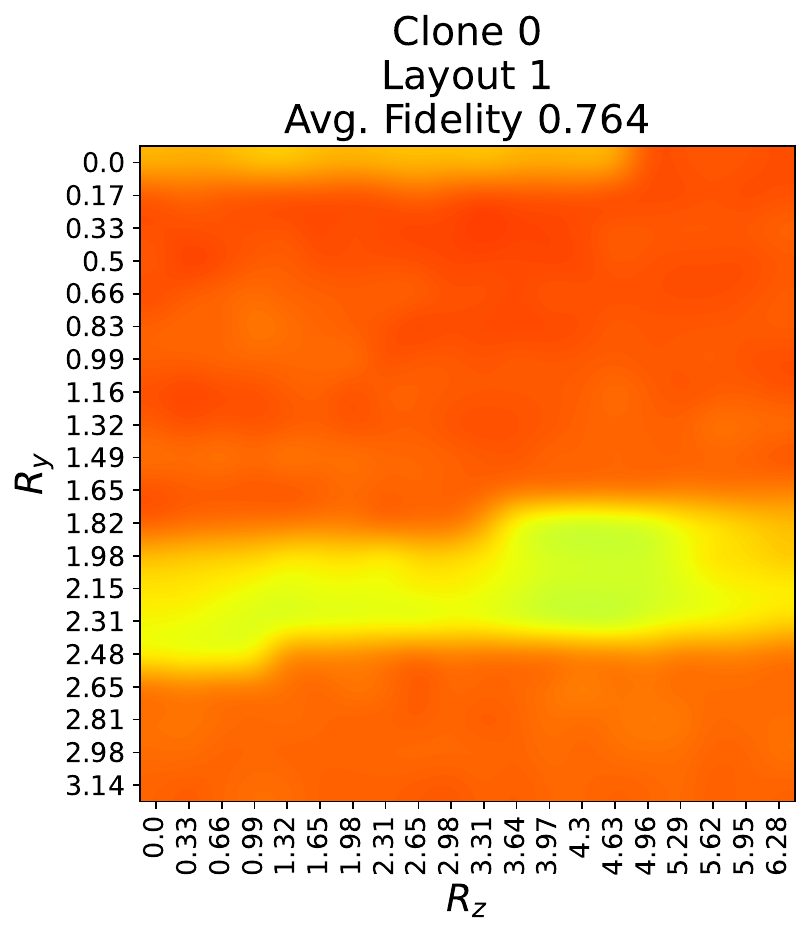}
    \includegraphics[width=0.13\textwidth]{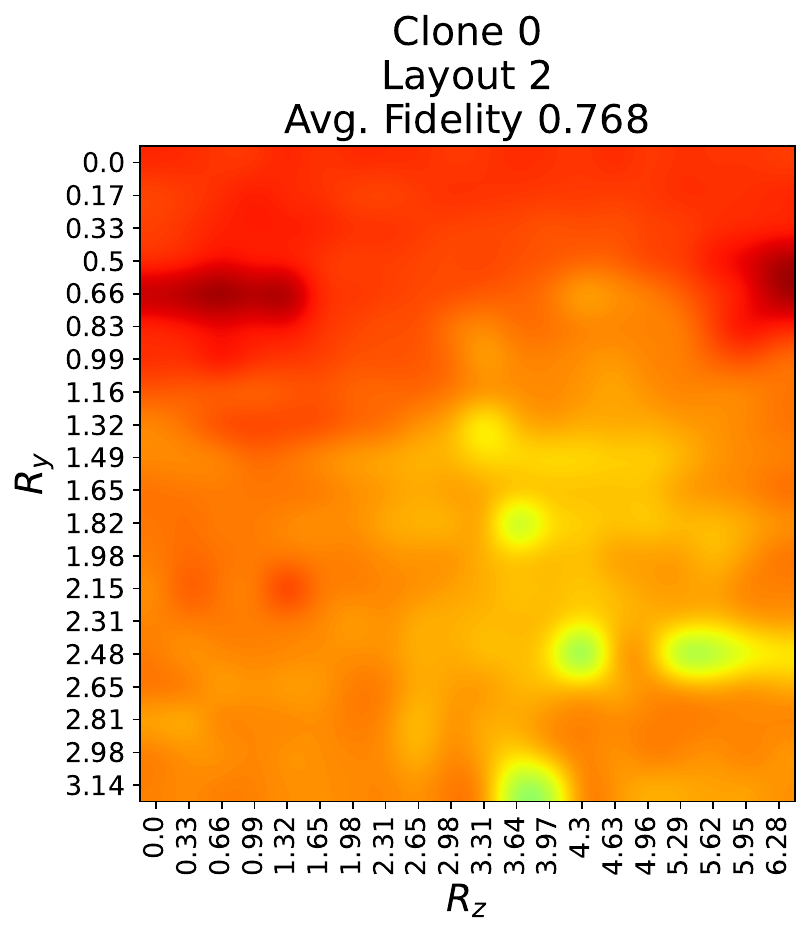}
    \includegraphics[width=0.13\textwidth]{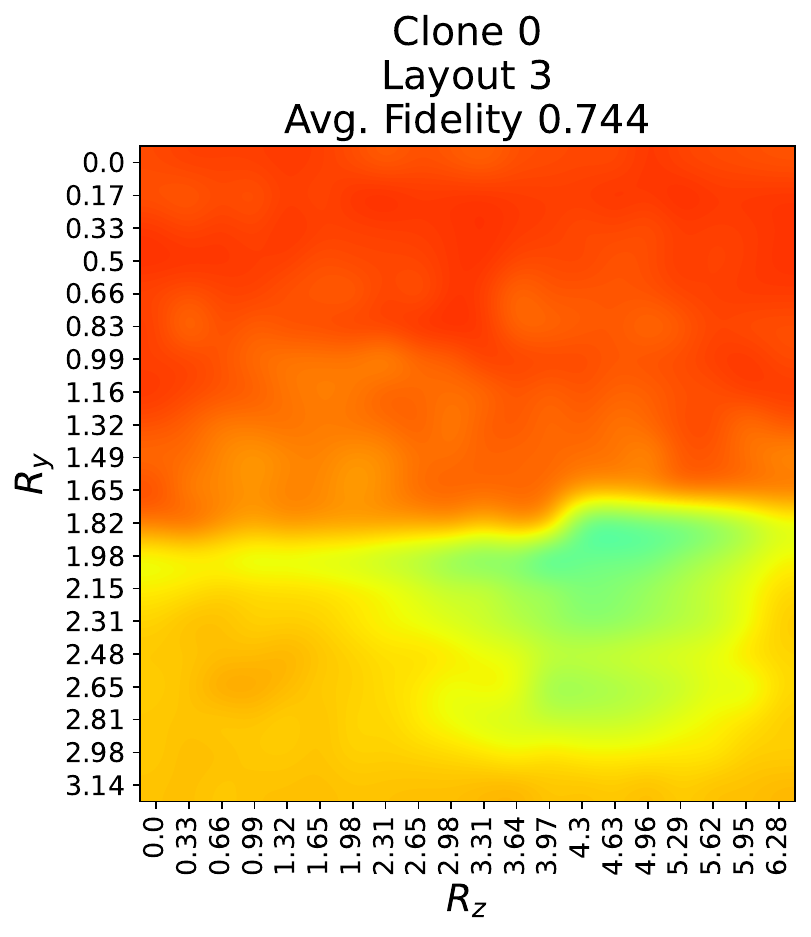}
    \includegraphics[width=0.13\textwidth]{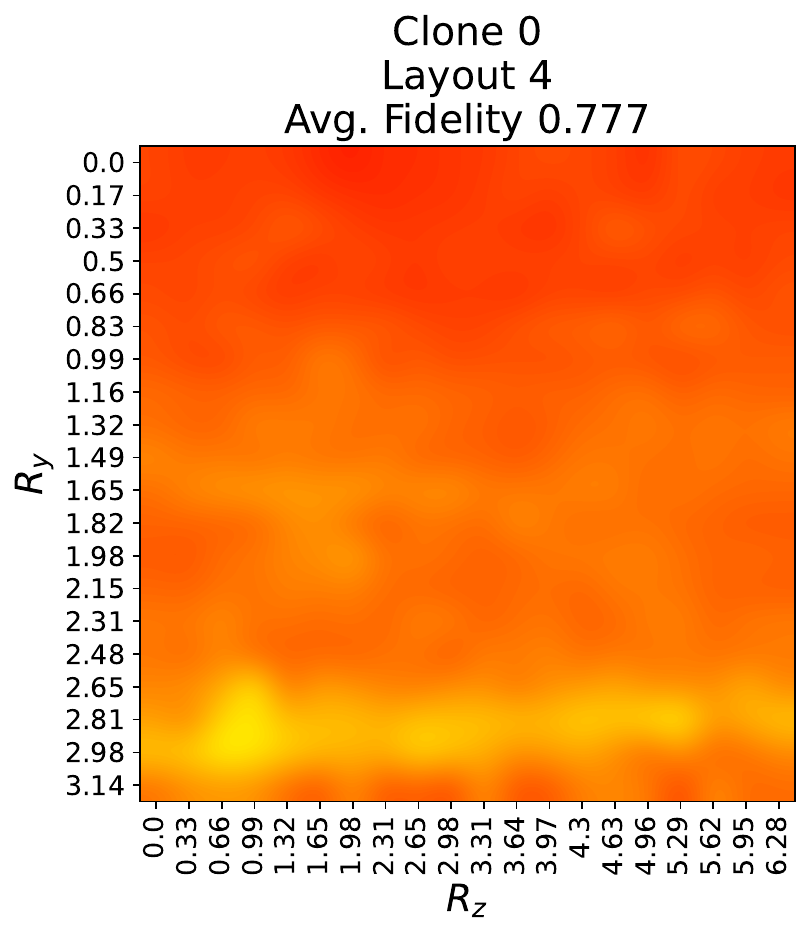}
    \includegraphics[width=0.13\textwidth]{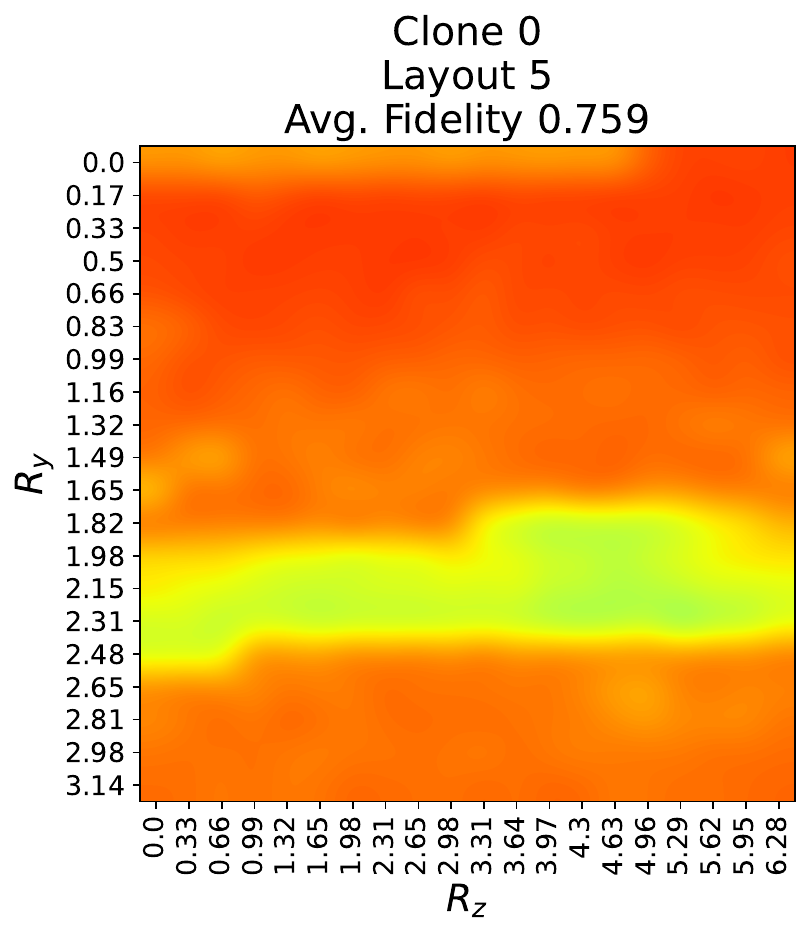}
    \includegraphics[width=0.13\textwidth]{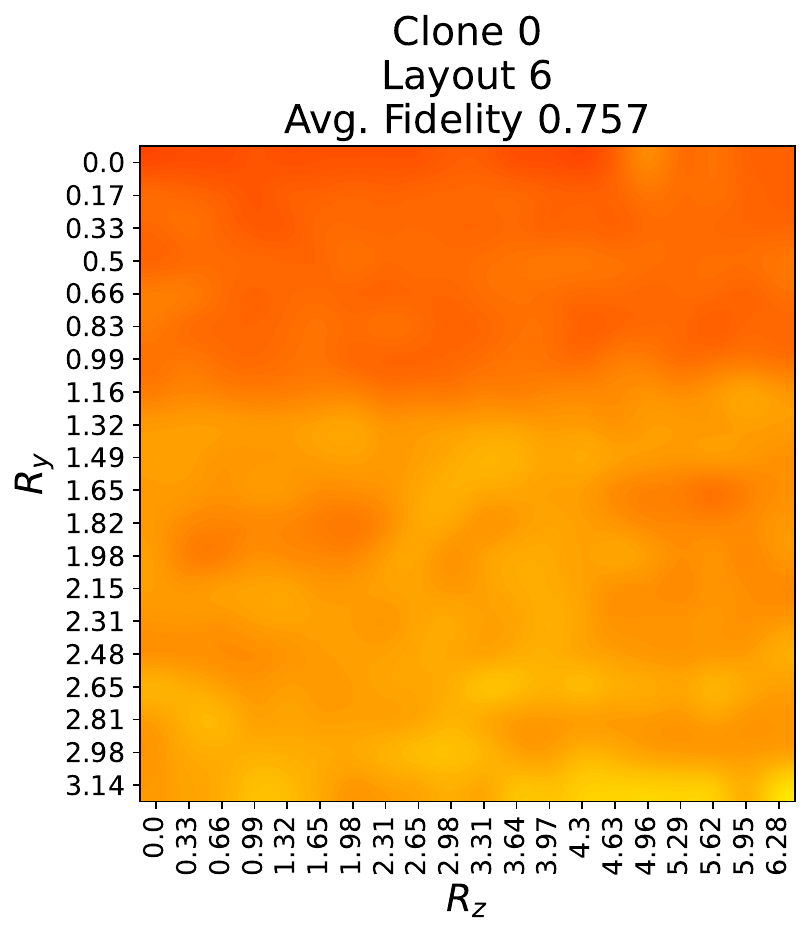}
    \includegraphics[width=0.13\textwidth]{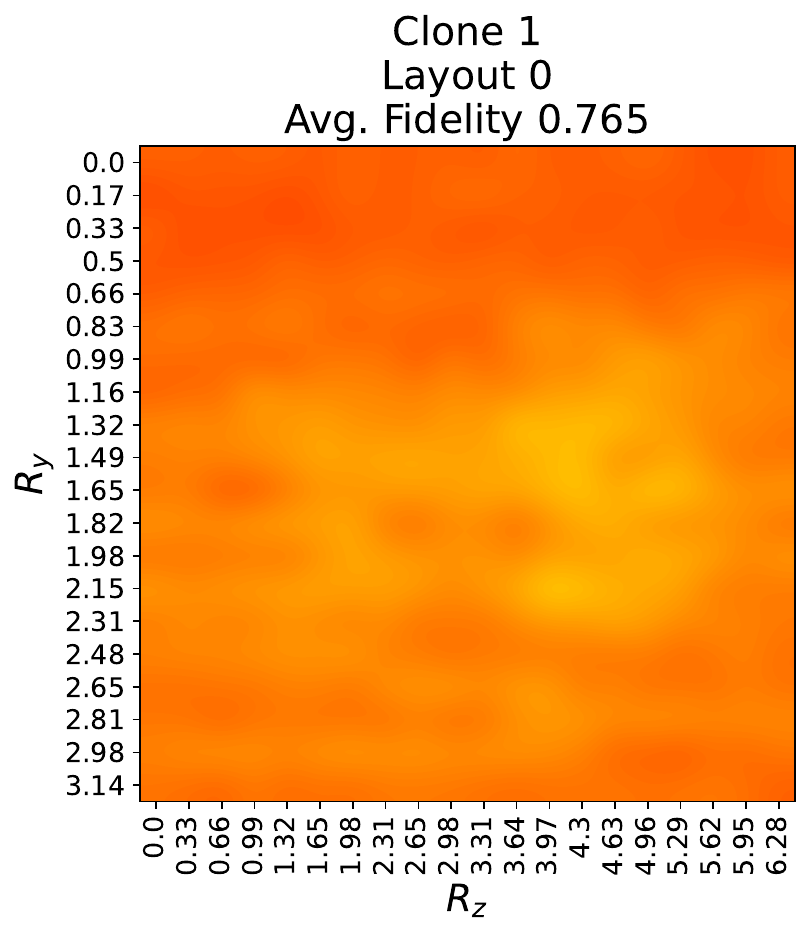}
    \includegraphics[width=0.13\textwidth]{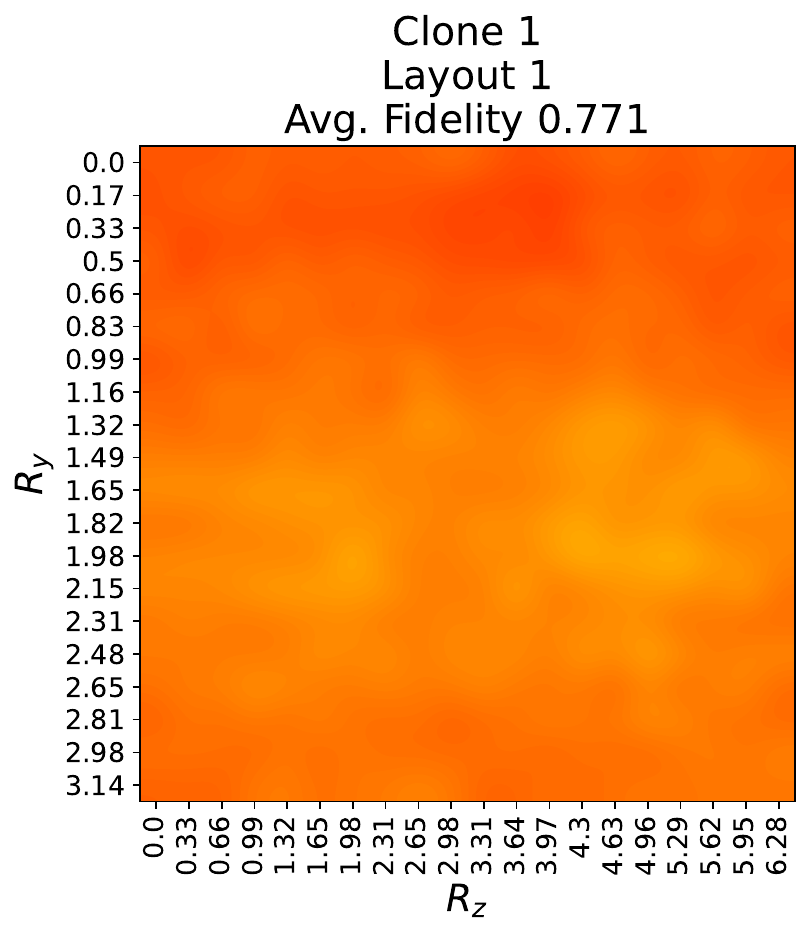}
    \includegraphics[width=0.13\textwidth]{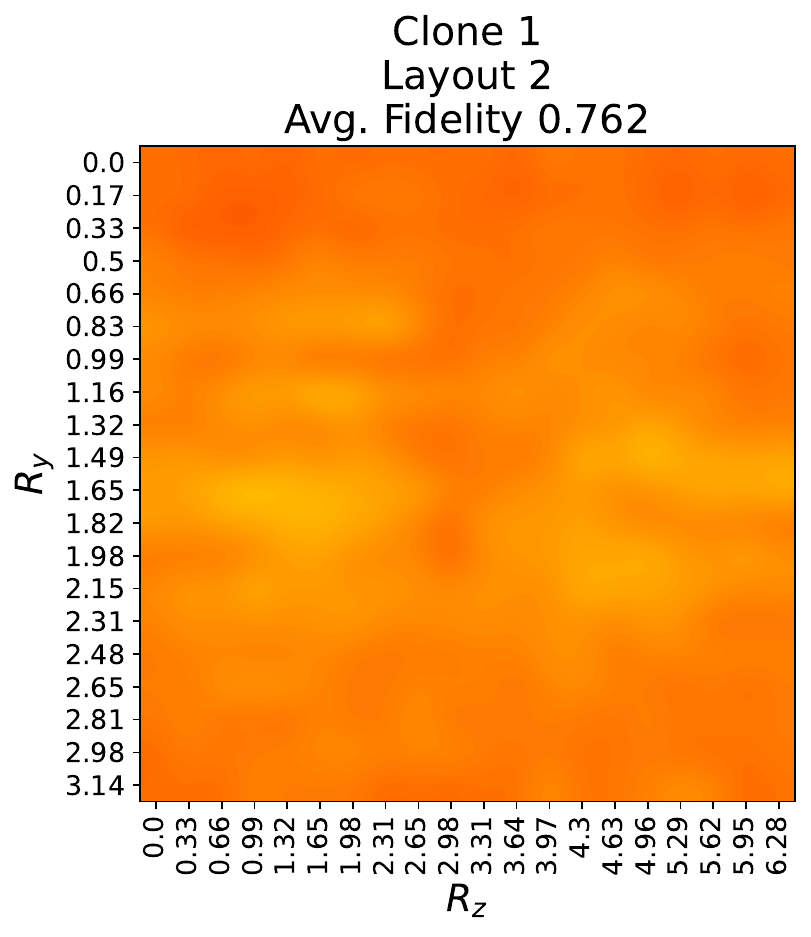}
    \includegraphics[width=0.13\textwidth]{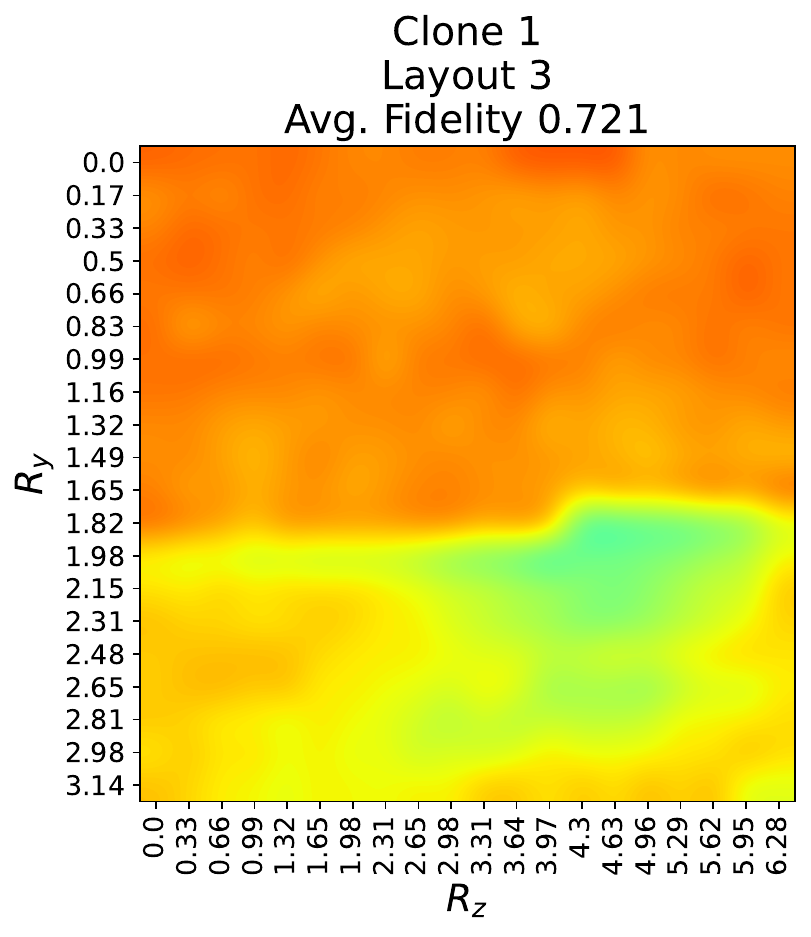}
    \includegraphics[width=0.13\textwidth]{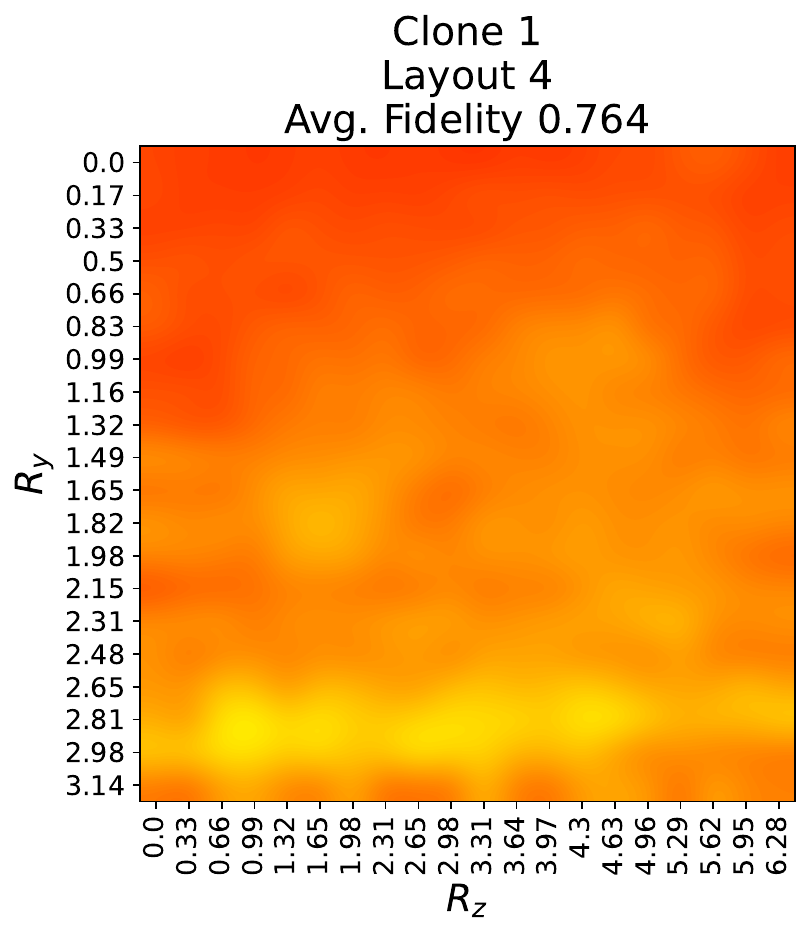}
    \includegraphics[width=0.13\textwidth]{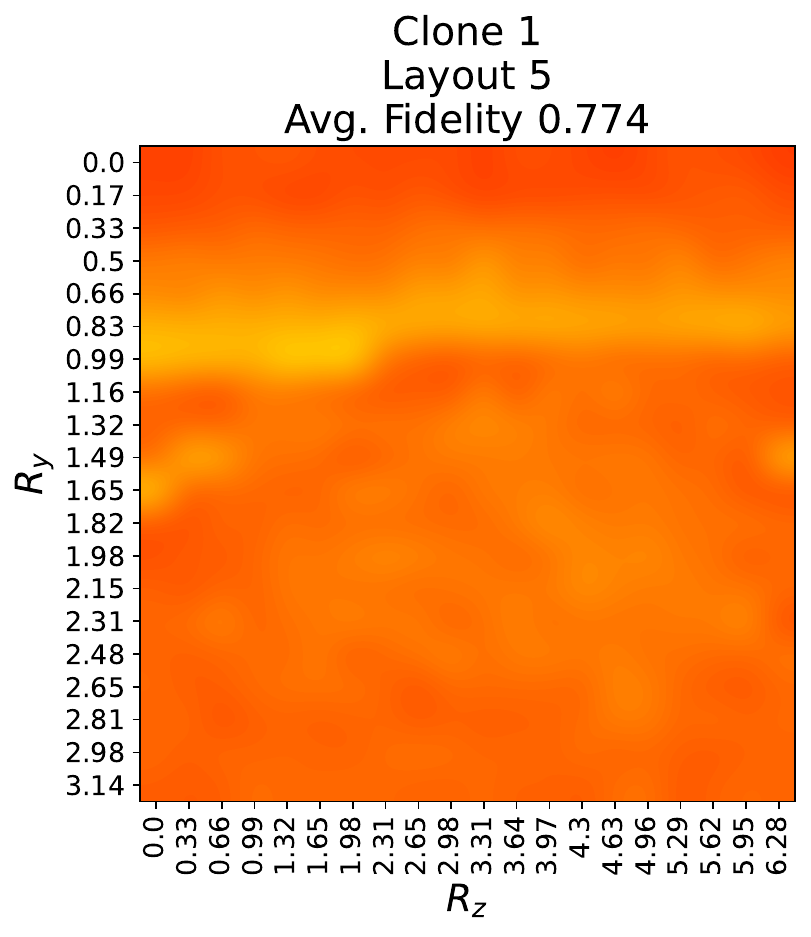}
    \includegraphics[width=0.13\textwidth]{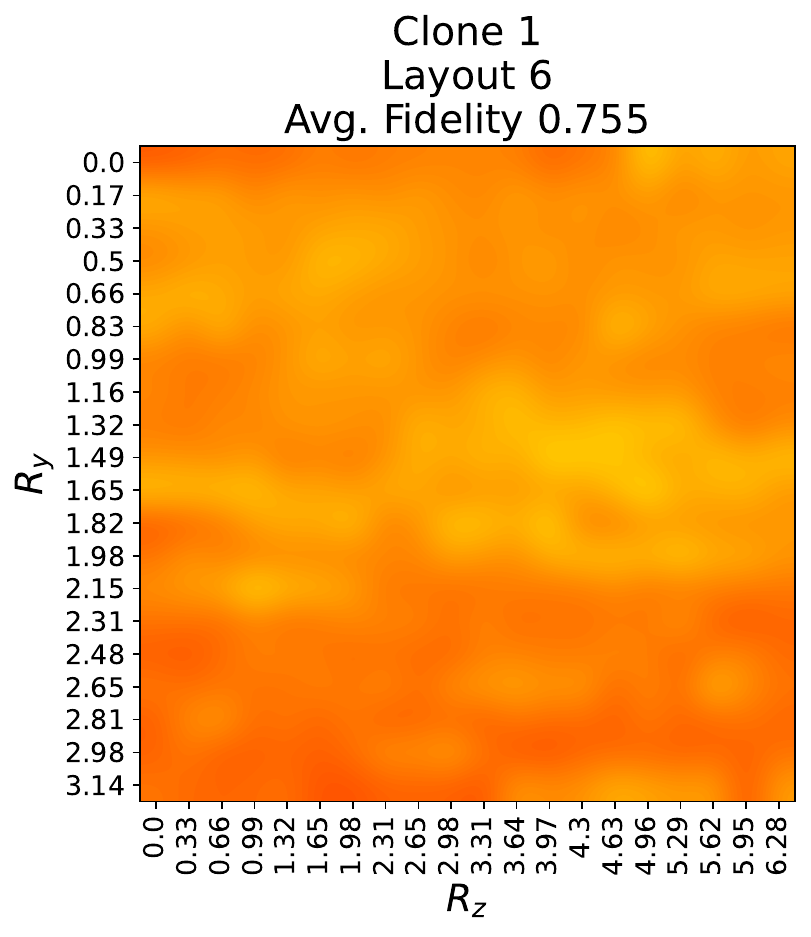}
    \includegraphics[width=0.43\textwidth]{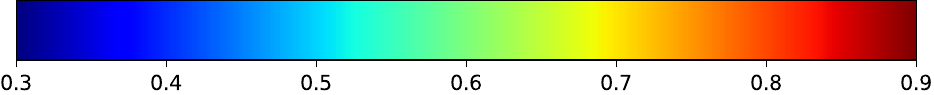}\\
    \caption{Single qubit cloning results for $M=2$ telecloning circuits with no ancilla qubits, executed with dynamical decoupling. Each column corresponds to the $7$ different compiled hardware layouts. Bottom two rows show the fidelity heatmaps of the $2$ single qubit clones, where the numerical fidelity is shown in the colormap legend below the figures. Each individual heatmap is plotting the $R_y$ and $R_z$ rotation angles to prepare the message qubit state (which is then fed into the quantum telecloning circuit). Top two rows show Bloch sphere vector representations of the single qubit state tomography computed density matrices. The x and y axis on the heatmpas in the top two rows encode the varying pure quantum states which are cloned. The heatmaps use bicubic interpolation, and each plot represents $400$ fidelity measures. The Bloch sphere vector representations and the fidelity heatmap representations share the same ordering with respect to column and row positions of the clone numbers and hardware layout positions. The average clone fidelity over all message qubit states for each clone qubit and hardware layout is shown in the fidelity heatmap plot titles. Data from \texttt{ibm\_auckland}. }
    \label{fig:fidelity_heatmaps_M2_no_ancilla_ALAP_DD_ibm_auckland}
\end{figure*}

\begin{figure*}[th!]
    \centering
    \includegraphics[width=0.13\textwidth]{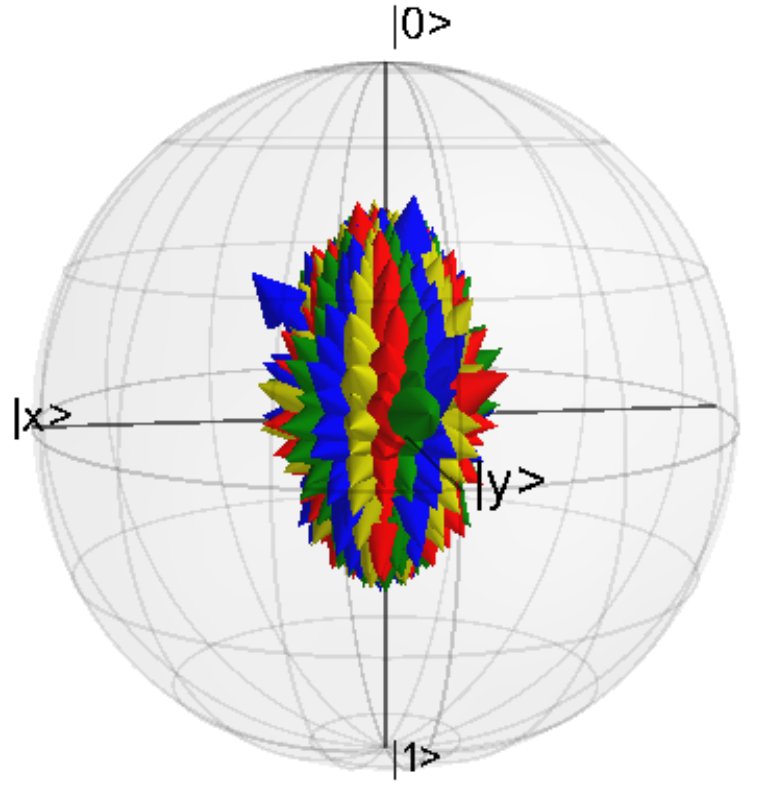}
    \includegraphics[width=0.13\textwidth]{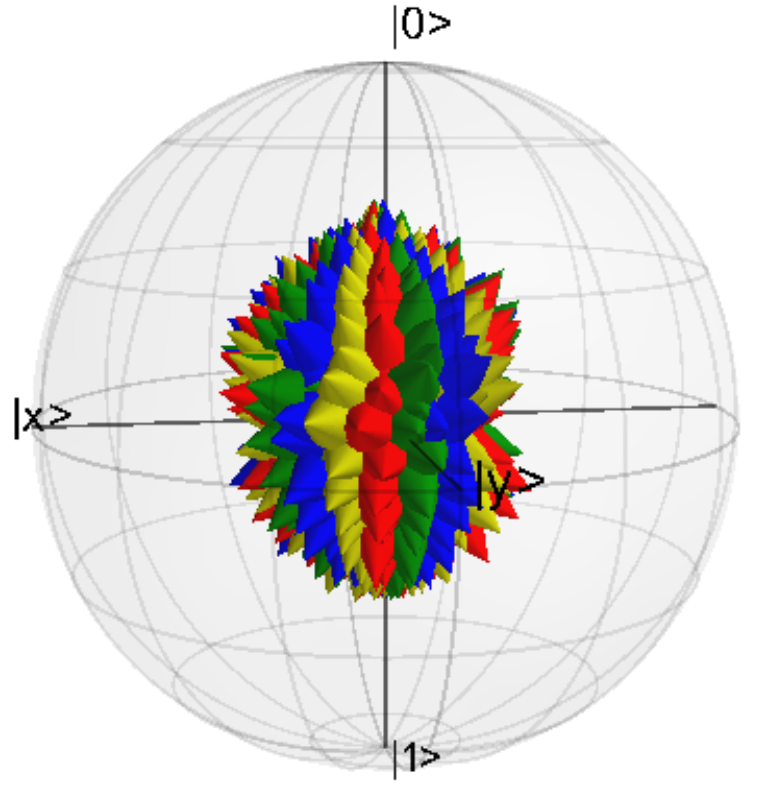}
    \includegraphics[width=0.13\textwidth]{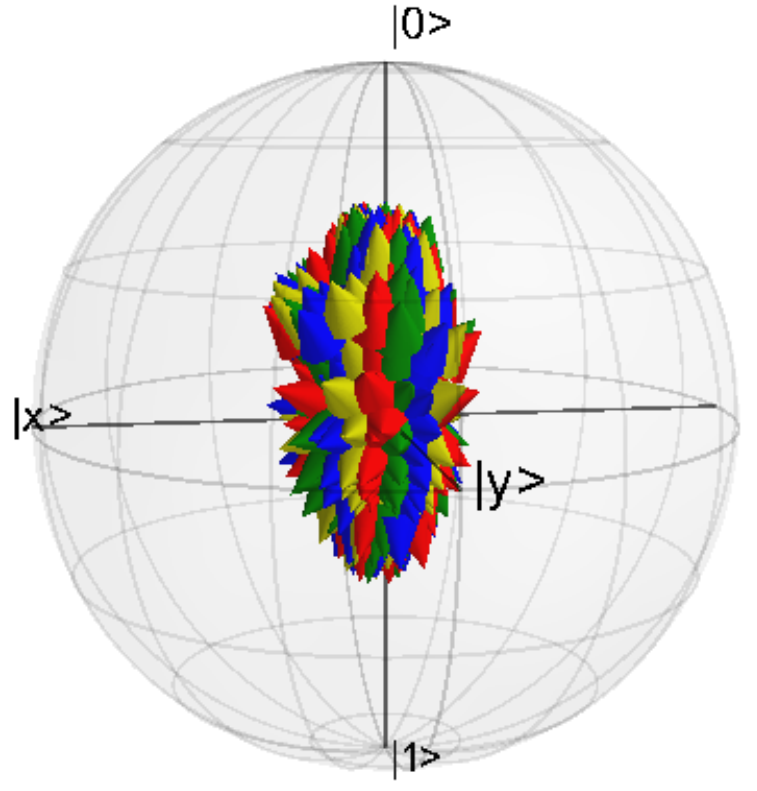}
    \includegraphics[width=0.13\textwidth]{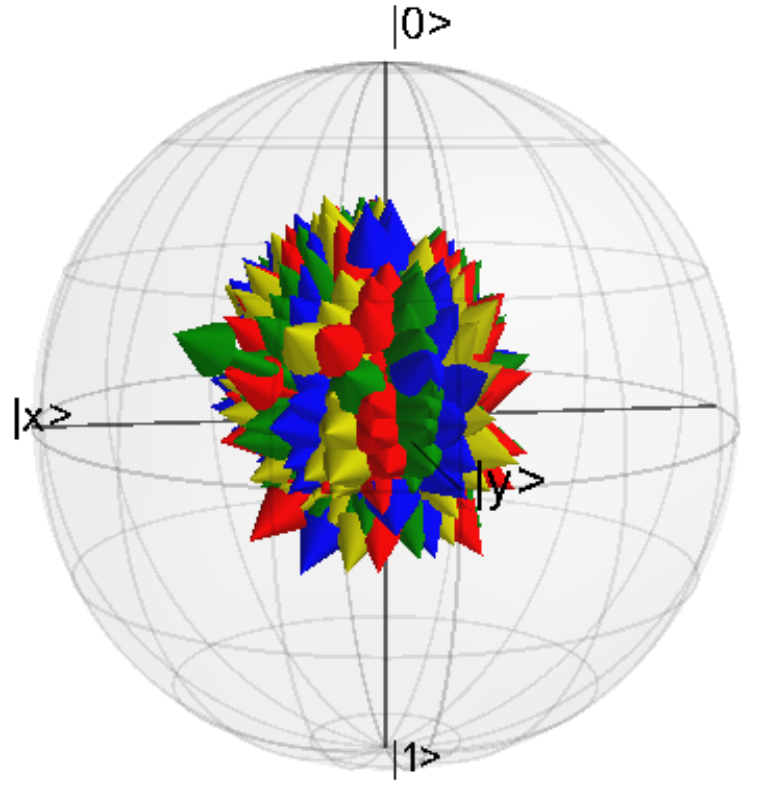}
    \includegraphics[width=0.13\textwidth]{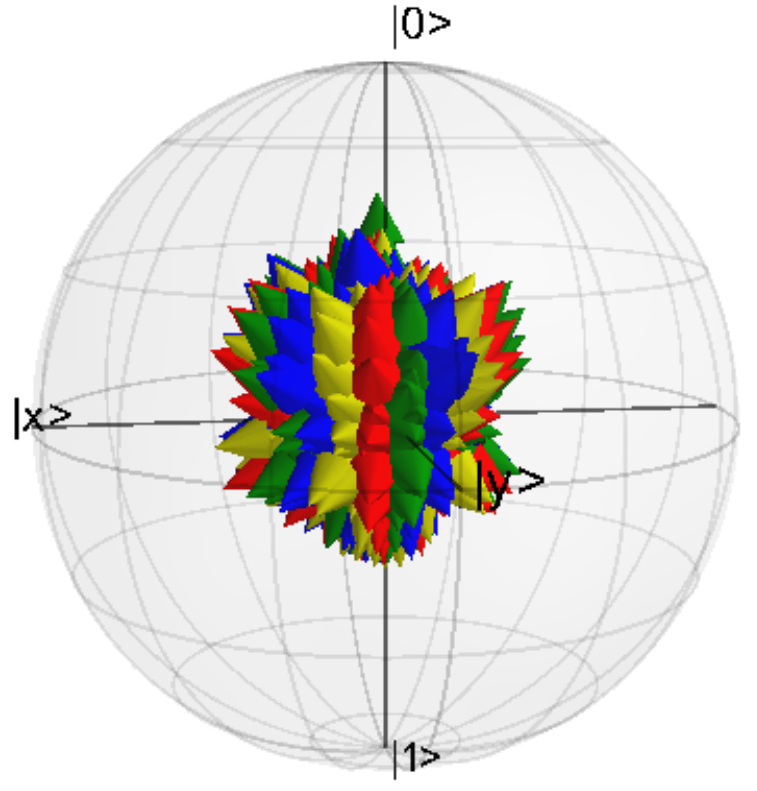}
    \includegraphics[width=0.13\textwidth]{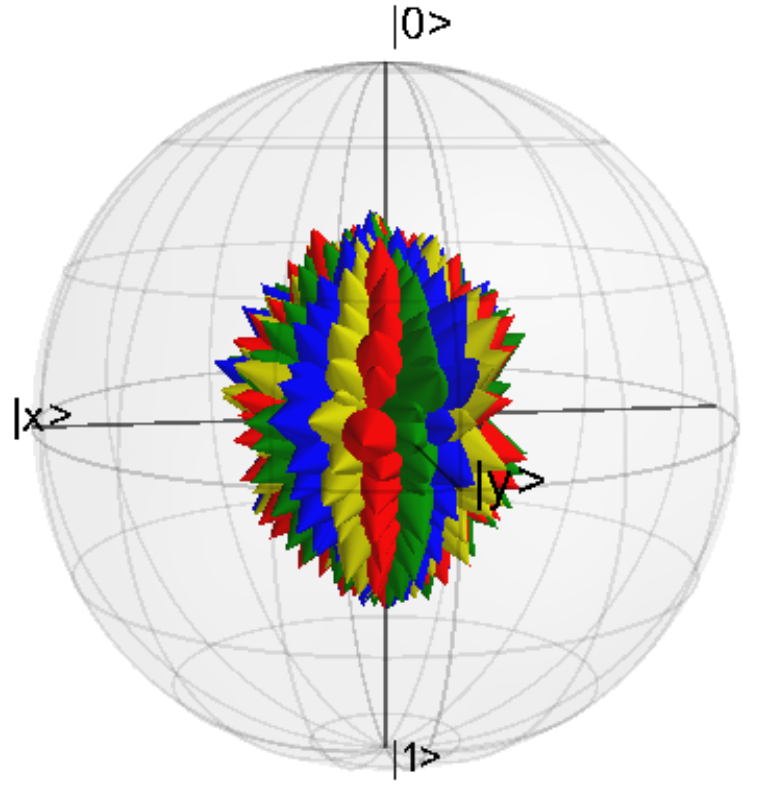}
    \includegraphics[width=0.13\textwidth]{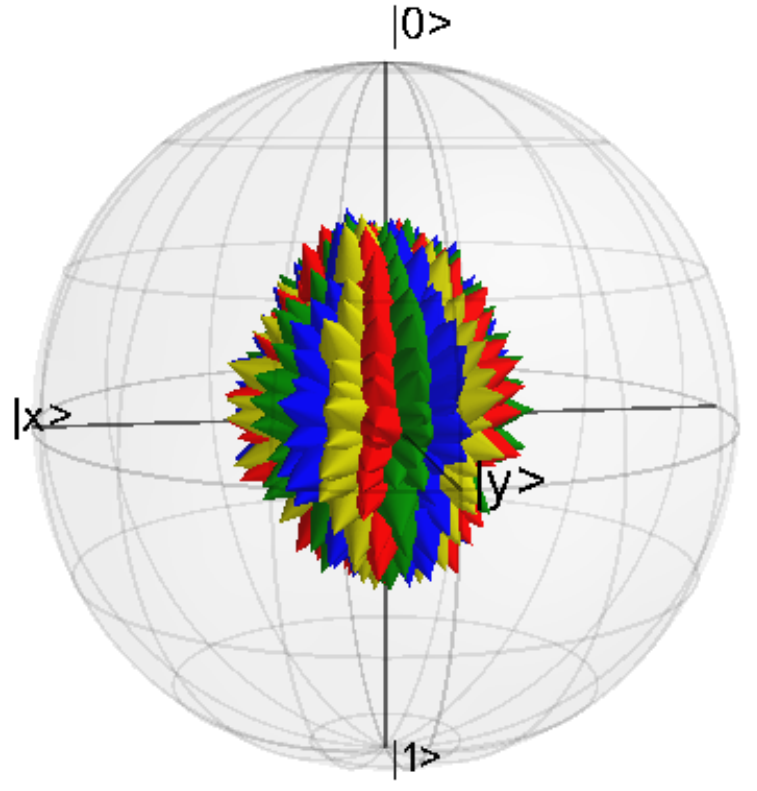}
    \includegraphics[width=0.13\textwidth]{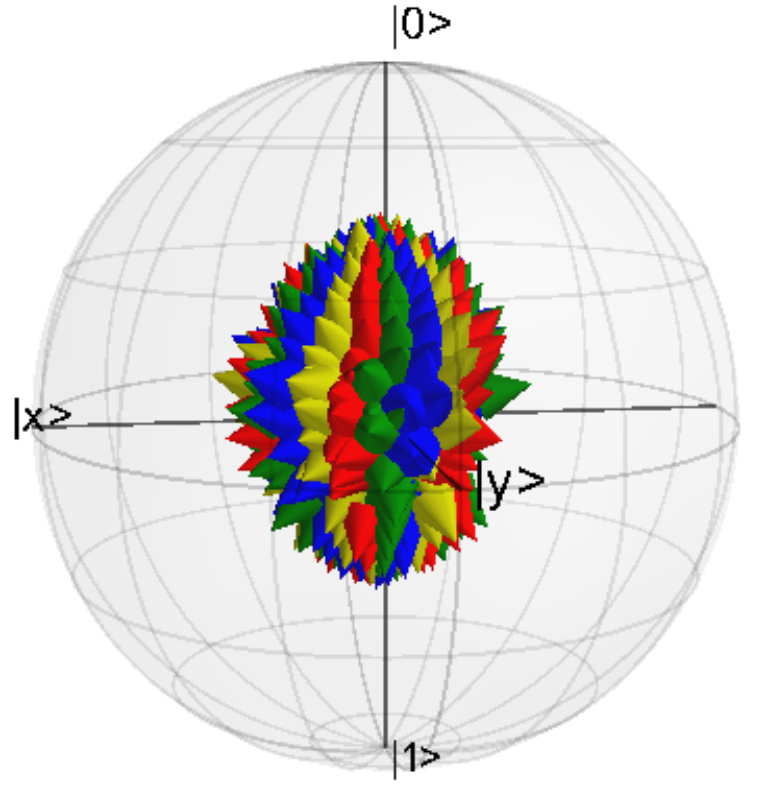}
    \includegraphics[width=0.13\textwidth]{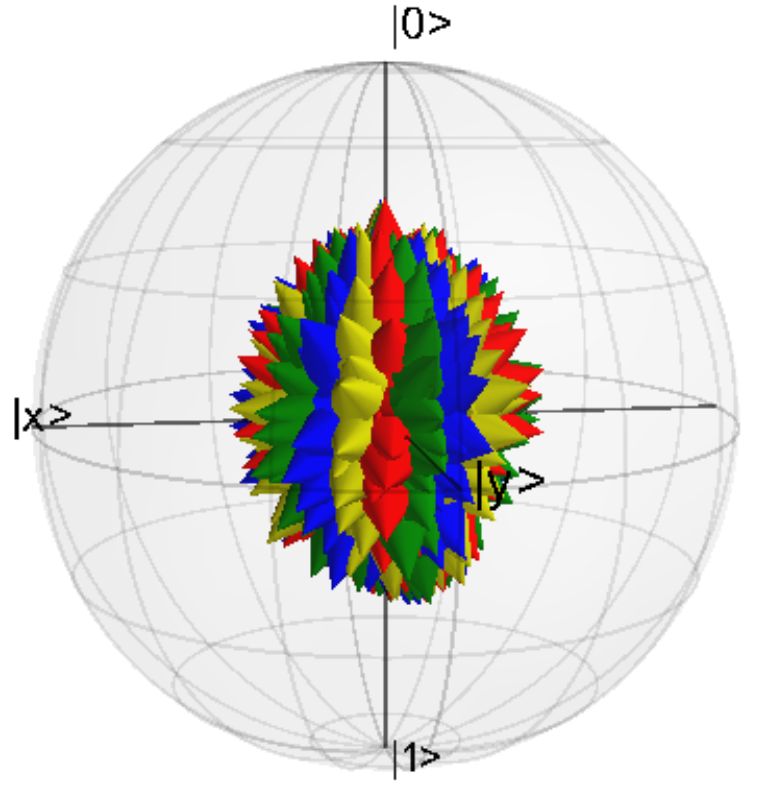}
    \includegraphics[width=0.13\textwidth]{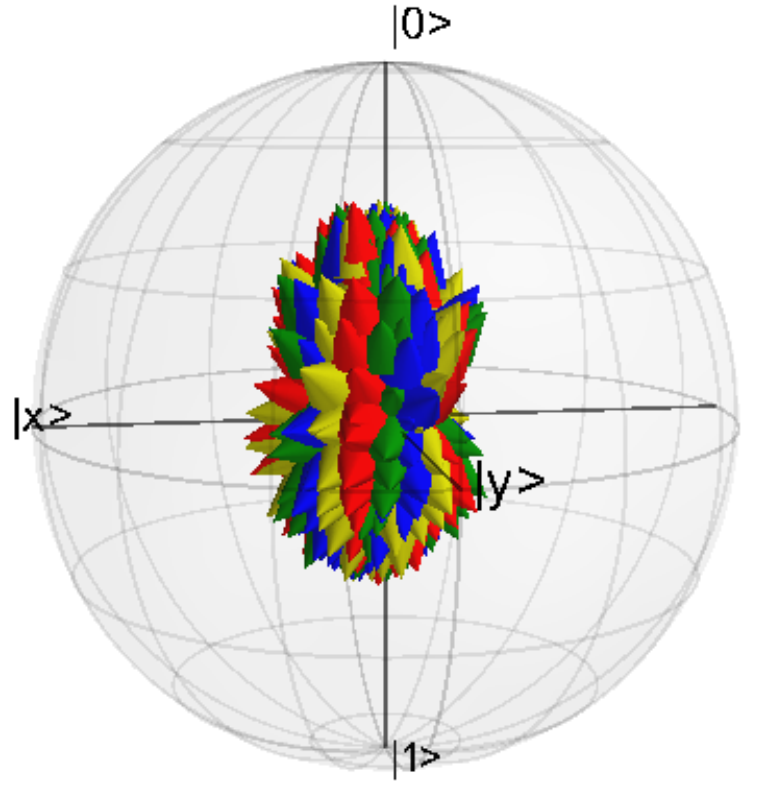}
    \includegraphics[width=0.13\textwidth]{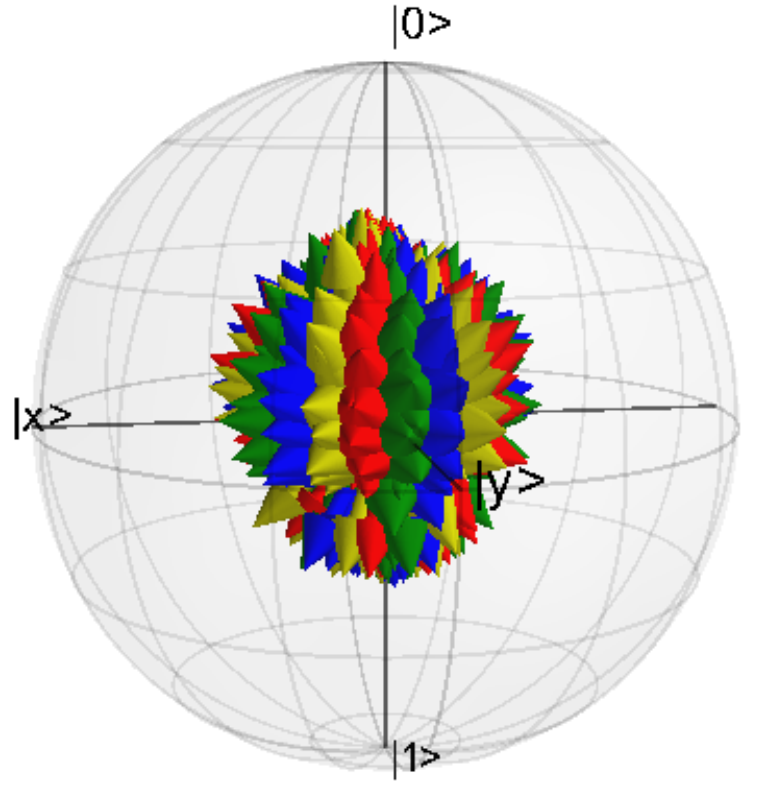}
    \includegraphics[width=0.13\textwidth]{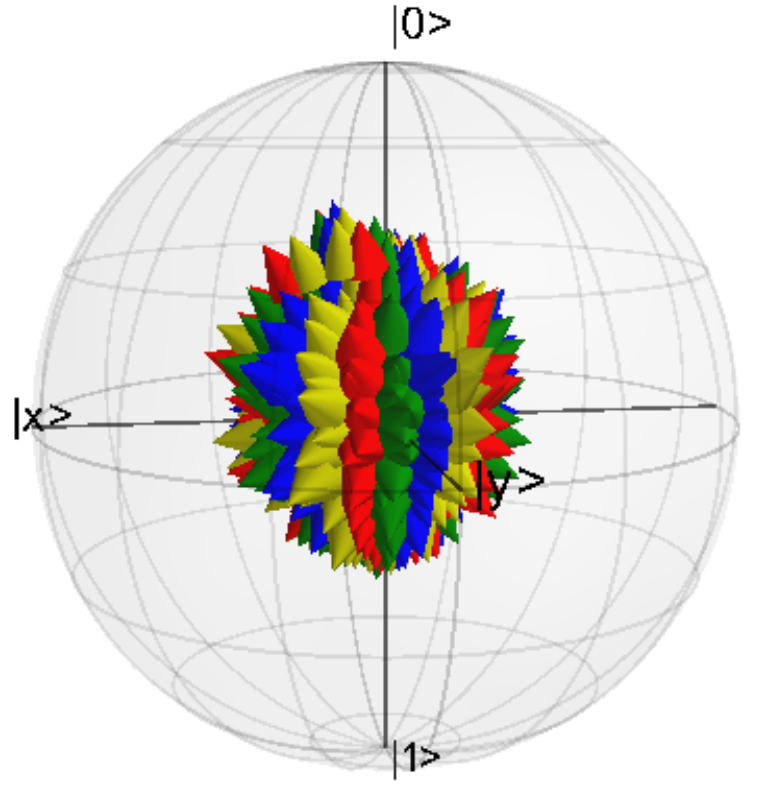}
    \includegraphics[width=0.13\textwidth]{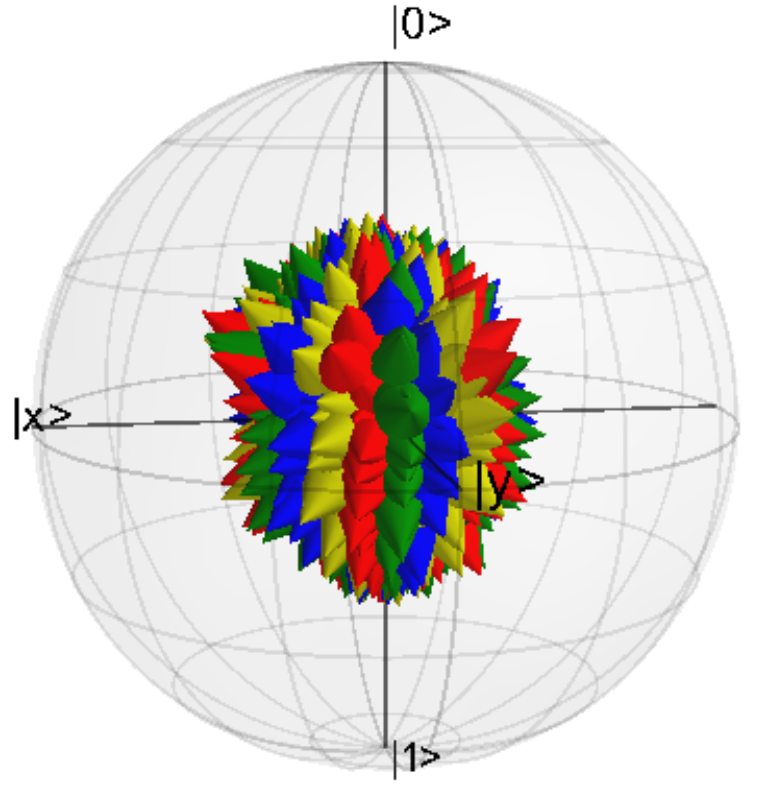}
    \includegraphics[width=0.13\textwidth]{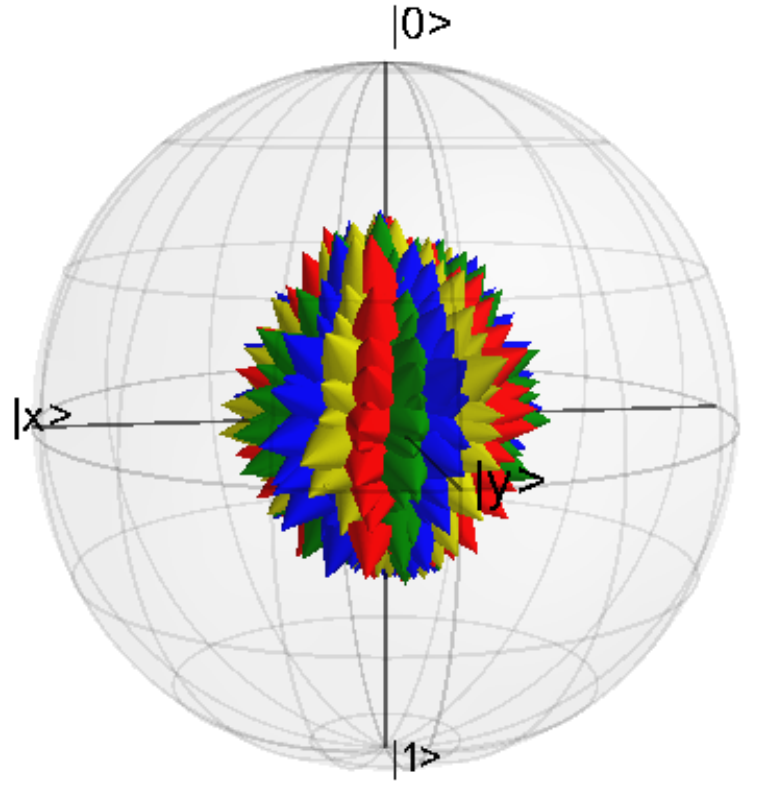}
    \includegraphics[width=0.13\textwidth]{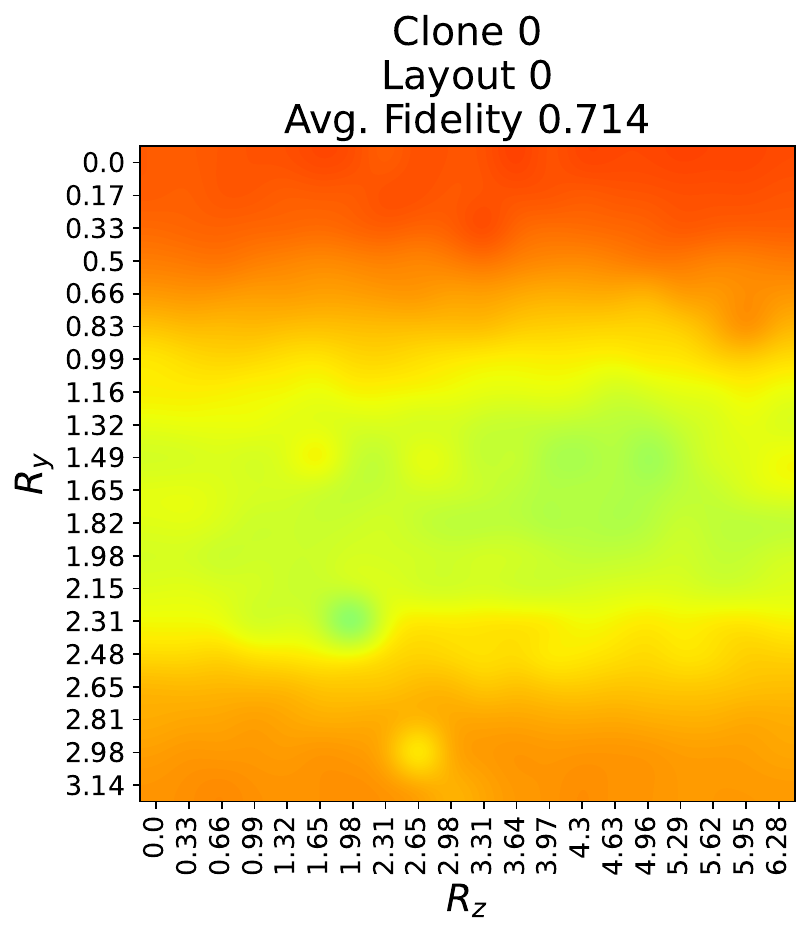}
    \includegraphics[width=0.13\textwidth]{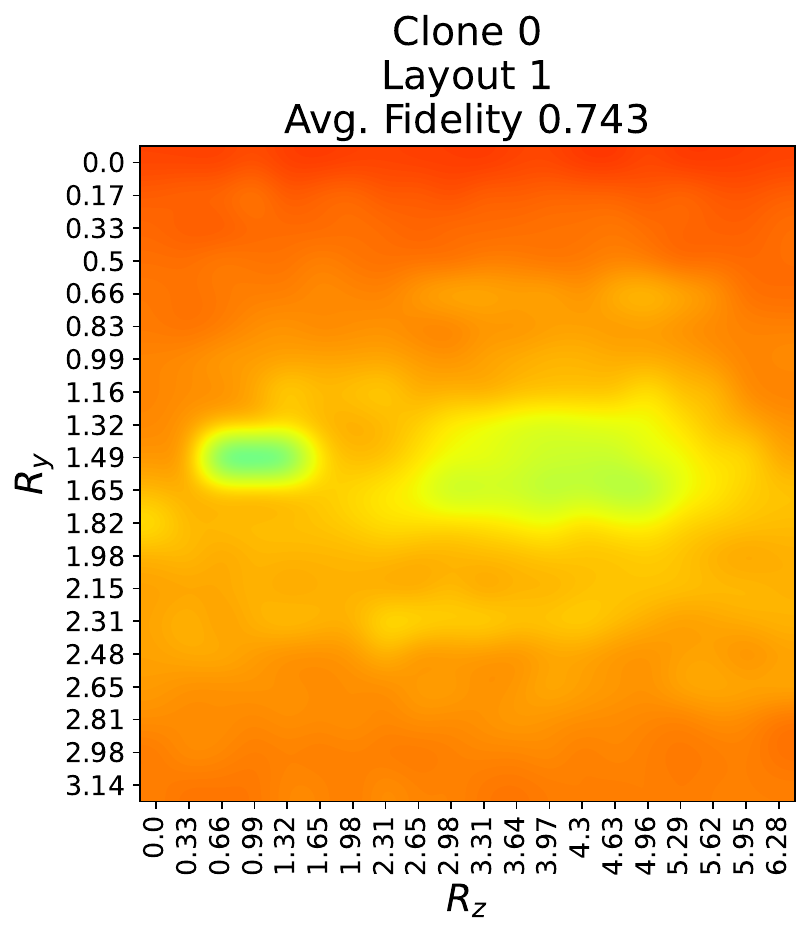}
    \includegraphics[width=0.13\textwidth]{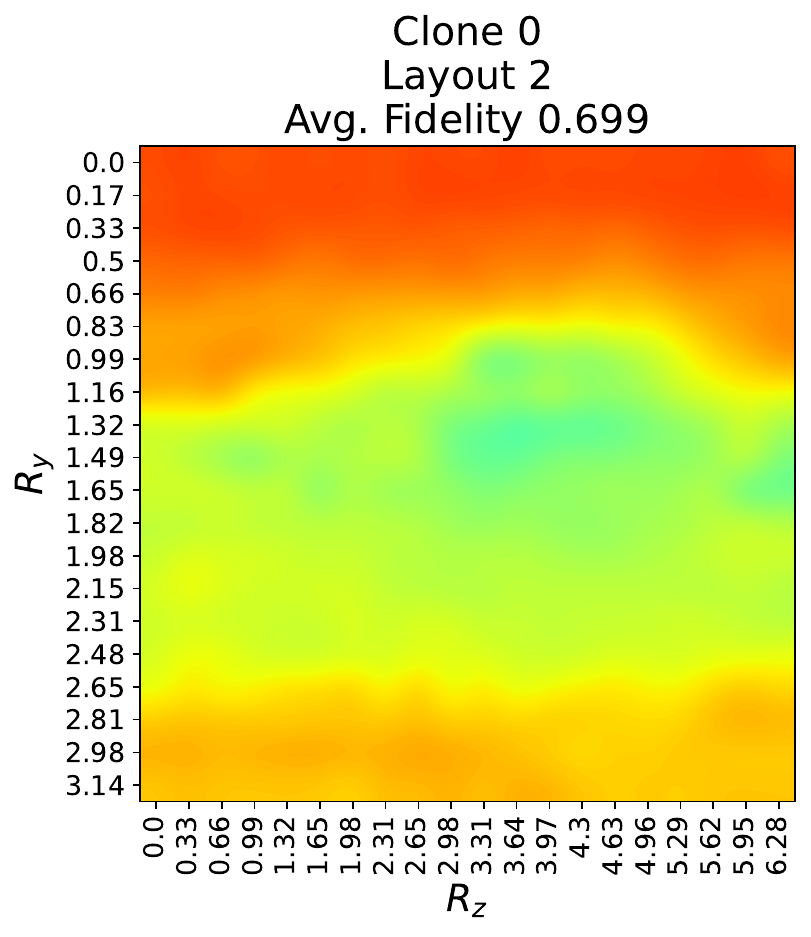}
    \includegraphics[width=0.13\textwidth]{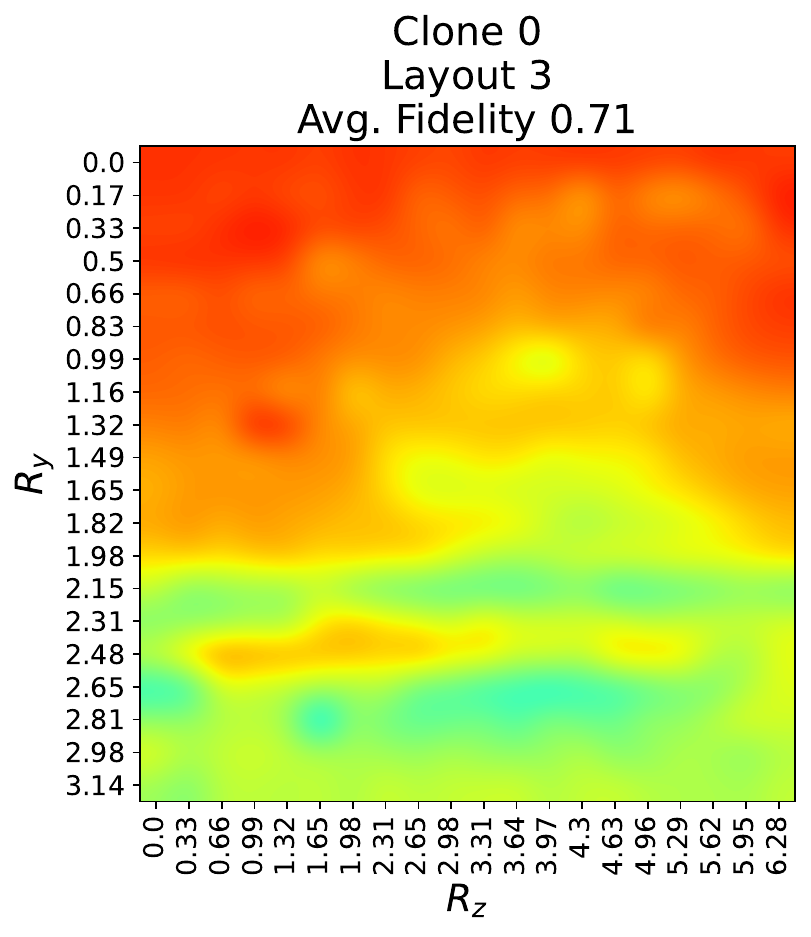}
    \includegraphics[width=0.13\textwidth]{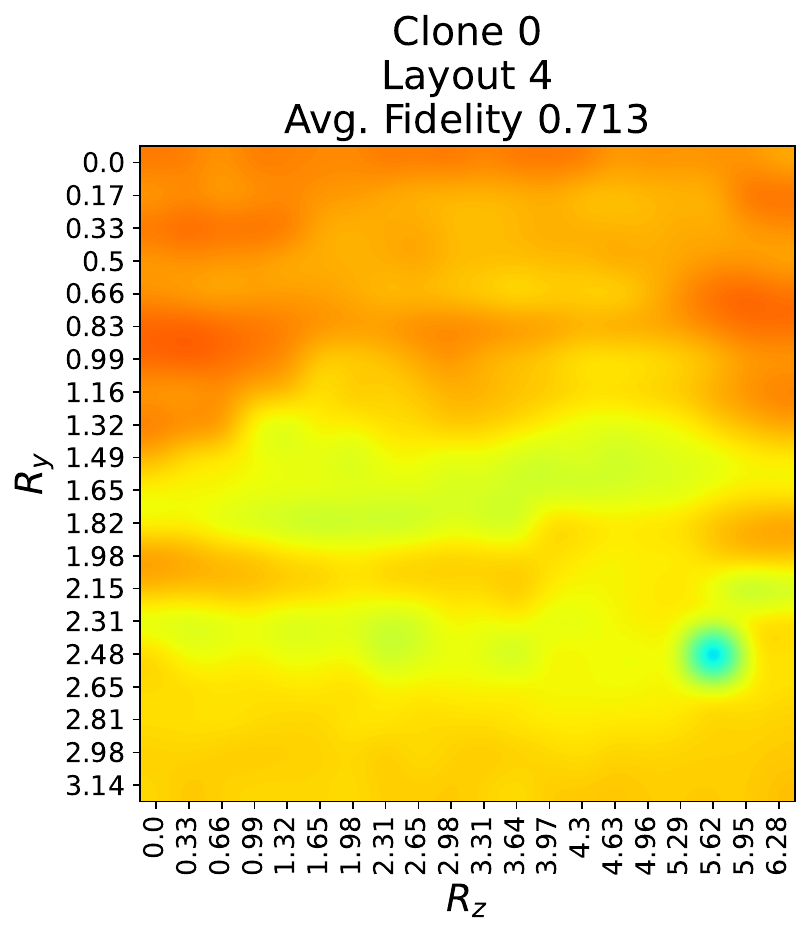}
    \includegraphics[width=0.13\textwidth]{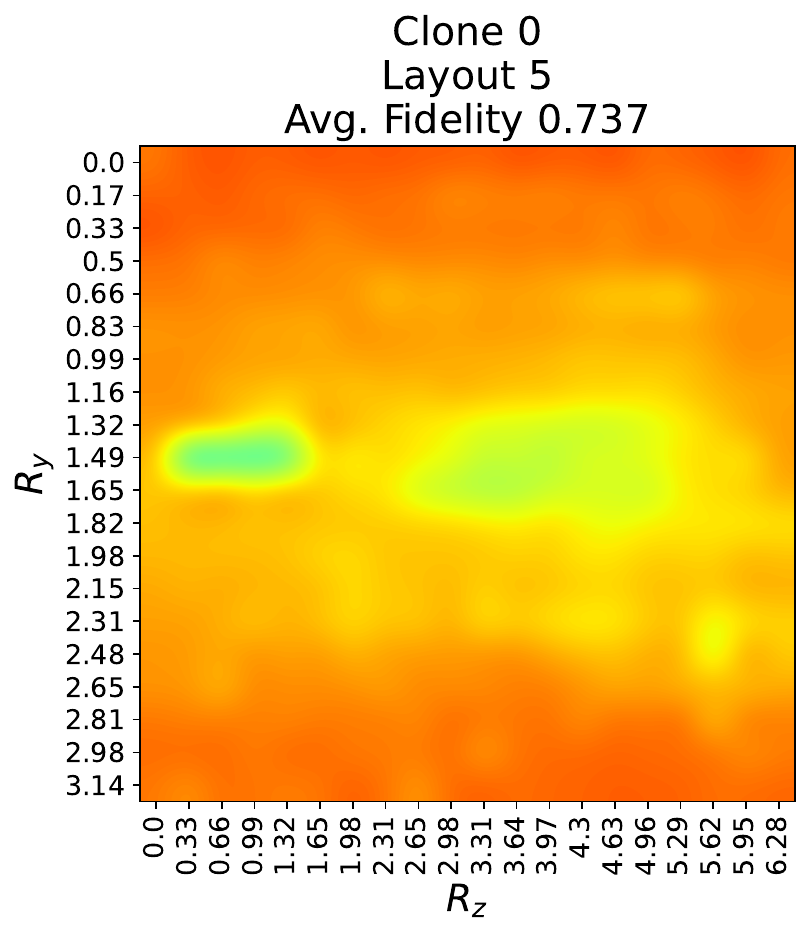}
    \includegraphics[width=0.13\textwidth]{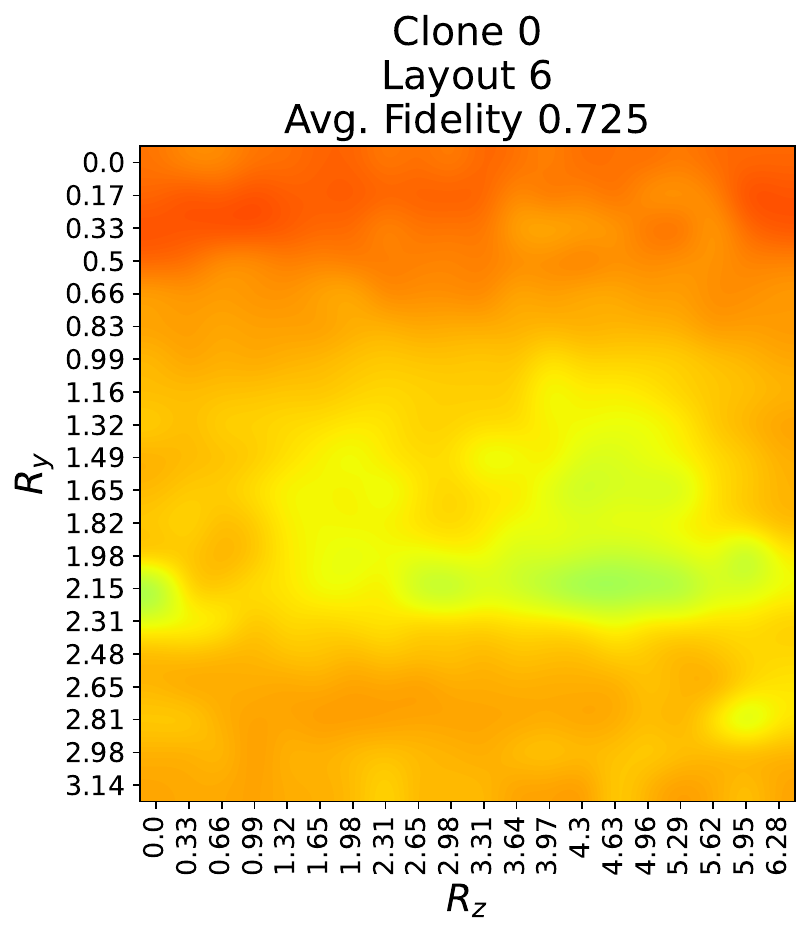}
    \includegraphics[width=0.13\textwidth]{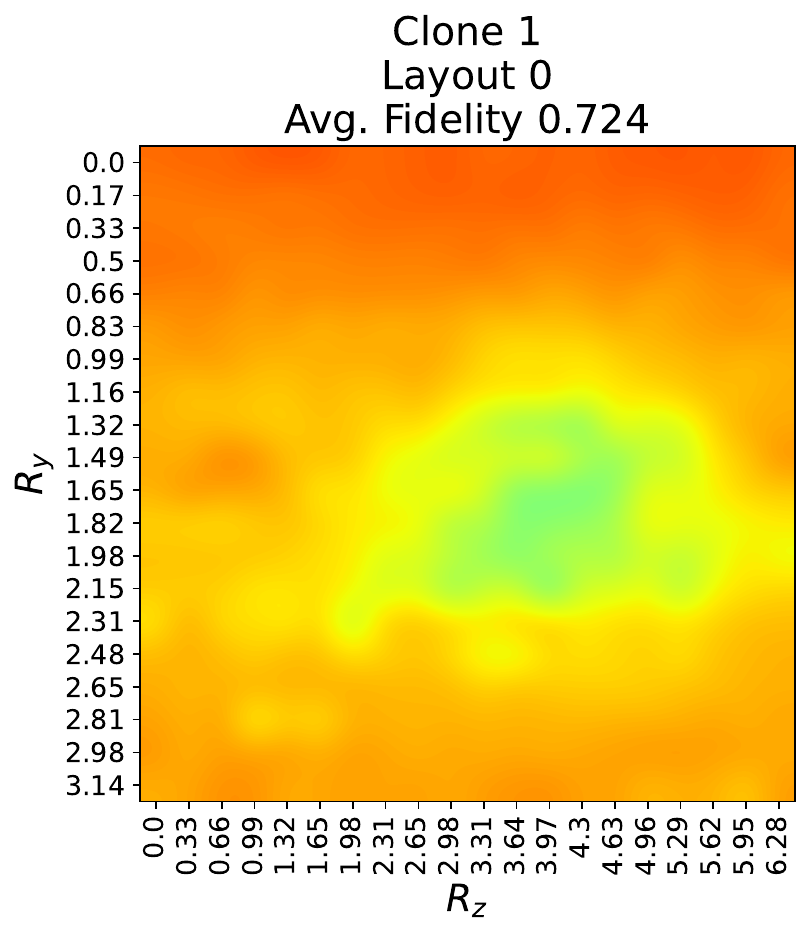}
    \includegraphics[width=0.13\textwidth]{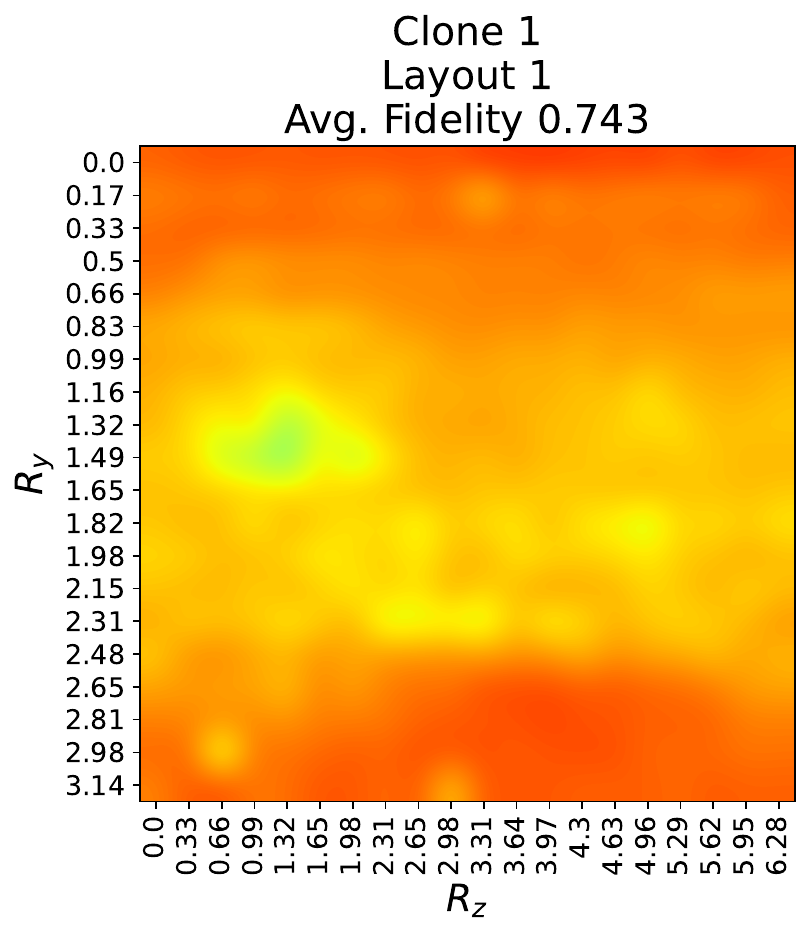}
    \includegraphics[width=0.13\textwidth]{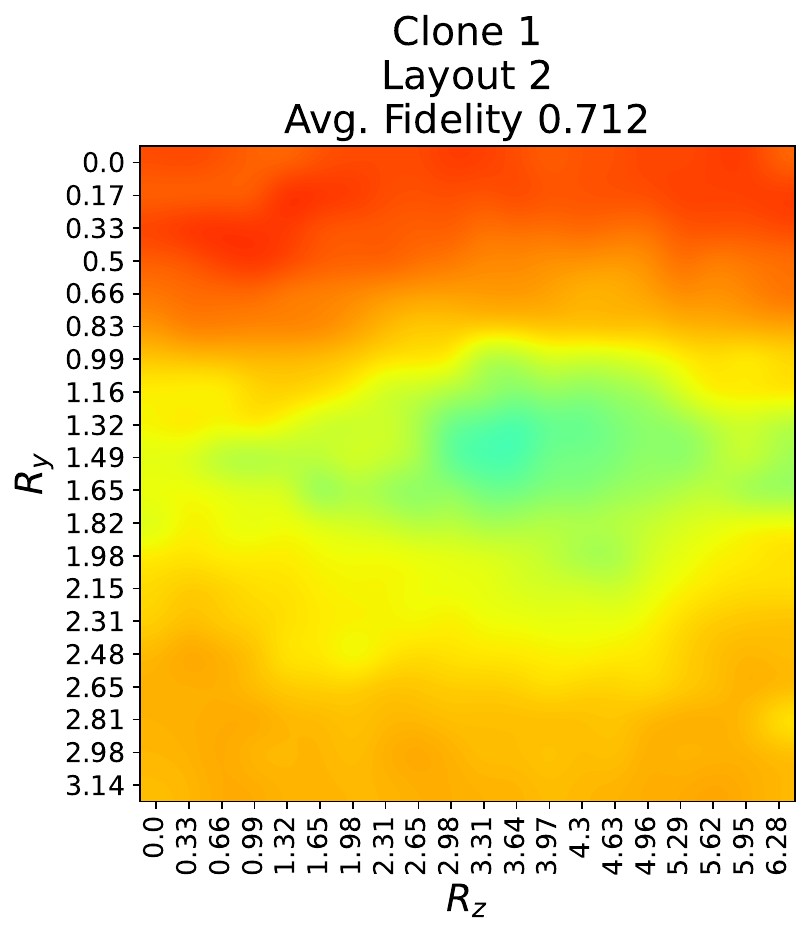}
    \includegraphics[width=0.13\textwidth]{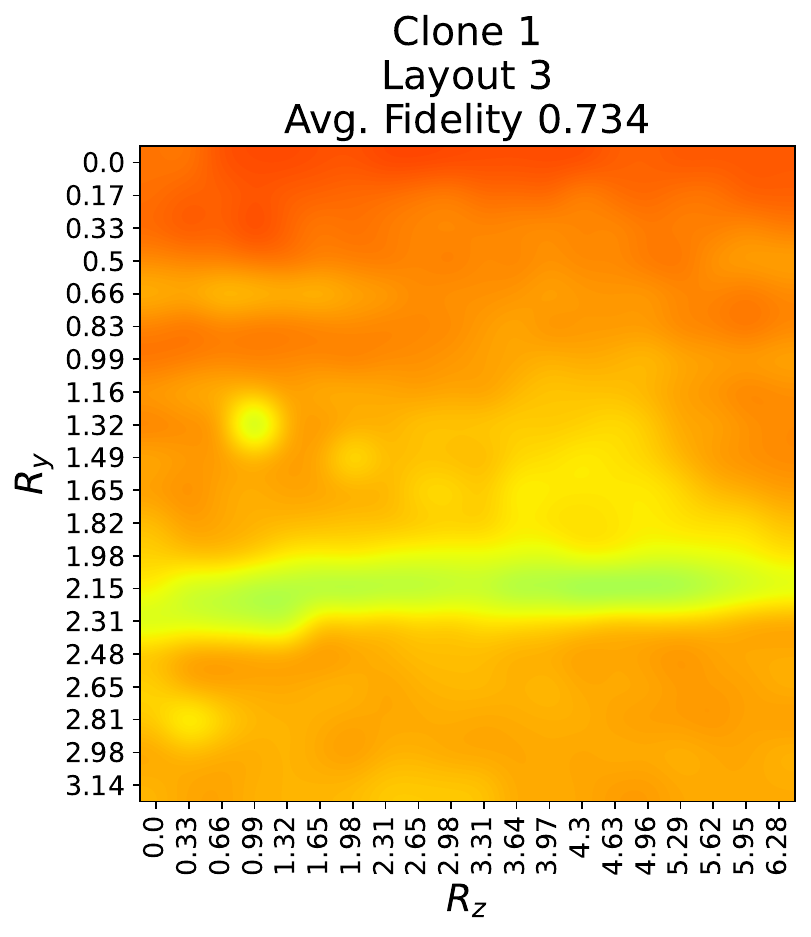}
    \includegraphics[width=0.13\textwidth]{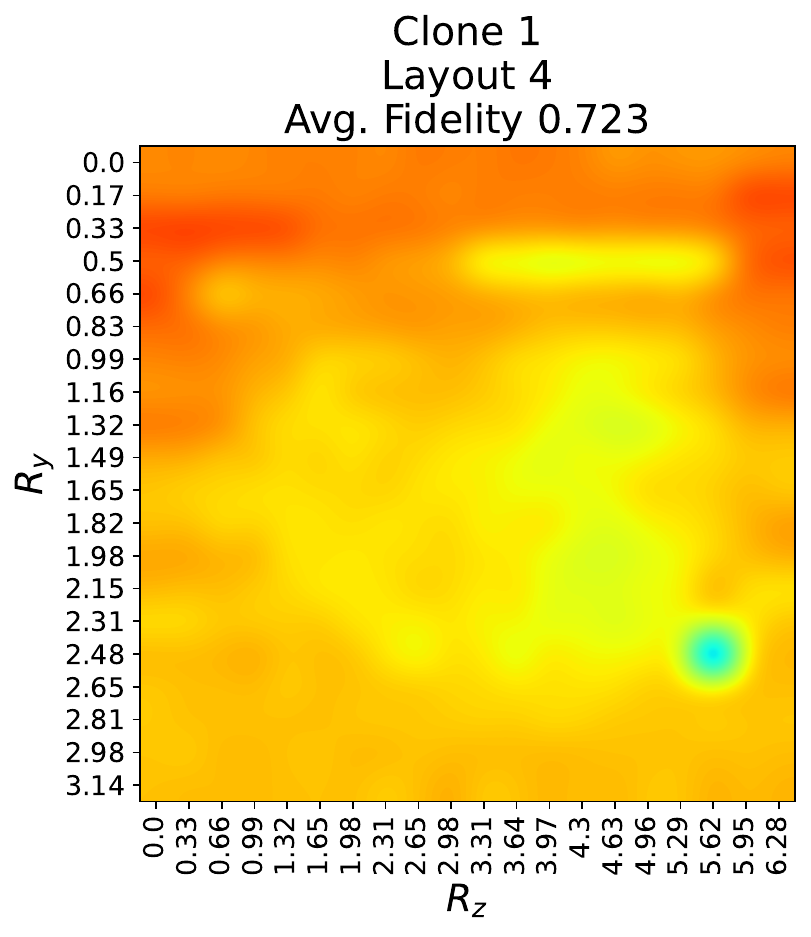}
    \includegraphics[width=0.13\textwidth]{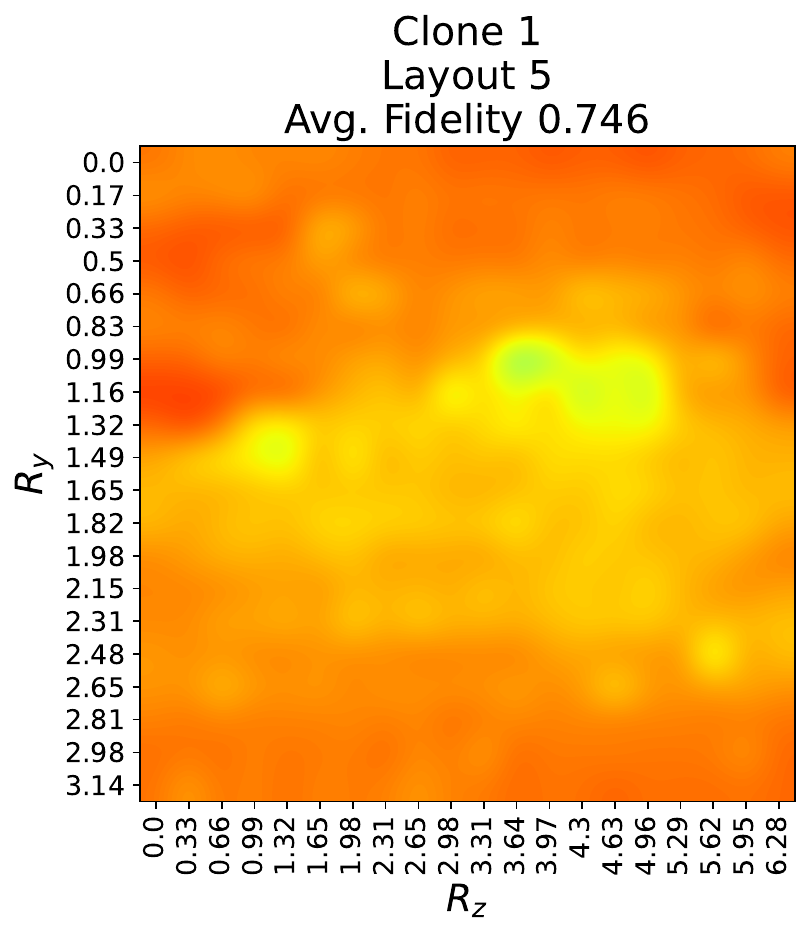}
    \includegraphics[width=0.13\textwidth]{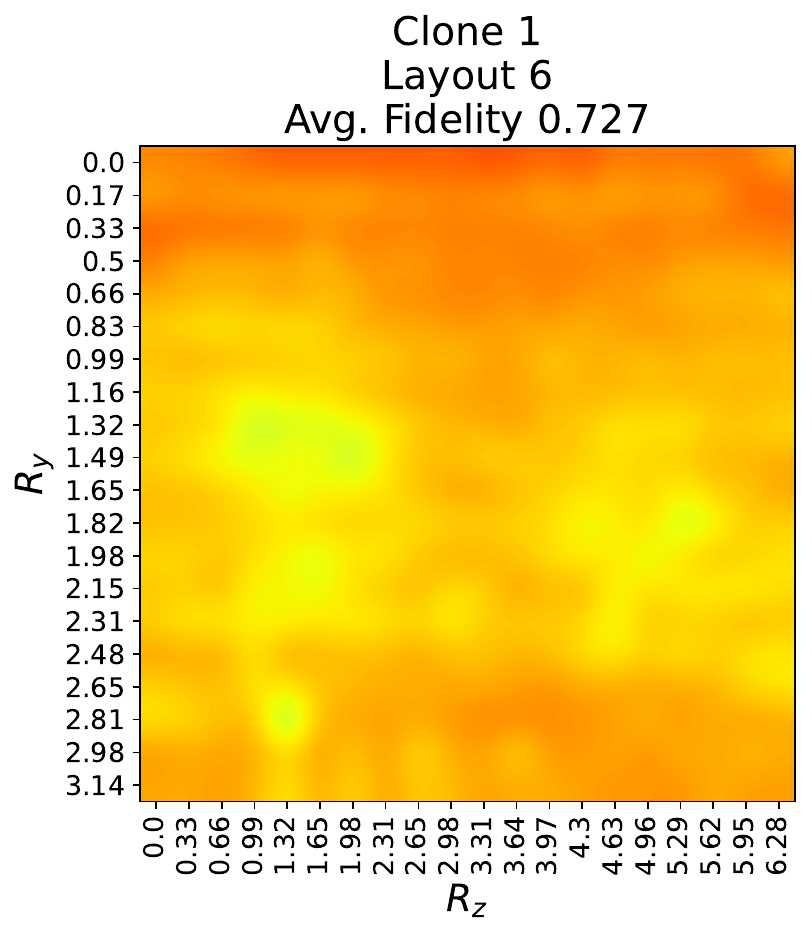}
    \includegraphics[width=0.43\textwidth]{figures/colorbar.pdf}\\
    \caption{Single qubit cloning results for $M=2$ telecloning circuits with no ancilla qubits, executed with dynamical decoupling. Each column corresponds to the $7$ different hardware layouts. Bottom two rows show the fidelity heatmaps of the $2$ single qubit clones, where the numerical fidelity is shown in the legend below the figures. Top two rows show Bloch sphere vector representations of the single qubit state tomography density matrices. The x and y axis on the heatmpas in the top two rows encode the varying pure quantum states which are cloned. All heatmaps use bicubic interpolation, and each sub-plot represents $400$ separate fidelity measures. Data from \texttt{ibmq\_mumbai}. }
    \label{fig:fidelity_heatmaps_M2_no_ancilla_ALAP_DD_ibmq_mumbai}
\end{figure*}

Table~\ref{table:summary_clone_fidelity} presents the mean fidelity measures of the experimentally computed clones, across $7$ IBM Quantum processors and for a variety of quantum telecloning circuit sizes and implementations (including X-X dynamical decoupling passes). Each data entry in Table~\ref{table:summary_clone_fidelity} represents a total of $8,400$ dynamic circuits having been executed on the quantum computer, which corresponds to $84,000,000$ dynamic circuit measurements for each entry. Appendix~\ref{section:appendix_job_execution_metadata} contains aggregated statistics on the QPU time and queue times required to gather these measurements, along with examples of the types of server-side errors that were encountered. The empty cells in the data table represent sets of experiments which were not completed or performed on the device. This incomplete data is due to a variety of factors, including devices being shut-down before more circuits could be executed, server-side errors requiring circuit re-submission and thus increasing compute time, and the overall compute and queuing time being quite high for this ensemble of quantum cloning fidelity characterizations. See Appendix~\ref{section:appendix_job_execution_metadata} for more detailed information. 

The fidelities in Table~\ref{table:summary_clone_fidelity} are the maximum mean (the mean is taken across the $400$ initial states) measured fidelity across the $7$ hardware layouts (see Figure~\ref{fig:qubit_layouts}). Table~\ref{table:summary_clone_fidelity} shows that the use of dynamical decoupling improves the best mean clone fidelity computations, with a single exception of $M=2$ on \texttt{ibm\_cairo}. Table~\ref{table:summary_clone_fidelity} shows that in general, it seems that the newer QPU generations resulted in improved clone fidelities compared to older generation QPUs. 

The clearest trend observed across all of the fidelity measures shown in Table~\ref{table:summary_clone_fidelity} is that there is a steep decrease of clone fidelity as $M$ increases, and without the circuit optimization that removes the need for ancilla qubits in the computation, the mean fidelities are lower. At $M=10$, the mean clone fidelity converges to the noise limit of $0.5$ for all devices on which that circuit size was tested. Additionally, the mean clone fidelity is typically far from the theoretical optimal fidelity. These results are consistent with the gate-level error rates of these devices. 

\begin{figure*}[th!]
    \centering
    \includegraphics[width=0.13\textwidth]{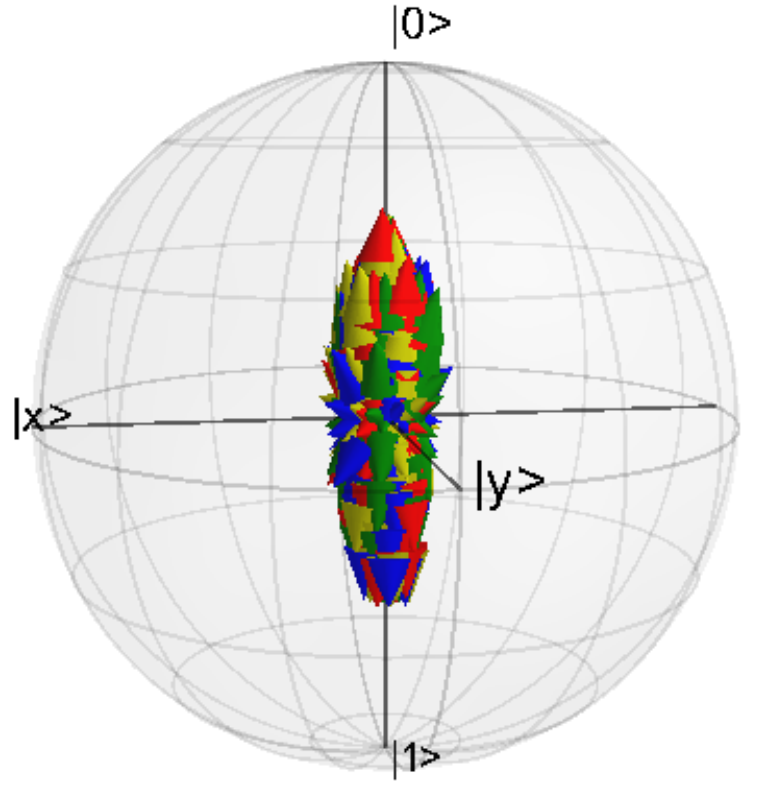}
    \includegraphics[width=0.13\textwidth]{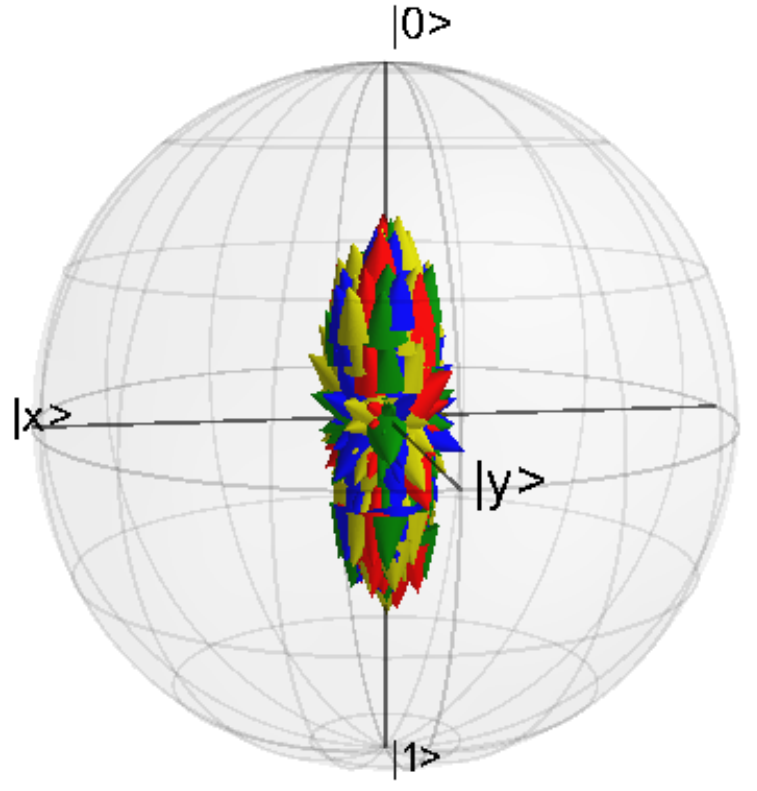}
    \includegraphics[width=0.13\textwidth]{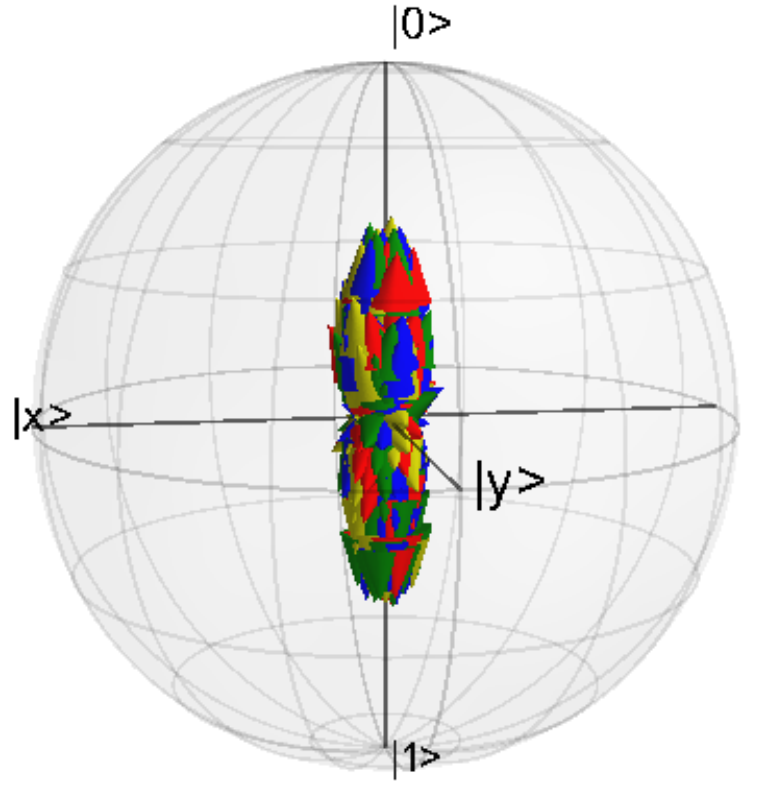}
    \includegraphics[width=0.13\textwidth]{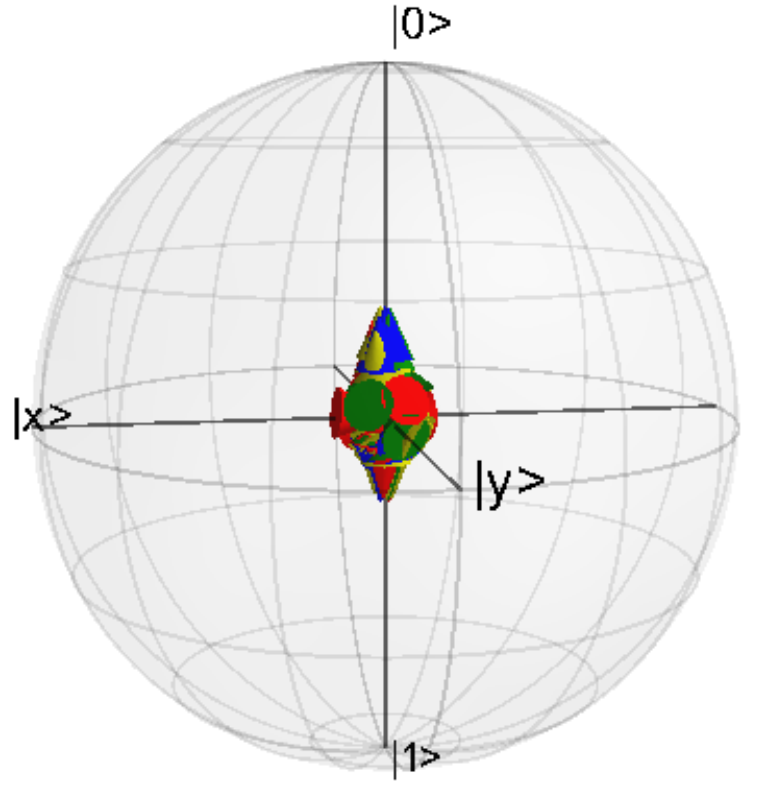}
    \includegraphics[width=0.13\textwidth]{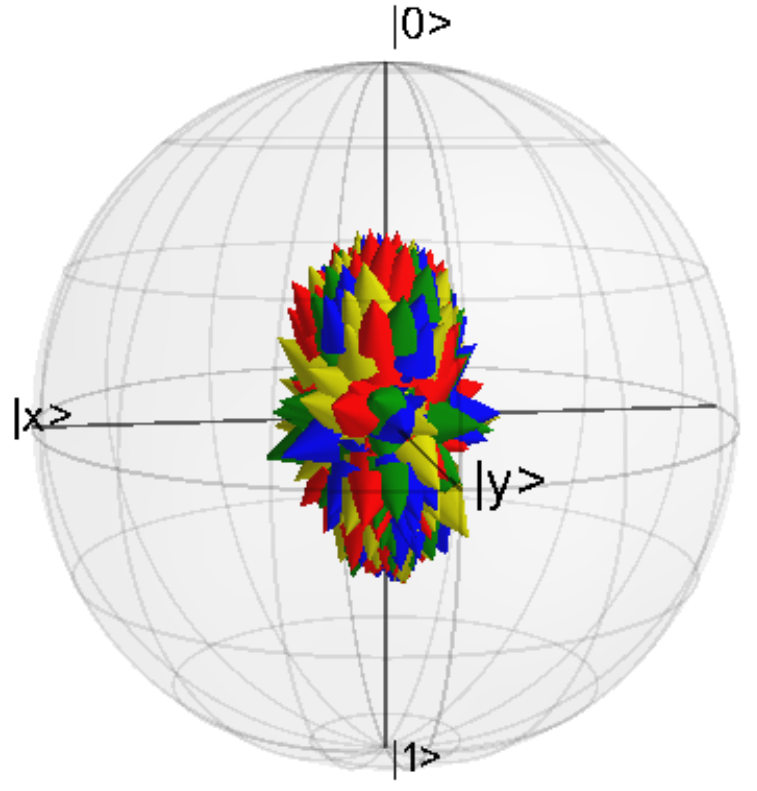}
    \includegraphics[width=0.13\textwidth]{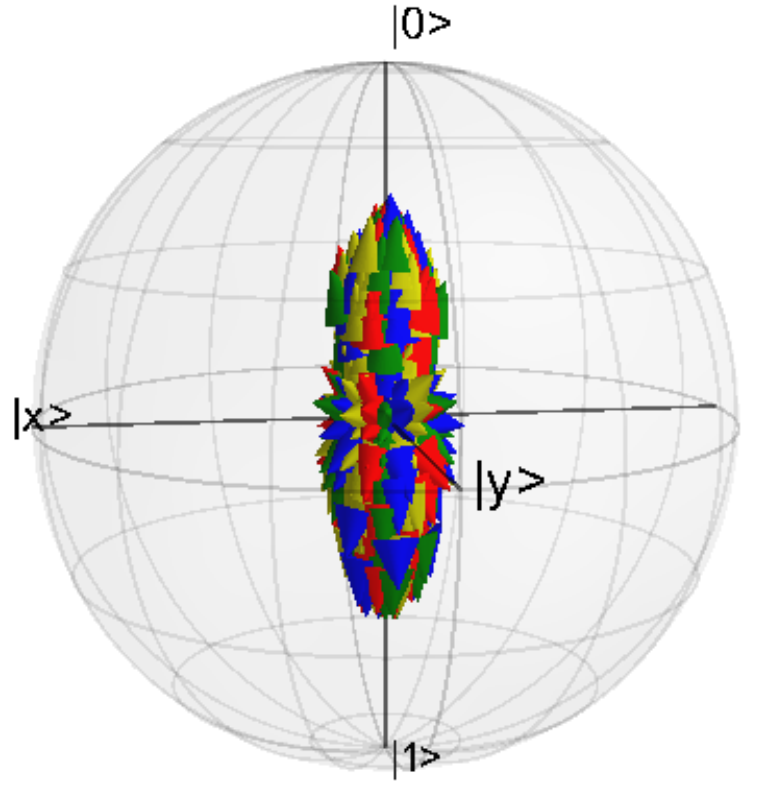}
    \includegraphics[width=0.13\textwidth]{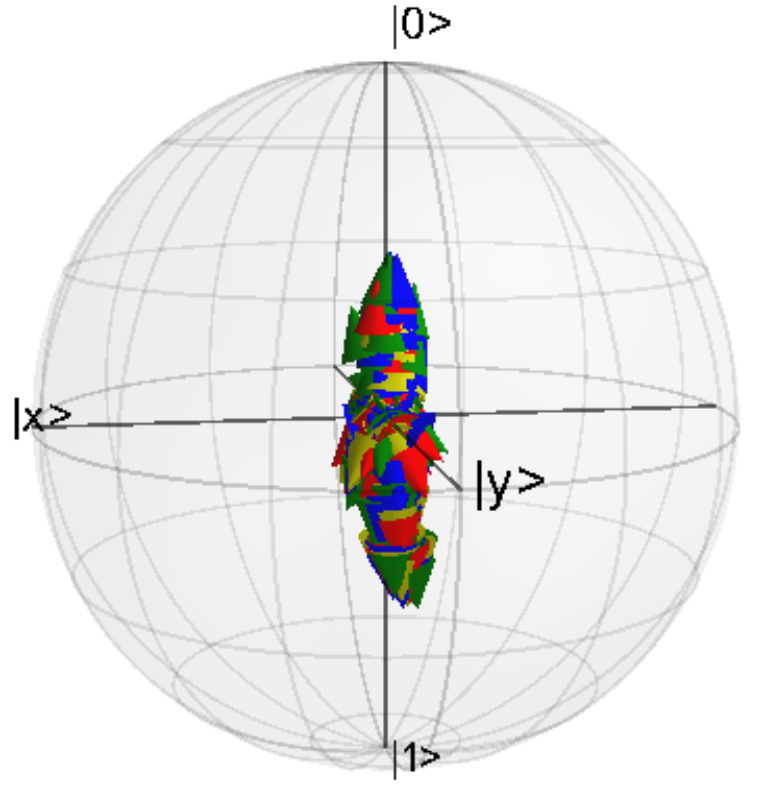}
    \includegraphics[width=0.13\textwidth]{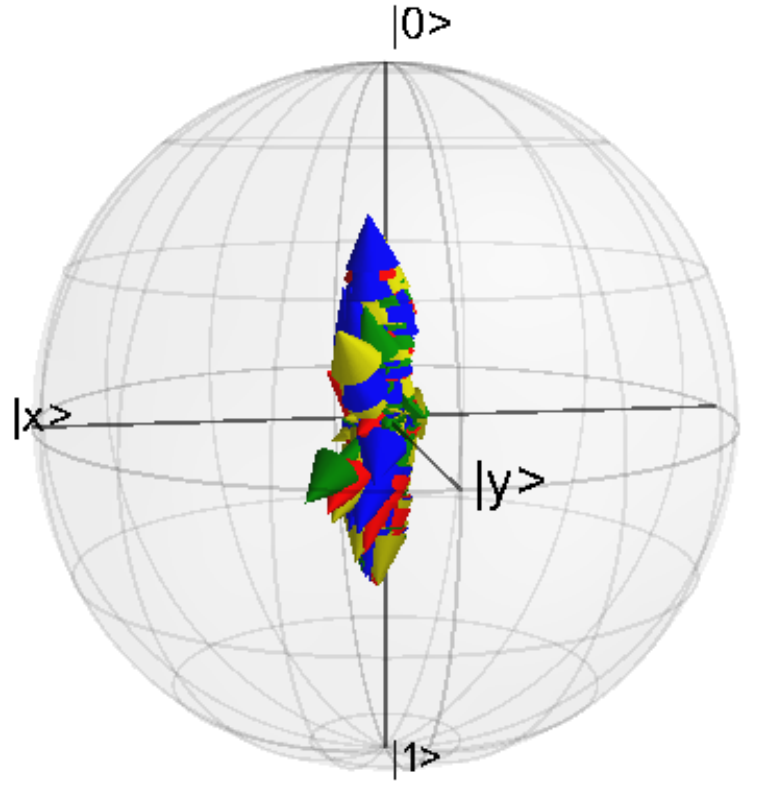}
    \includegraphics[width=0.13\textwidth]{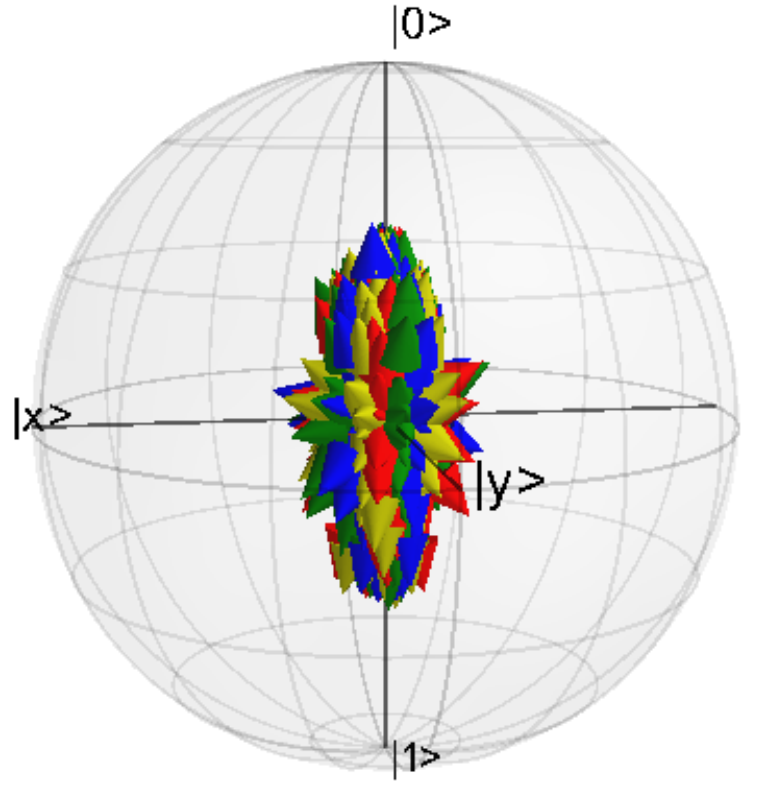}
    \includegraphics[width=0.13\textwidth]{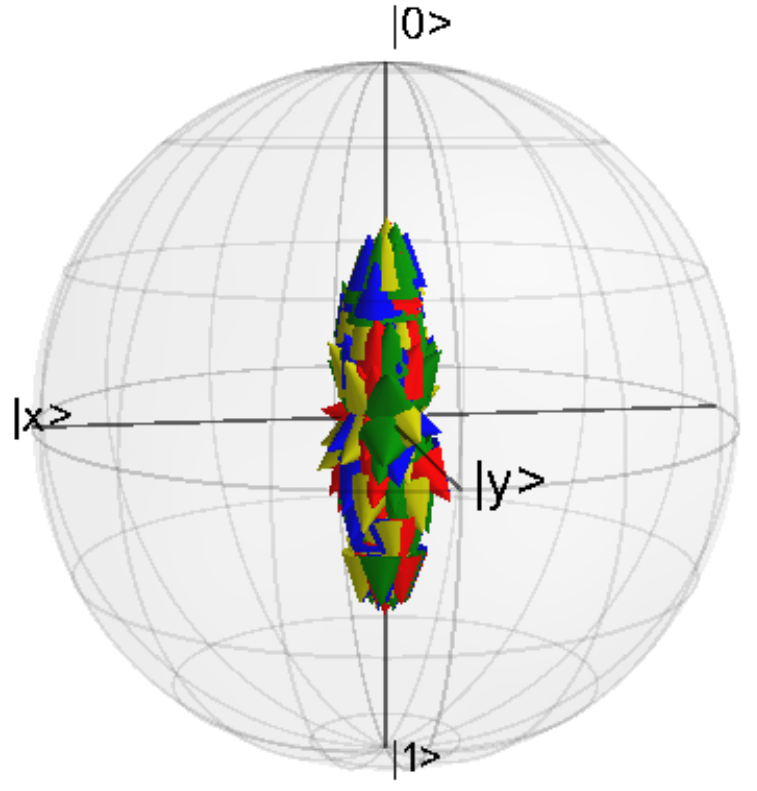}
    \includegraphics[width=0.13\textwidth]{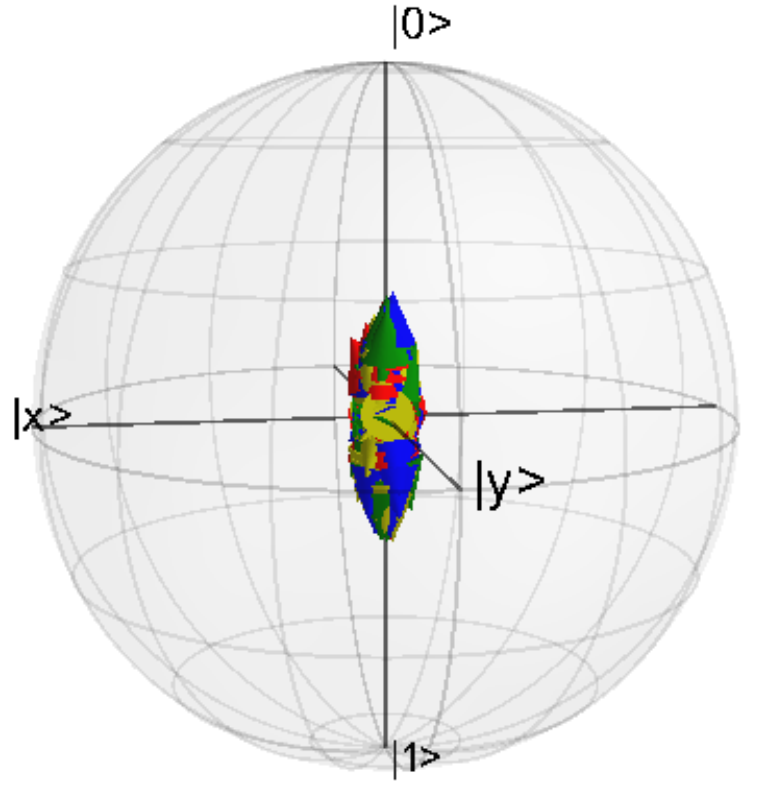}
    \includegraphics[width=0.13\textwidth]{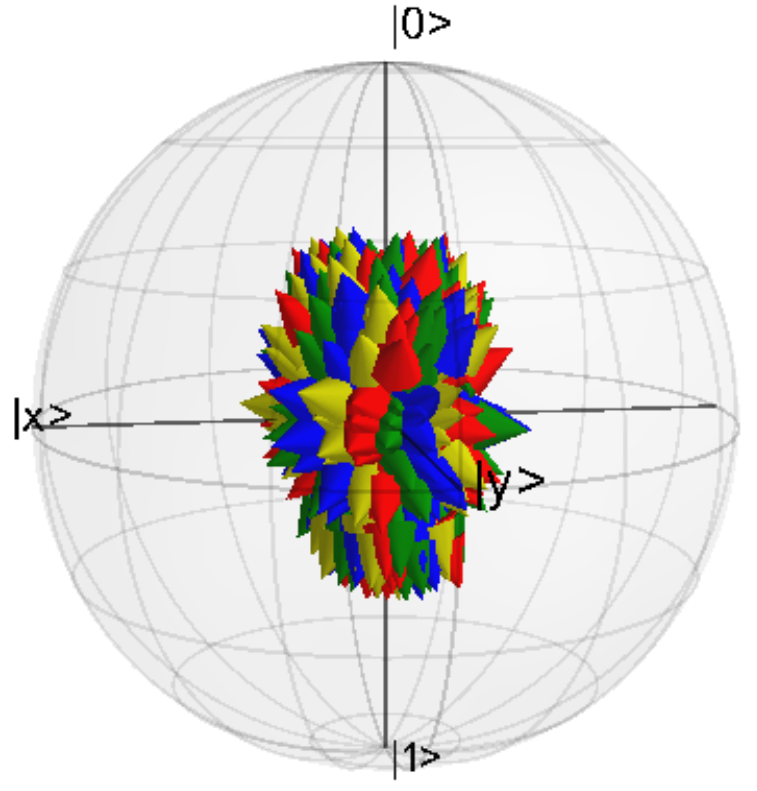}
    \includegraphics[width=0.13\textwidth]{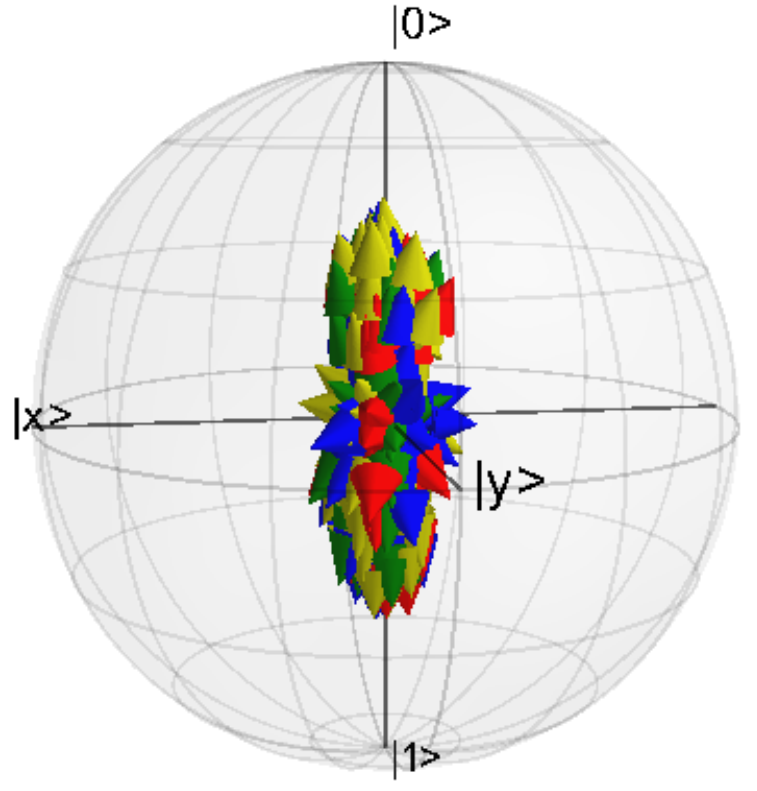}
    \includegraphics[width=0.13\textwidth]{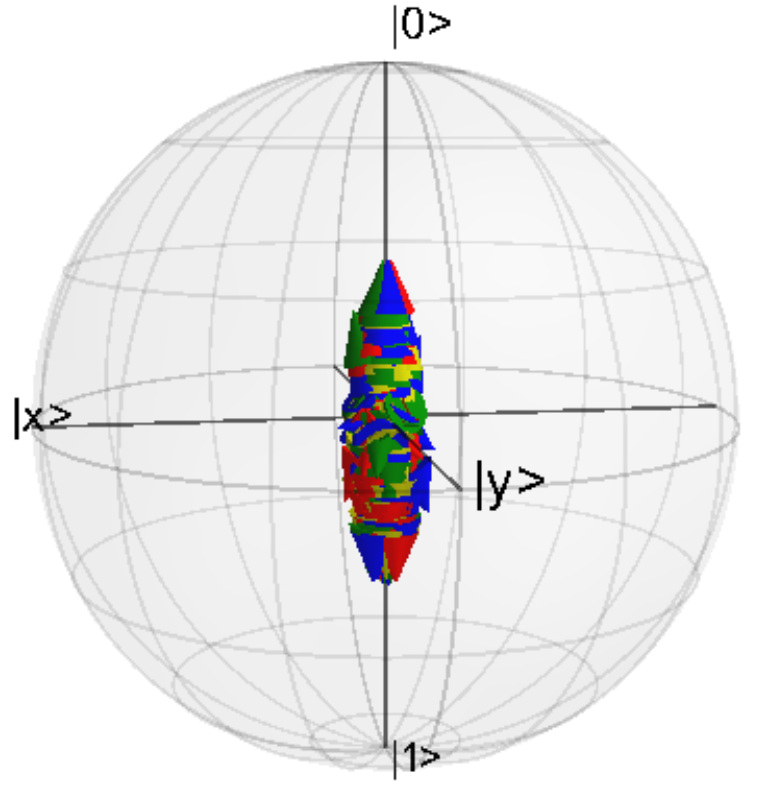}
    \includegraphics[width=0.13\textwidth]{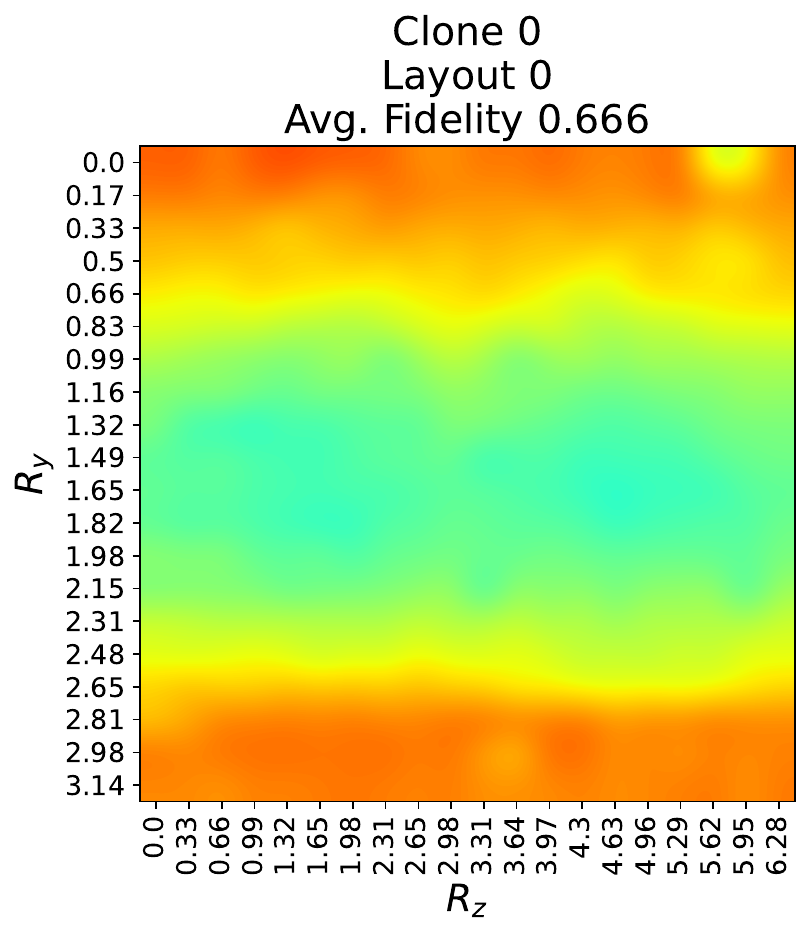}
    \includegraphics[width=0.13\textwidth]{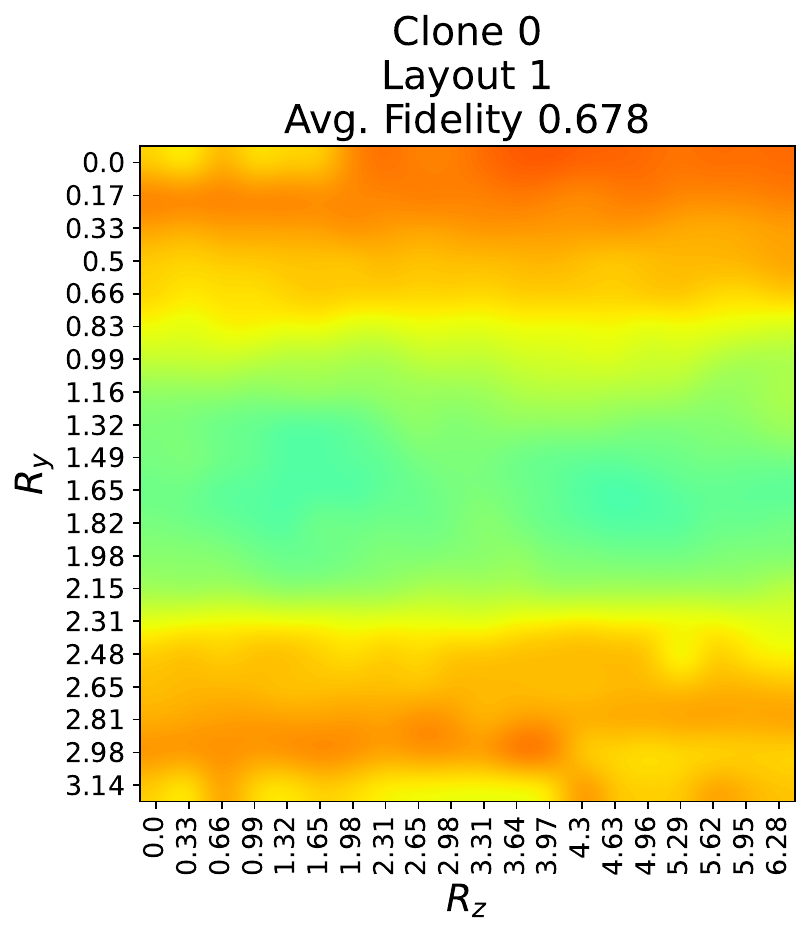}
    \includegraphics[width=0.13\textwidth]{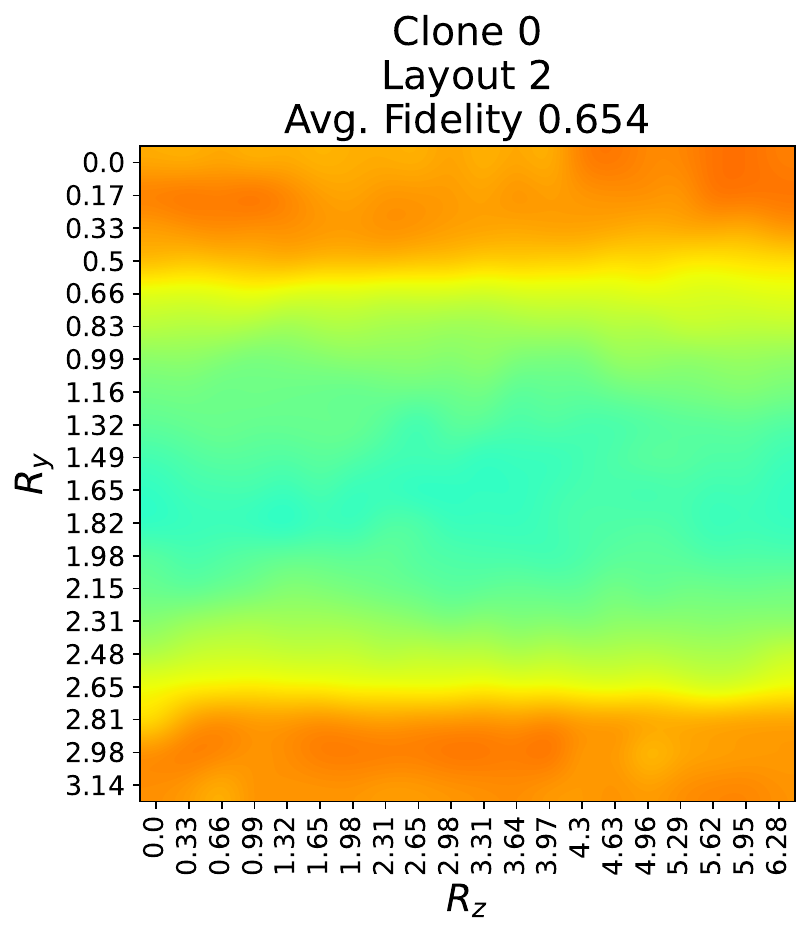}
    \includegraphics[width=0.13\textwidth]{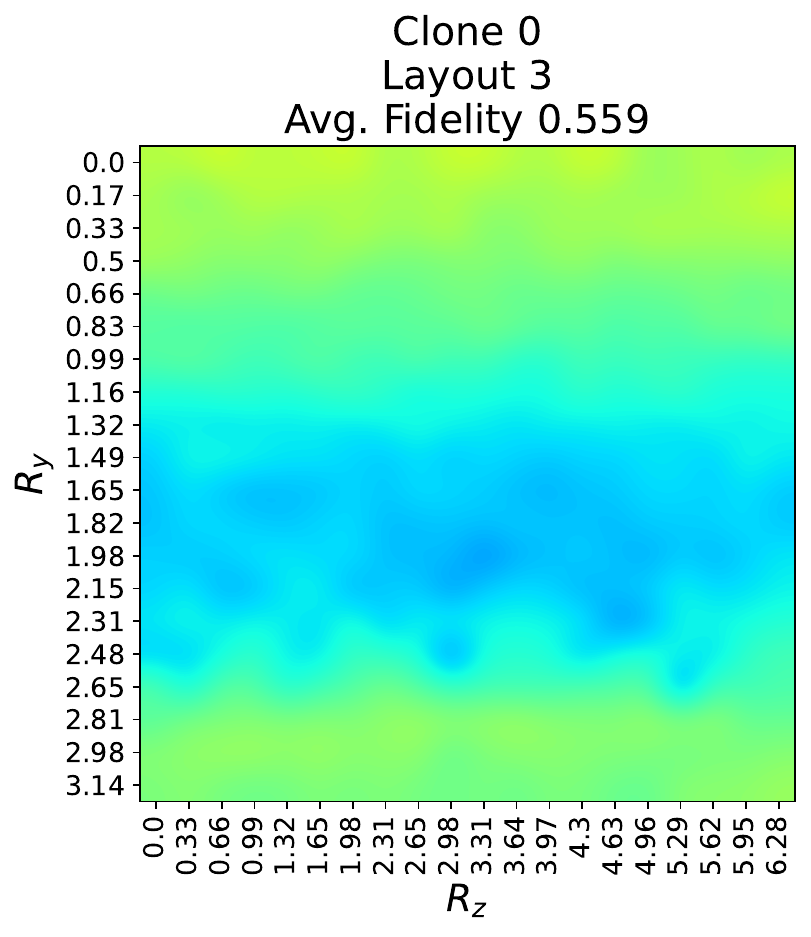}
    \includegraphics[width=0.13\textwidth]{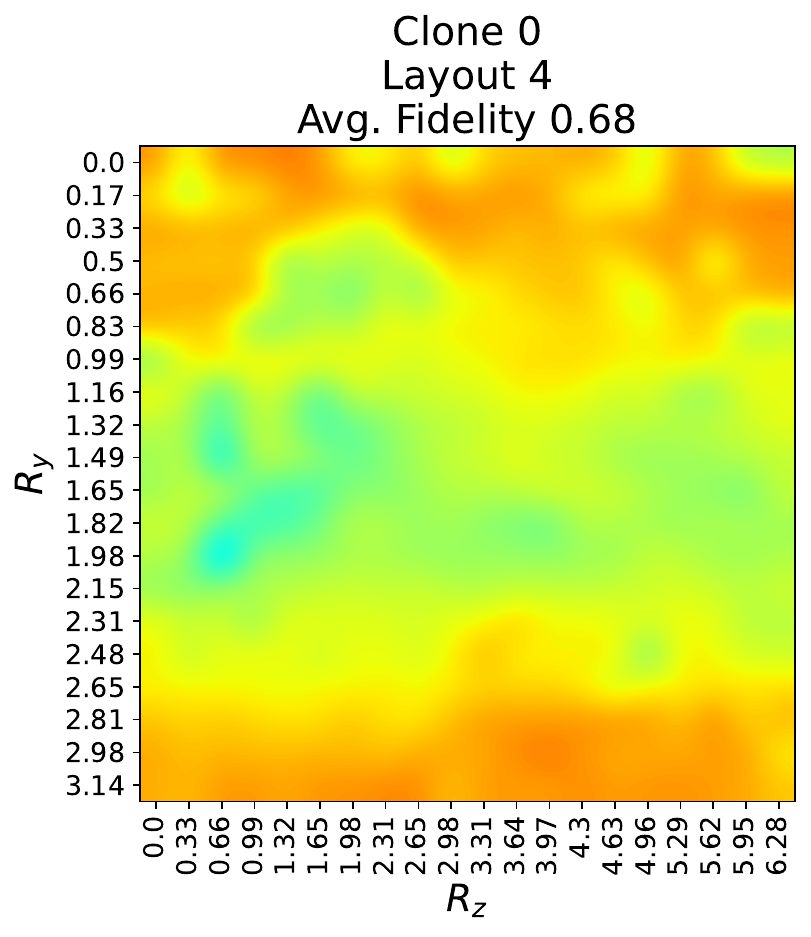}
    \includegraphics[width=0.13\textwidth]{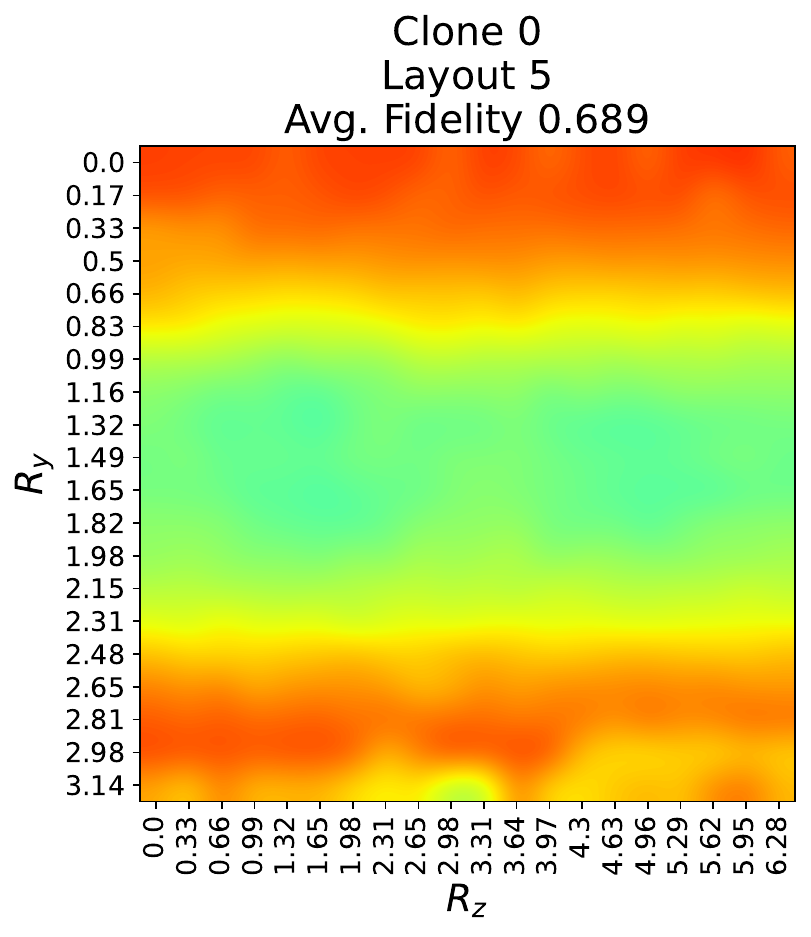}
    \includegraphics[width=0.13\textwidth]{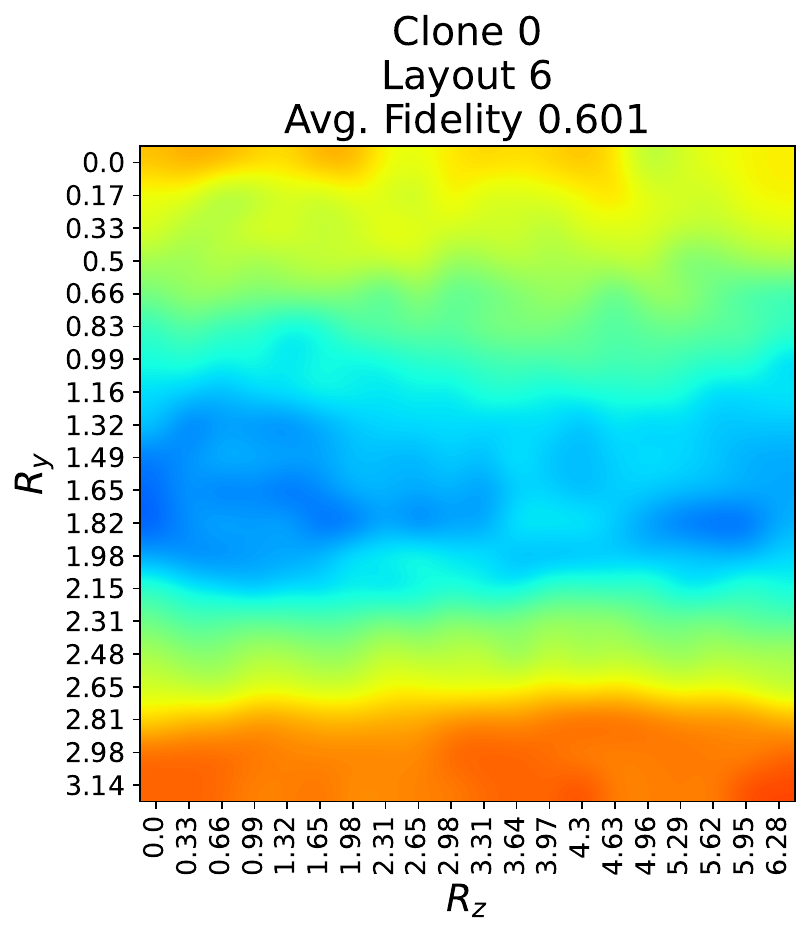}
    \includegraphics[width=0.13\textwidth]{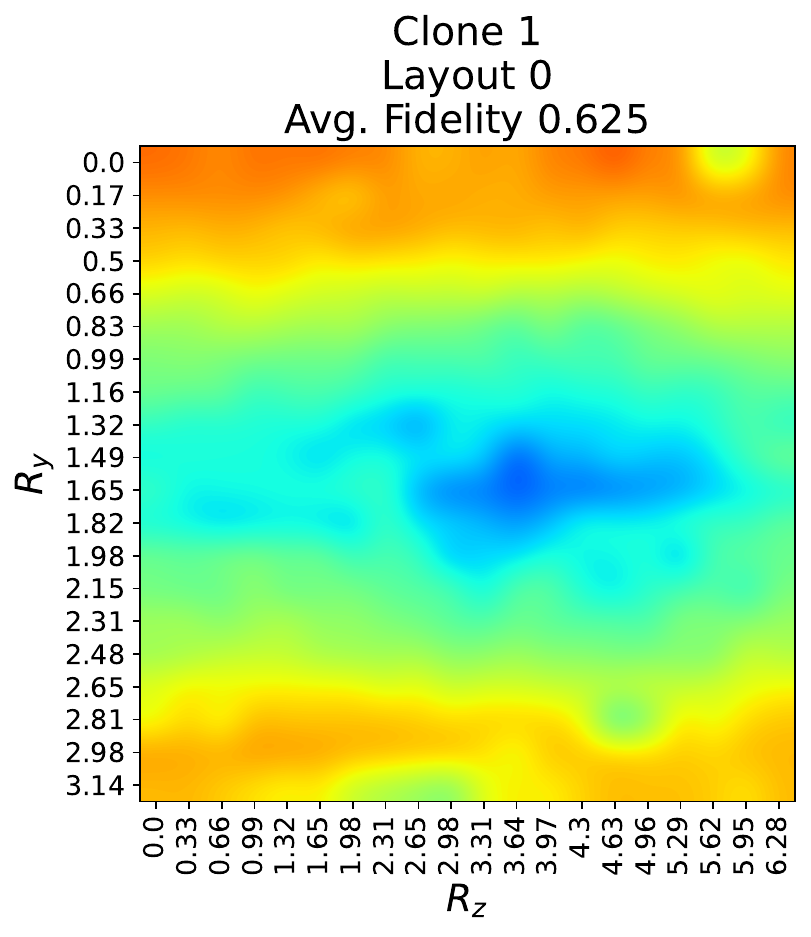}
    \includegraphics[width=0.13\textwidth]{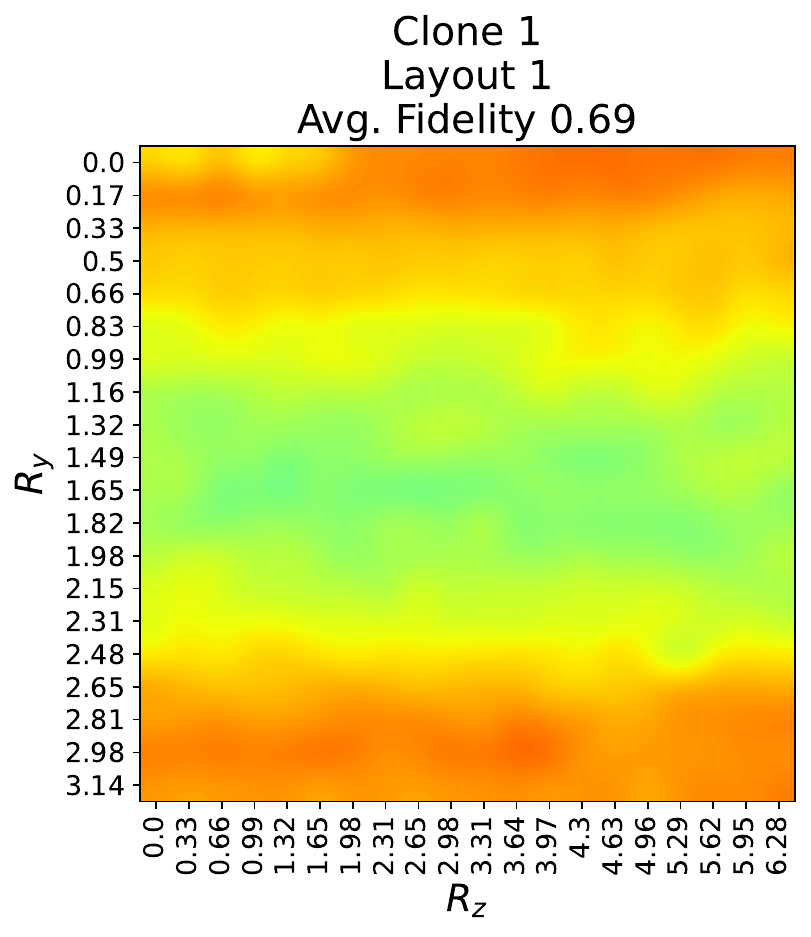}
    \includegraphics[width=0.13\textwidth]{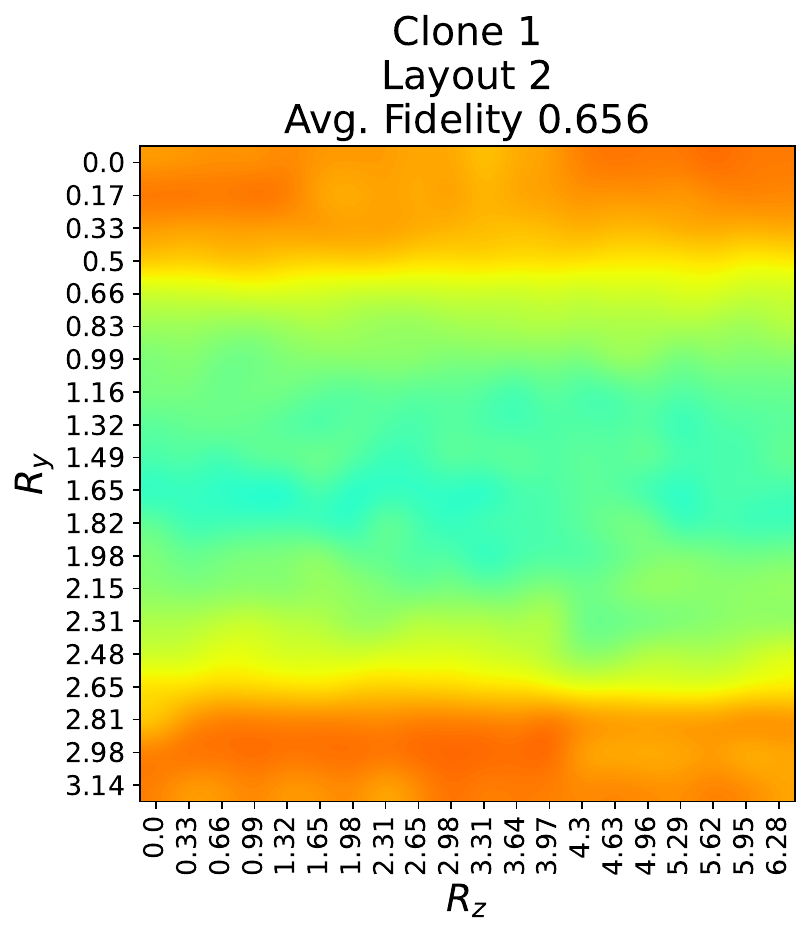}
    \includegraphics[width=0.13\textwidth]{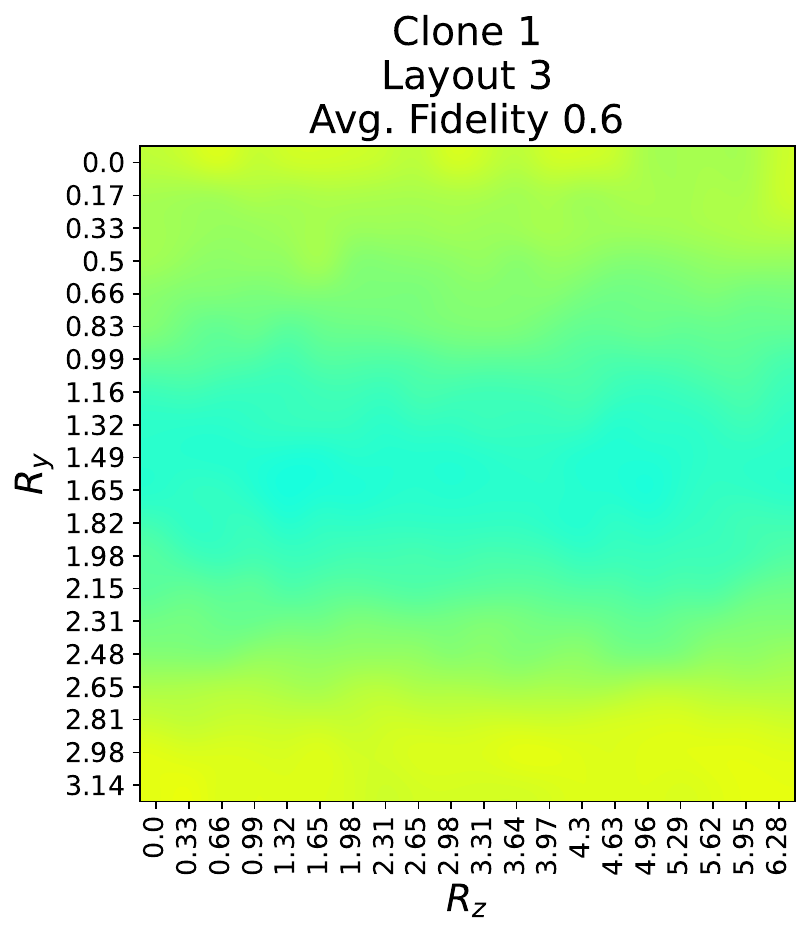}
    \includegraphics[width=0.13\textwidth]{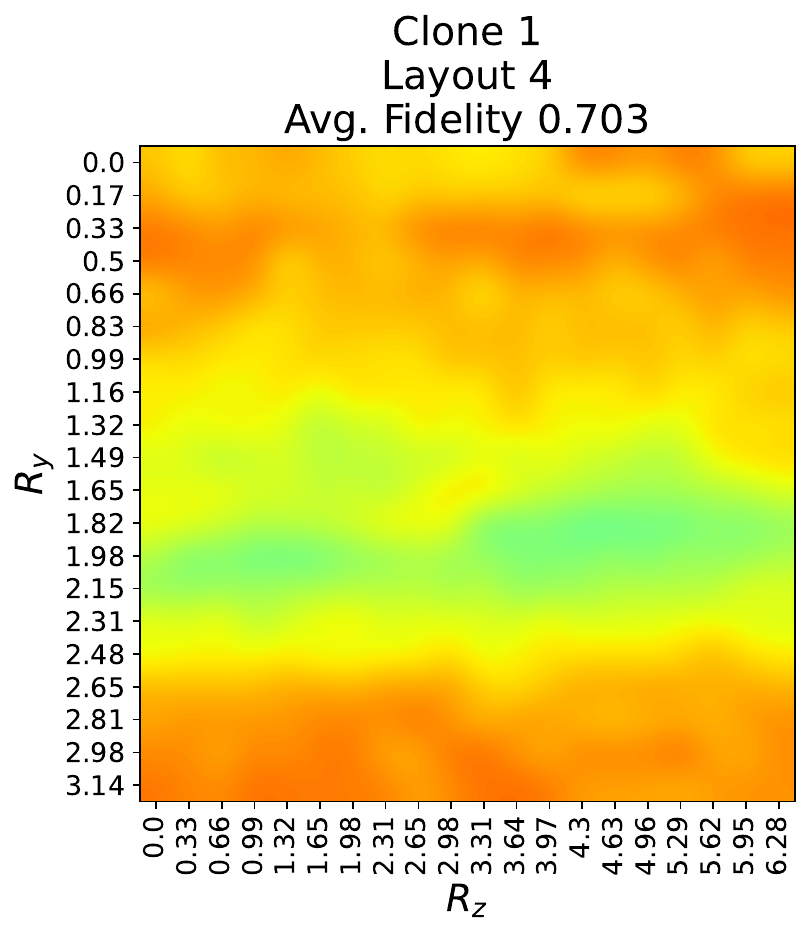}
    \includegraphics[width=0.13\textwidth]{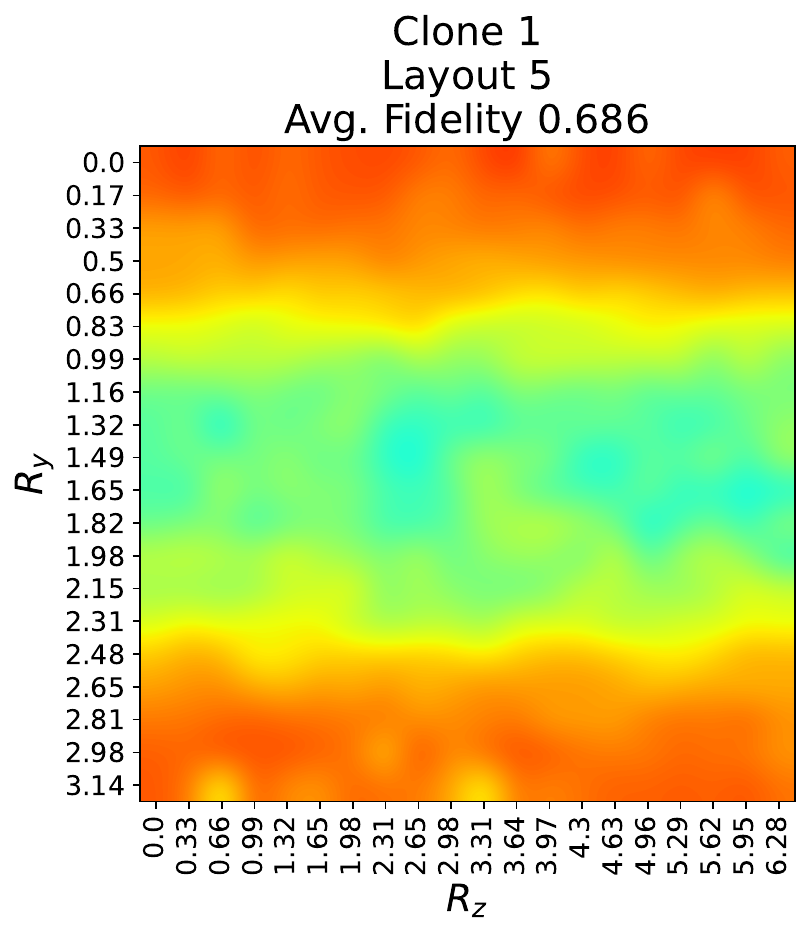}
    \includegraphics[width=0.13\textwidth]{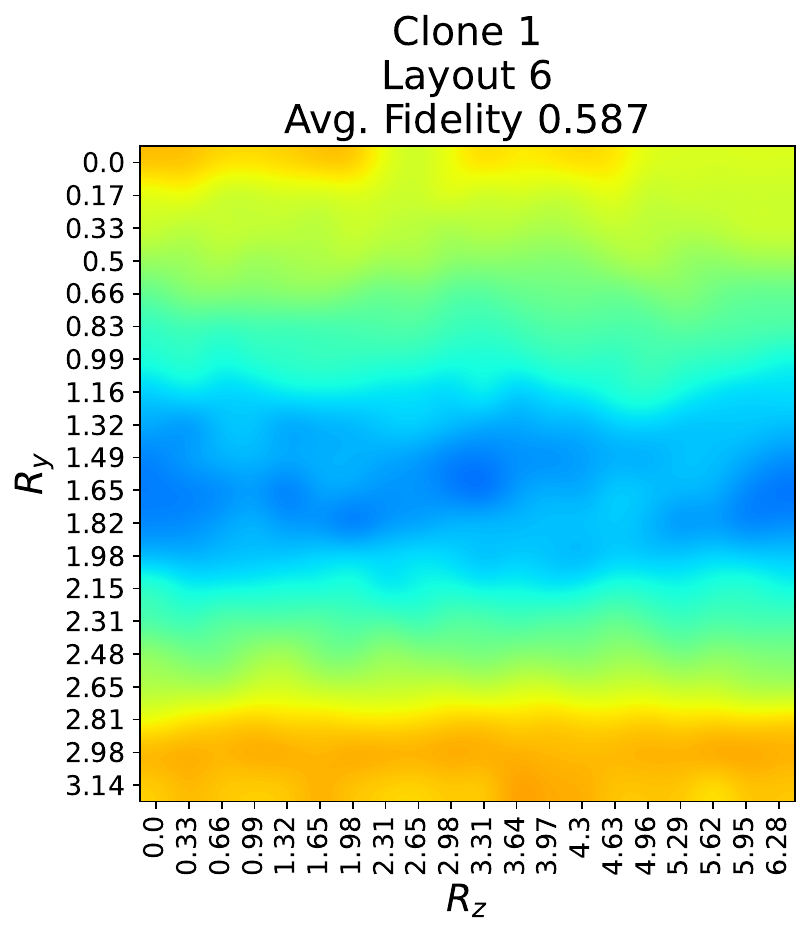}
    \includegraphics[width=0.43\textwidth]{figures/colorbar.pdf}\\
    \caption{Single qubit clone fidelity heatmaps for $M=2$ quantum telecloning circuits with ancilla qubits, executed without dynamical decoupling pulses. Each column corresponds to the $7$ different compiled hardware layouts. Bottom two rows show the fidelities of the $2$ single qubit clones. Top two rows show Bloch sphere vector representations of the single qubit state tomography computed density matrices. Data from \texttt{ibmq\_kolkata}.  }
    \label{fig:fidelity_heatmaps_M2_ibmq_kolkata_with_ancilla}
\end{figure*}

\begin{figure*}[th!]
    \centering
    \includegraphics[width=0.13\textwidth]{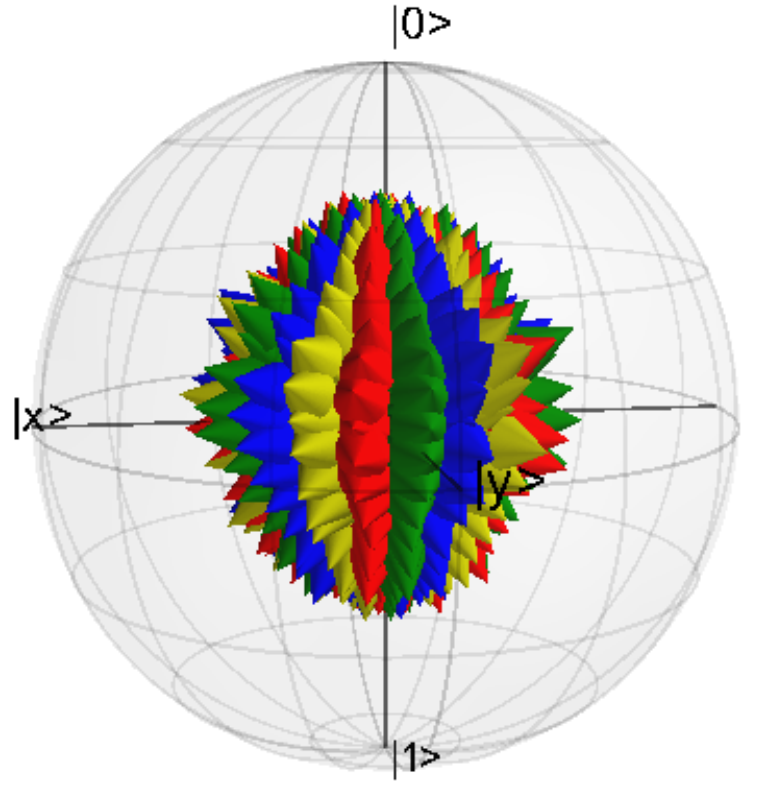}
    \includegraphics[width=0.13\textwidth]{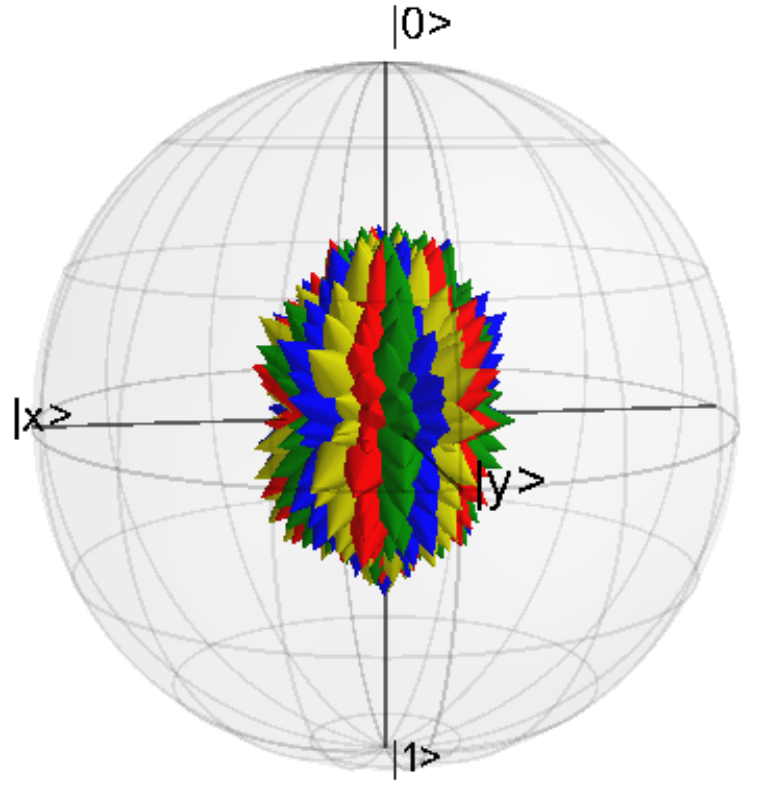}
    \includegraphics[width=0.13\textwidth]{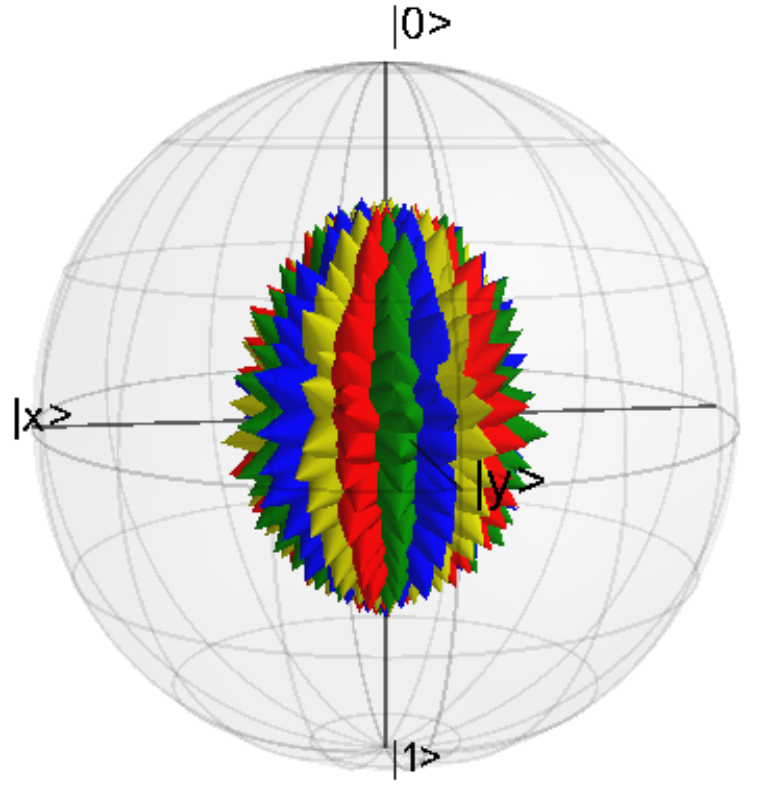}
    \includegraphics[width=0.13\textwidth]{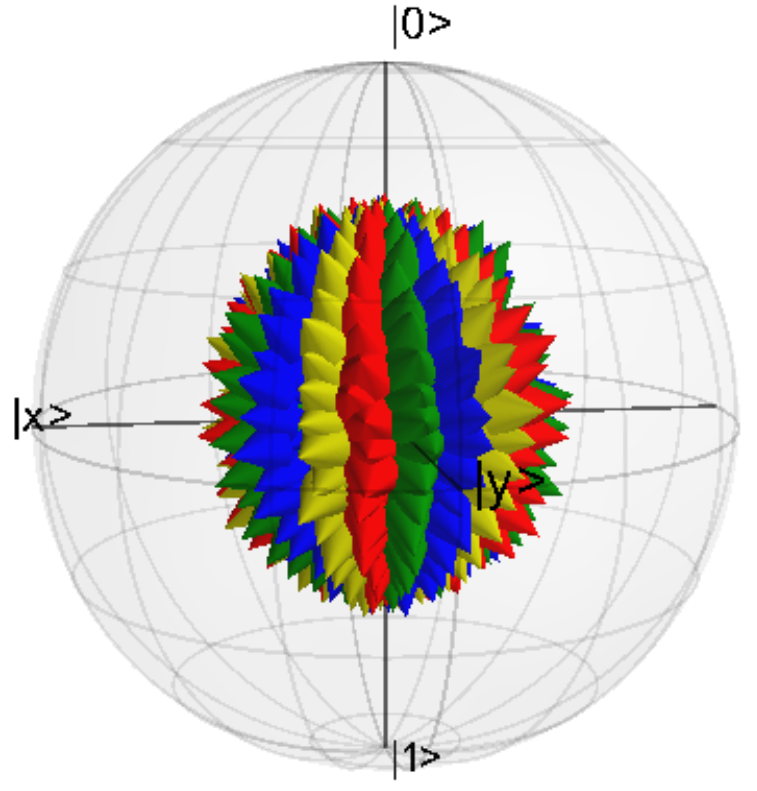}
    \includegraphics[width=0.13\textwidth]{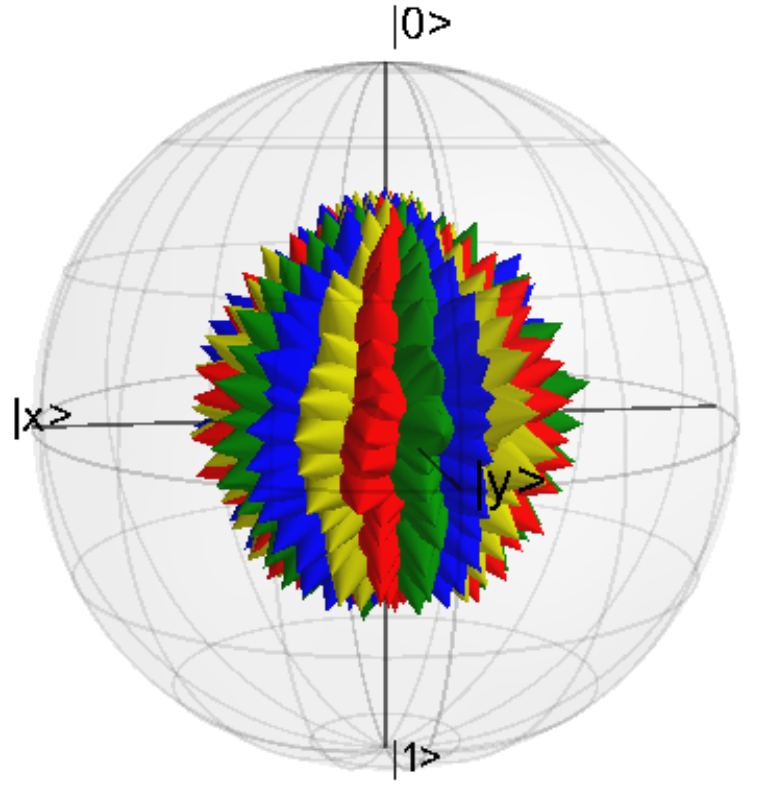}
    \includegraphics[width=0.13\textwidth]{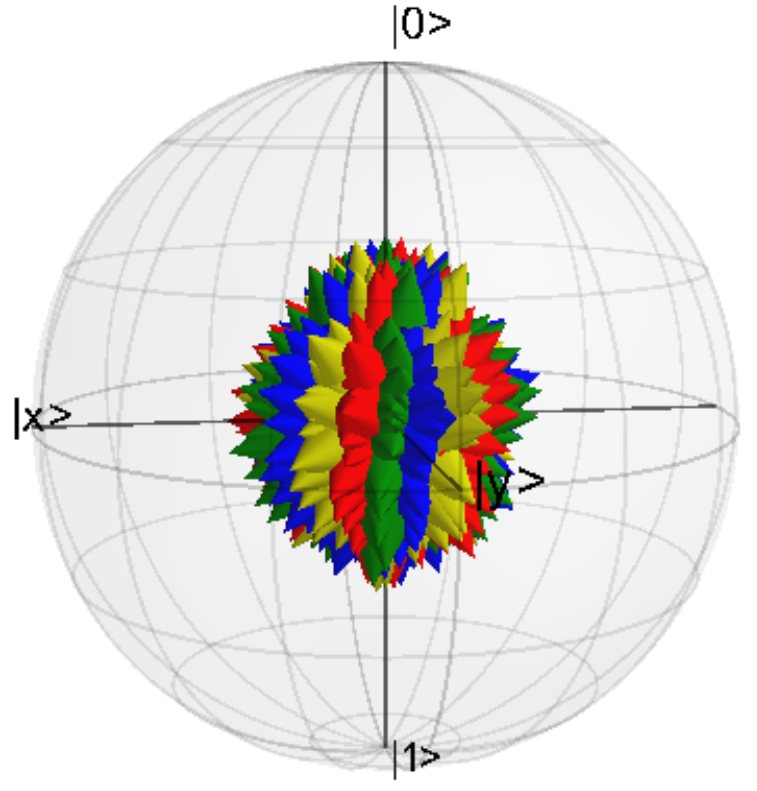}
    \includegraphics[width=0.13\textwidth]{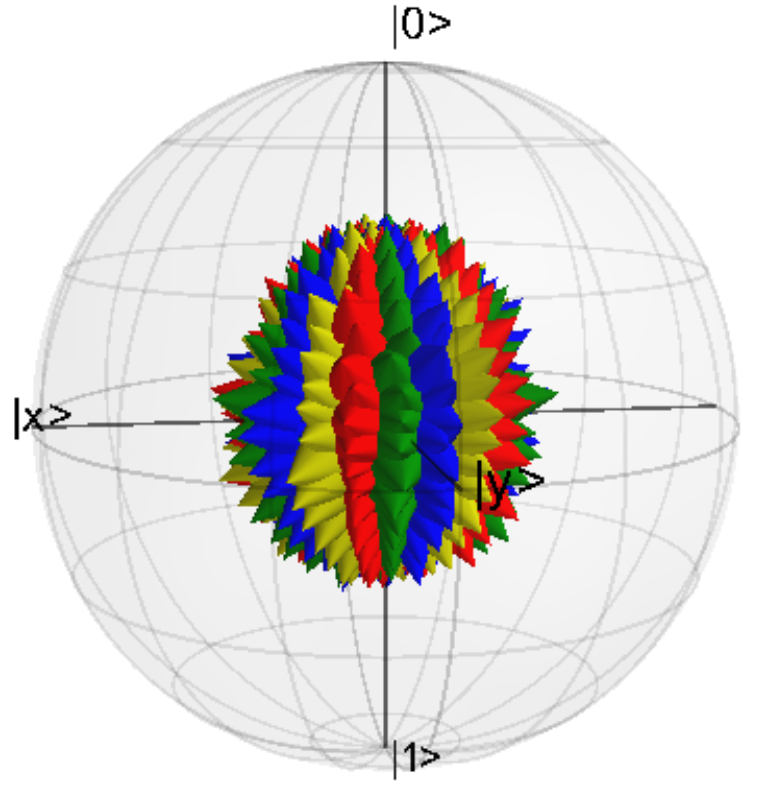}
    \includegraphics[width=0.13\textwidth]{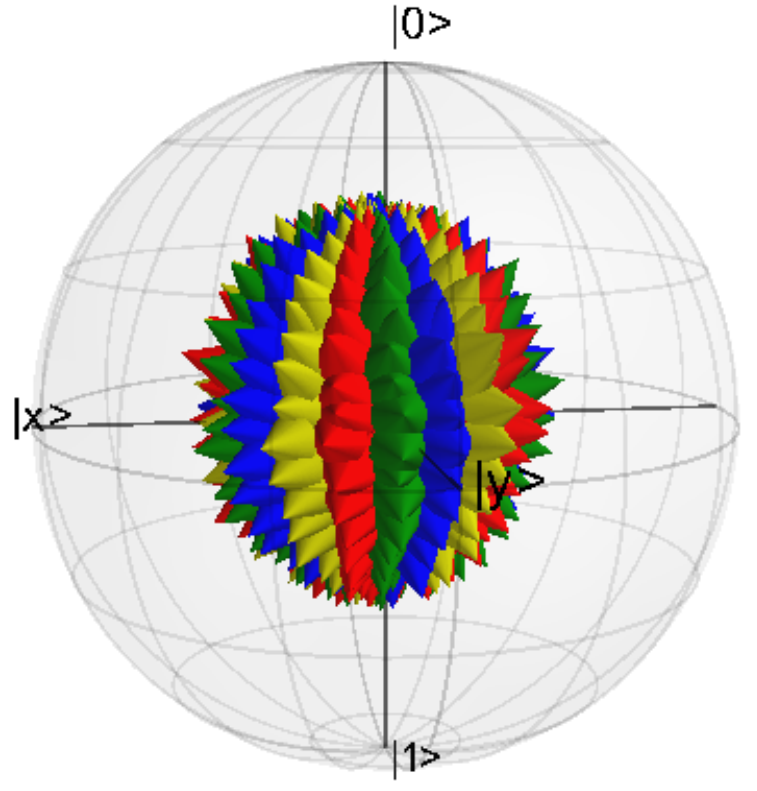}
    \includegraphics[width=0.13\textwidth]{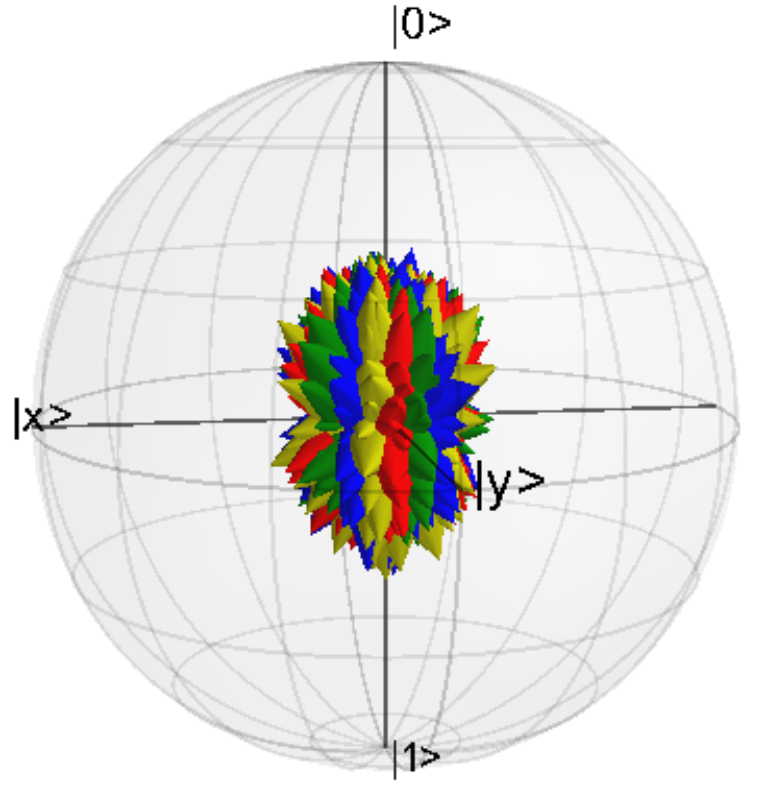}
    \includegraphics[width=0.13\textwidth]{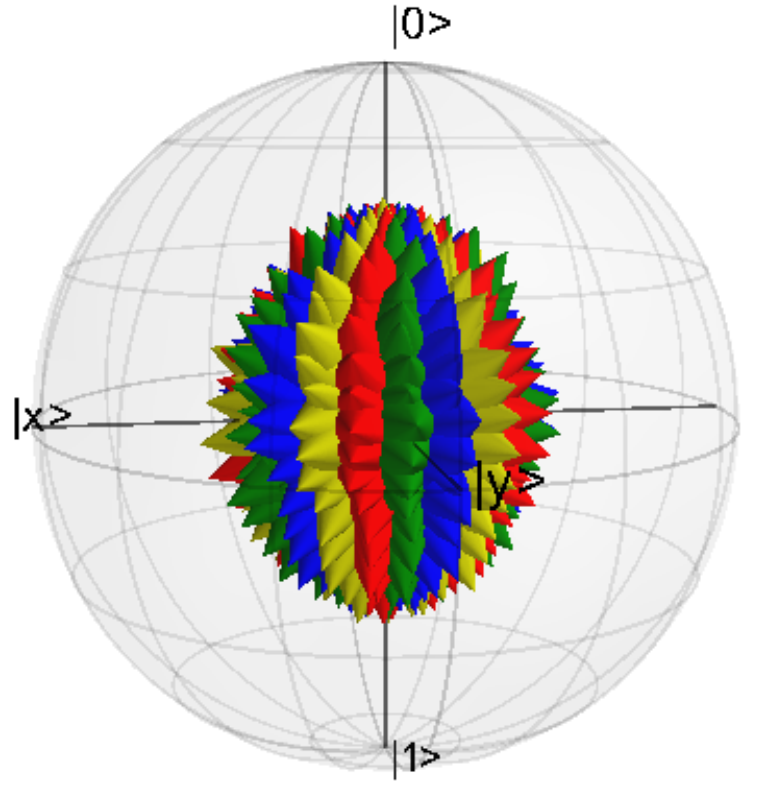}
    \includegraphics[width=0.13\textwidth]{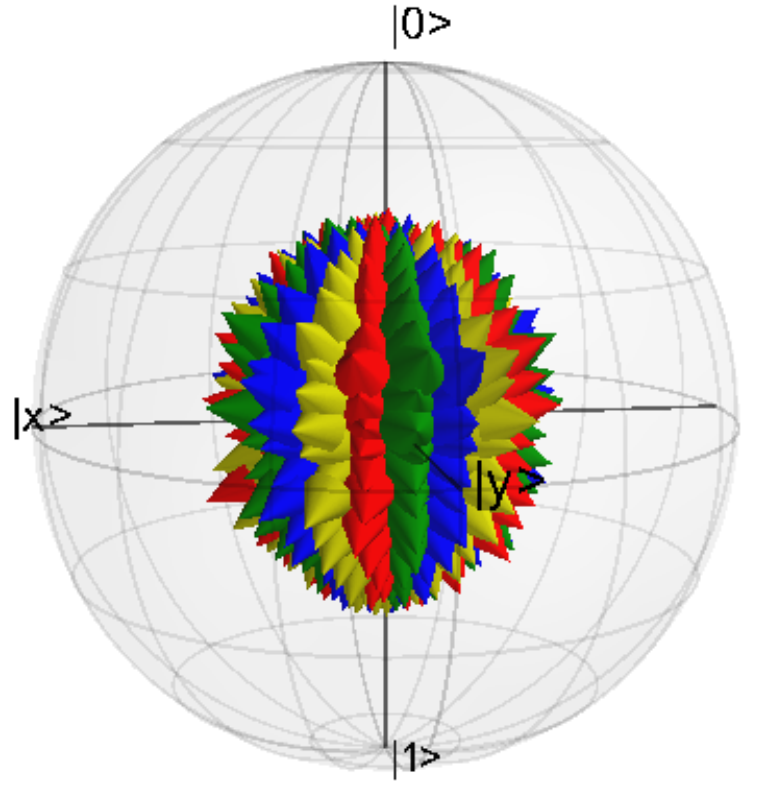}
    \includegraphics[width=0.13\textwidth]{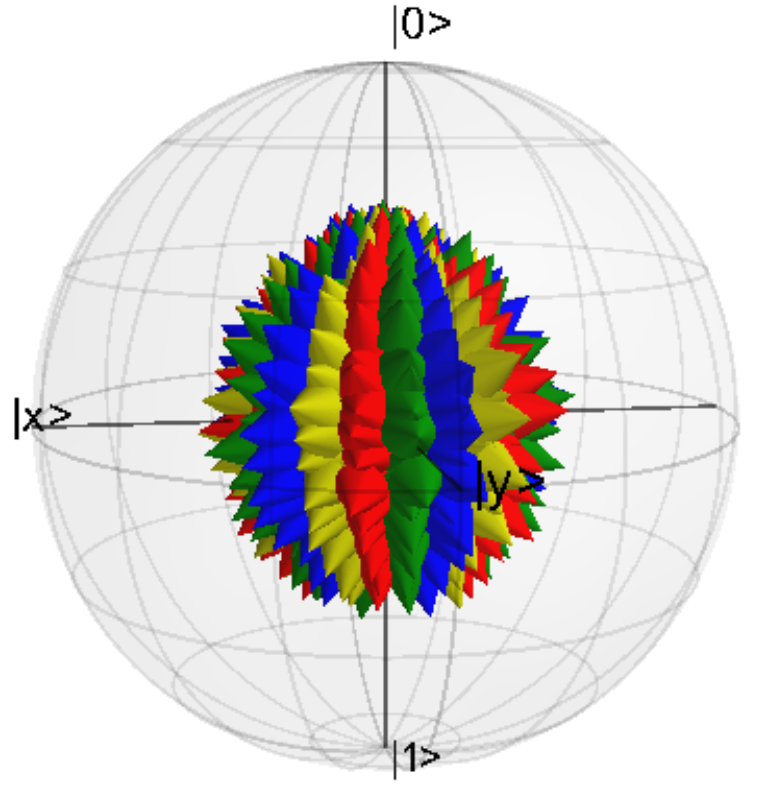}
    \includegraphics[width=0.13\textwidth]{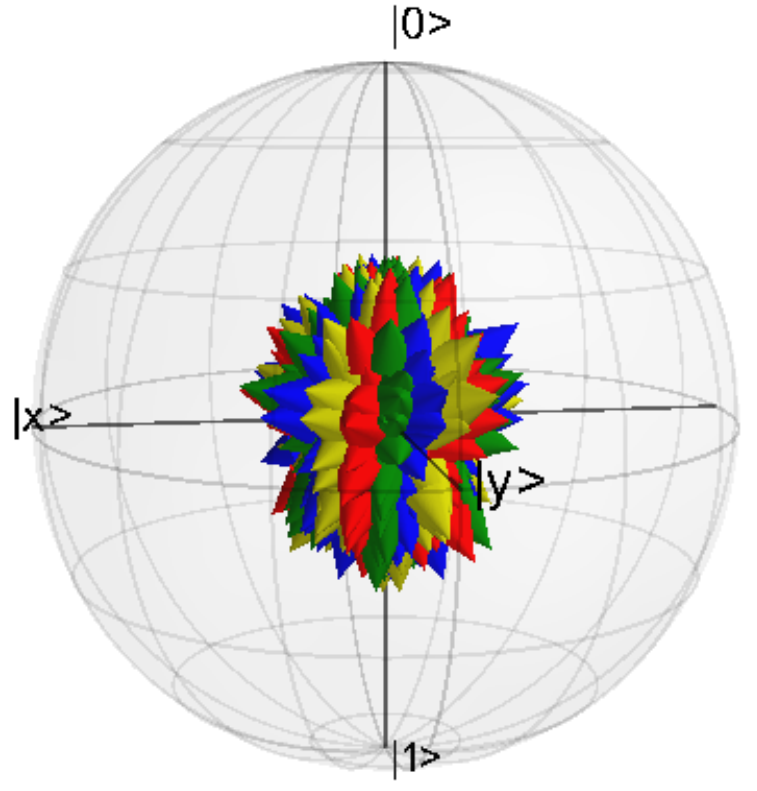}
    \includegraphics[width=0.13\textwidth]{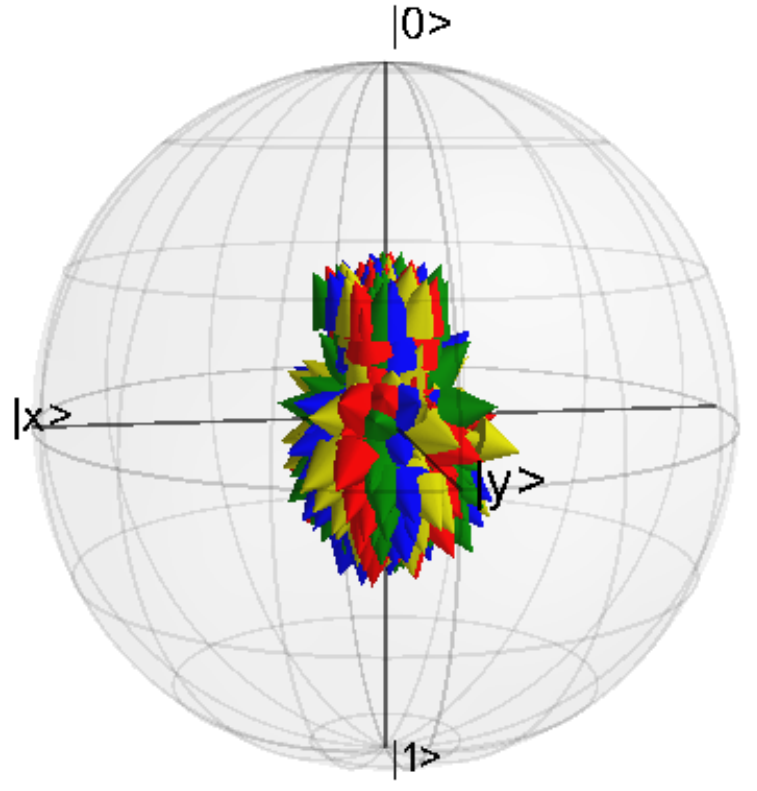}
    \includegraphics[width=0.13\textwidth]{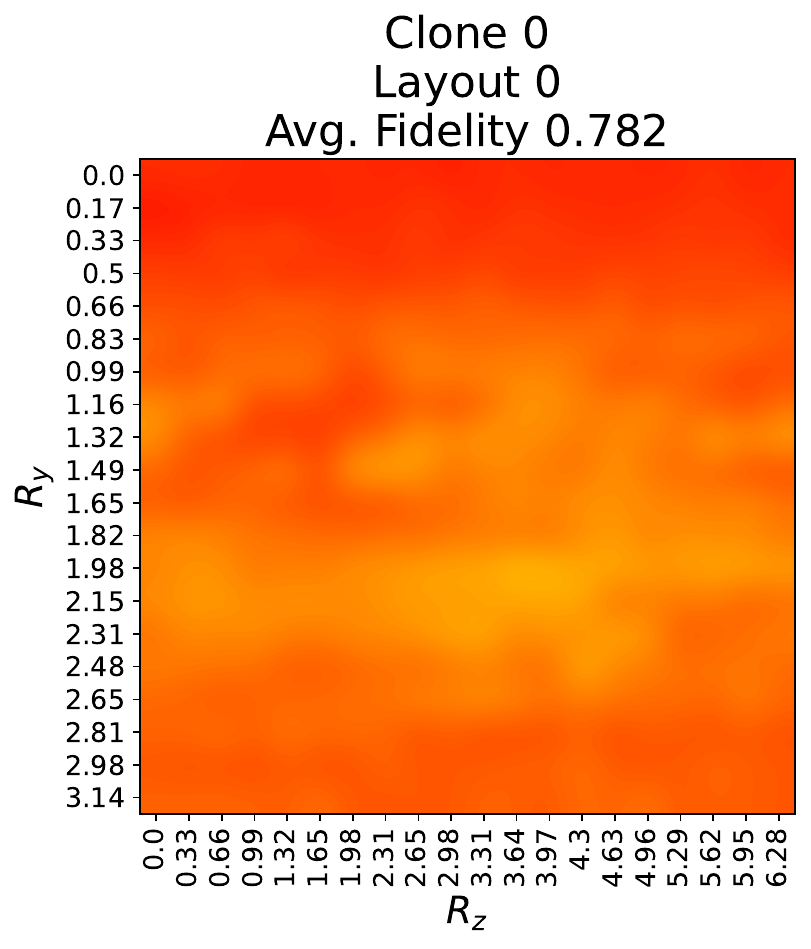}
    \includegraphics[width=0.13\textwidth]{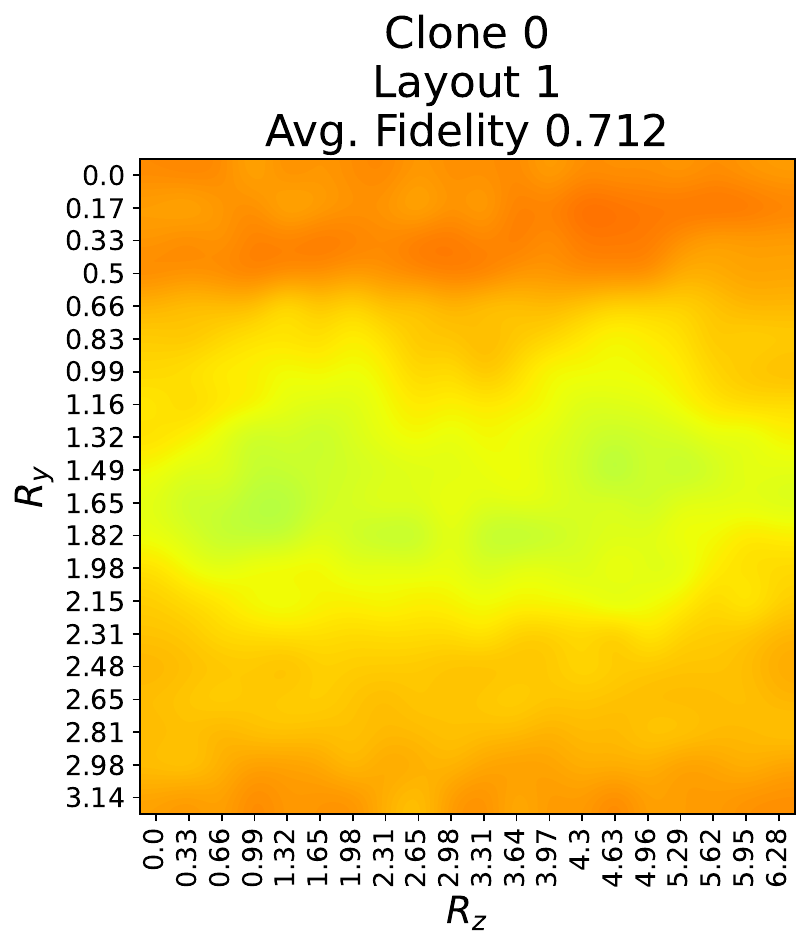}
    \includegraphics[width=0.13\textwidth]{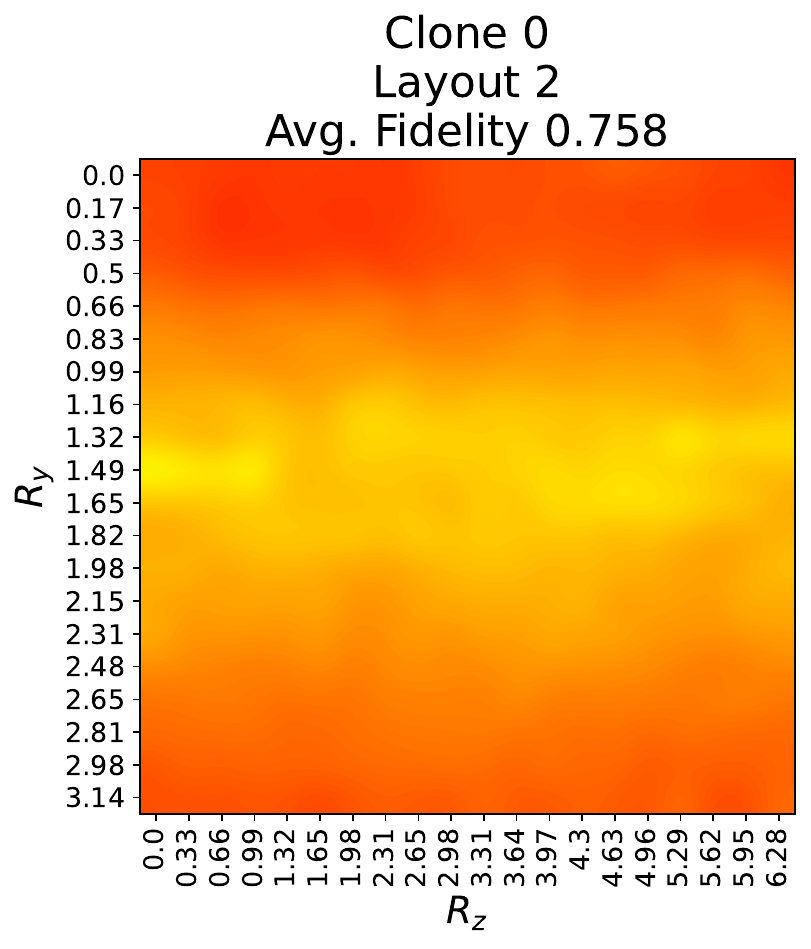}
    \includegraphics[width=0.13\textwidth]{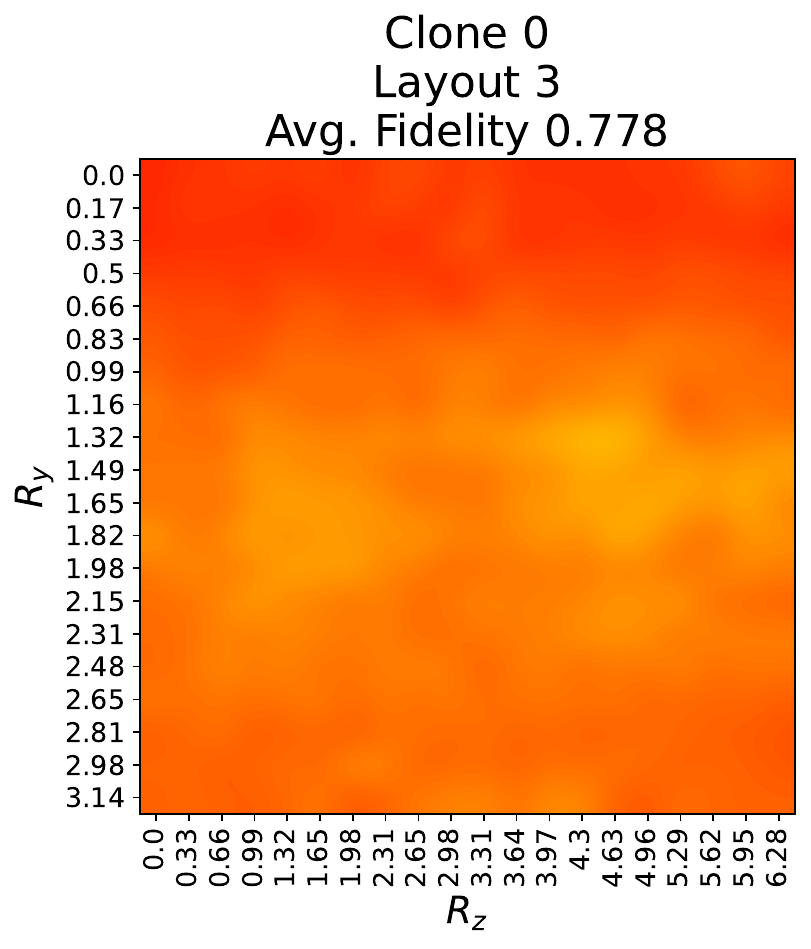}
    \includegraphics[width=0.13\textwidth]{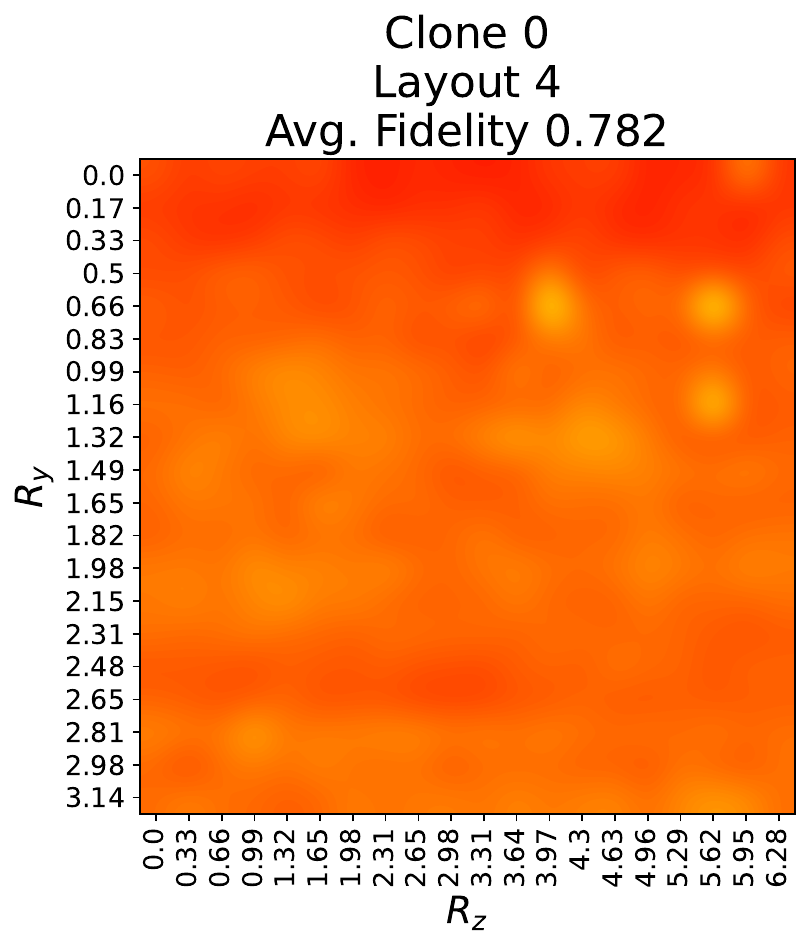}
    \includegraphics[width=0.13\textwidth]{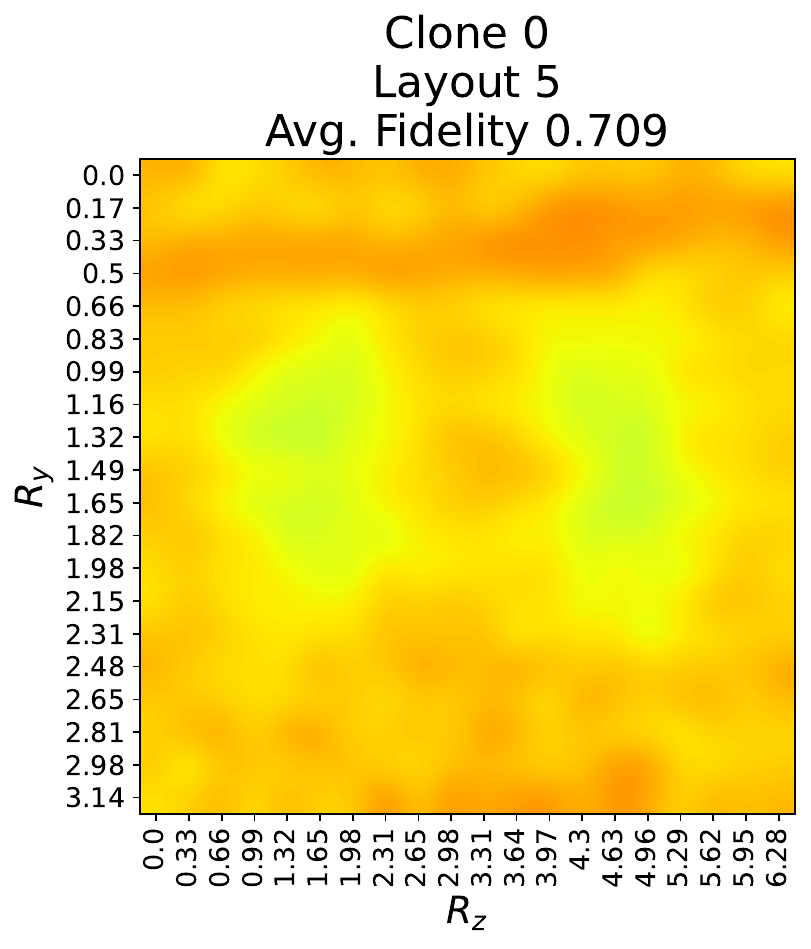}
    \includegraphics[width=0.13\textwidth]{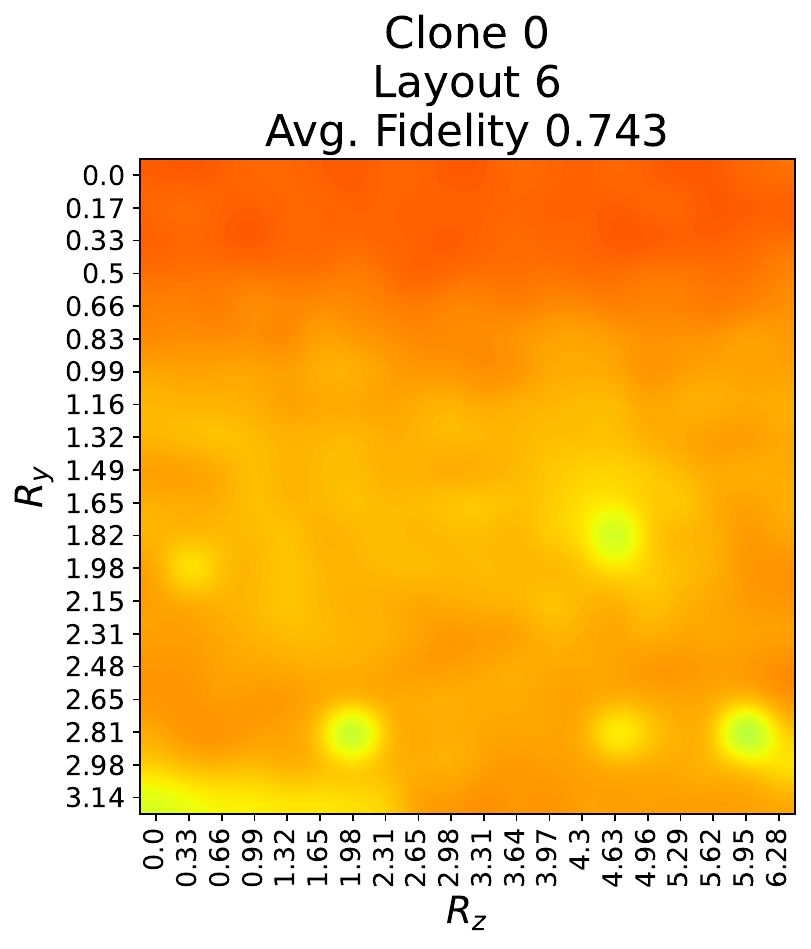}
    \includegraphics[width=0.13\textwidth]{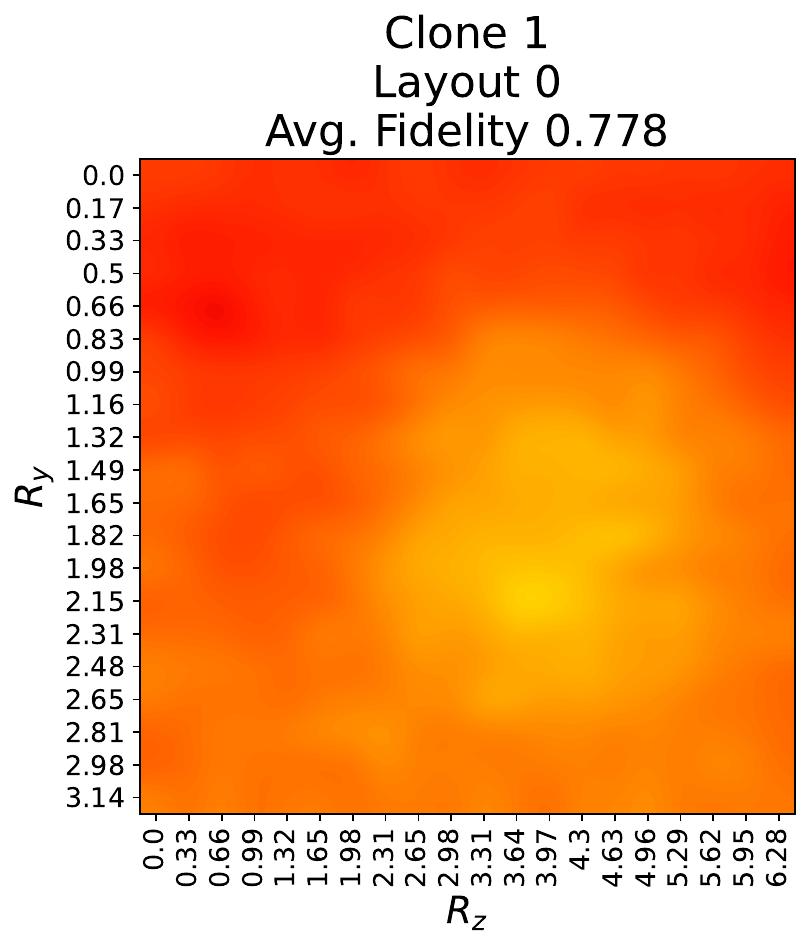}
    \includegraphics[width=0.13\textwidth]{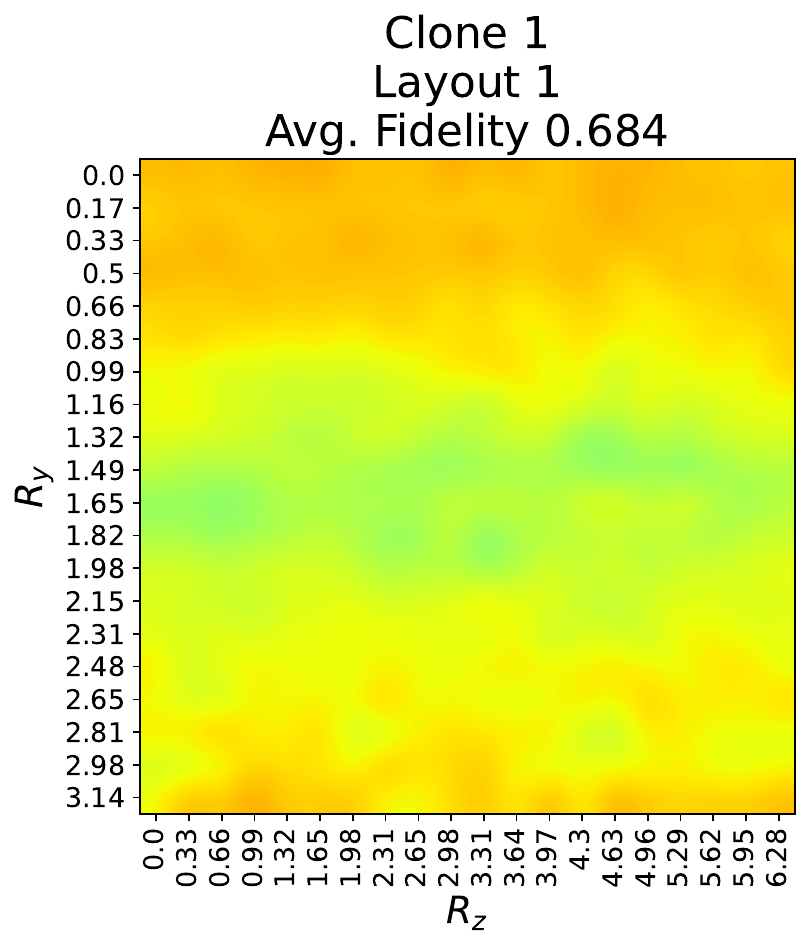}
    \includegraphics[width=0.13\textwidth]{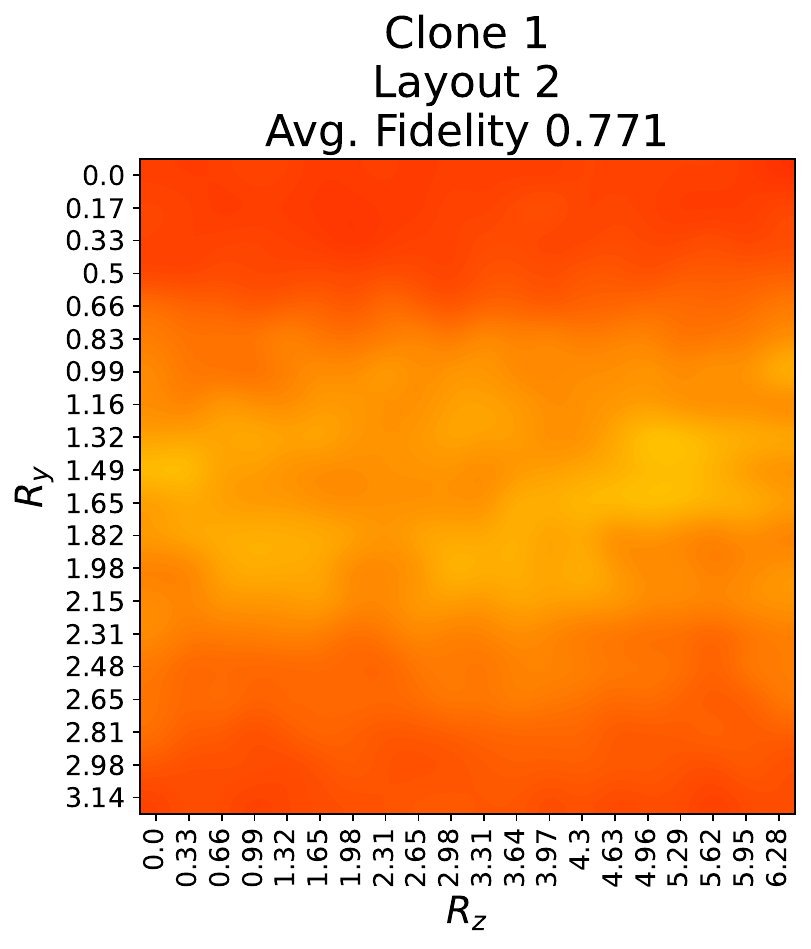}
    \includegraphics[width=0.13\textwidth]{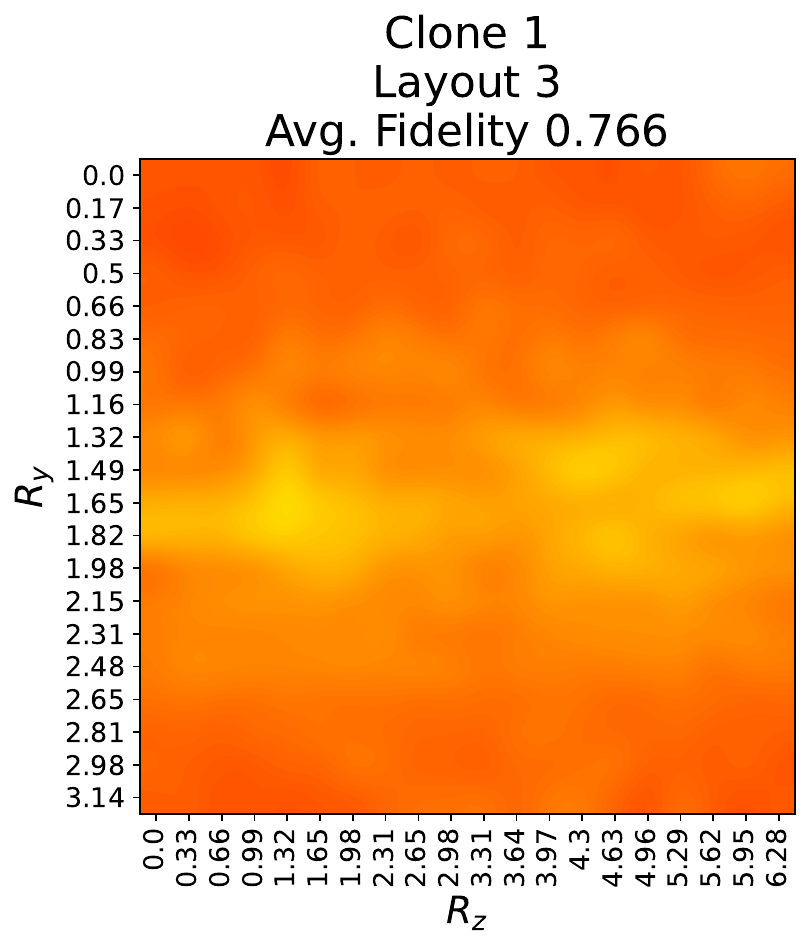}
    \includegraphics[width=0.13\textwidth]{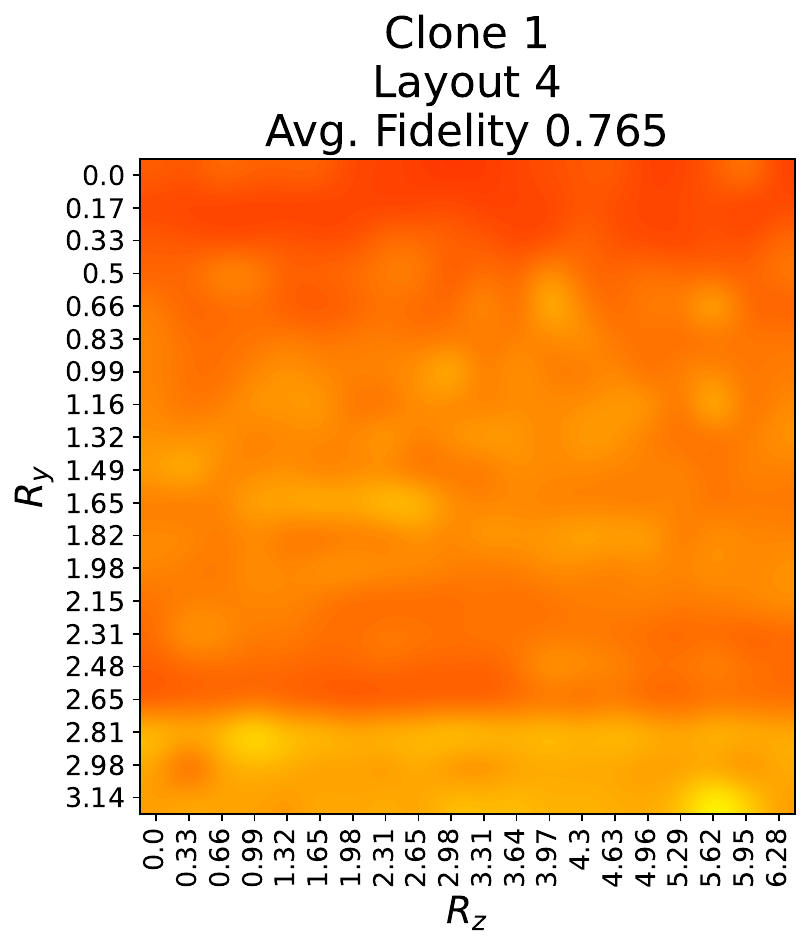}
    \includegraphics[width=0.13\textwidth]{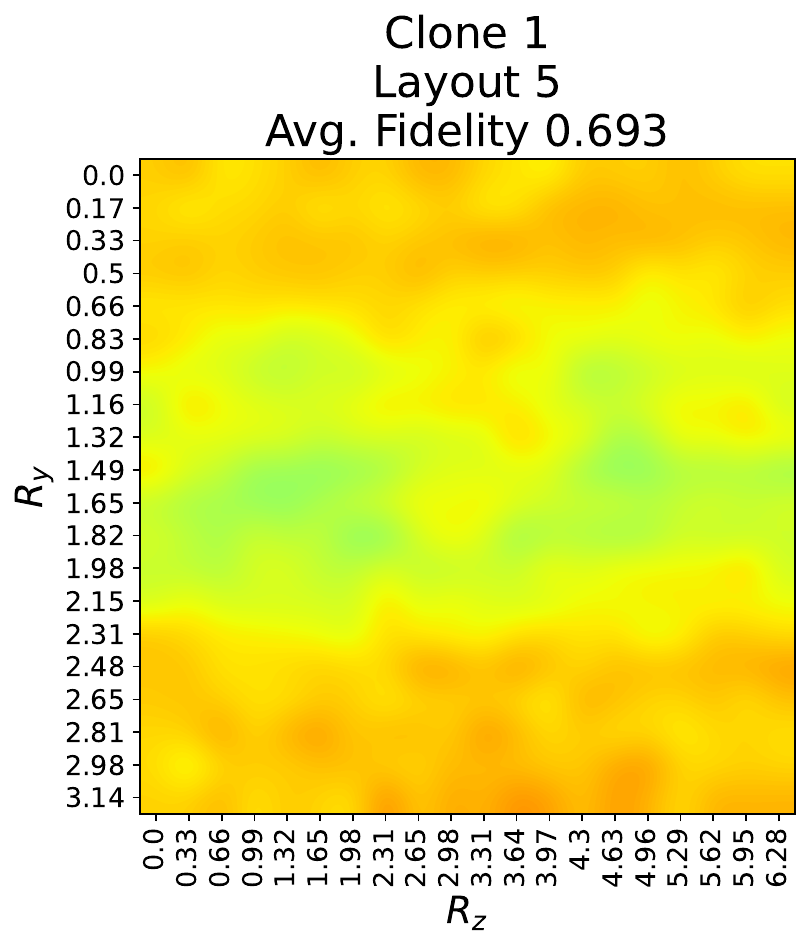}
    \includegraphics[width=0.13\textwidth]{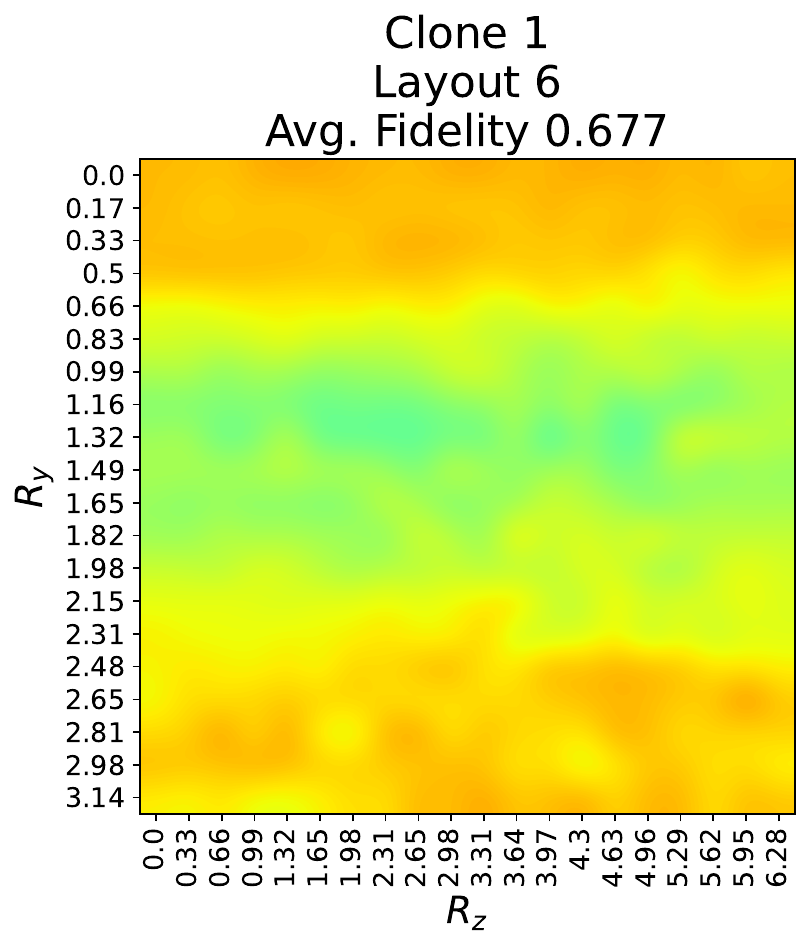}
    \includegraphics[width=0.43\textwidth]{figures/colorbar.pdf}\\
    \caption{Single qubit clone fidelity heatmaps for $M=2$ quantum telecloning circuits with no ancilla qubits, executed without dynamical decoupling sequences. Each column corresponds to the $7$ different compiled hardware layouts. Bottom two rows show the fidelities of the $2$ single qubit clones. Top two rows show Bloch sphere vector representations of the single qubit state tomography computed density matrices. Data from \texttt{ibm\_cairo}.  }
    \label{fig:fidelity_heatmaps_M2_ibm_cairo_with_ancilla}
\end{figure*}

Figure~\ref{fig:fidelity_heatmaps_M2_no_ancilla_ALAP_DD_ibm_auckland} shows the computed clone fidelity, and the computed single qubit clone density matrices (represented as aggregated vectors plotted on a Bloch sphere) for $M=2$, with the optimization of no ancilla qubits and with dynamical decoupling sequences having been applied, run on \texttt{ibm\_auckland}. Figure~\ref{fig:fidelity_heatmaps_M2_no_ancilla_ALAP_DD_ibmq_mumbai} shows the results for the same experimental setup as Figure~\ref{fig:fidelity_heatmaps_M2_no_ancilla_ALAP_DD_ibmq_mumbai} ($M=2$ no ancilla qubit telecloning circuits, and X-X dynamical decoupling sequences added) but on \texttt{ibmq\_mumbai}. Figure~\ref{fig:fidelity_heatmaps_M2_ibmq_kolkata_with_ancilla} shows the computed clone fidelity representation for $M=2$, but with ancilla qubits used and no dynamical decoupling used, run on \texttt{ibmq\_kolkata}. Figure~\ref{fig:fidelity_heatmaps_M2_ibm_cairo_with_ancilla} shows the fidelity results for $M=2$ run on \texttt{ibm\_cairo} with the removed ancilla qubit optimization, and no dynamical decoupling. 

\begin{figure*}[th!]
    \centering
    \includegraphics[width=0.13\textwidth]{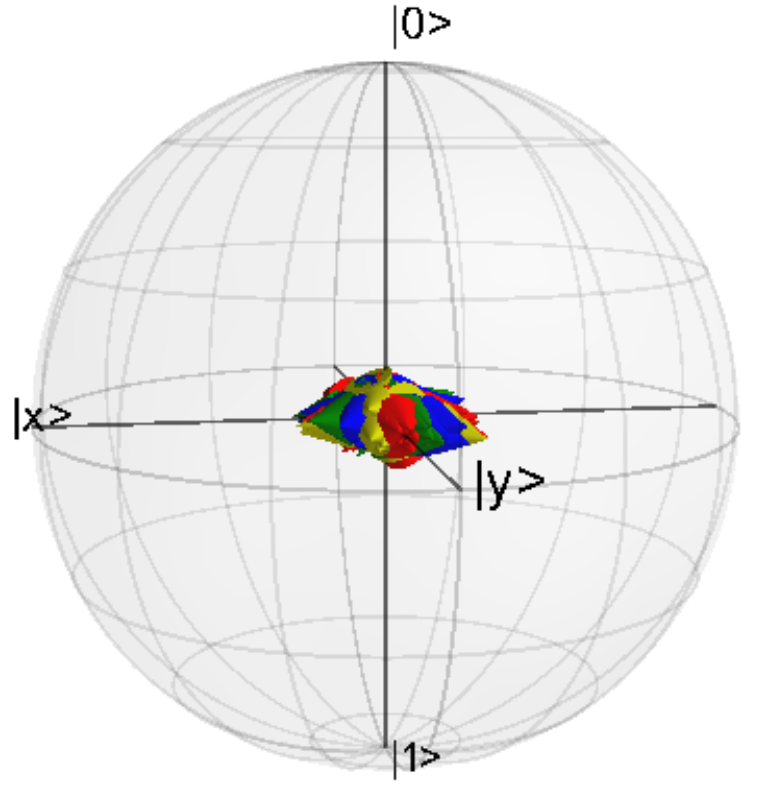}
    \includegraphics[width=0.13\textwidth]{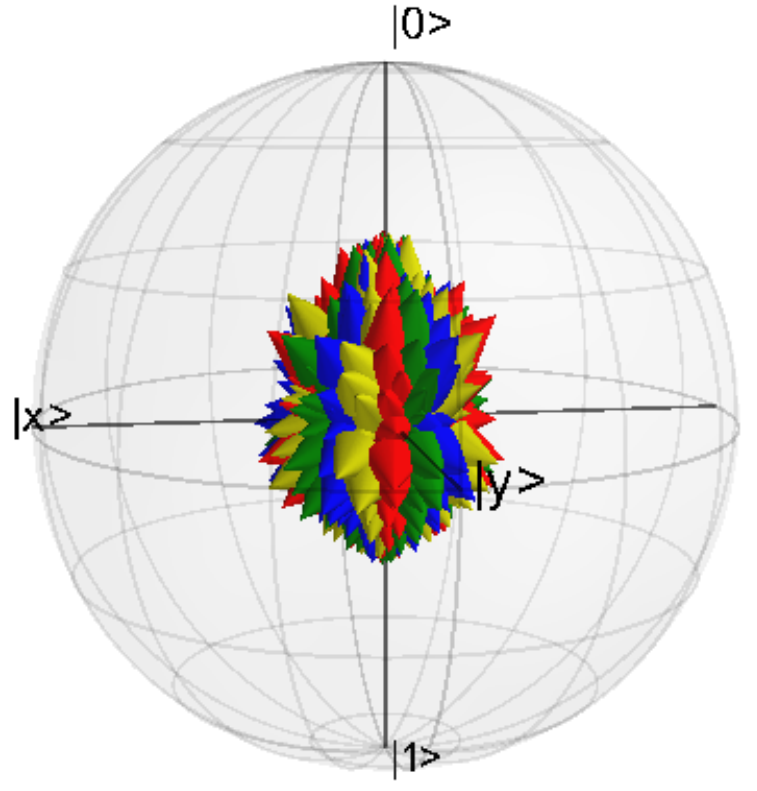}
    \includegraphics[width=0.13\textwidth]{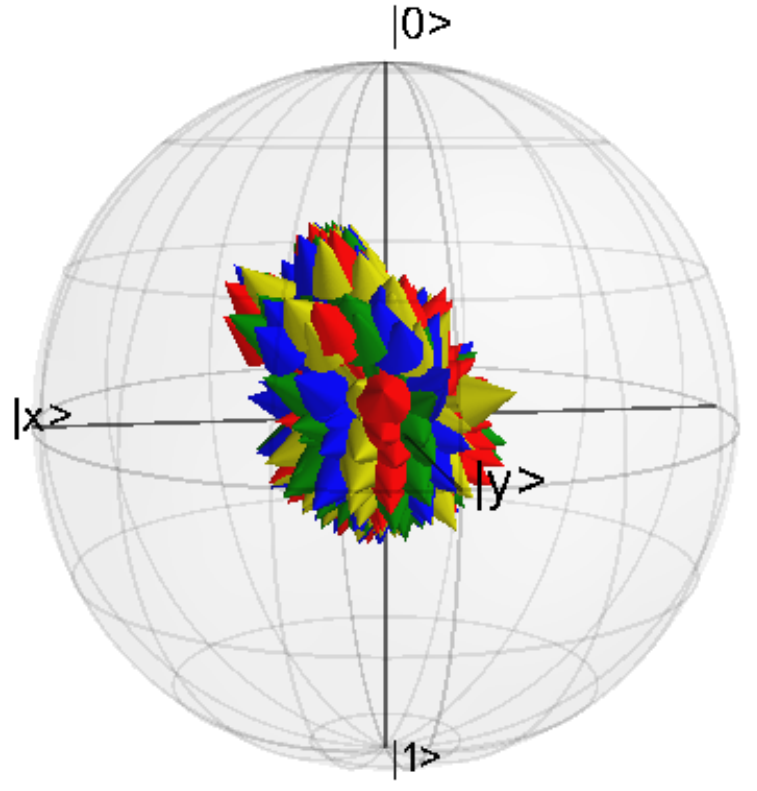}
    \includegraphics[width=0.13\textwidth]{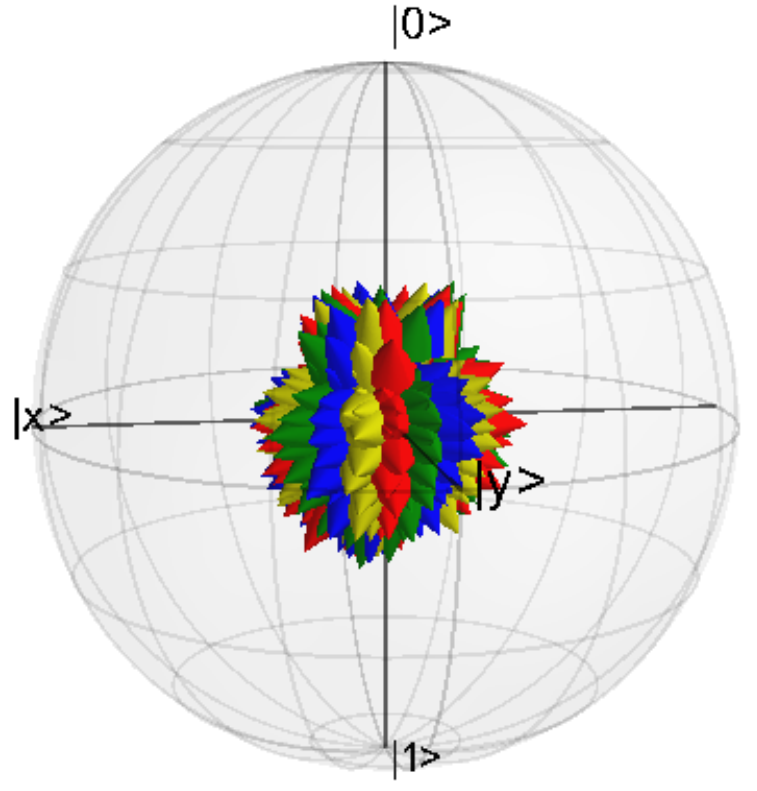}
    \includegraphics[width=0.13\textwidth]{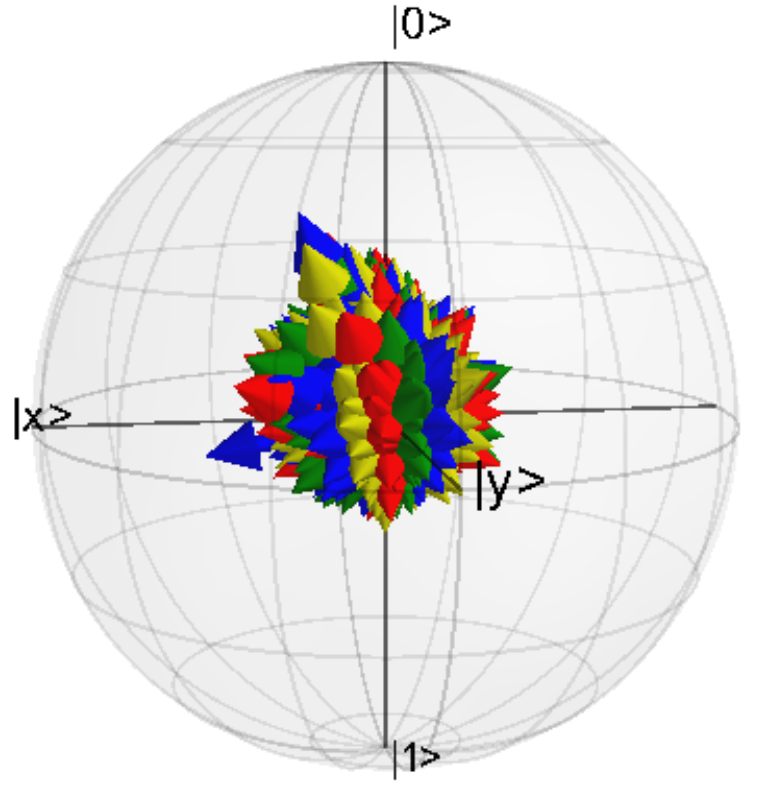}
    \includegraphics[width=0.13\textwidth]{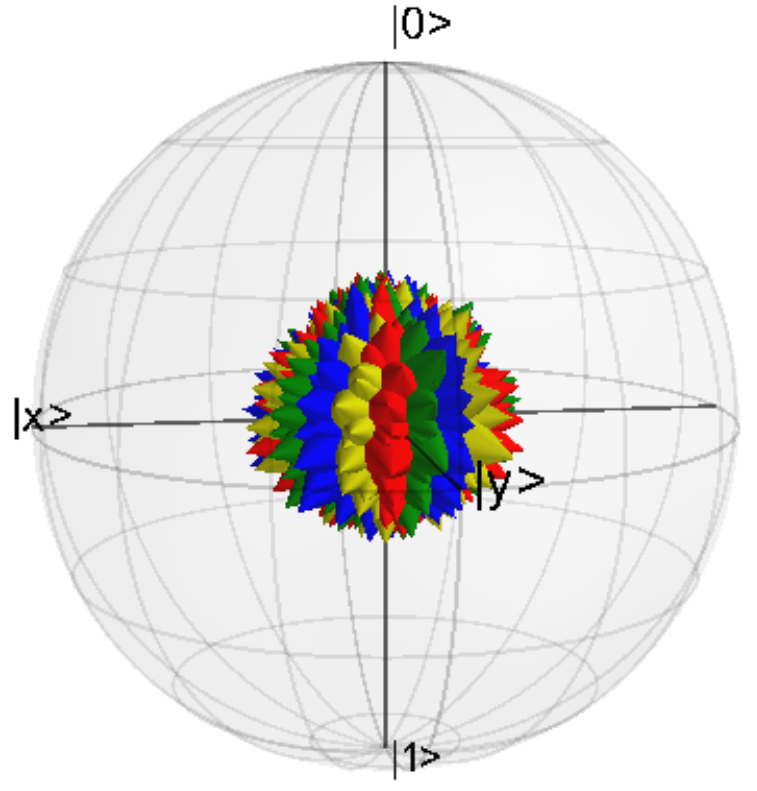}
    \includegraphics[width=0.13\textwidth]{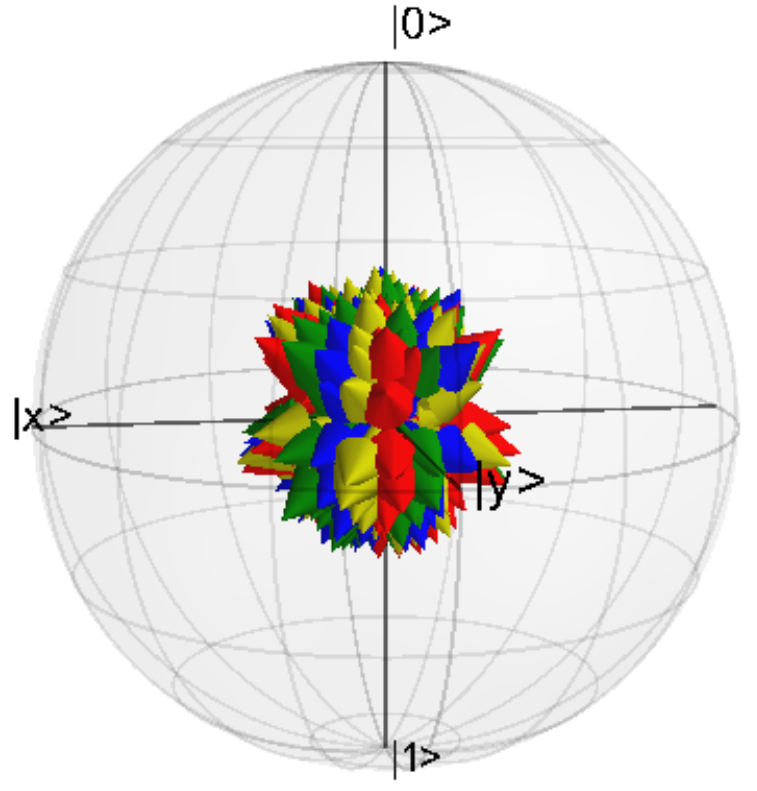}
    \includegraphics[width=0.13\textwidth]{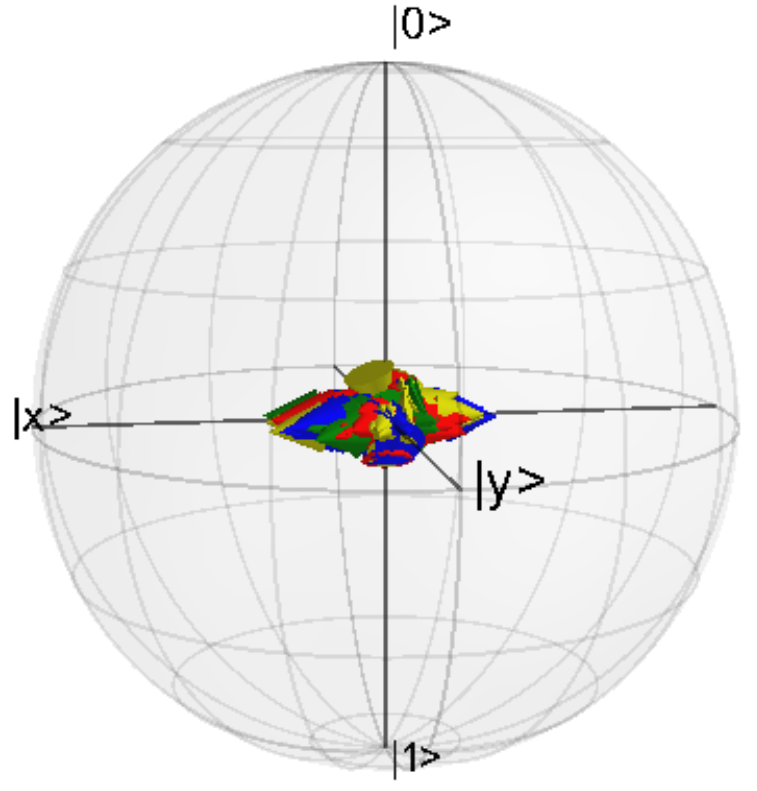}
    \includegraphics[width=0.13\textwidth]{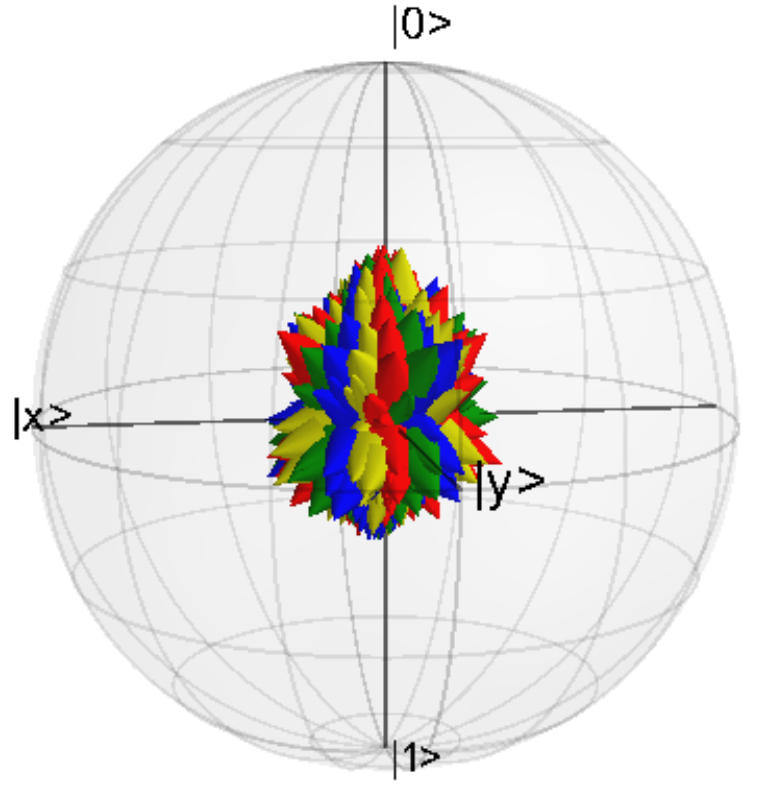}
    \includegraphics[width=0.13\textwidth]{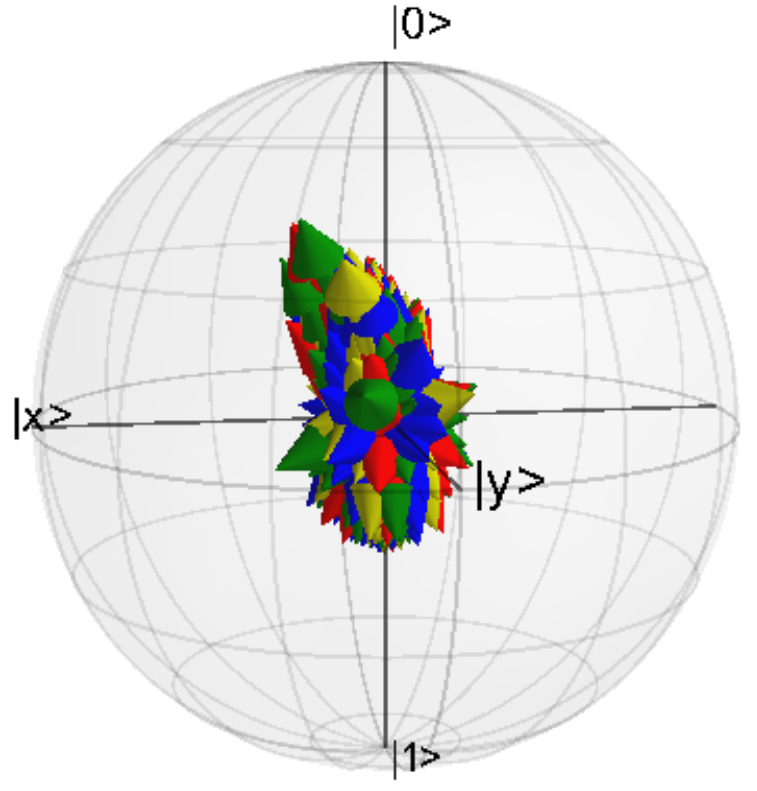}
    \includegraphics[width=0.13\textwidth]{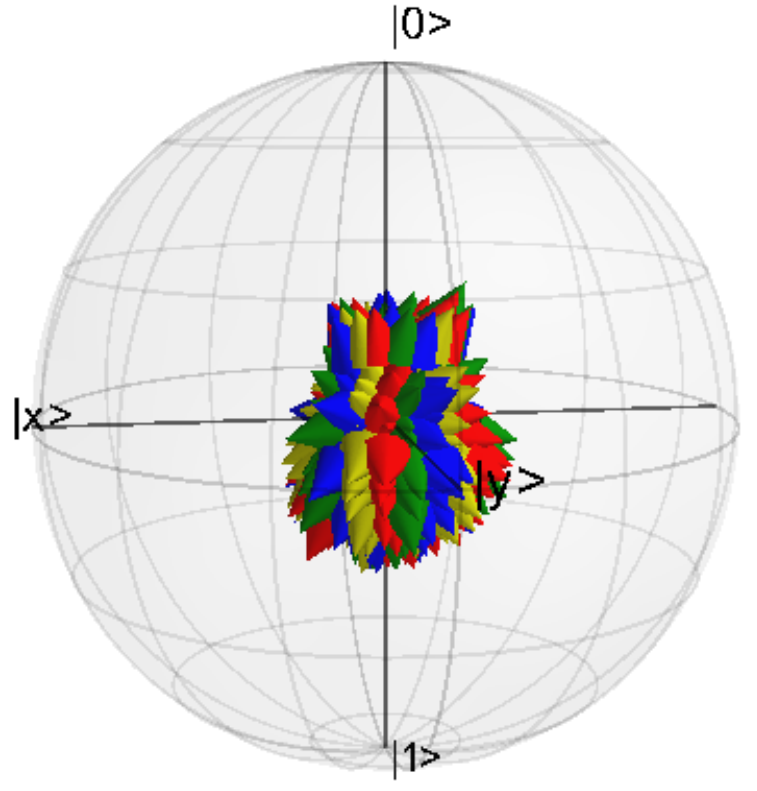}
    \includegraphics[width=0.13\textwidth]{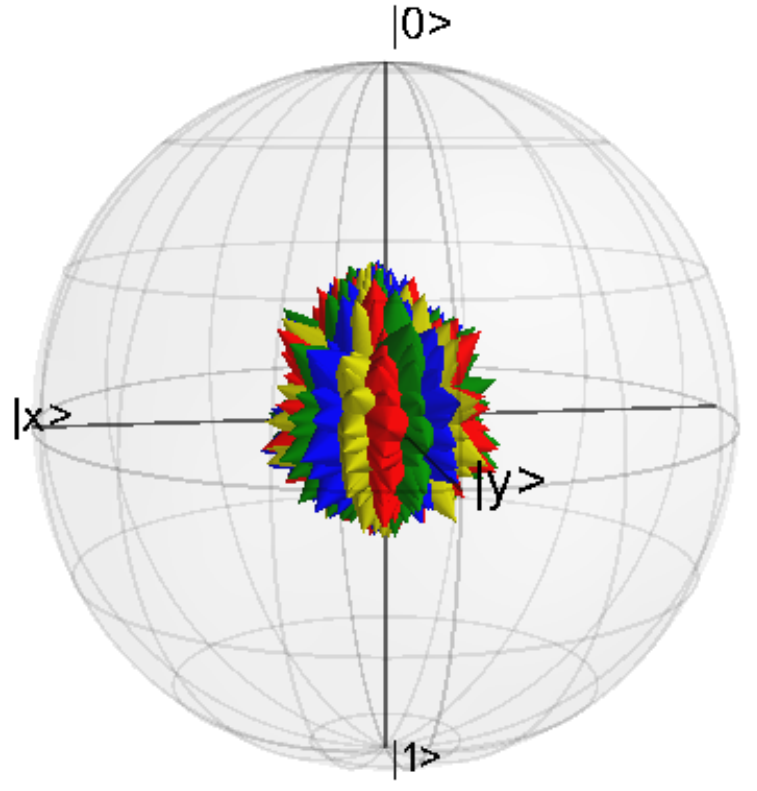}
    \includegraphics[width=0.13\textwidth]{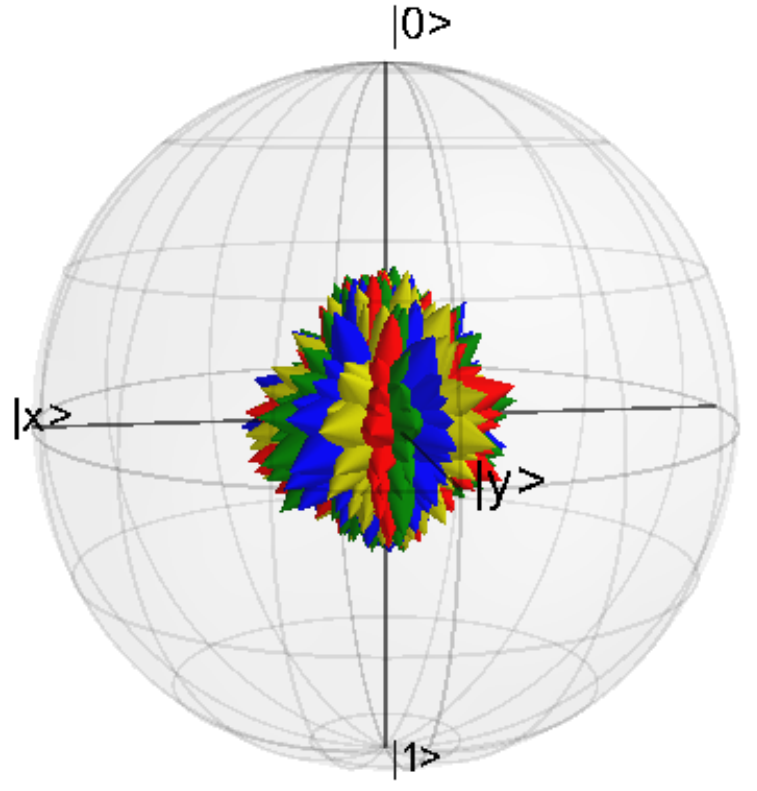}
    \includegraphics[width=0.13\textwidth]{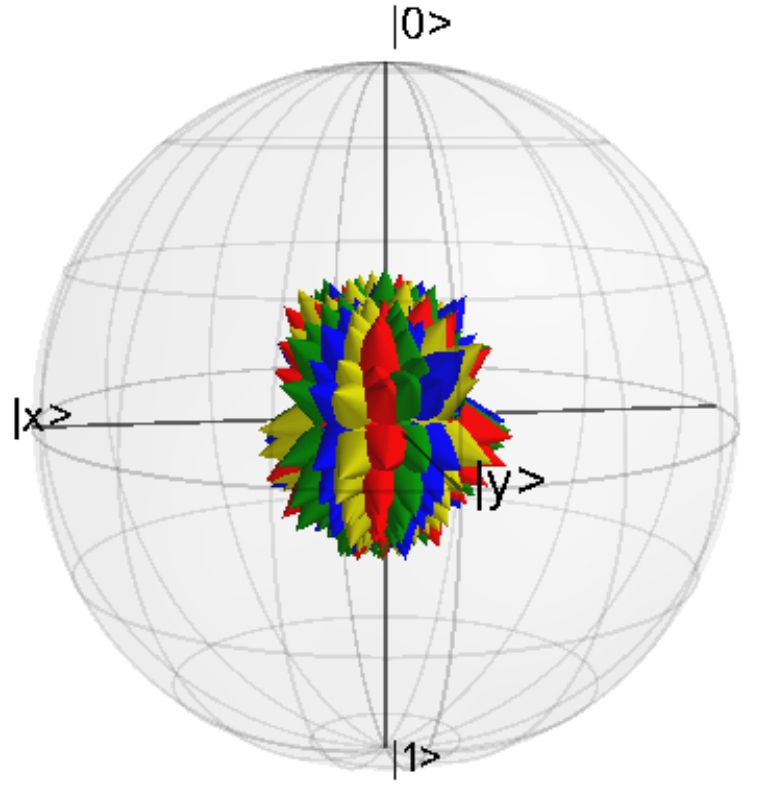}
    \includegraphics[width=0.13\textwidth]{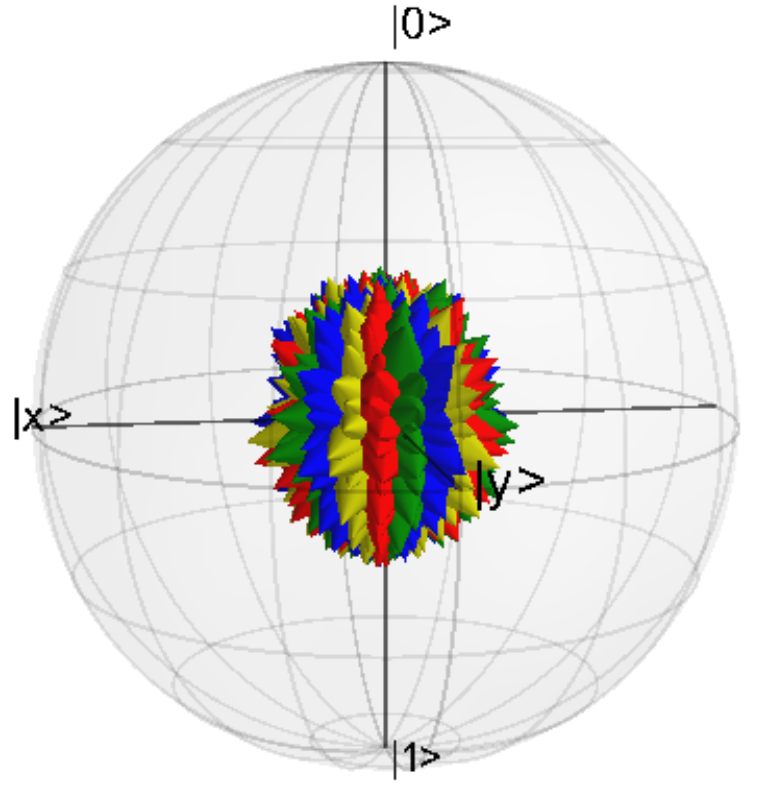}
    \includegraphics[width=0.13\textwidth]{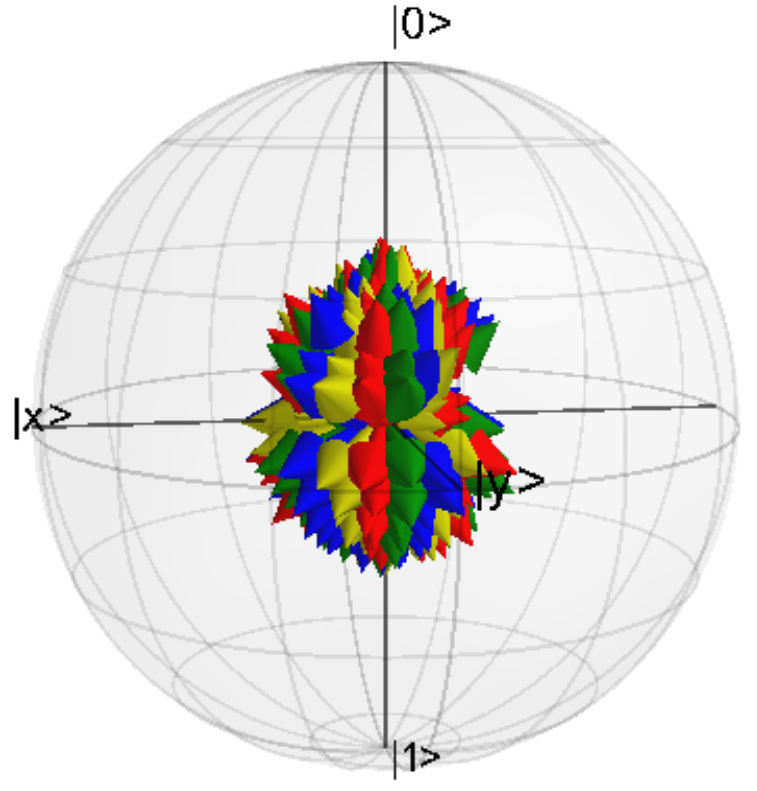}
    \includegraphics[width=0.13\textwidth]{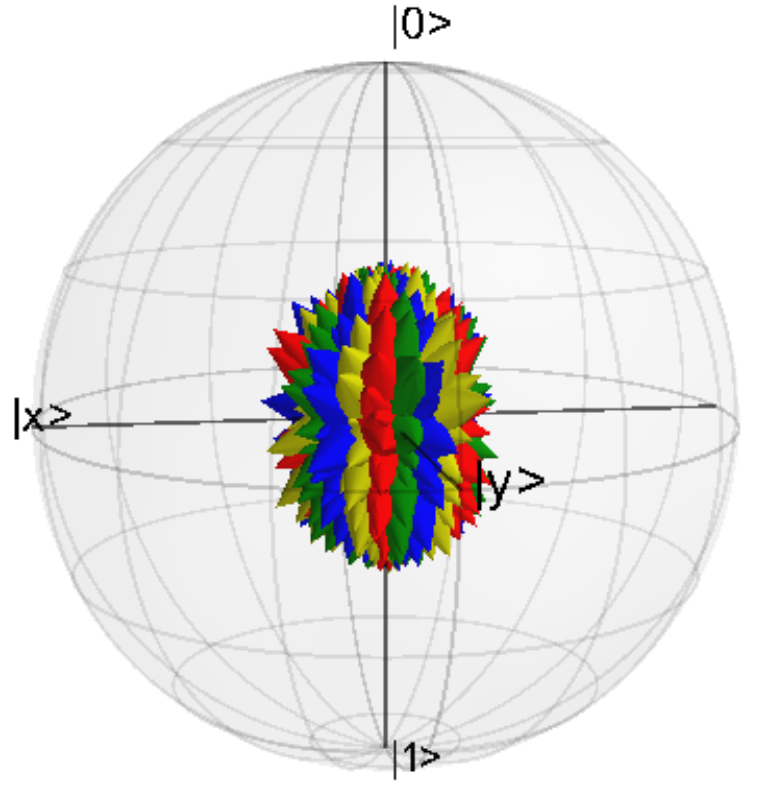}
    \includegraphics[width=0.13\textwidth]{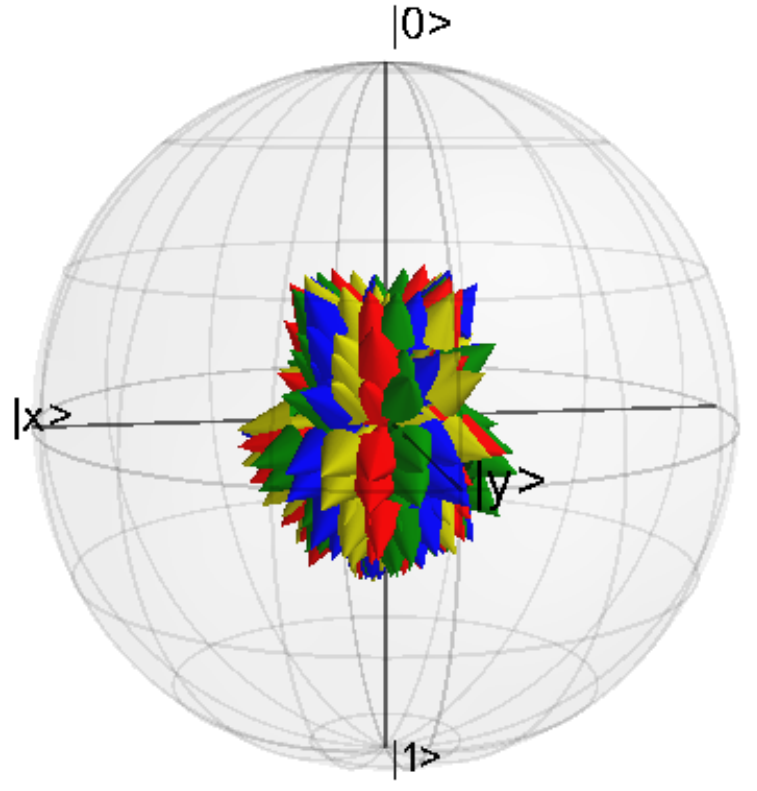}
    \includegraphics[width=0.13\textwidth]{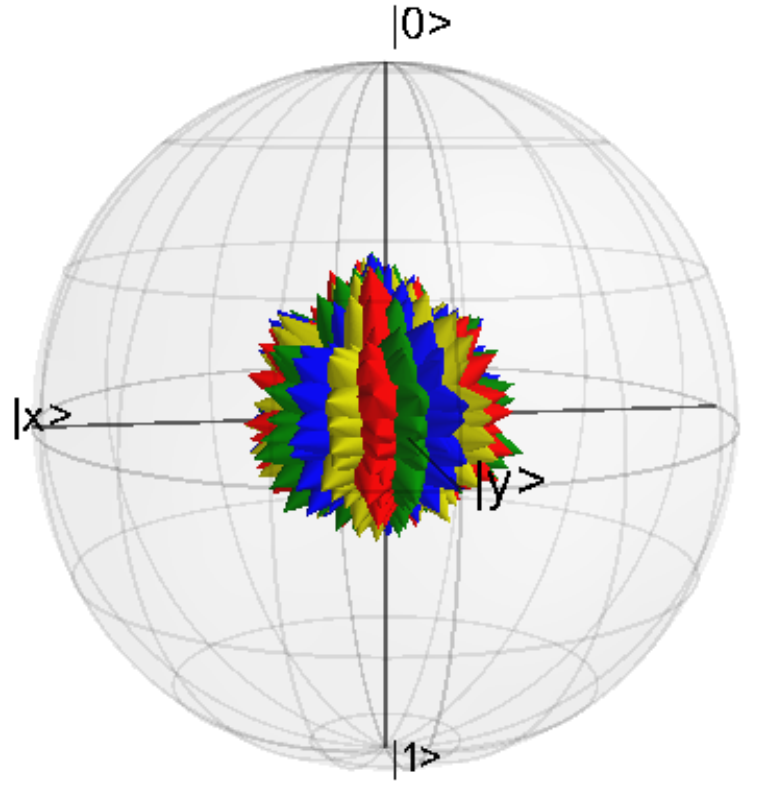}
    \includegraphics[width=0.13\textwidth]{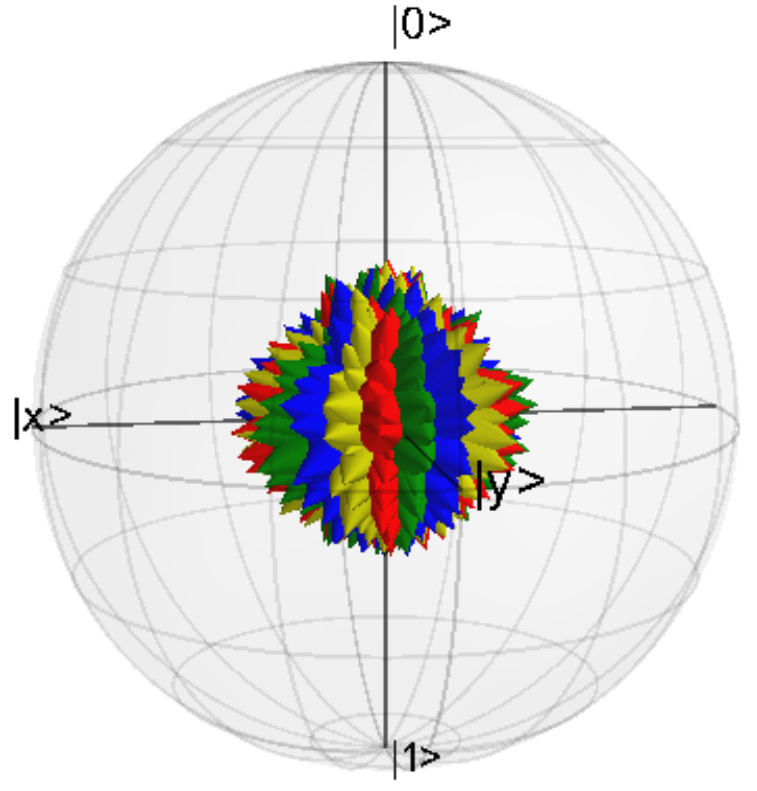}
    \includegraphics[width=0.13\textwidth]{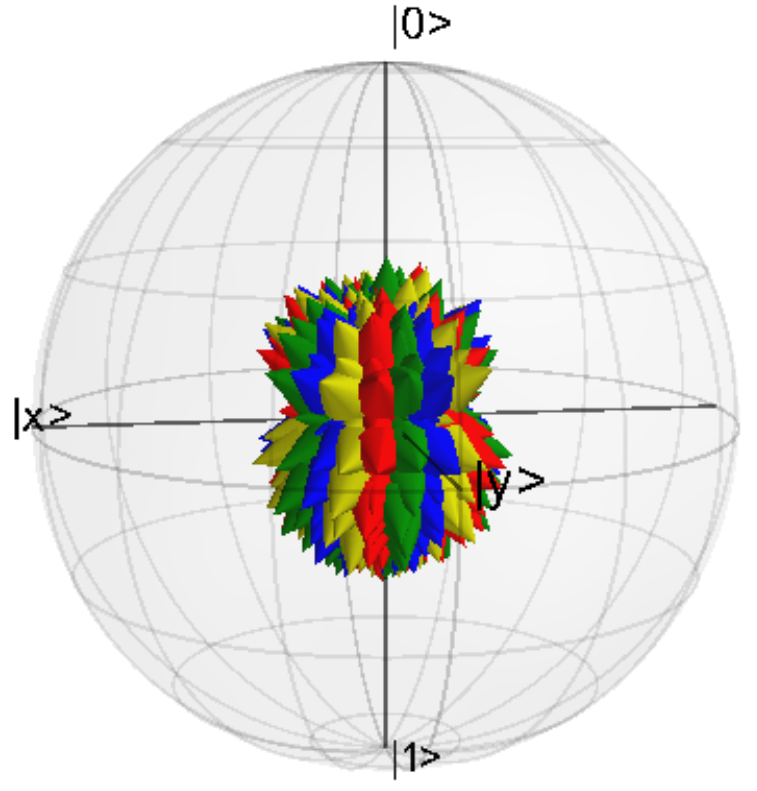}
    \includegraphics[width=0.13\textwidth]{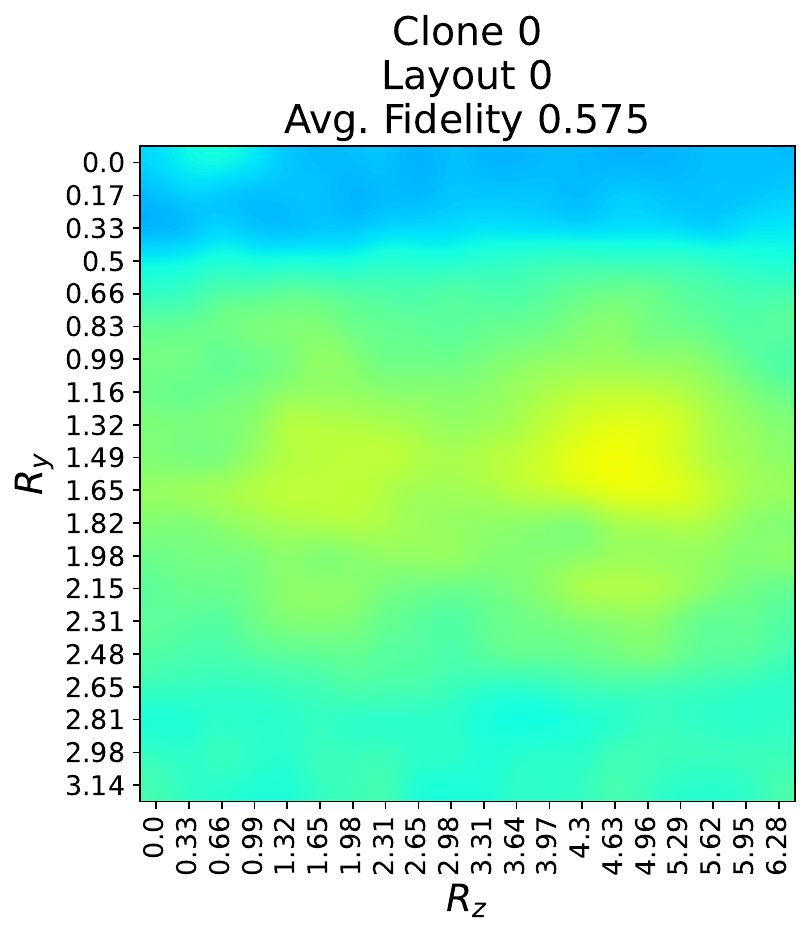}
    \includegraphics[width=0.13\textwidth]{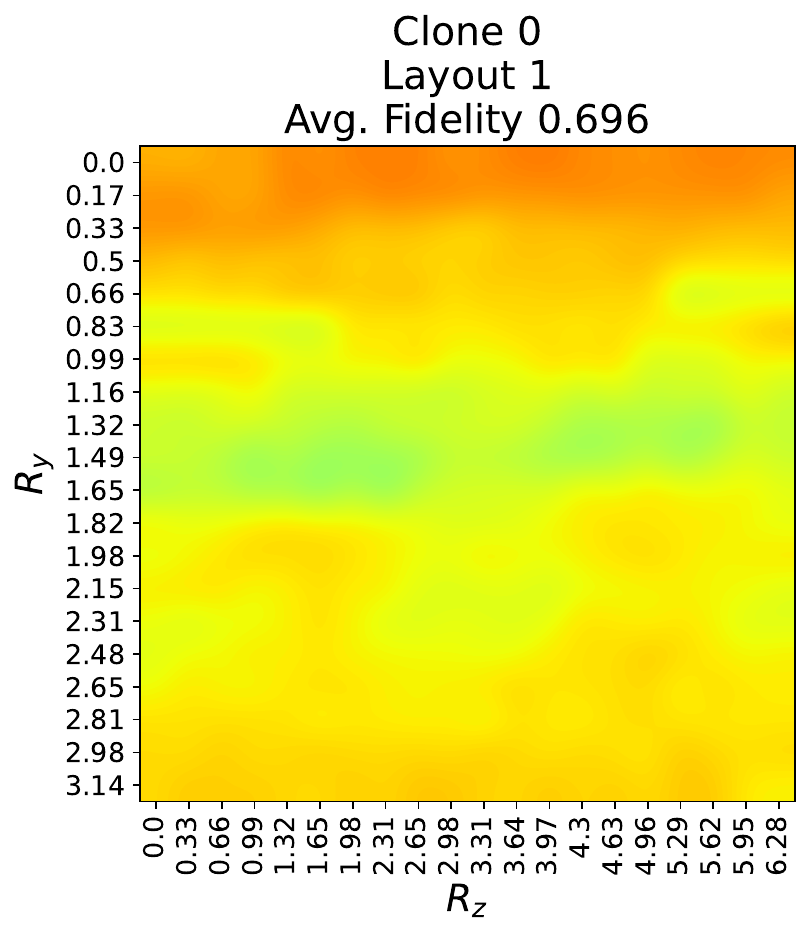}
    \includegraphics[width=0.13\textwidth]{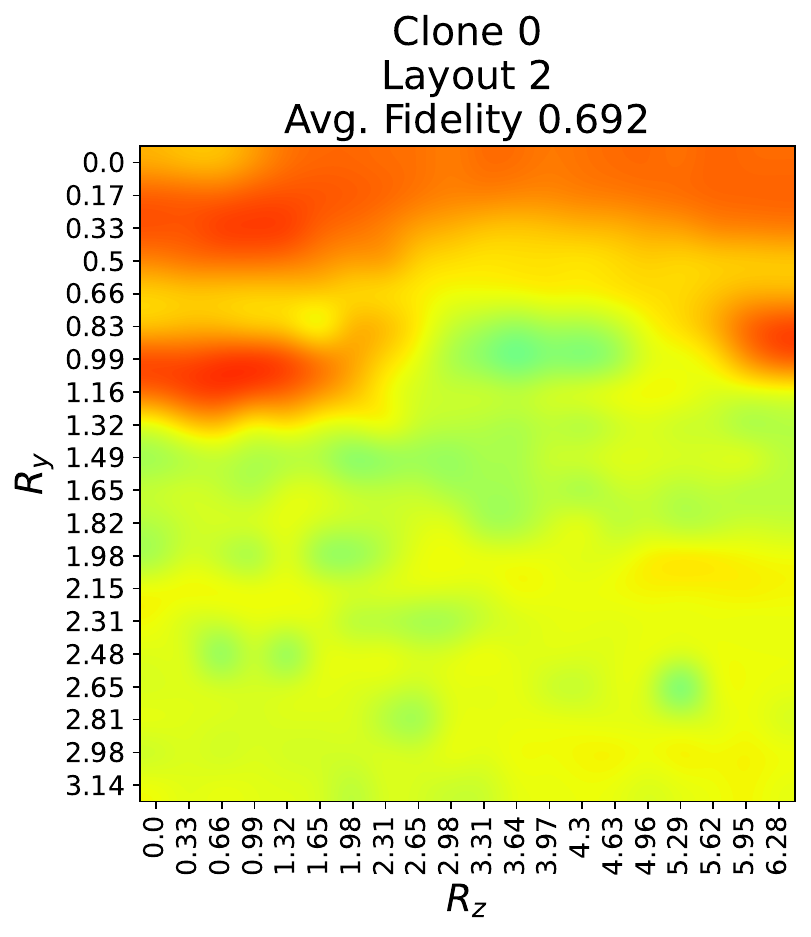}
    \includegraphics[width=0.13\textwidth]{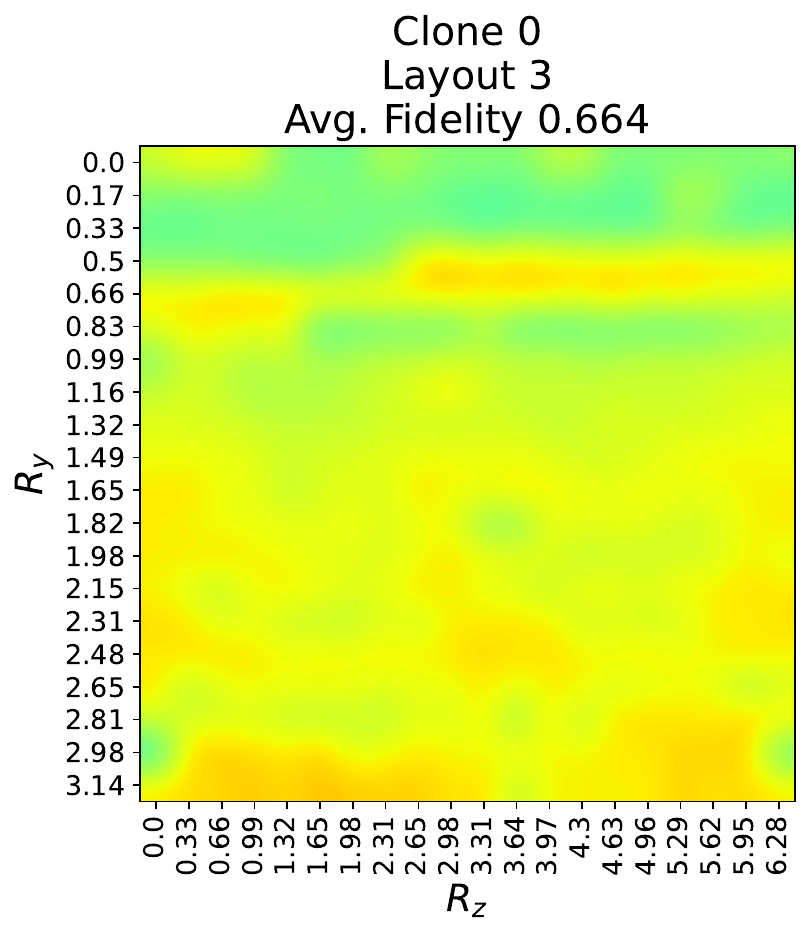}
    \includegraphics[width=0.13\textwidth]{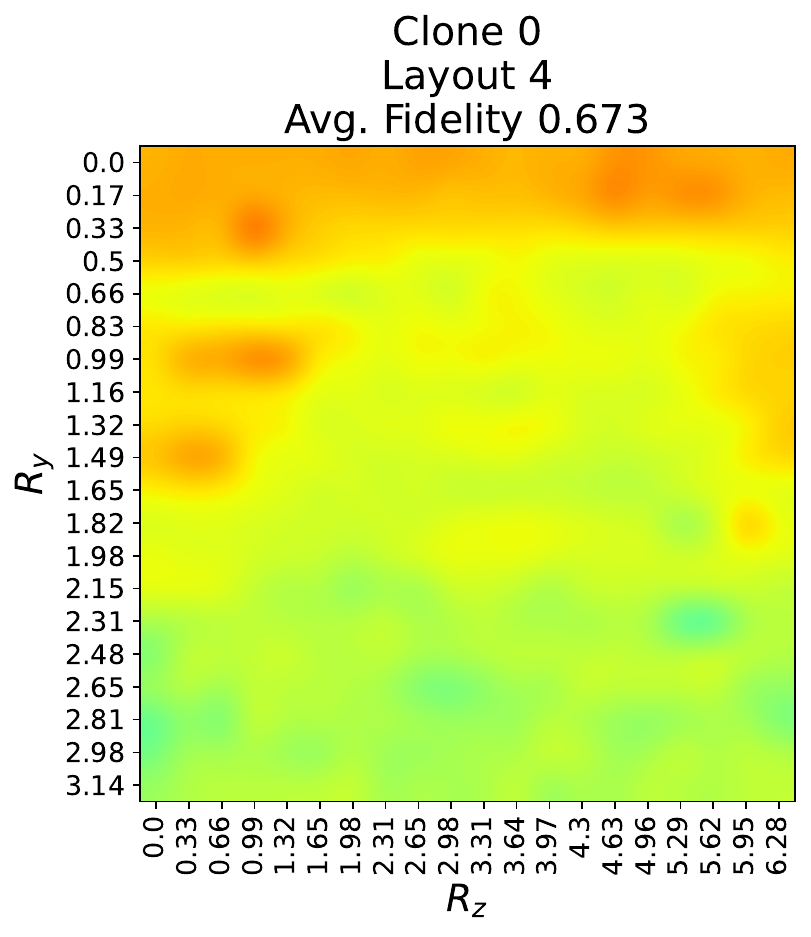}
    \includegraphics[width=0.13\textwidth]{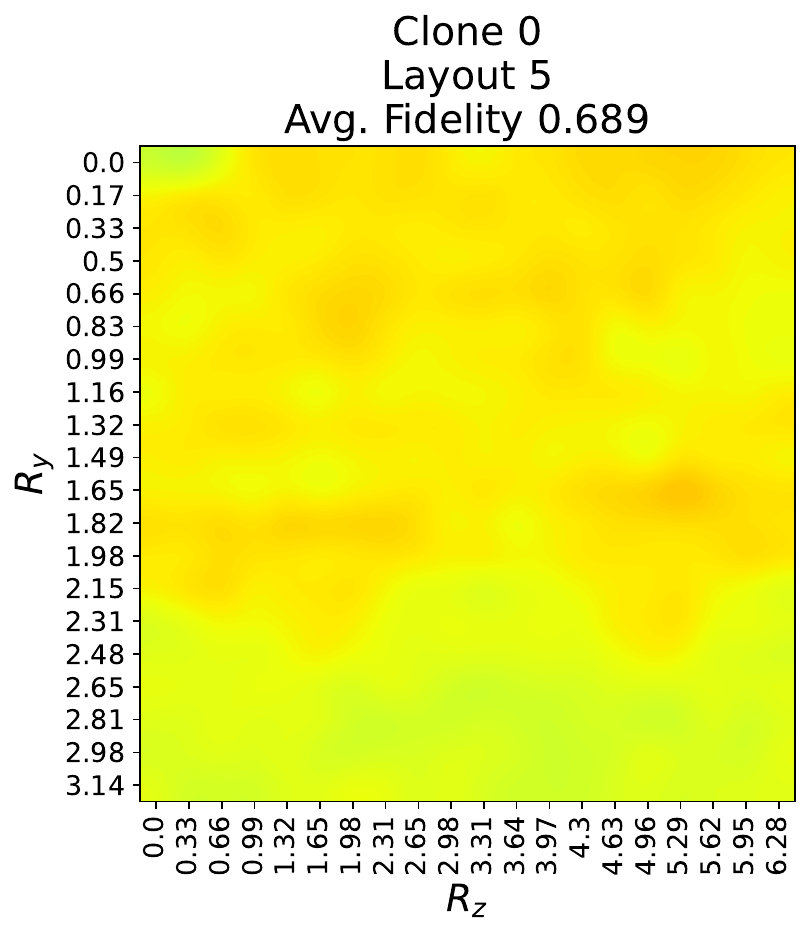}
    \includegraphics[width=0.13\textwidth]{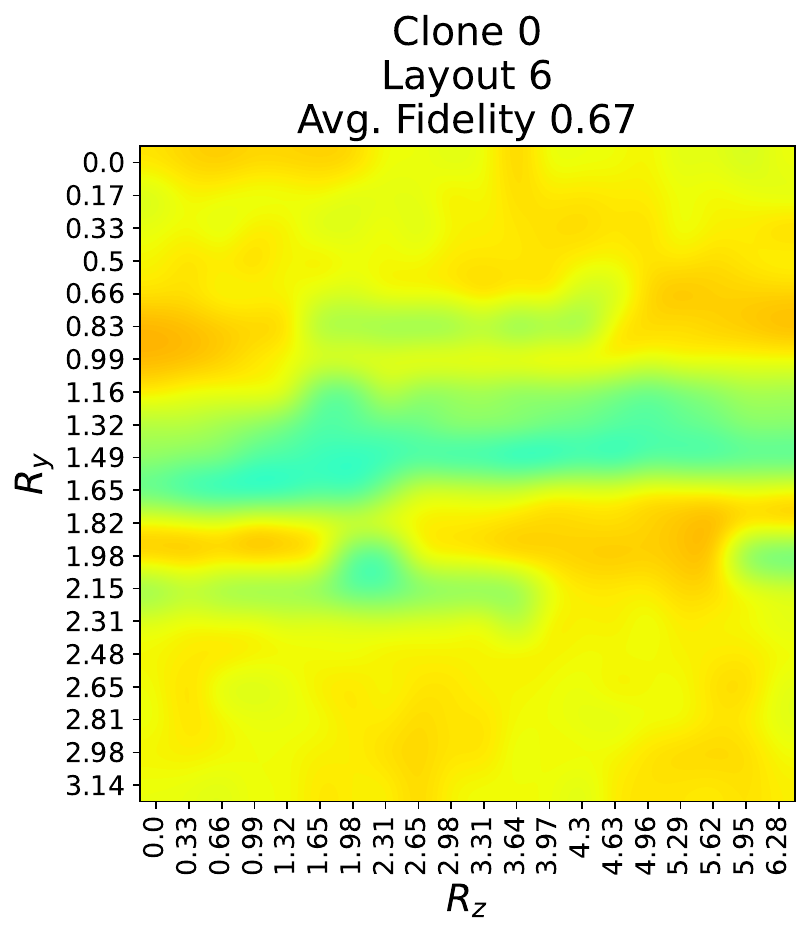}
    \includegraphics[width=0.13\textwidth]{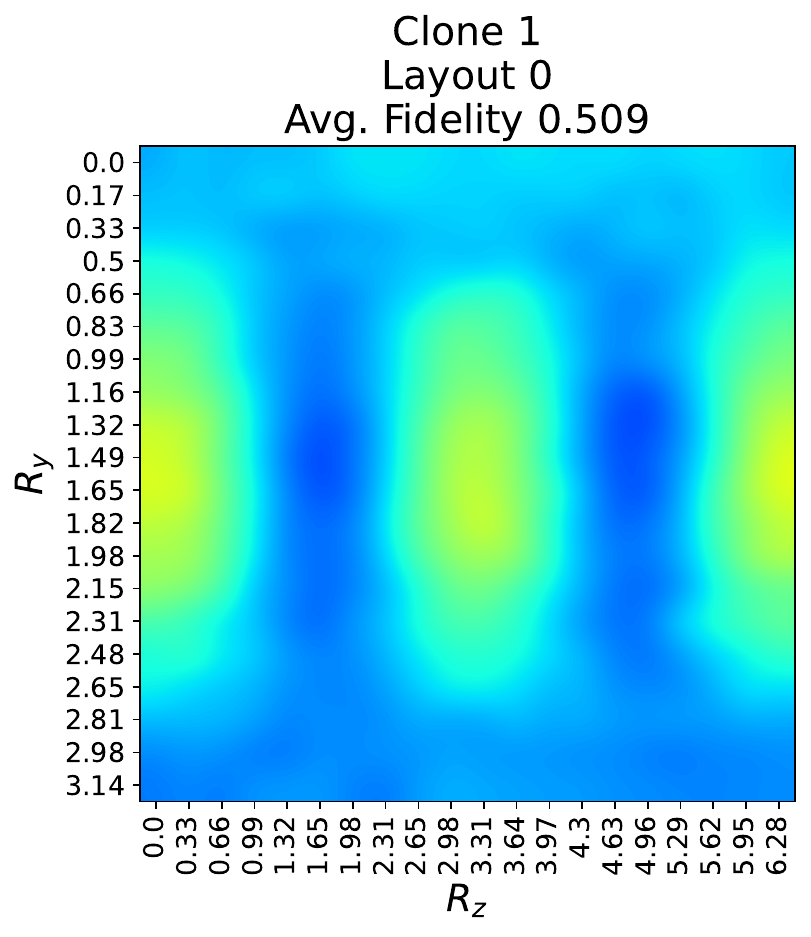}
    \includegraphics[width=0.13\textwidth]{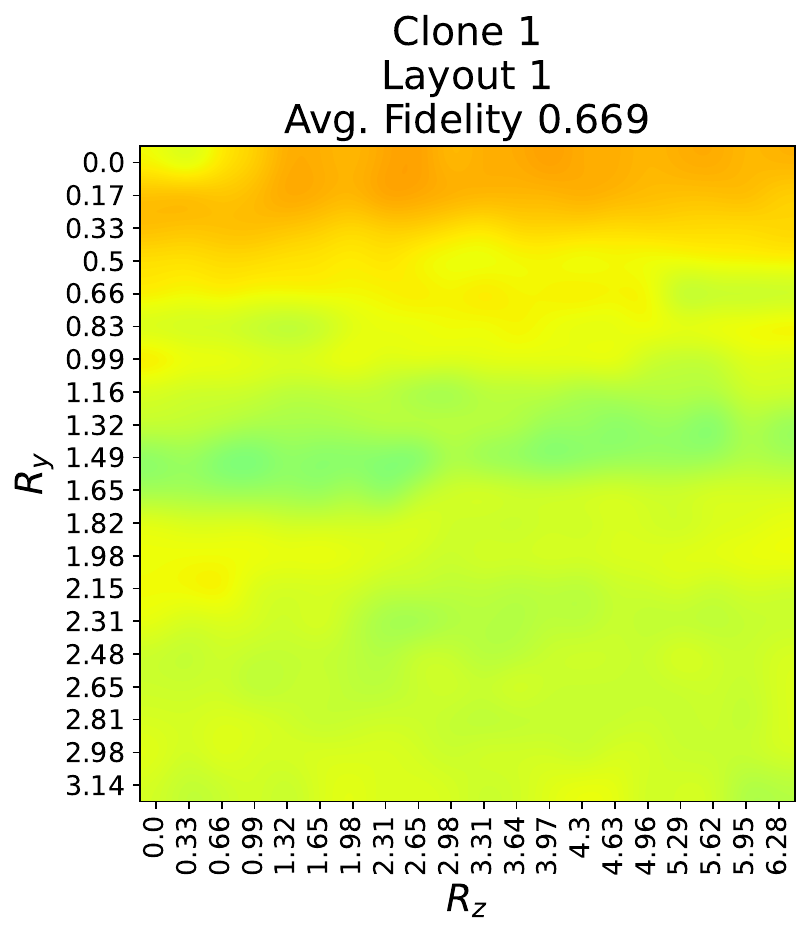}
    \includegraphics[width=0.13\textwidth]{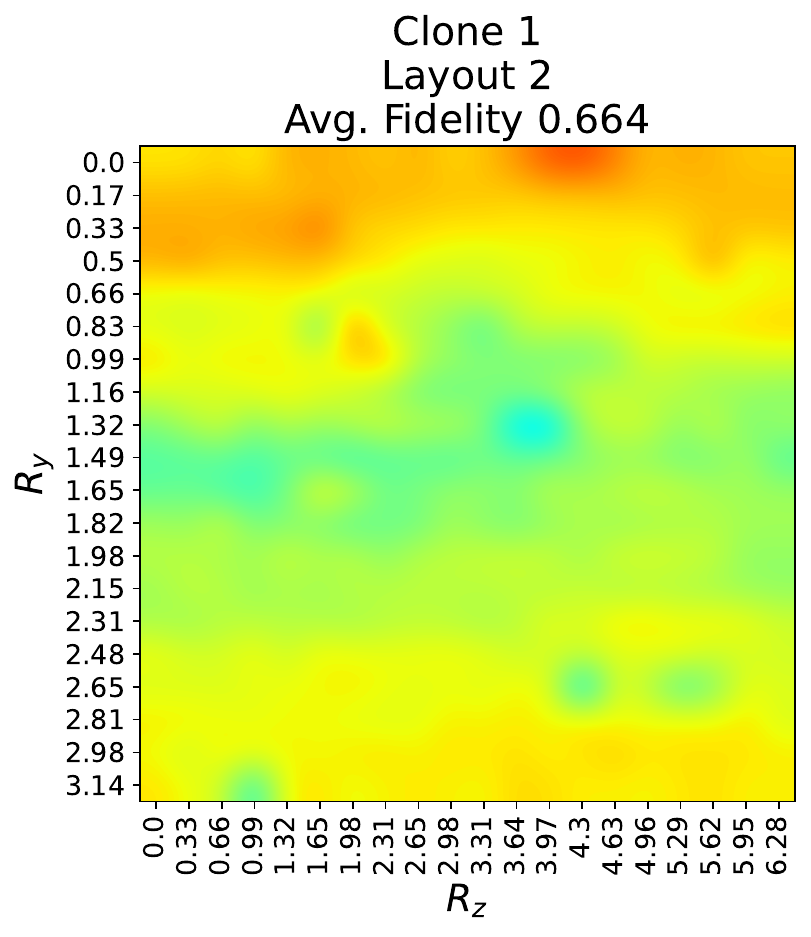}
    \includegraphics[width=0.13\textwidth]{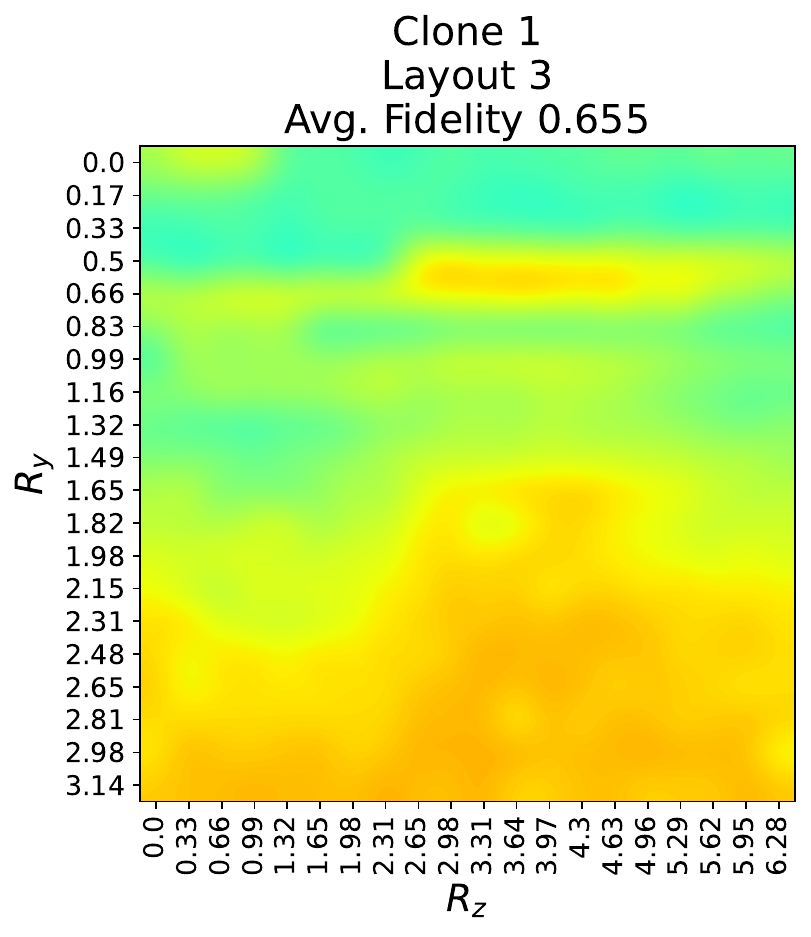}
    \includegraphics[width=0.13\textwidth]{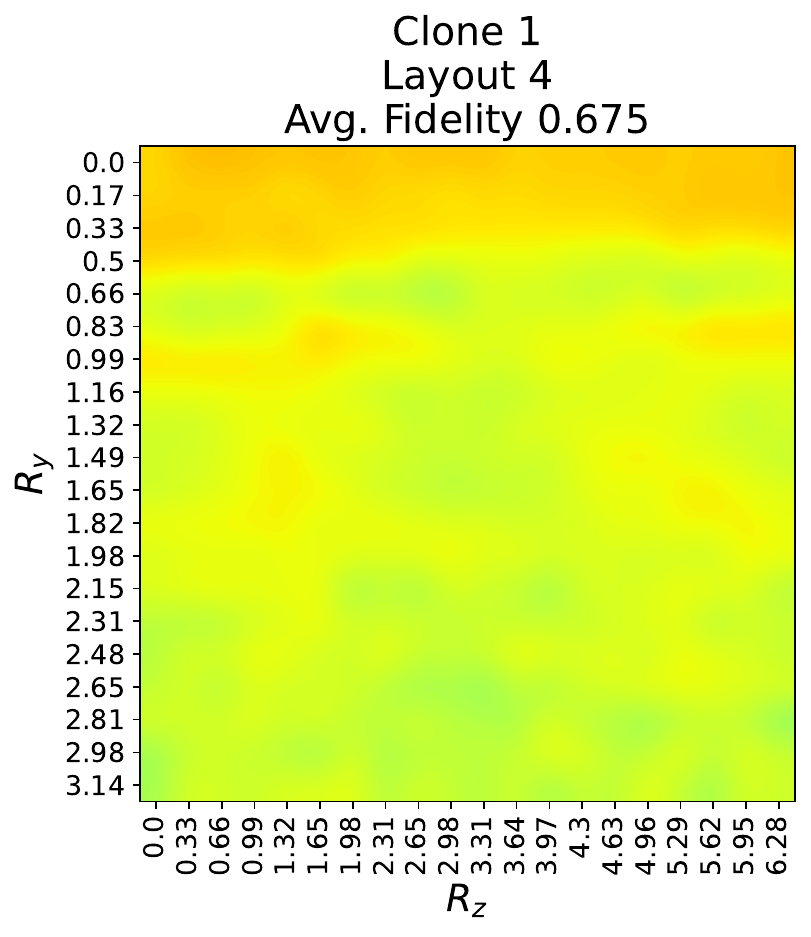}
    \includegraphics[width=0.13\textwidth]{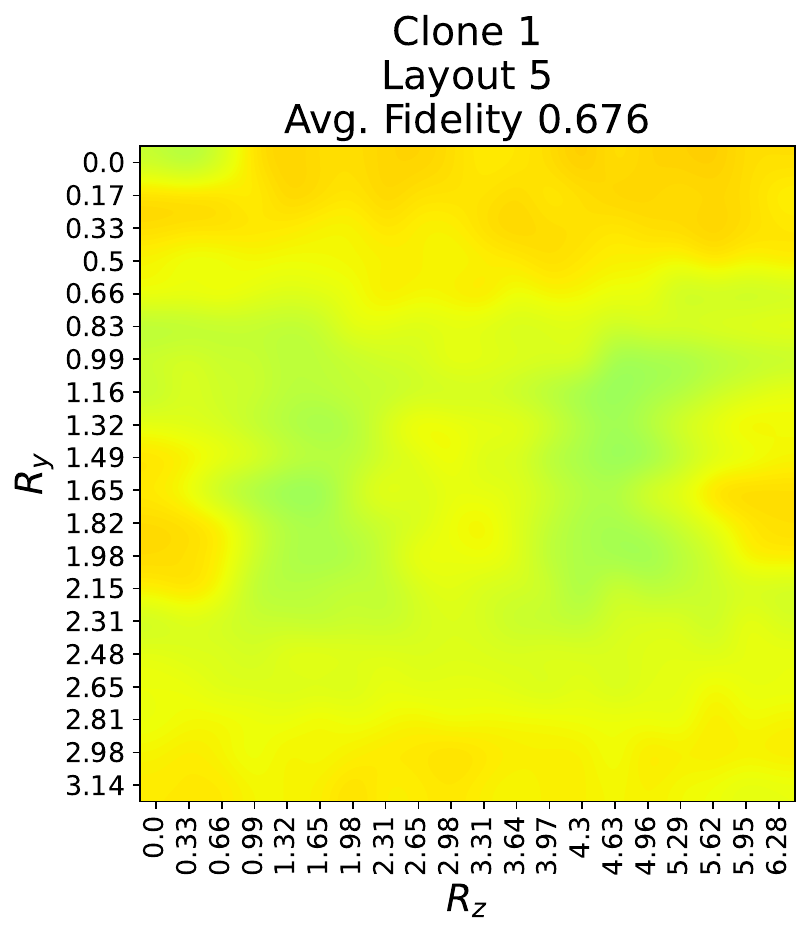}
    \includegraphics[width=0.13\textwidth]{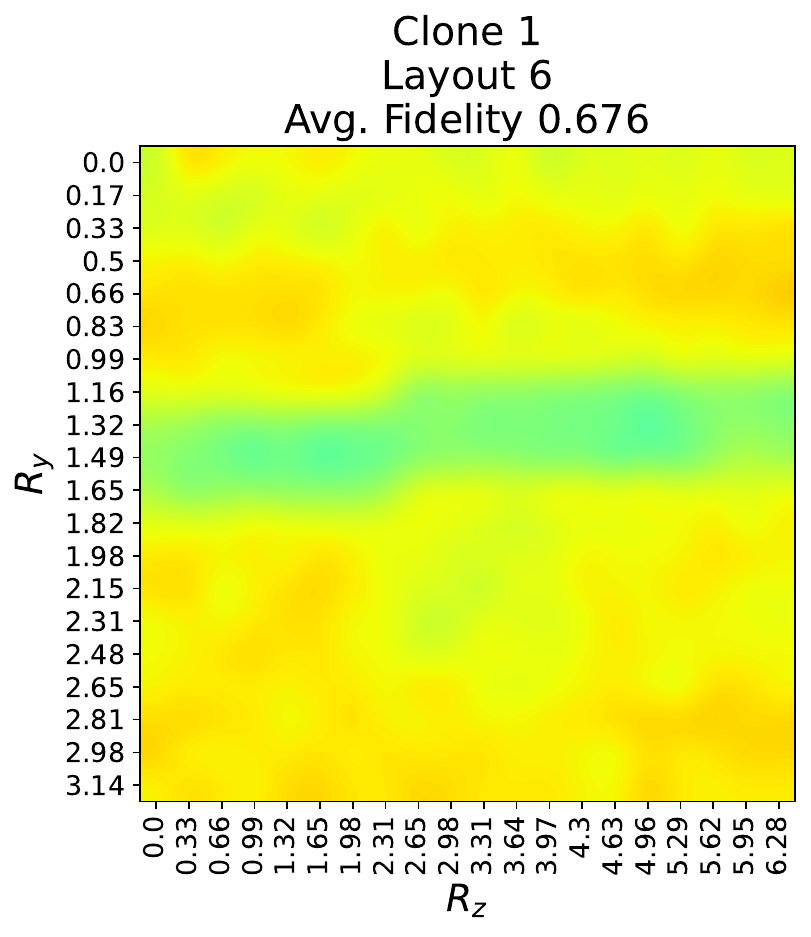}
    \includegraphics[width=0.13\textwidth]{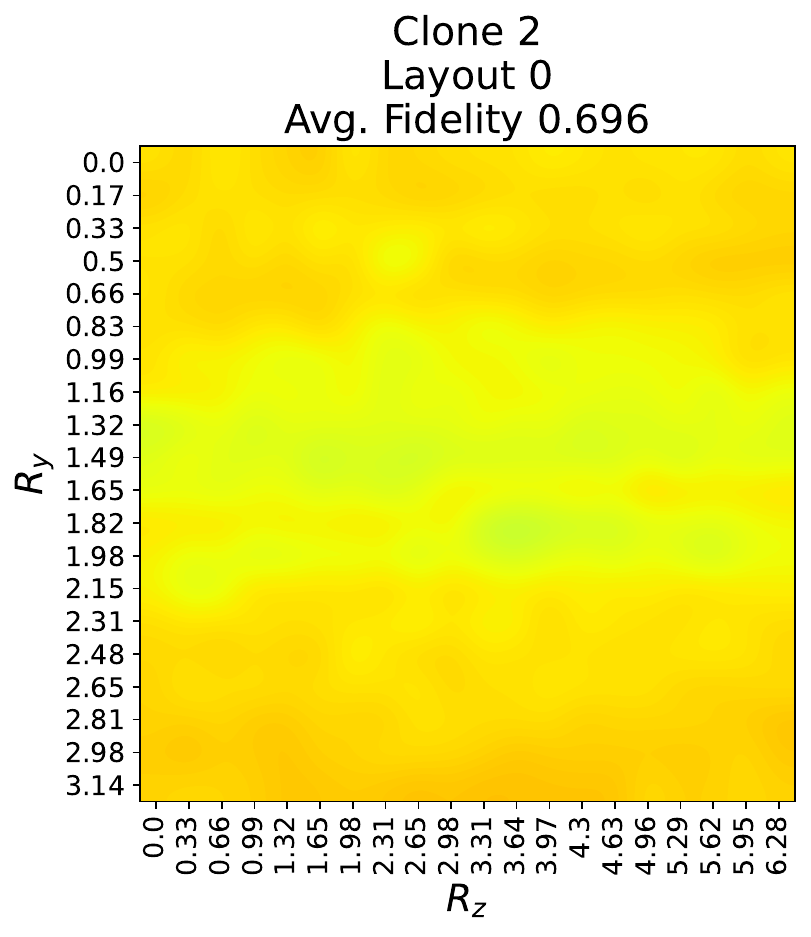}
    \includegraphics[width=0.13\textwidth]{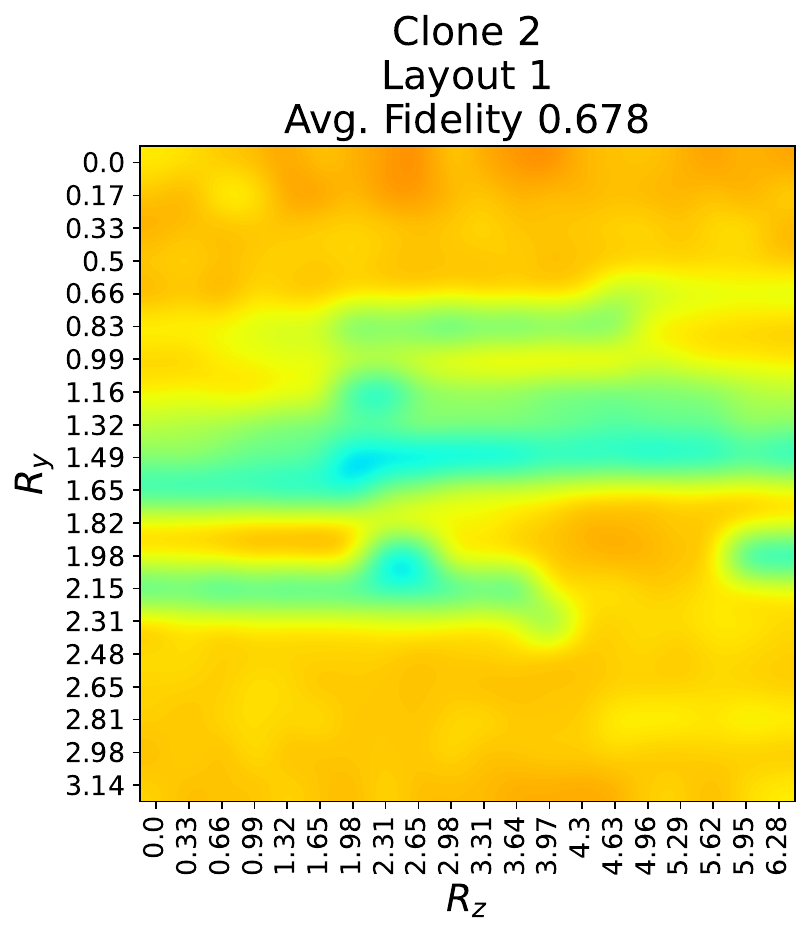}
    \includegraphics[width=0.13\textwidth]{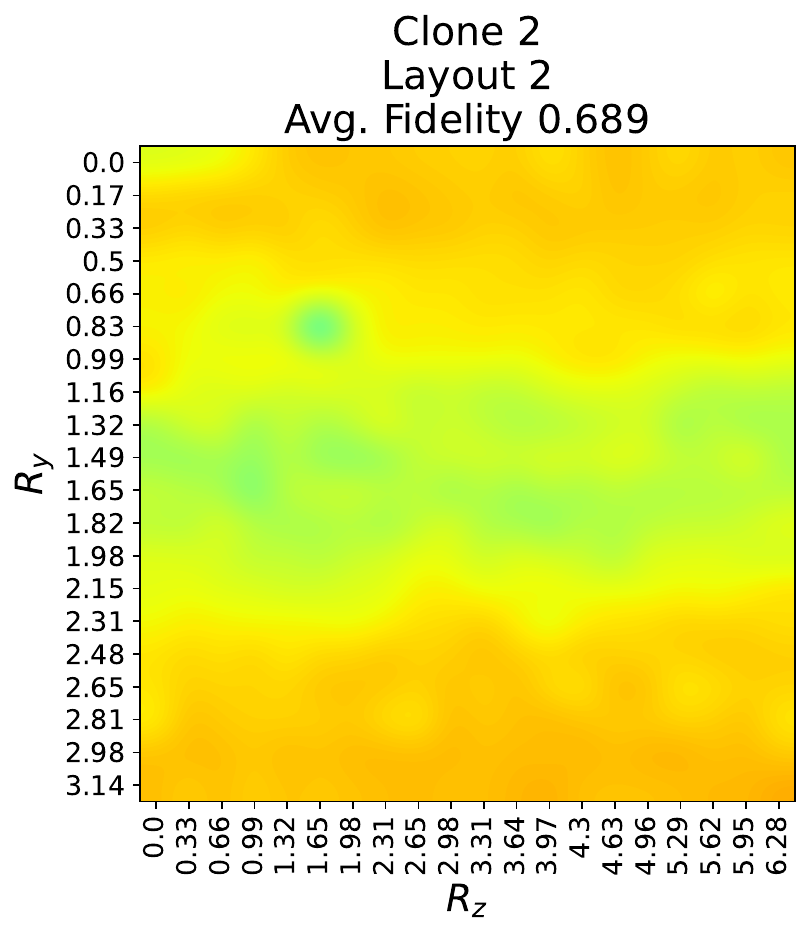}
    \includegraphics[width=0.13\textwidth]{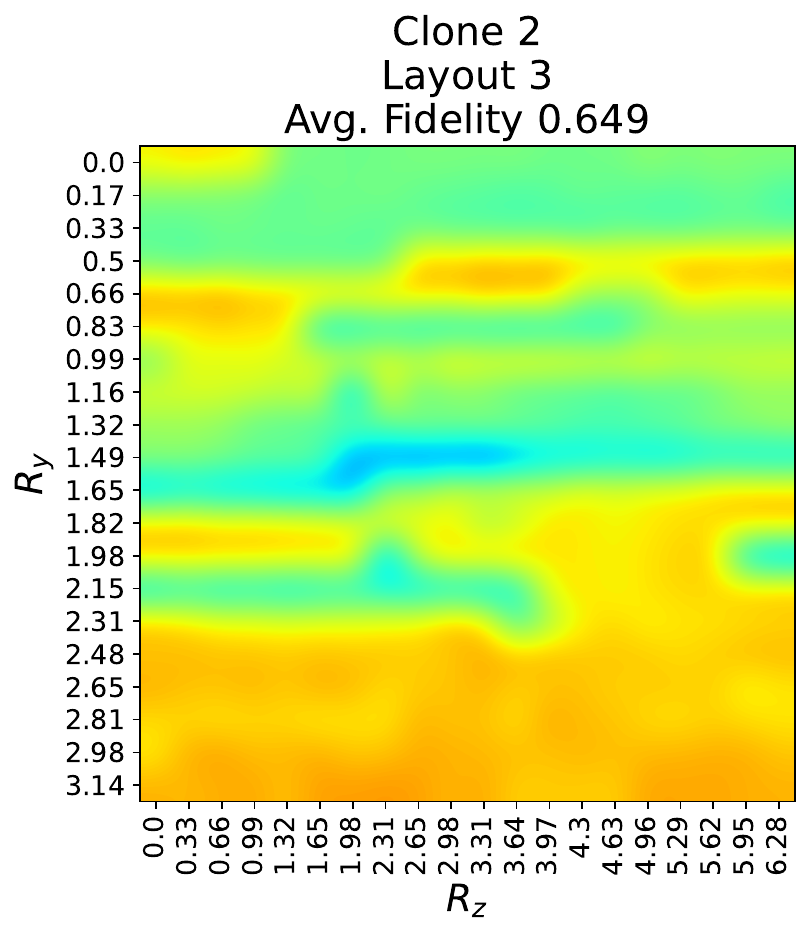}
    \includegraphics[width=0.13\textwidth]{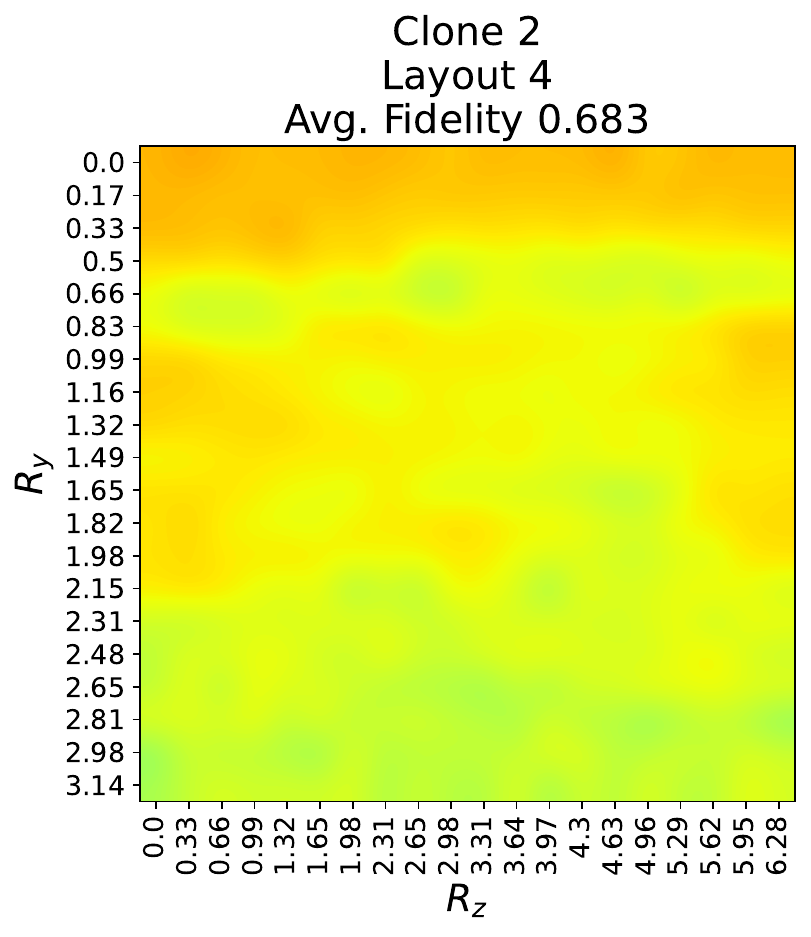}
    \includegraphics[width=0.13\textwidth]{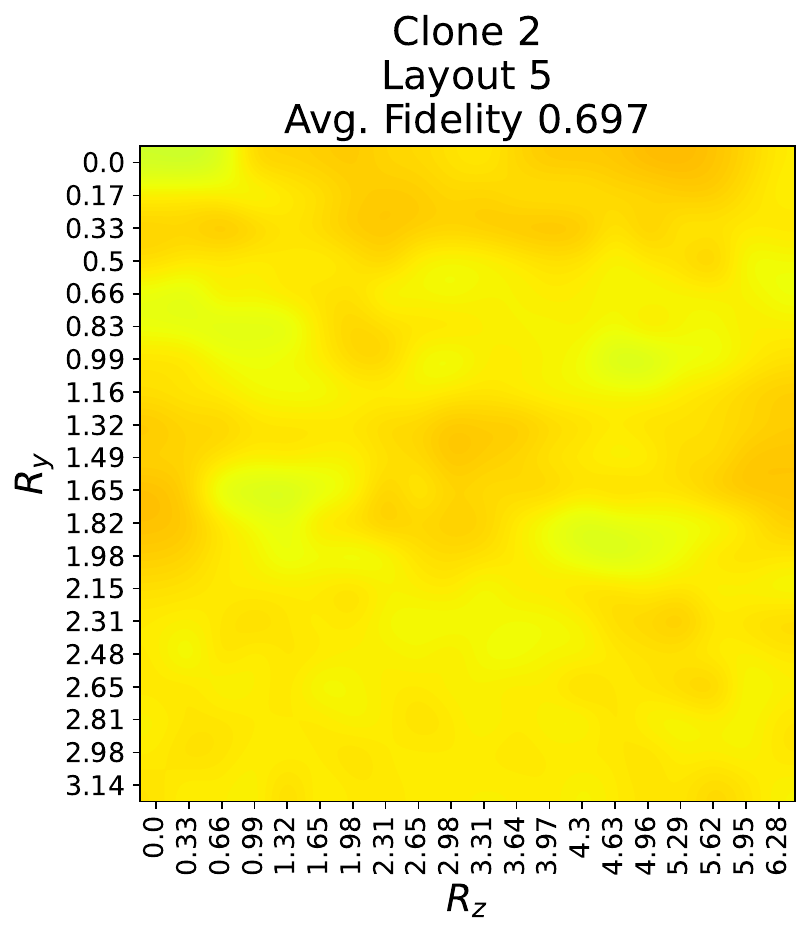}
    \includegraphics[width=0.13\textwidth]{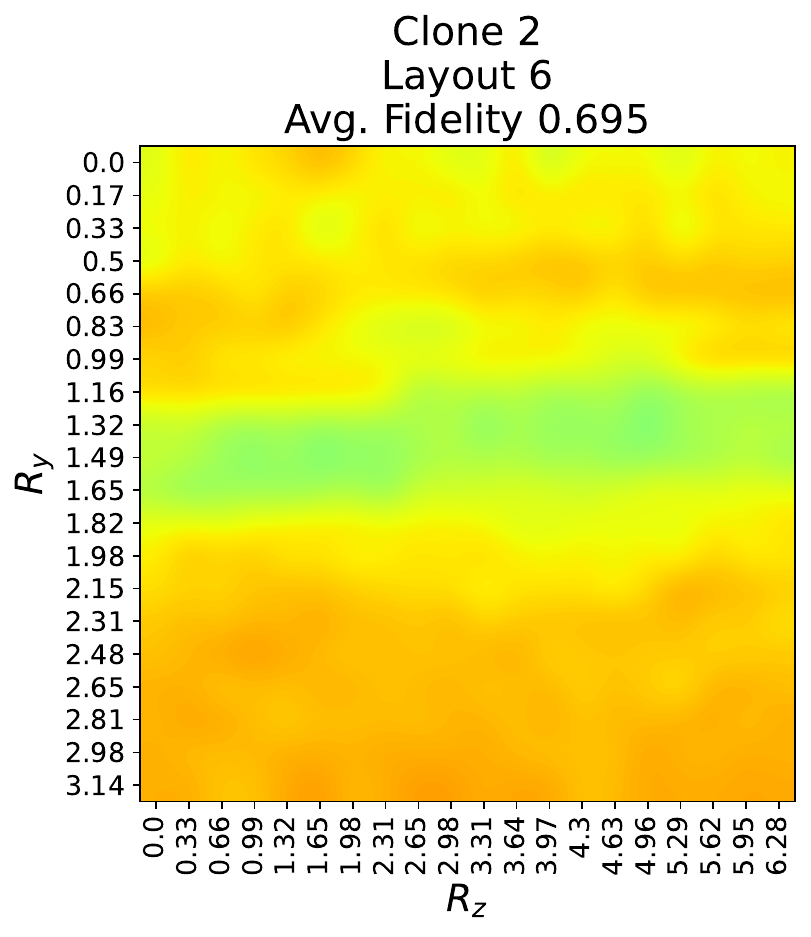}
    \includegraphics[width=0.43\textwidth]{figures/colorbar.pdf}\\
    \caption{Bloch sphere vector representations of the computed density matrices (top $3$ rows) and single qubit clone fidelity heatmaps (bottom $3$ rows) of the single qubit clones for $M=3$ with no ancilla, executed with dynamical decoupling. Each column corresponds to the $7$ different compiled hardware layouts. Each row corresponds to the $3$ different single qubit clones. The rows and column ordering of the sub-figures is the same between the Bloch sphere vector representations and the clone fidelity heatmaps. Data from \texttt{ibm\_hanoi}.  }
    \label{fig:fidelity_heatmaps_M3_ibm_hanoi_no_ancilla_DD}
\end{figure*}

\begin{figure*}[th!]
    \centering
    \includegraphics[width=0.13\textwidth]{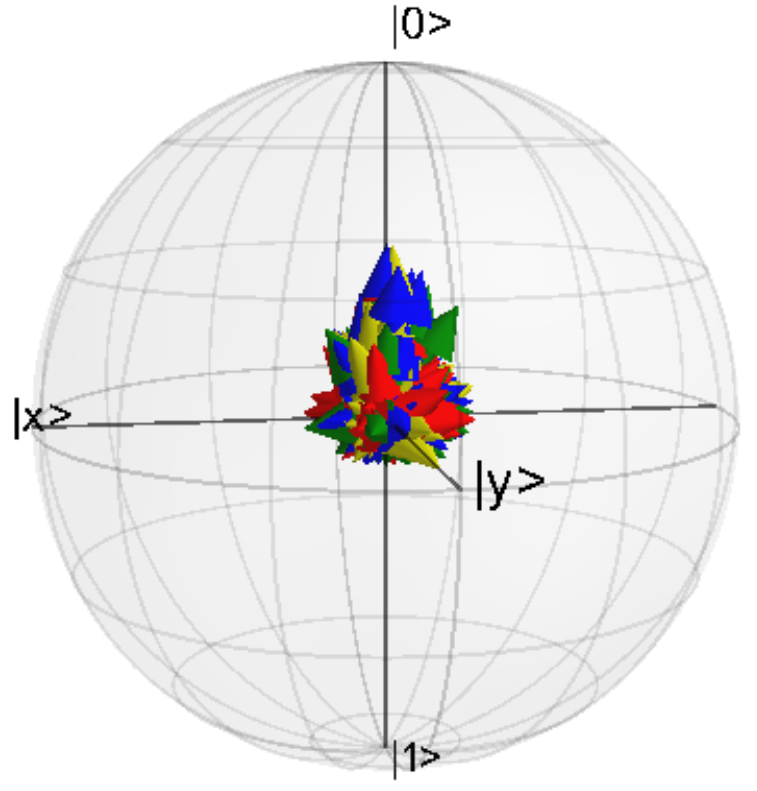}
    \includegraphics[width=0.13\textwidth]{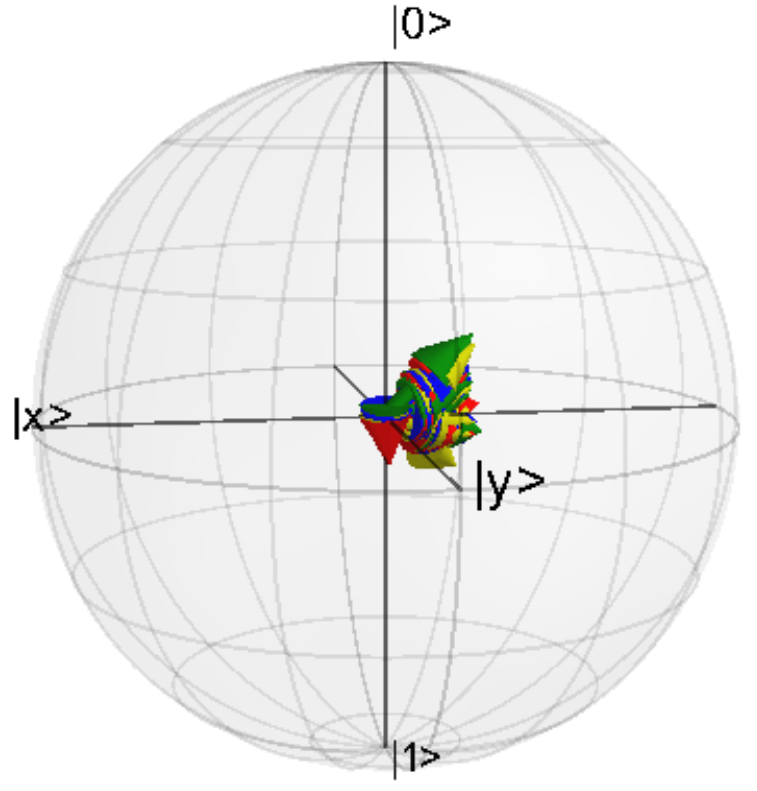}
    \includegraphics[width=0.13\textwidth]{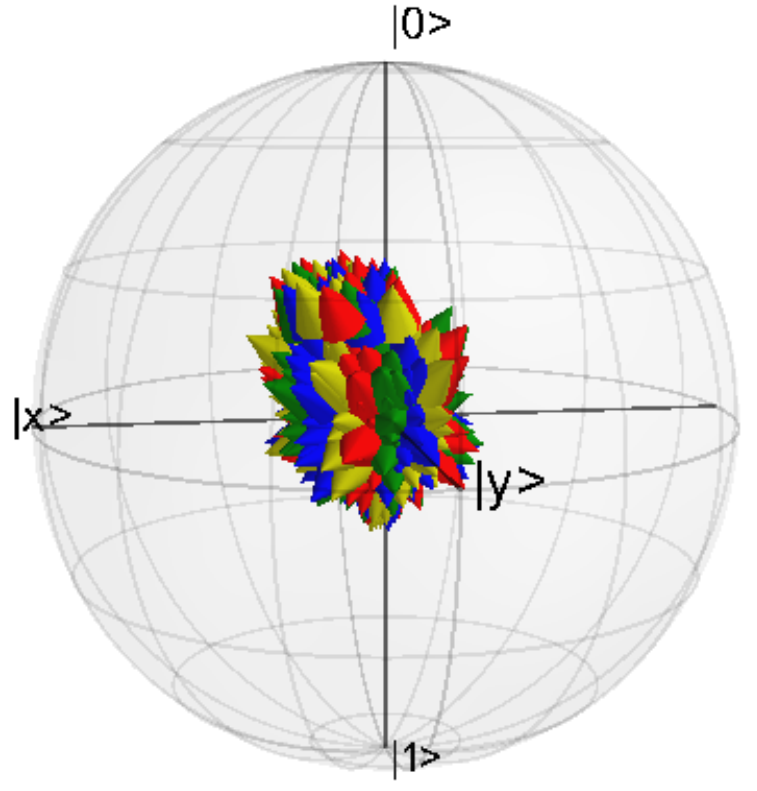}
    \includegraphics[width=0.13\textwidth]{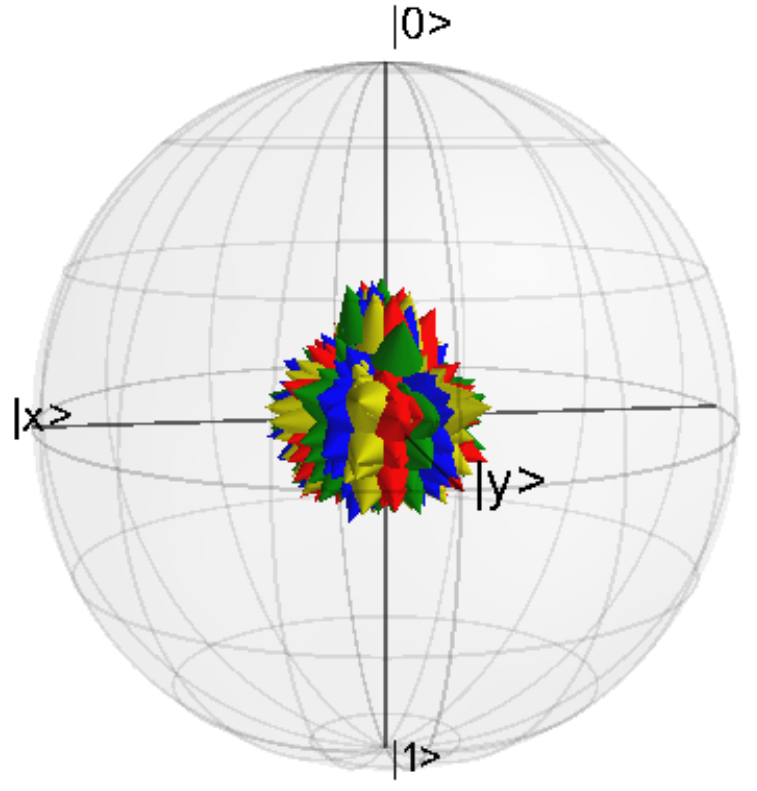}
    \includegraphics[width=0.13\textwidth]{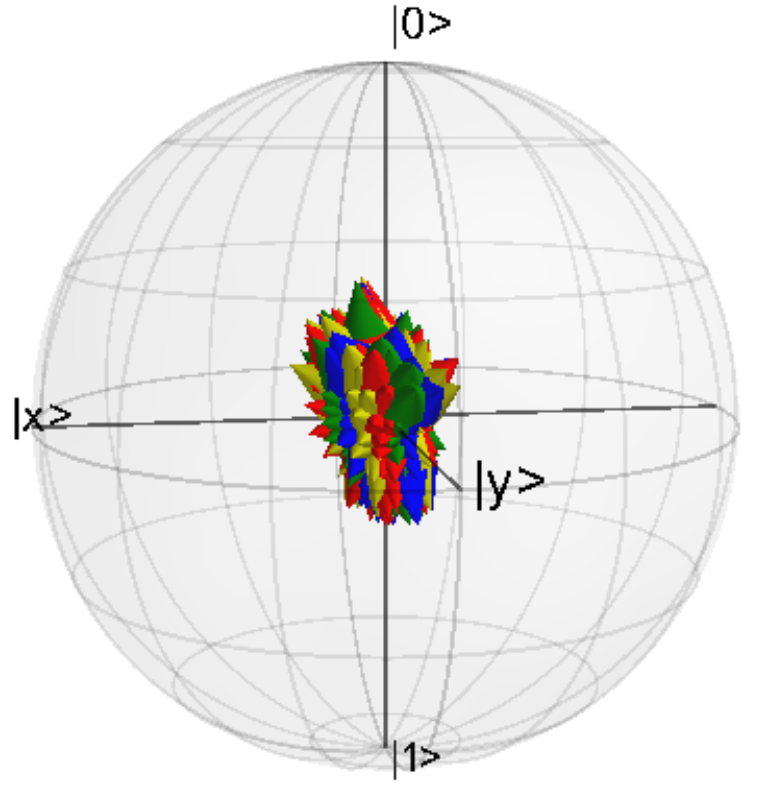}
    \includegraphics[width=0.13\textwidth]{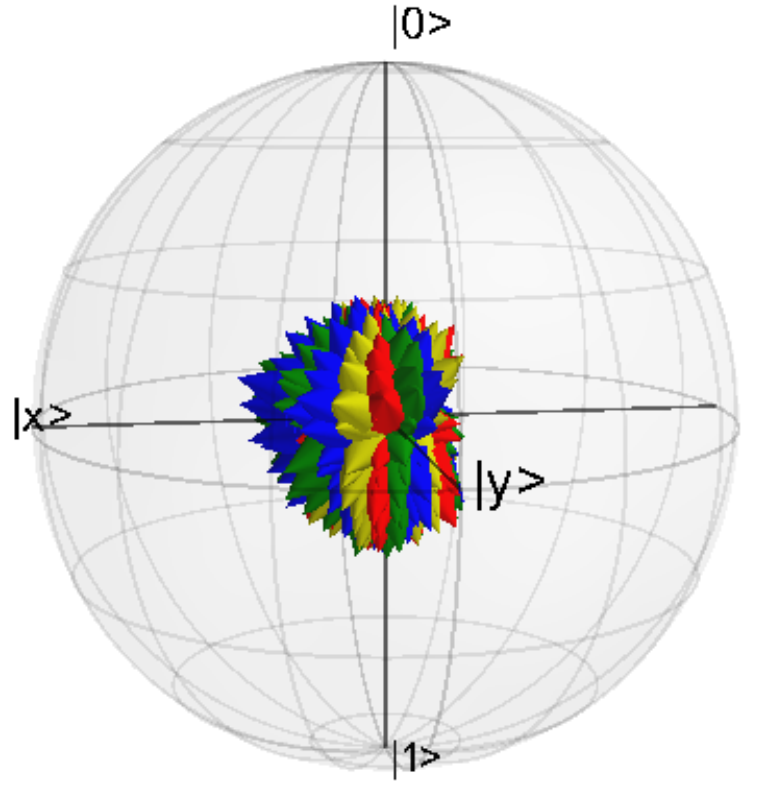}
    \includegraphics[width=0.13\textwidth]{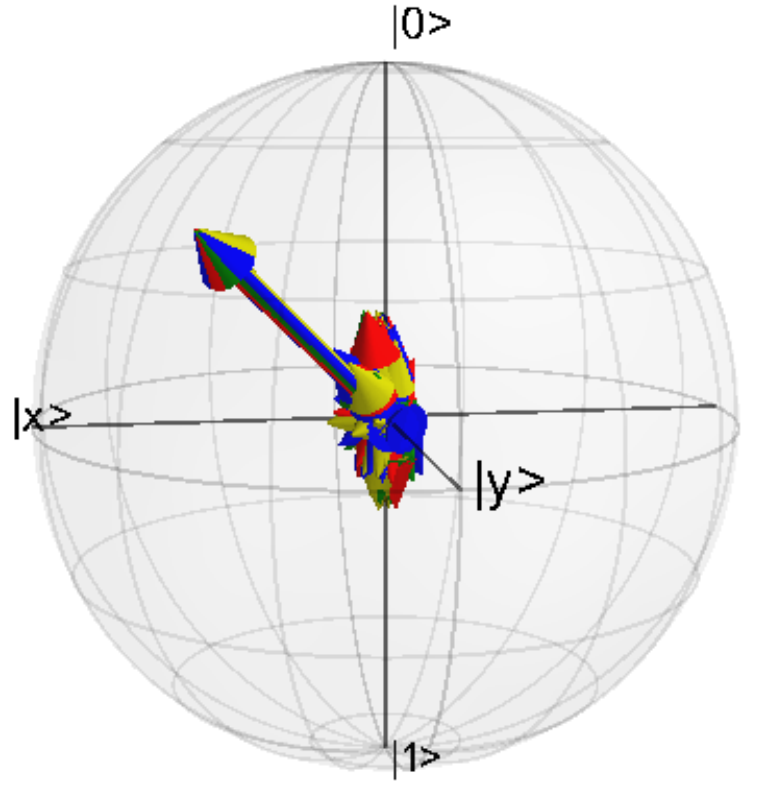}
    \includegraphics[width=0.13\textwidth]{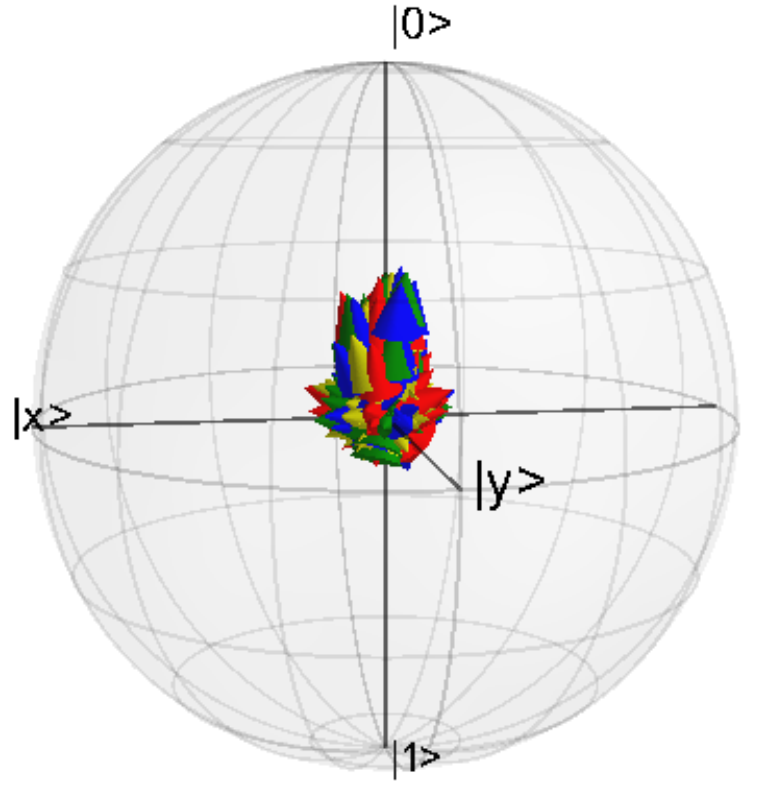}
    \includegraphics[width=0.13\textwidth]{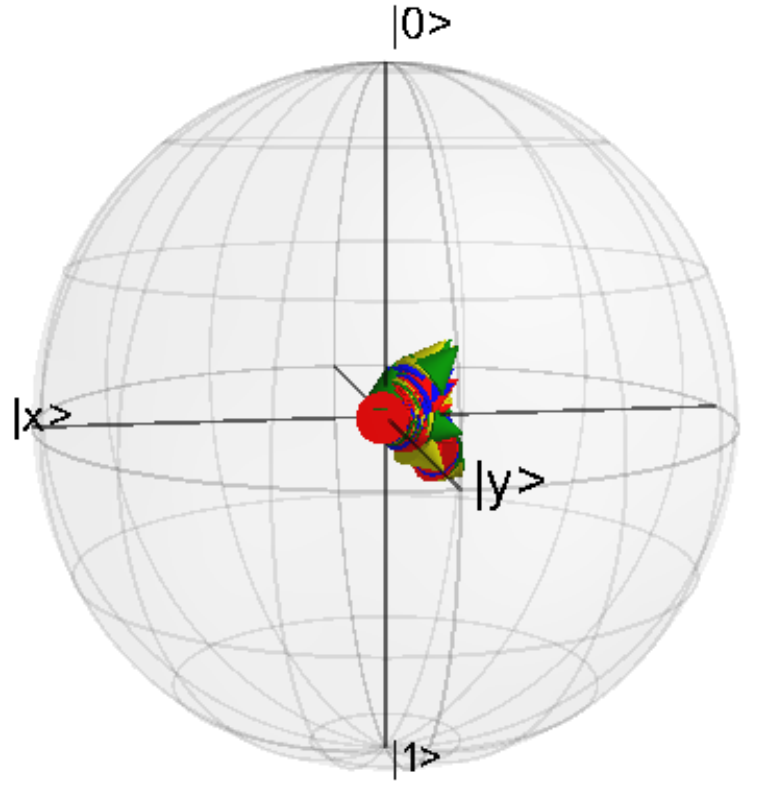}
    \includegraphics[width=0.13\textwidth]{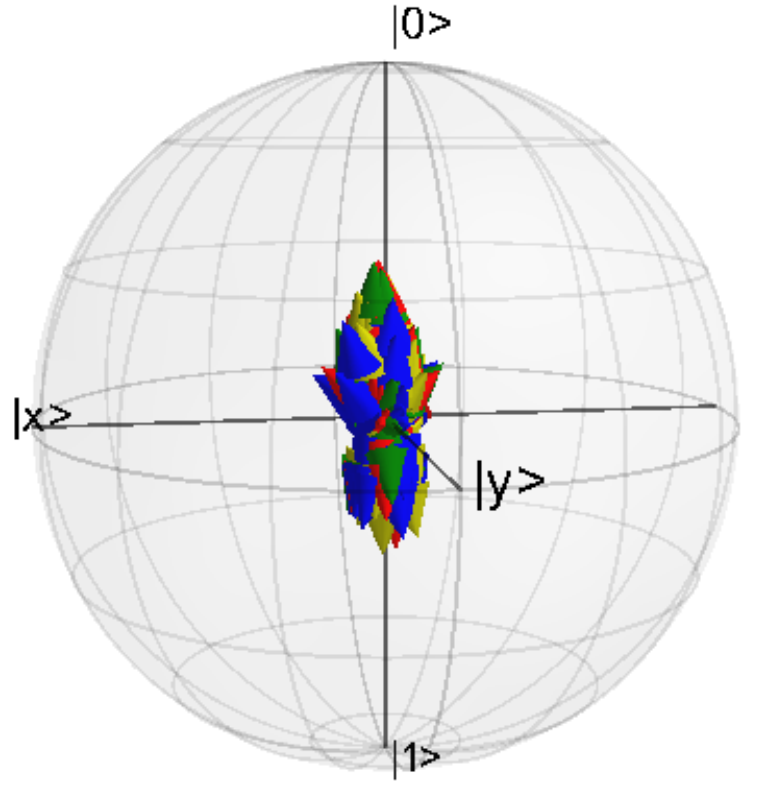}
    \includegraphics[width=0.13\textwidth]{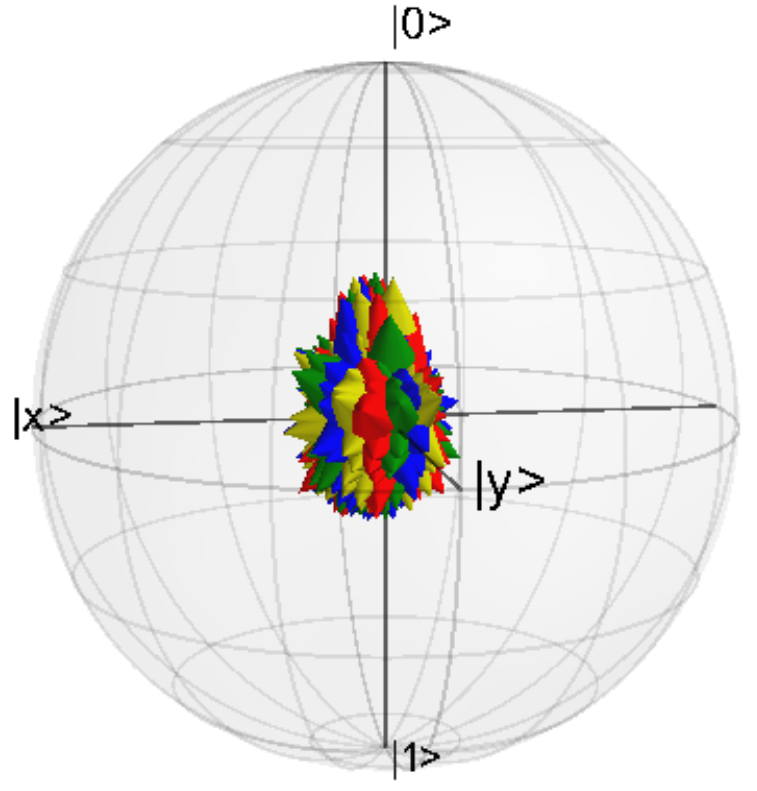}
    \includegraphics[width=0.13\textwidth]{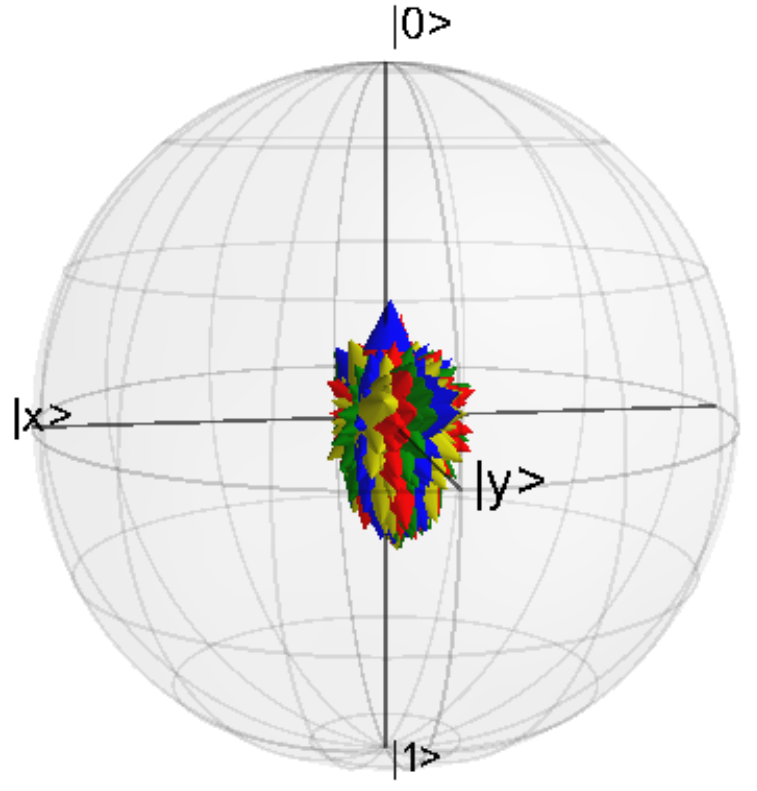}
    \includegraphics[width=0.13\textwidth]{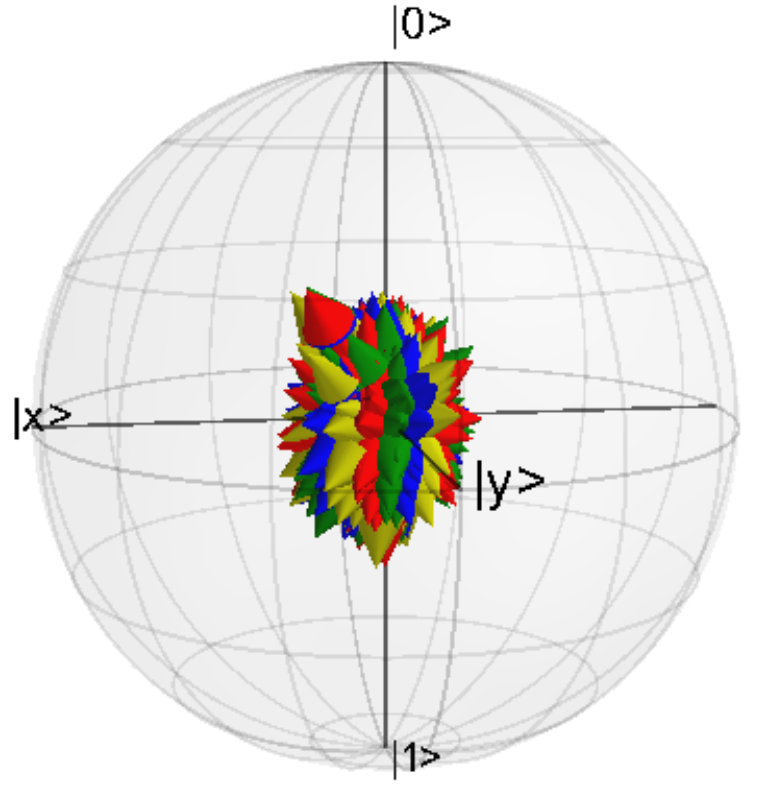}
    \includegraphics[width=0.13\textwidth]{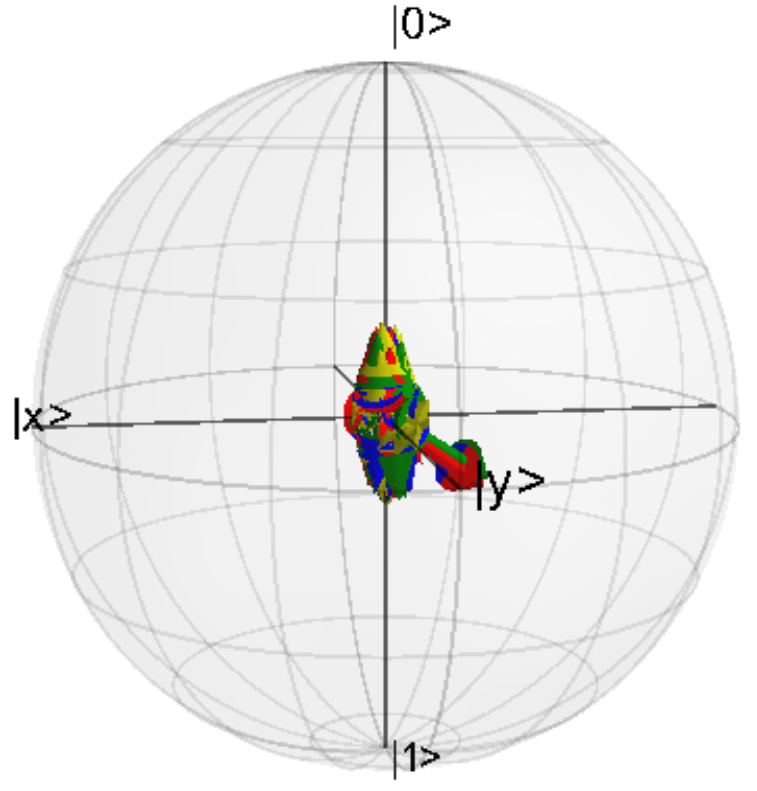}
    \includegraphics[width=0.13\textwidth]{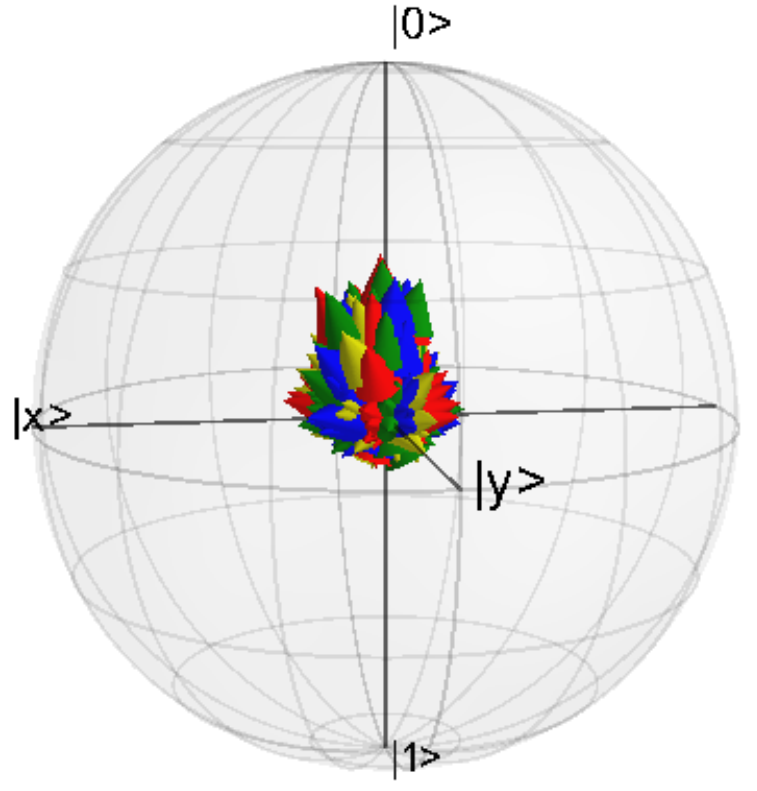}
    \includegraphics[width=0.13\textwidth]{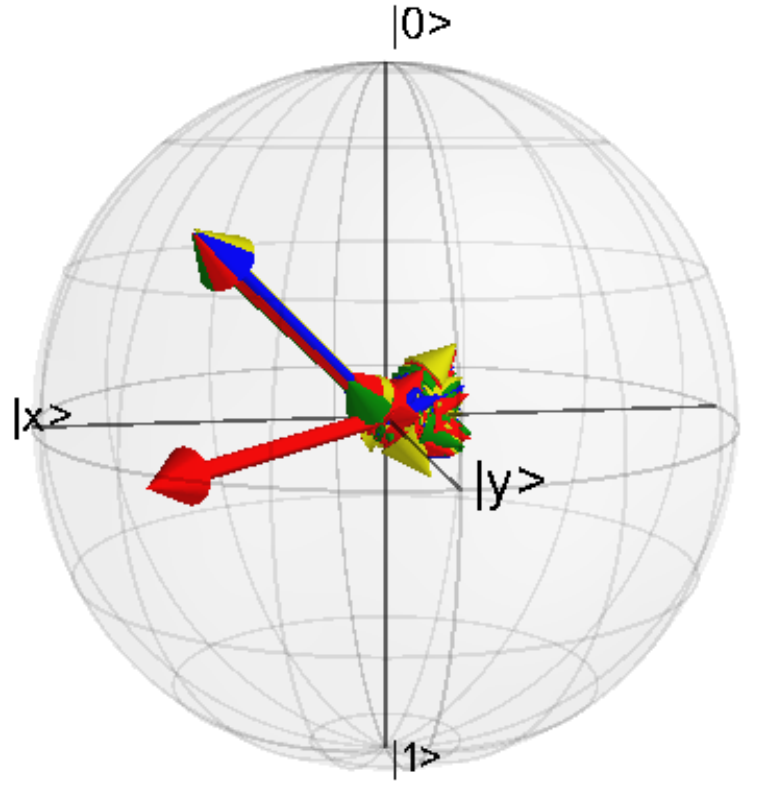}
    \includegraphics[width=0.13\textwidth]{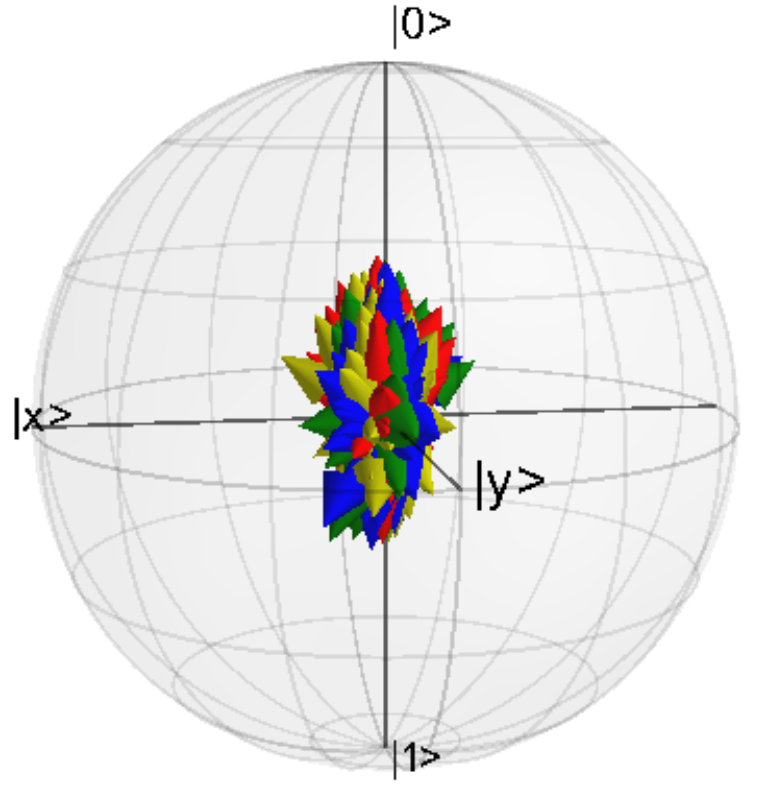}
    \includegraphics[width=0.13\textwidth]{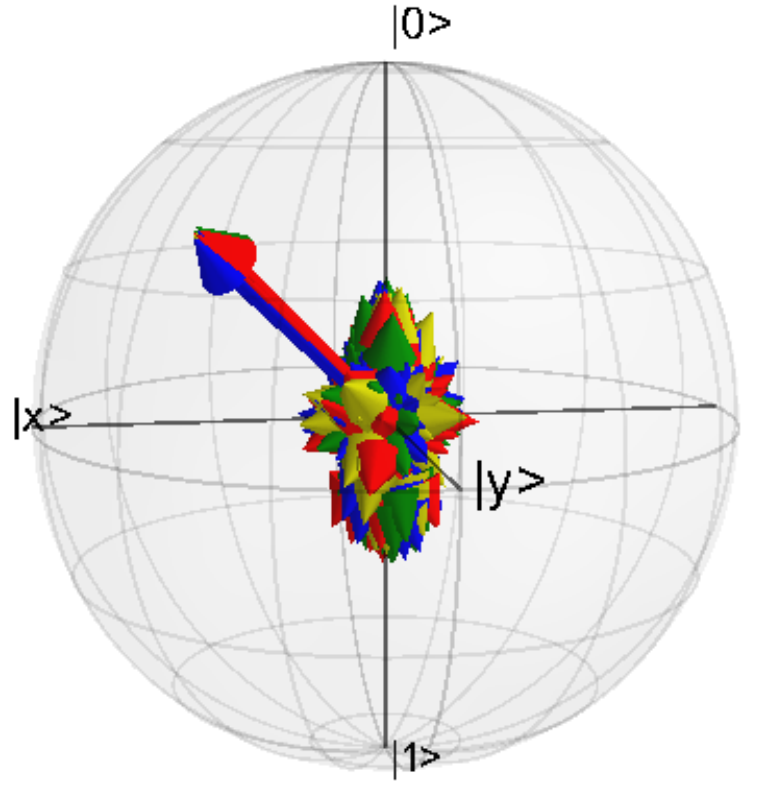}
    \includegraphics[width=0.13\textwidth]{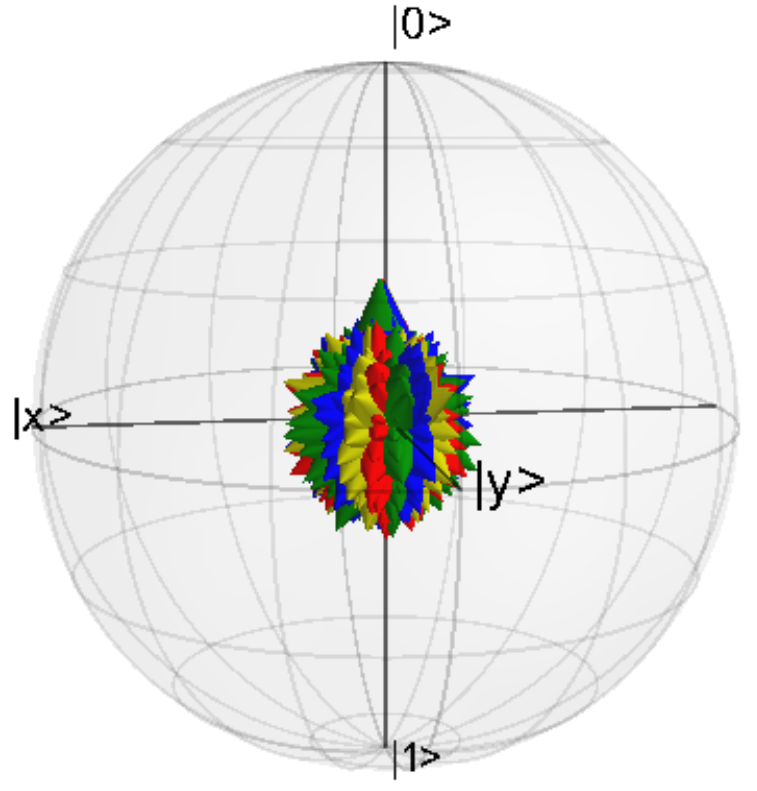}
    \includegraphics[width=0.13\textwidth]{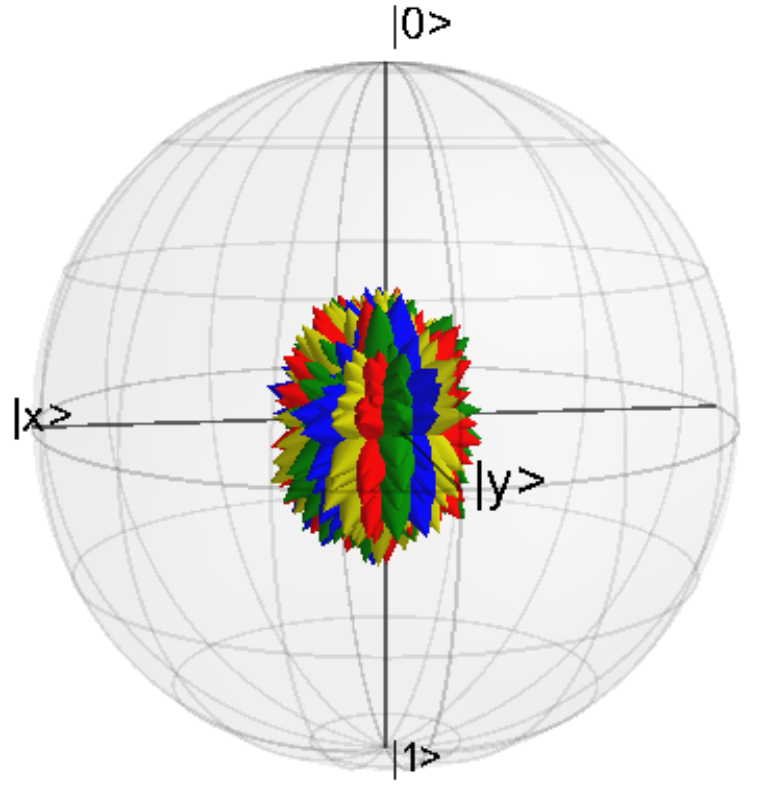}
    \includegraphics[width=0.13\textwidth]{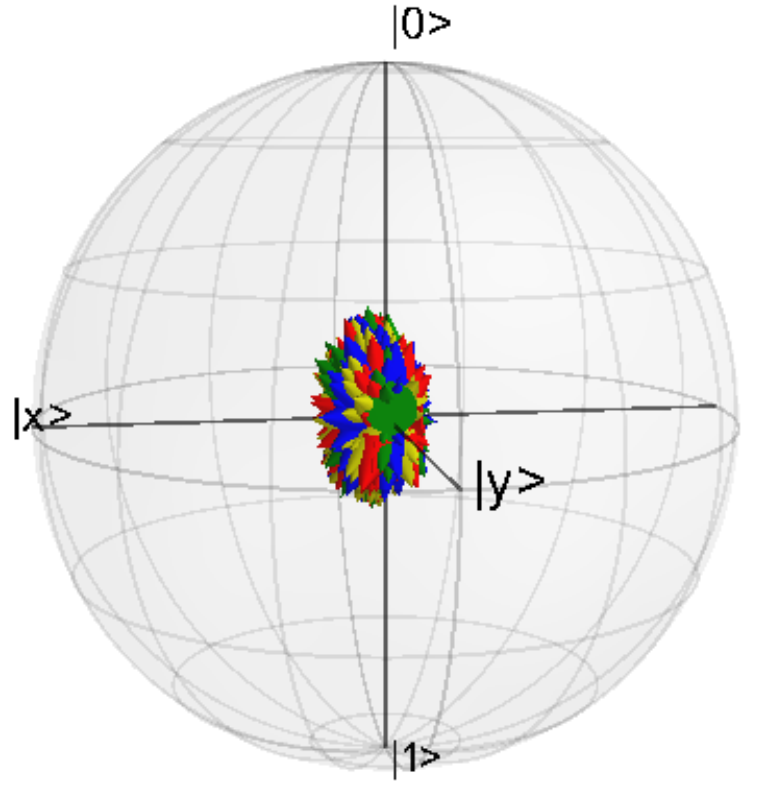}
    \includegraphics[width=0.13\textwidth]{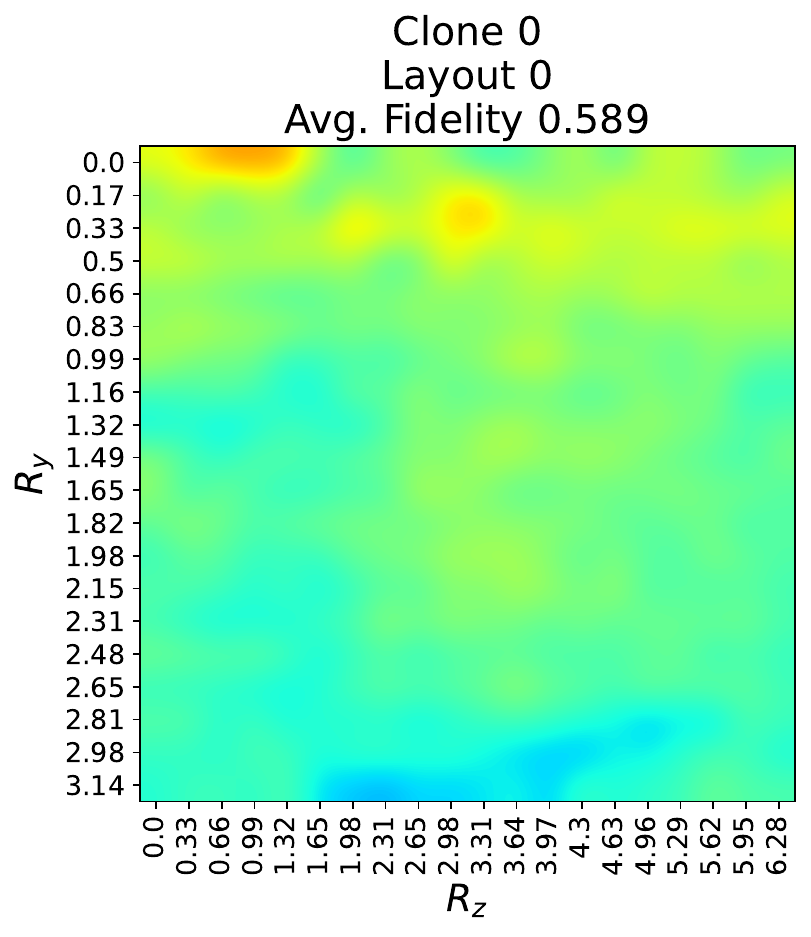}
    \includegraphics[width=0.13\textwidth]{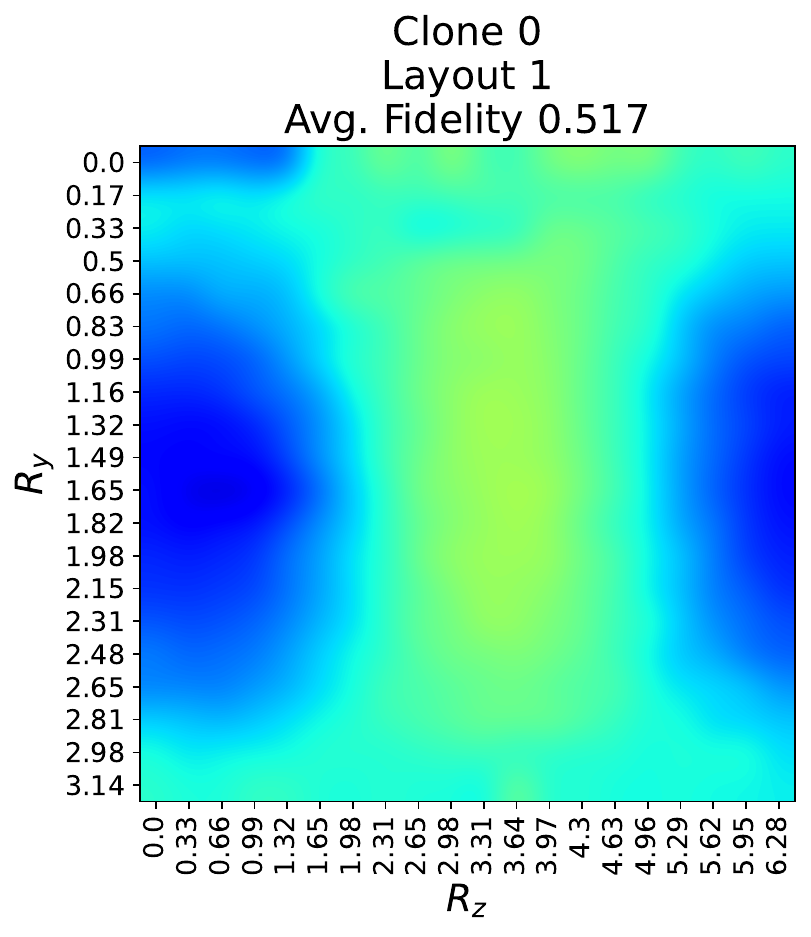}
    \includegraphics[width=0.13\textwidth]{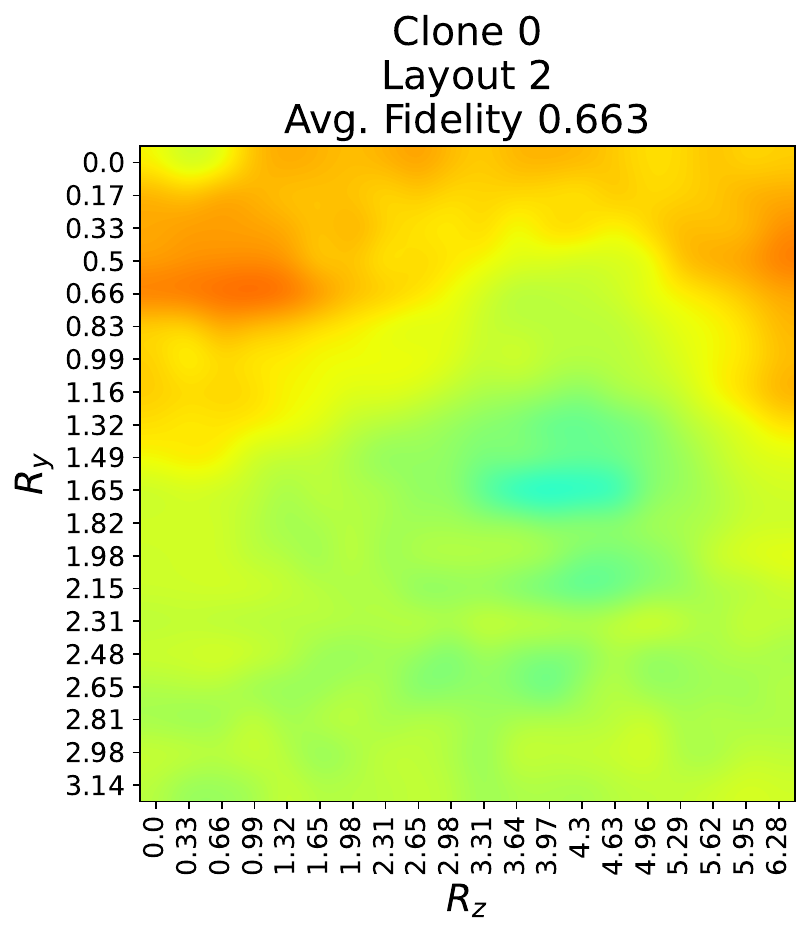}
    \includegraphics[width=0.13\textwidth]{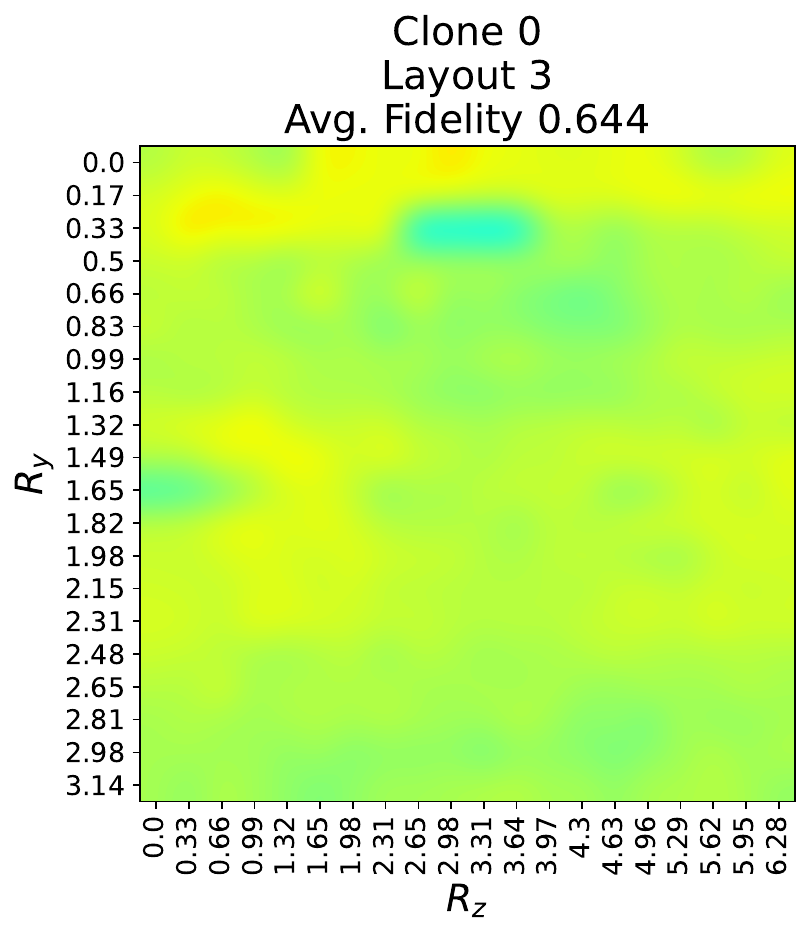}
    \includegraphics[width=0.13\textwidth]{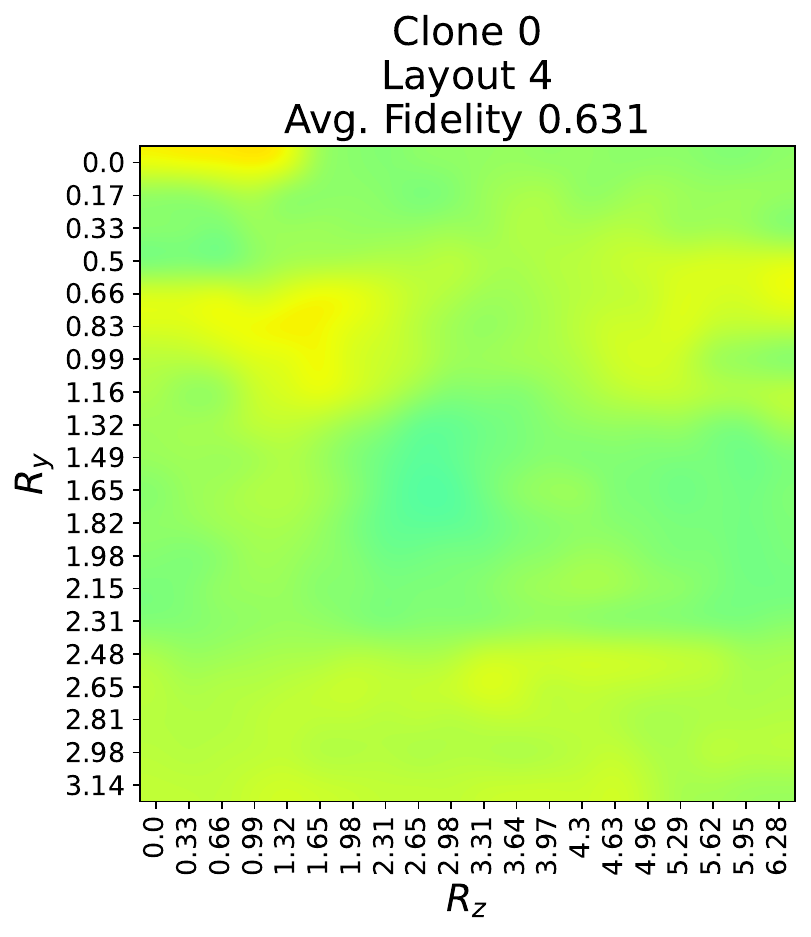}
    \includegraphics[width=0.13\textwidth]{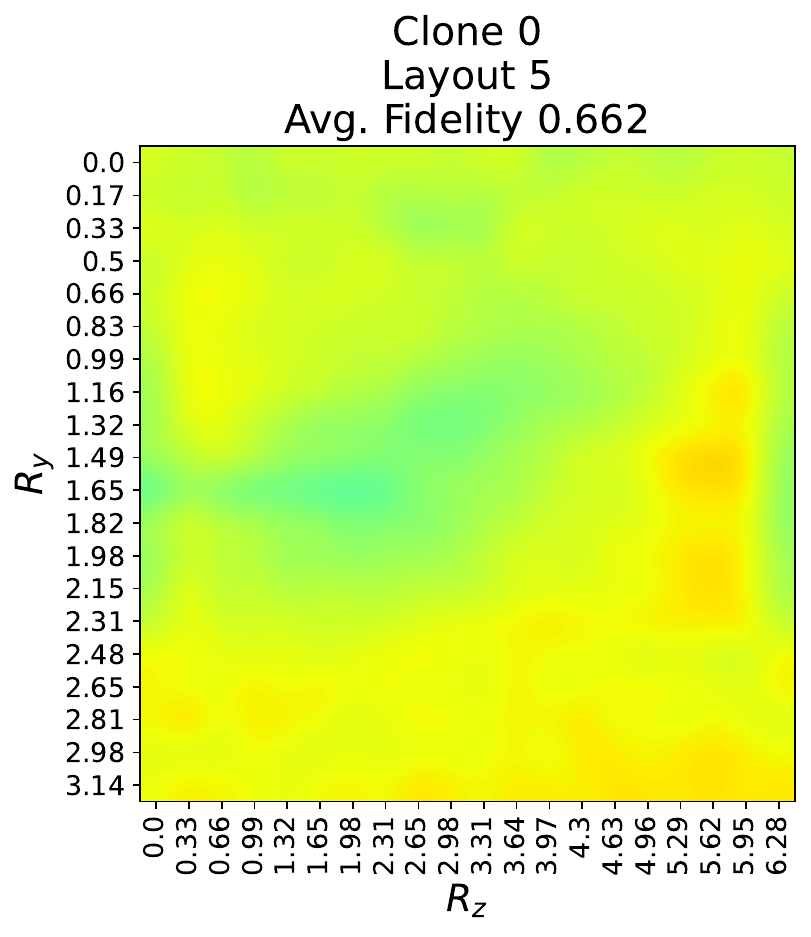}
    \includegraphics[width=0.13\textwidth]{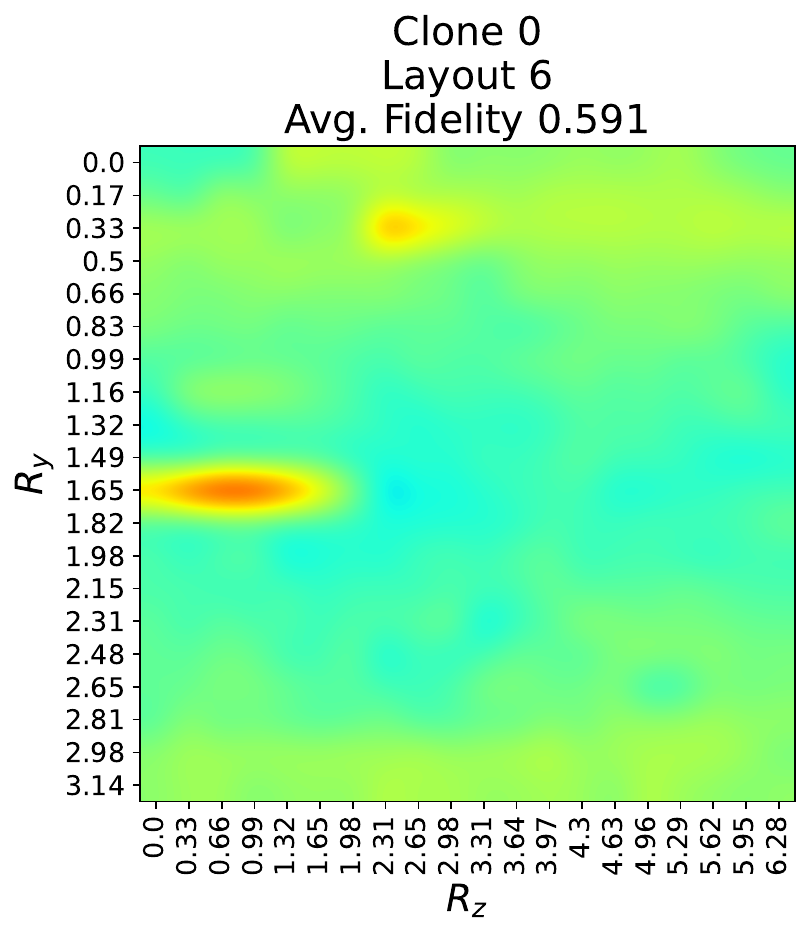}
    \includegraphics[width=0.13\textwidth]{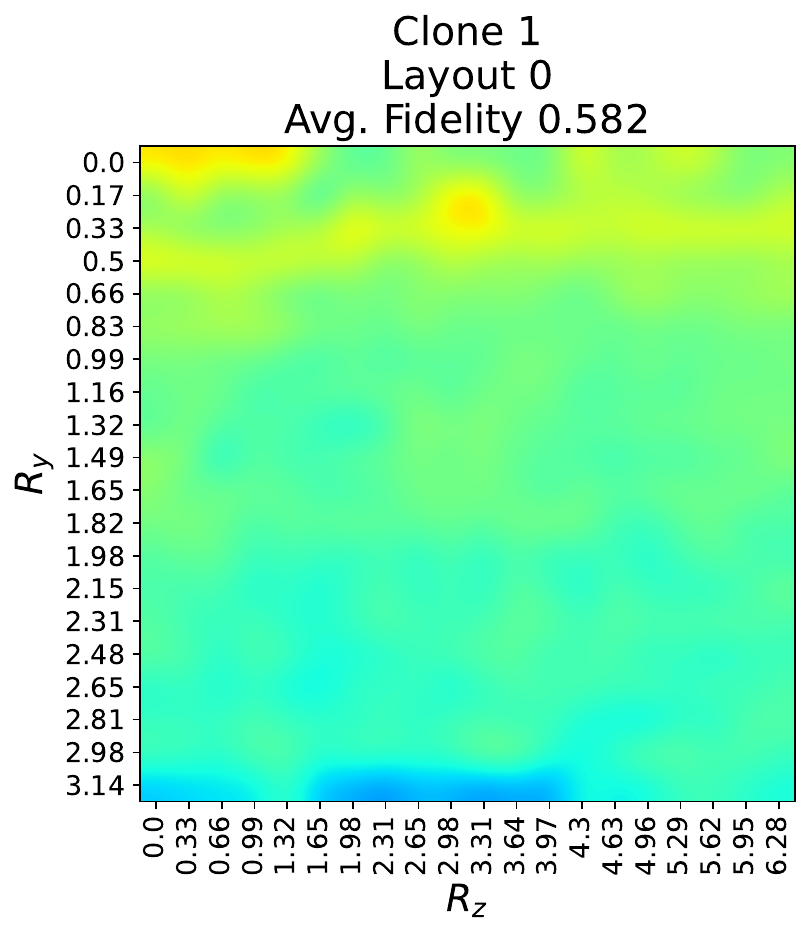}
    \includegraphics[width=0.13\textwidth]{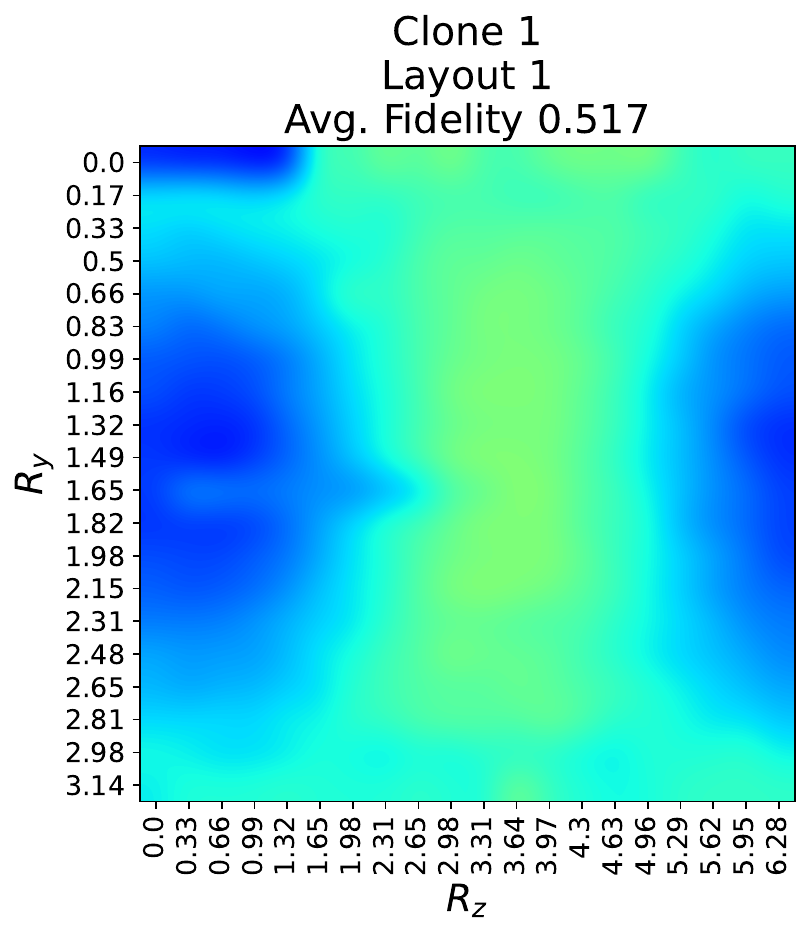}
    \includegraphics[width=0.13\textwidth]{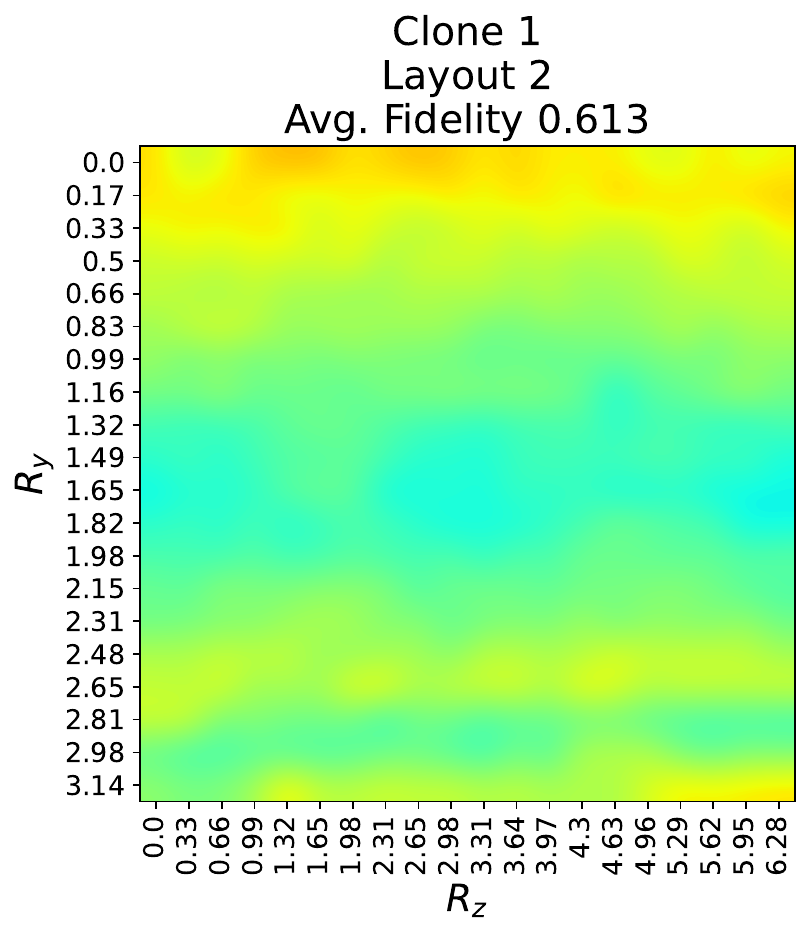}
    \includegraphics[width=0.13\textwidth]{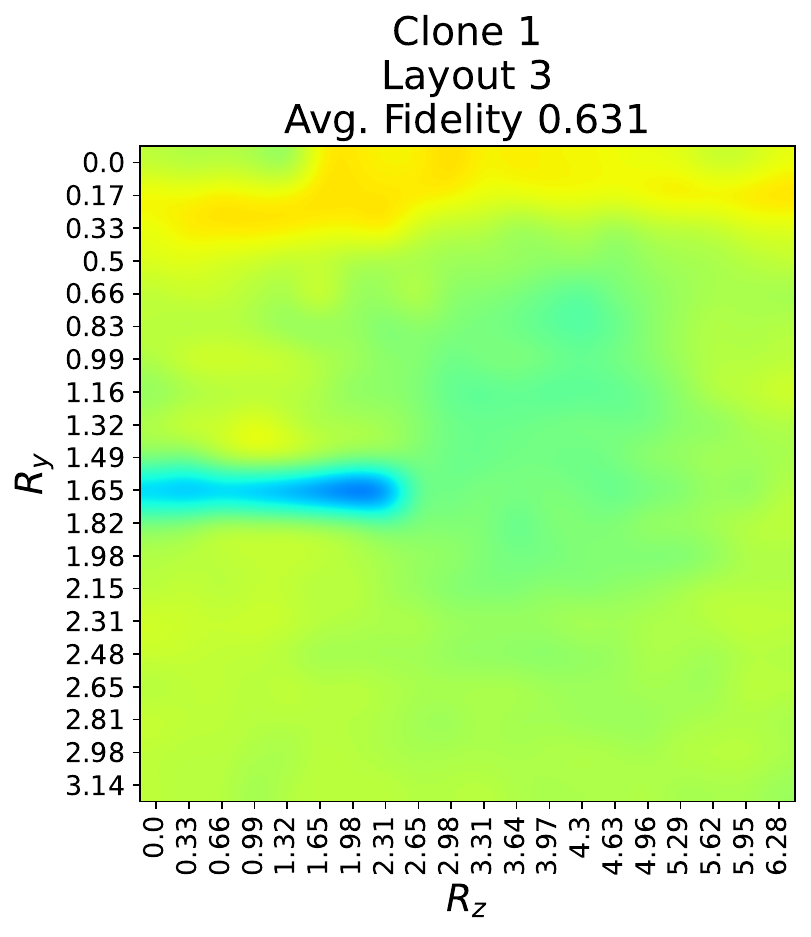}
    \includegraphics[width=0.13\textwidth]{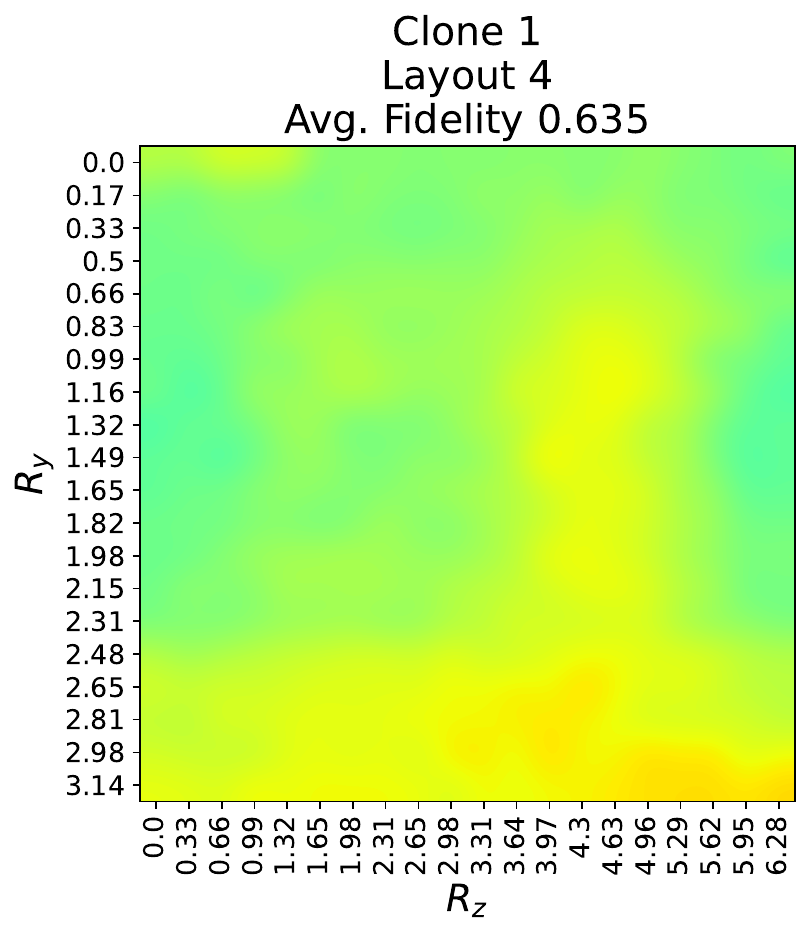}
    \includegraphics[width=0.13\textwidth]{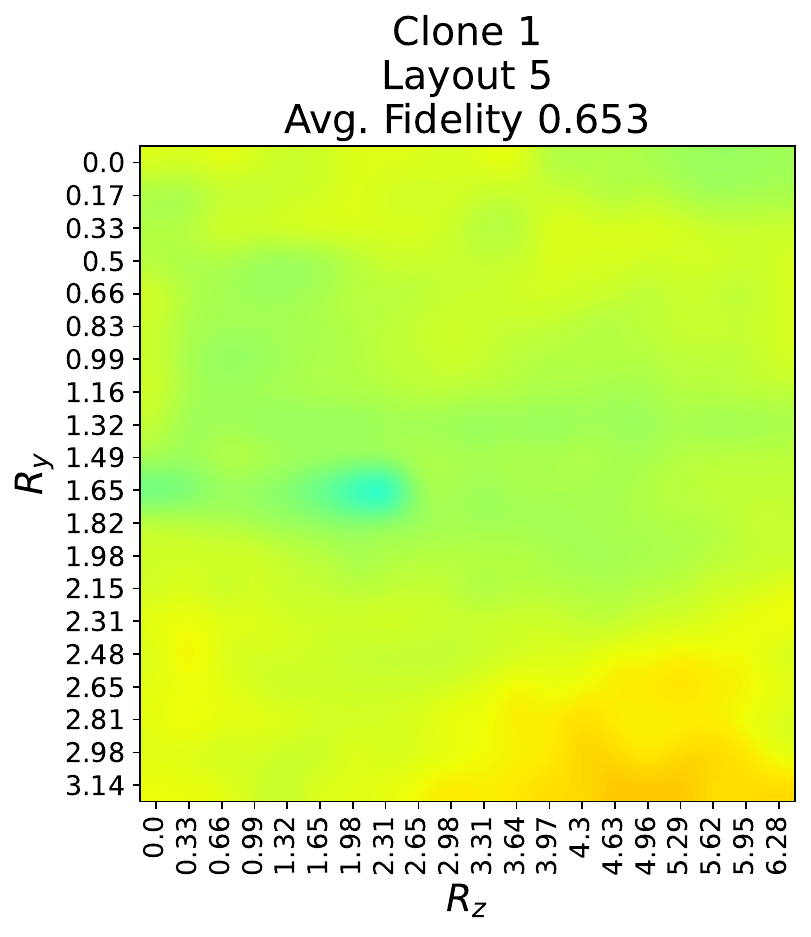}
    \includegraphics[width=0.13\textwidth]{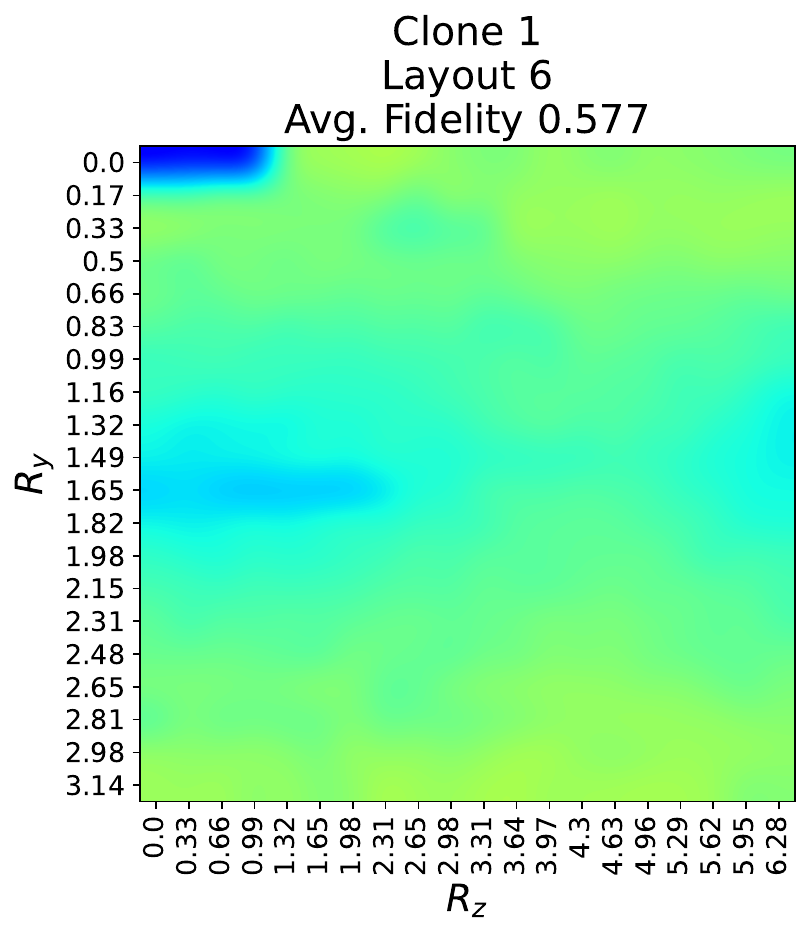}
    \includegraphics[width=0.13\textwidth]{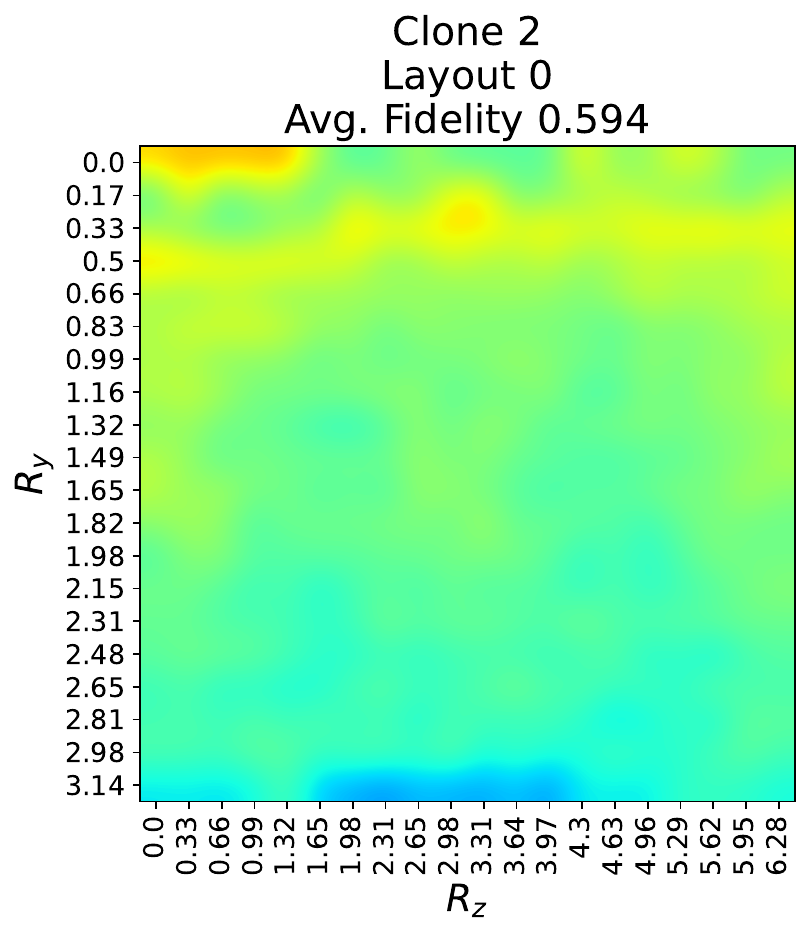}
    \includegraphics[width=0.13\textwidth]{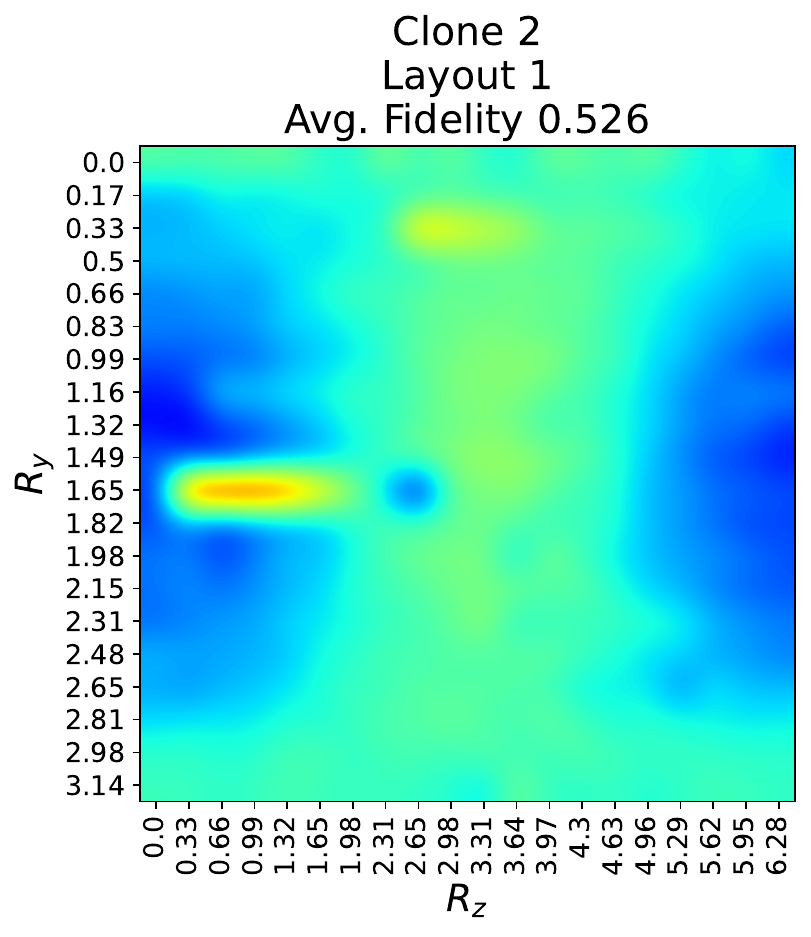}
    \includegraphics[width=0.13\textwidth]{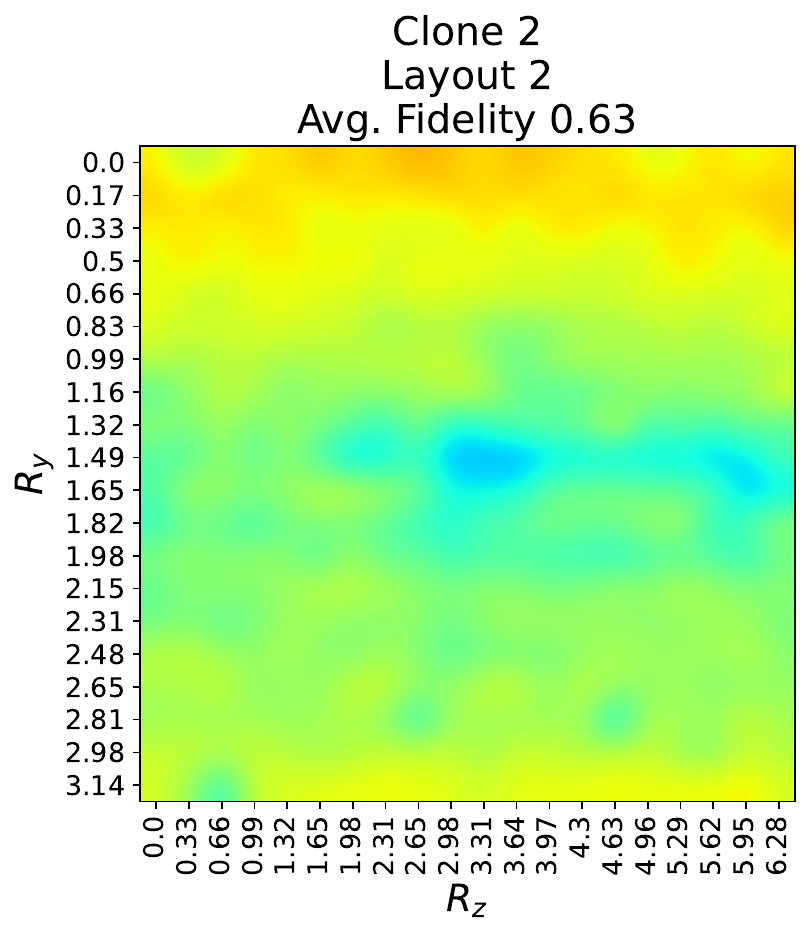}
    \includegraphics[width=0.13\textwidth]{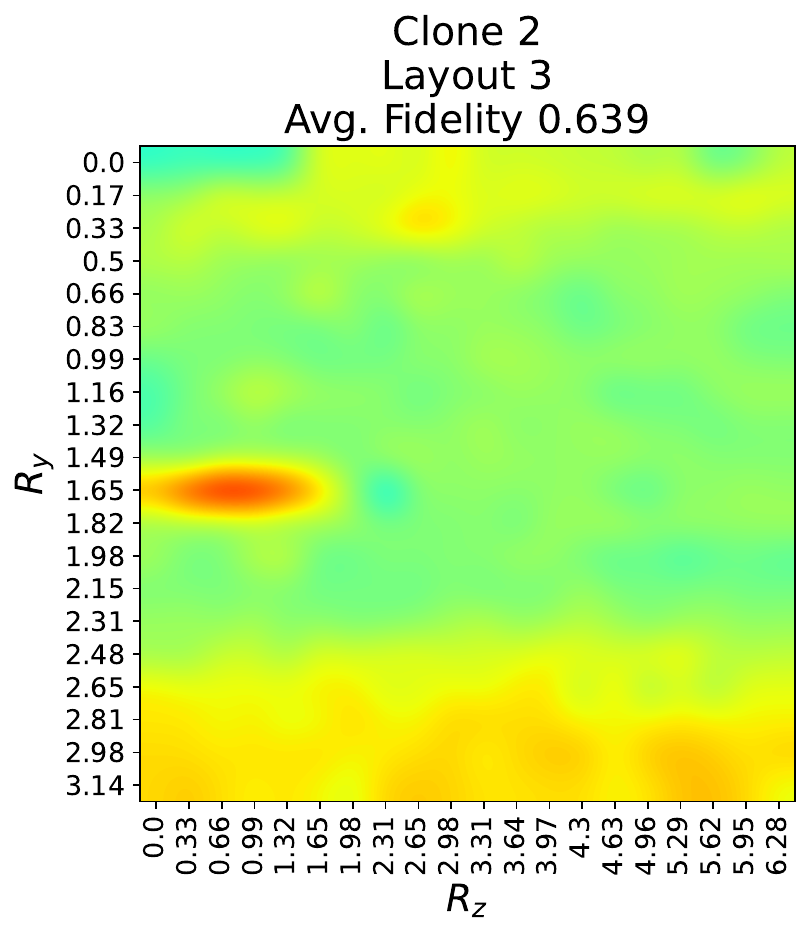}
    \includegraphics[width=0.13\textwidth]{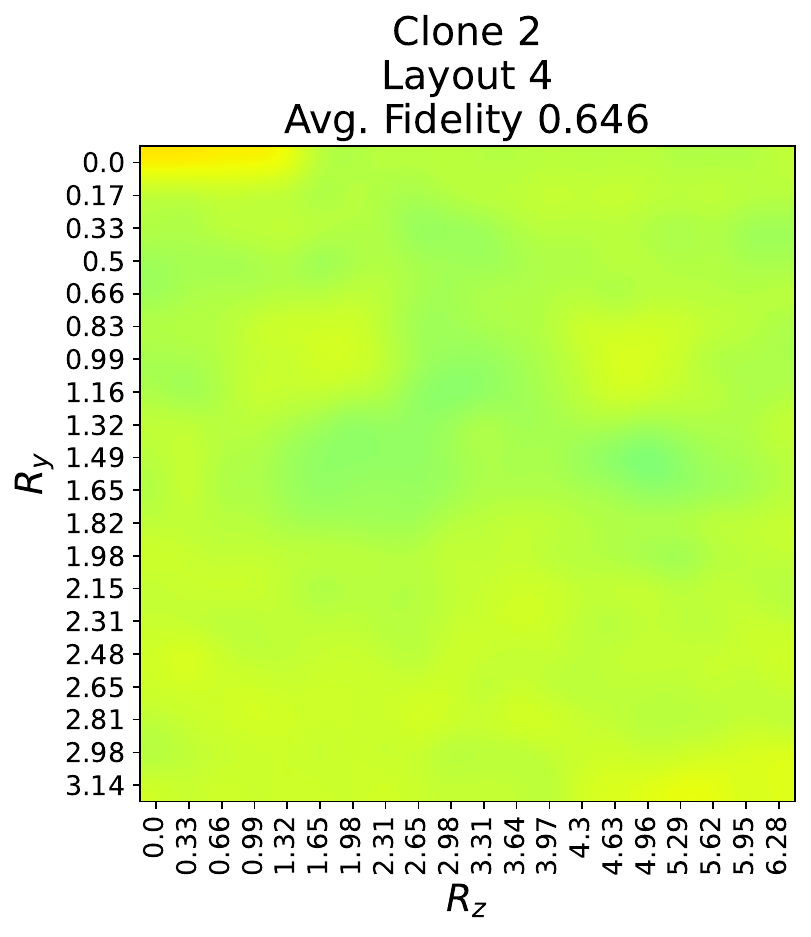}
    \includegraphics[width=0.13\textwidth]{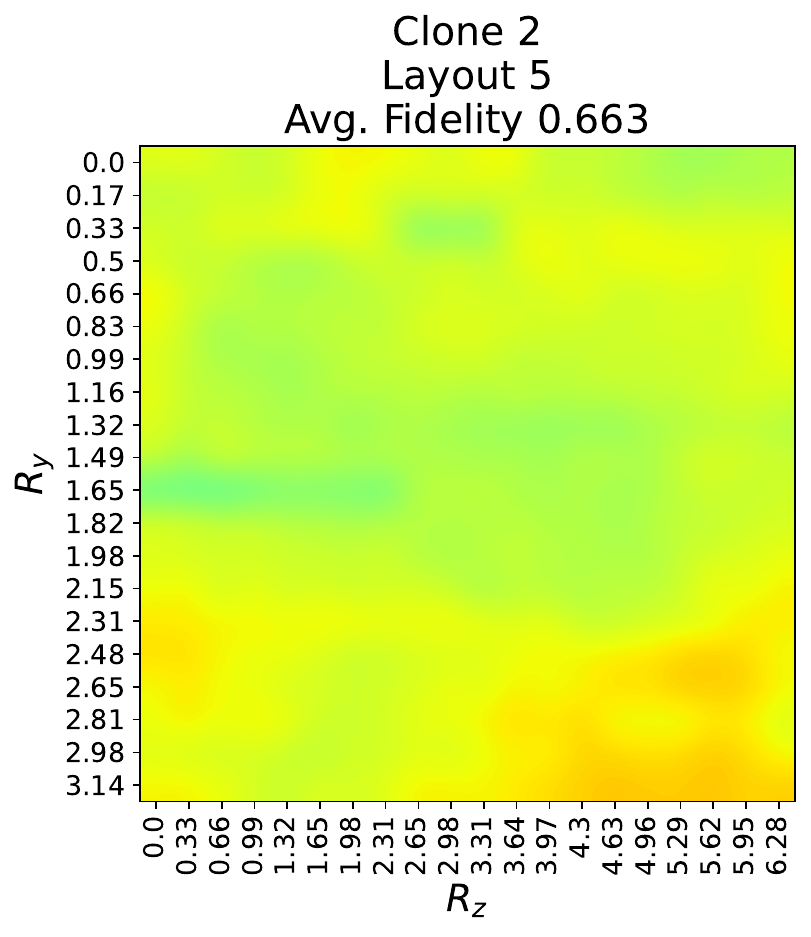}
    \includegraphics[width=0.13\textwidth]{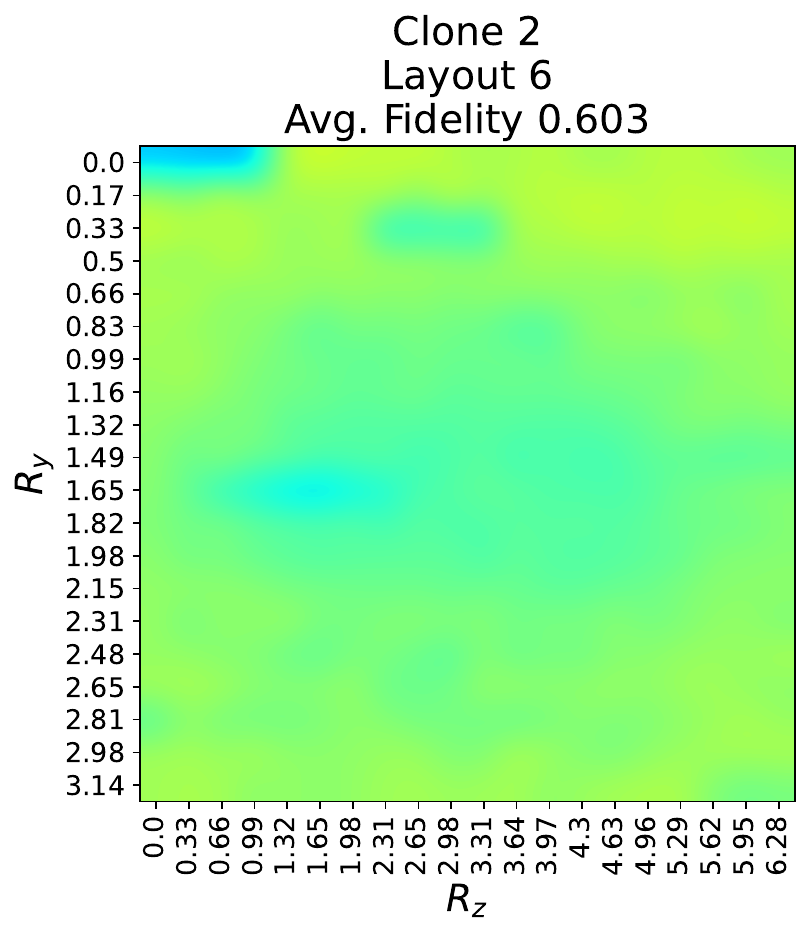}
    \includegraphics[width=0.43\textwidth]{figures/colorbar.pdf}\\
    \caption{Bloch sphere vector representations of the computed density matrices (top $3$ rows) and single qubit clone fidelity heatmaps (bottom $3$ rows) of the single qubit clones for $M=3$ with no ancilla, executed with dynamical decoupling. Each column corresponds to the $7$ different compiled hardware layouts. Each row corresponds to the $3$ different single qubit clones. The rows and column ordering of the sub-figures is the same between the Bloch sphere vector representations and the clone fidelity heatmaps. Notice that several of the Bloch vectors are clearly outlier density matrices, due to noise drift on the device. Data from \texttt{ibm\_algiers}. }
    \label{fig:fidelity_heatmaps_M3_ibm_algiers_no_ancilla_DD}
\end{figure*}

Figure~\ref{fig:fidelity_heatmaps_M3_ibm_hanoi_no_ancilla_DD} shows the computed clone fidelity for $M=3$ circuits, with the removed ancilla qubit optimization, run on \texttt{ibm\_hanoi} with added dynamical decoupling sequences. Figure~\ref{fig:fidelity_heatmaps_M3_ibm_algiers_no_ancilla_DD} shows the results for the same experimental settings ($M=3$, no ancilla qubits optimization used, and with dynamical decoupling), but run on \texttt{ibm\_algiers}.

\begin{figure*}[th!]
    \centering
    \includegraphics[width=0.13\textwidth]{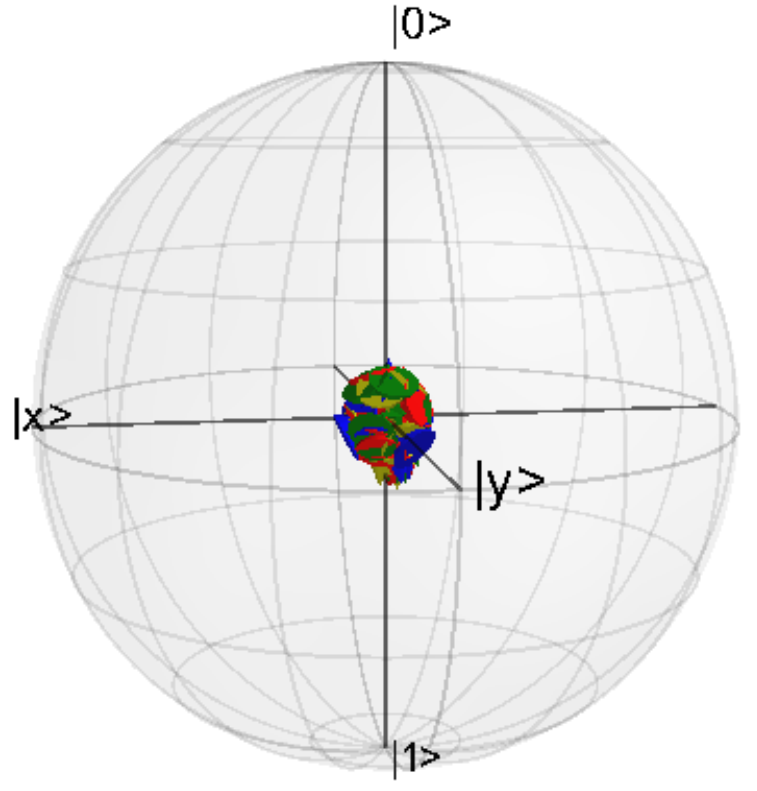}
    \includegraphics[width=0.13\textwidth]{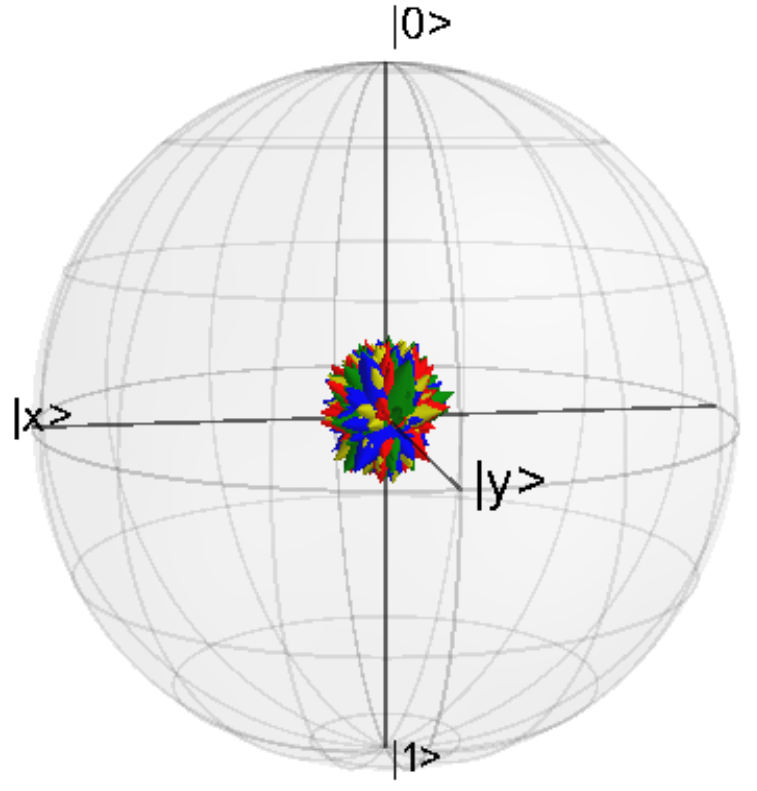}
    \includegraphics[width=0.13\textwidth]{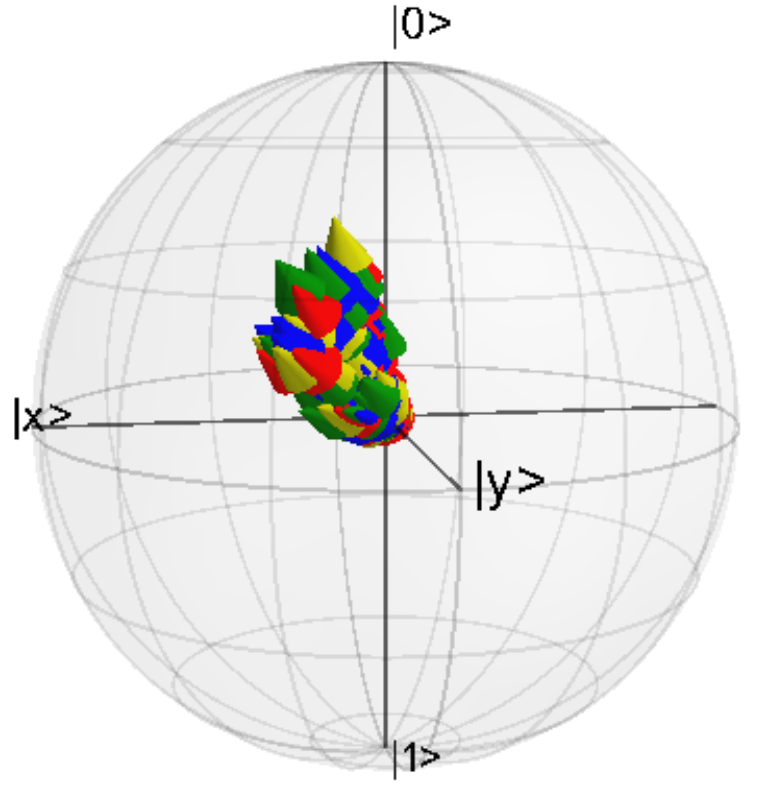}
    \includegraphics[width=0.13\textwidth]{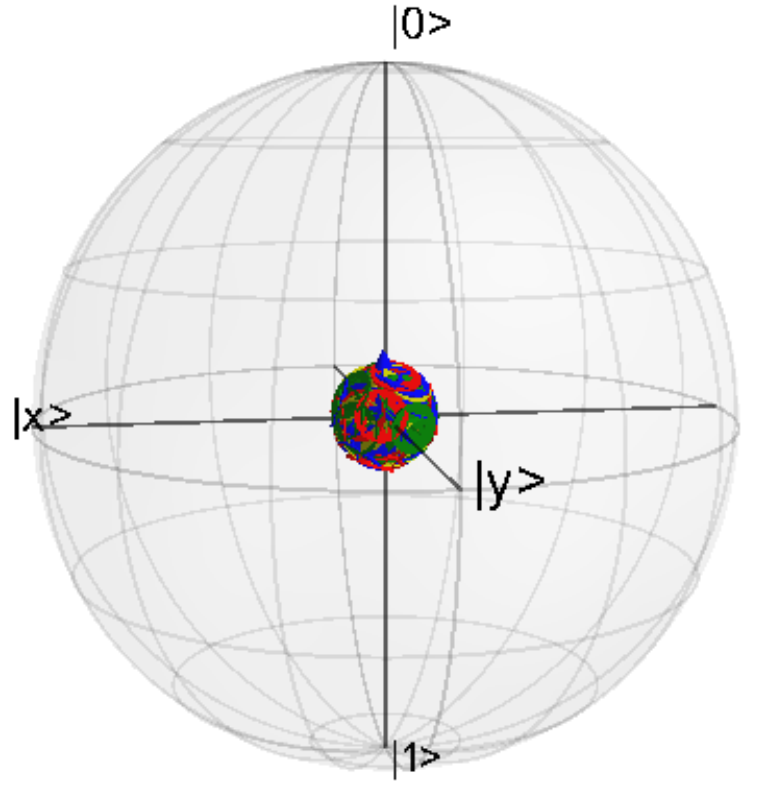}
    \includegraphics[width=0.13\textwidth]{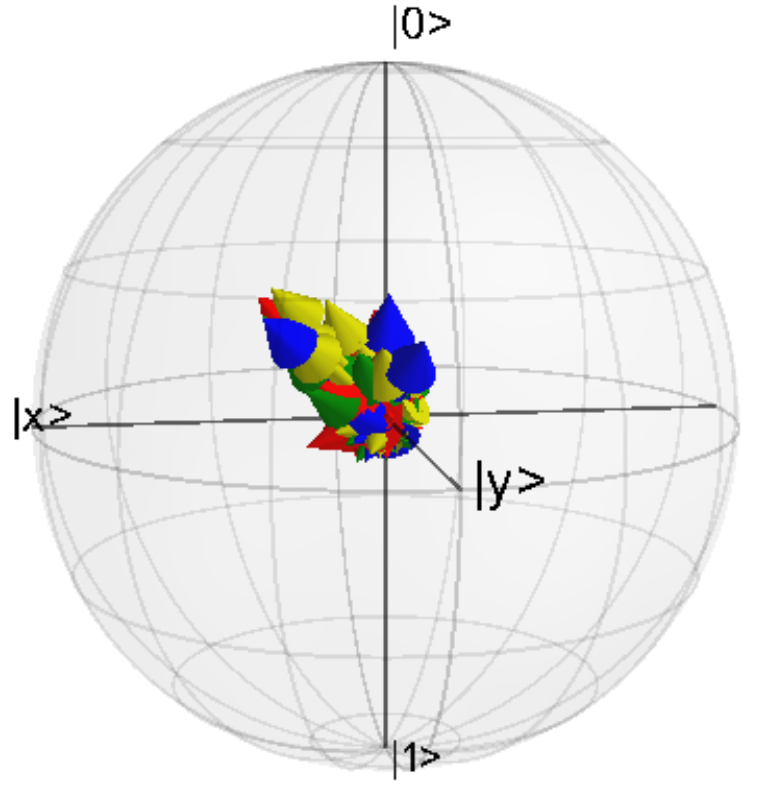}
    \includegraphics[width=0.13\textwidth]{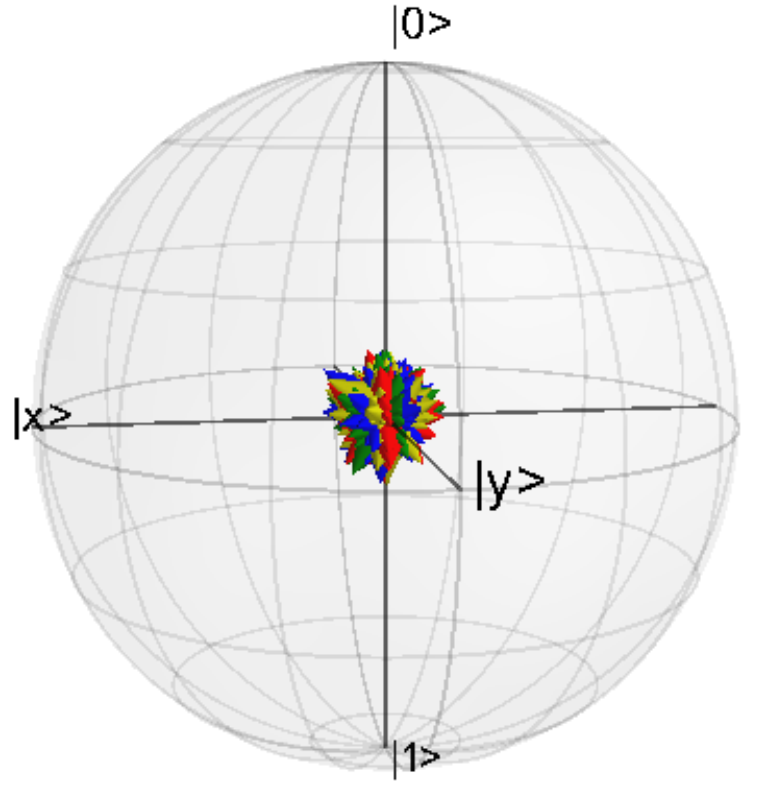}
    \includegraphics[width=0.13\textwidth]{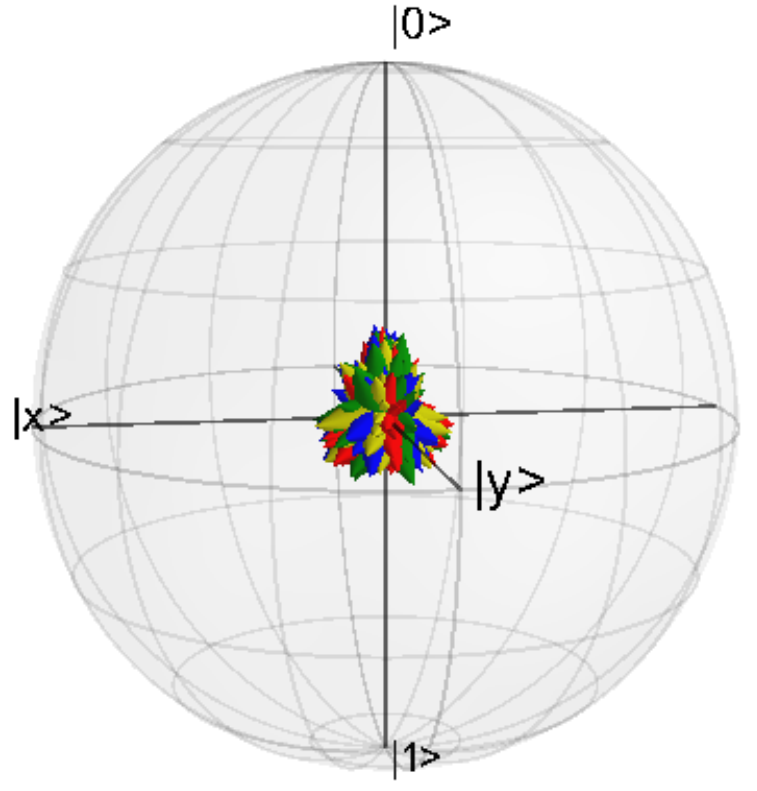}
    \includegraphics[width=0.13\textwidth]{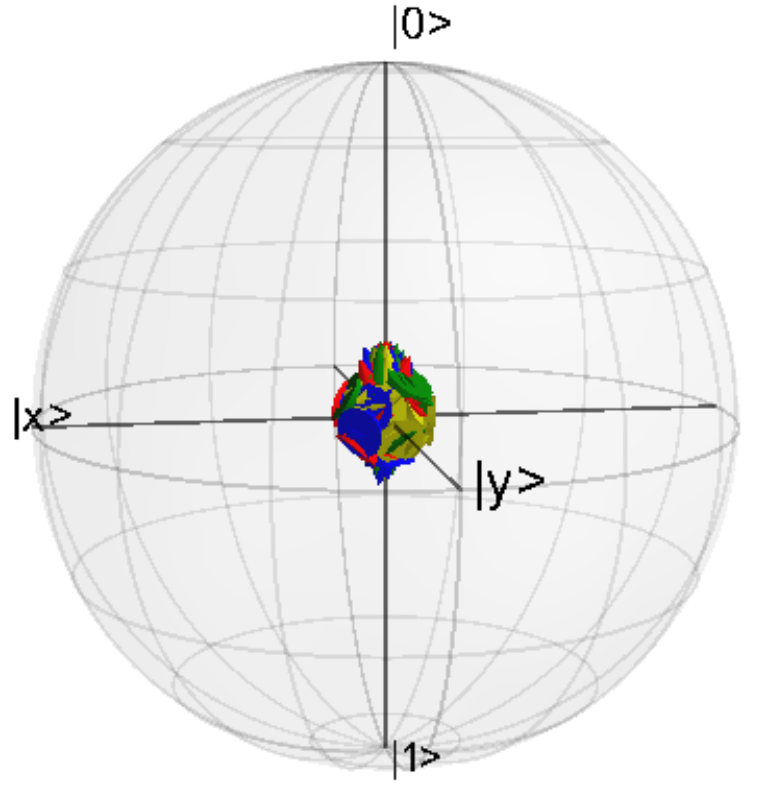}
    \includegraphics[width=0.13\textwidth]{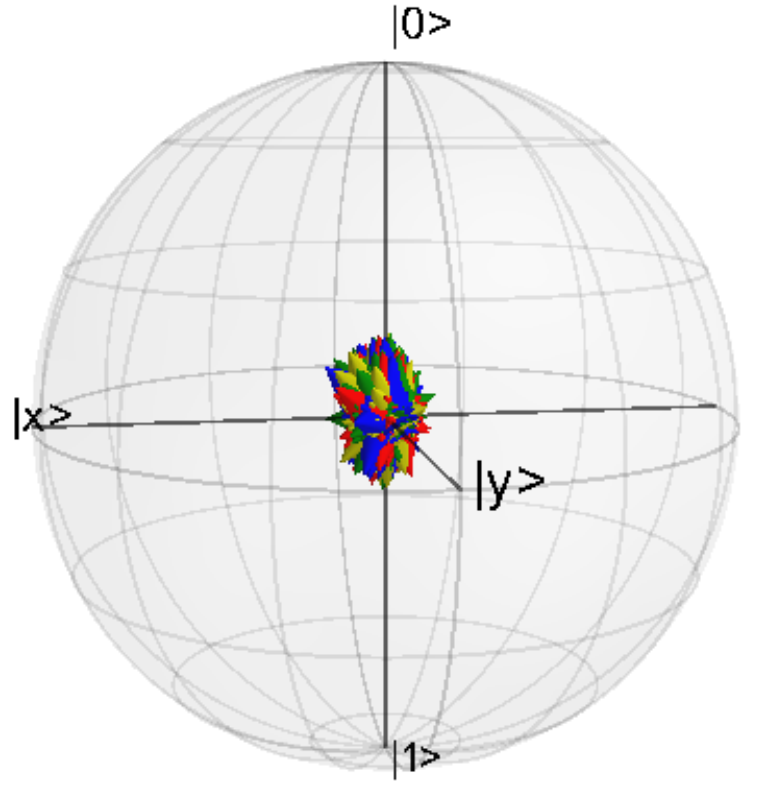}
    \includegraphics[width=0.13\textwidth]{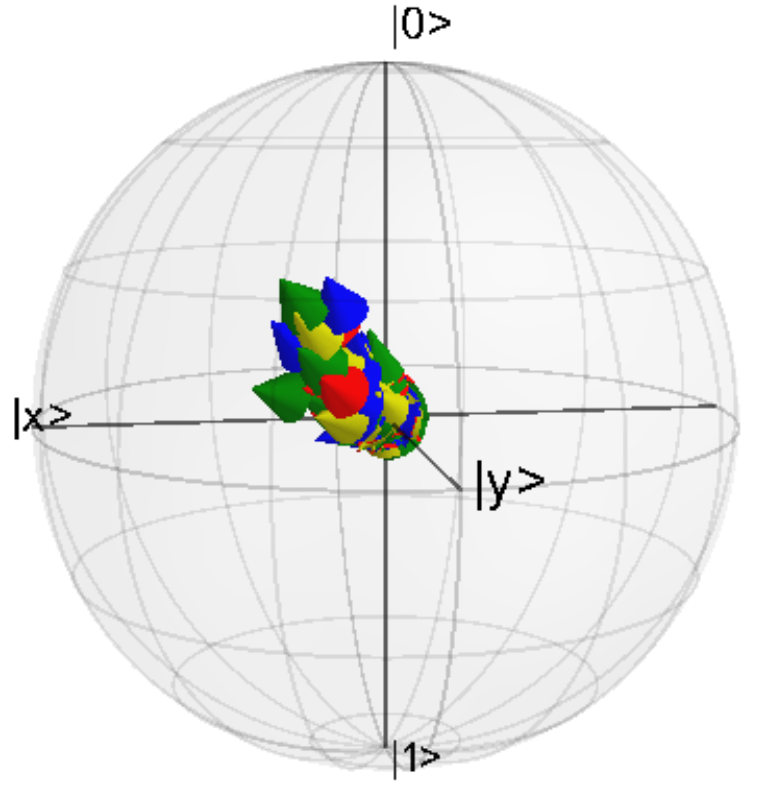}
    \includegraphics[width=0.13\textwidth]{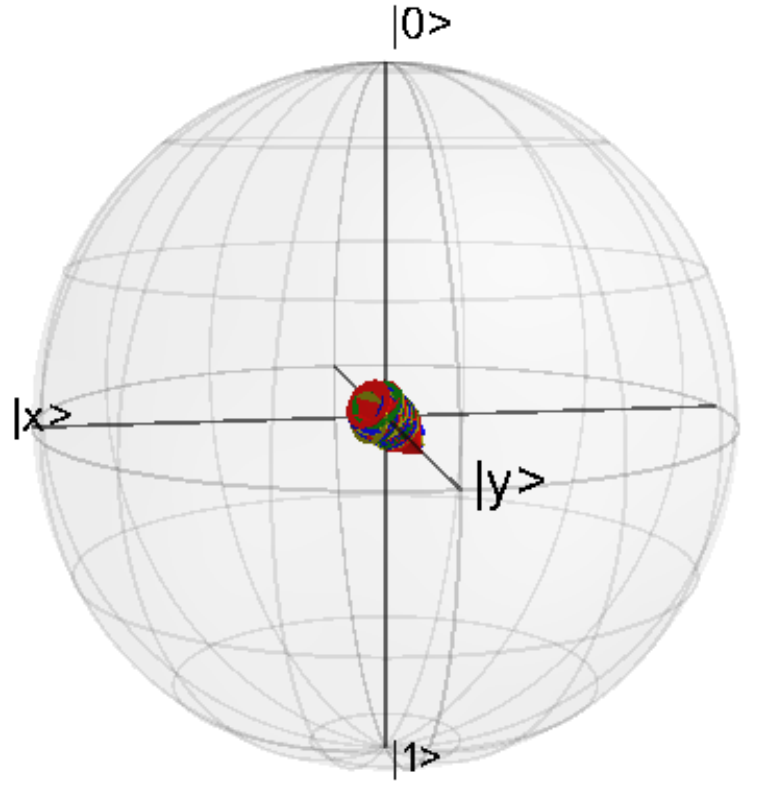}
    \includegraphics[width=0.13\textwidth]{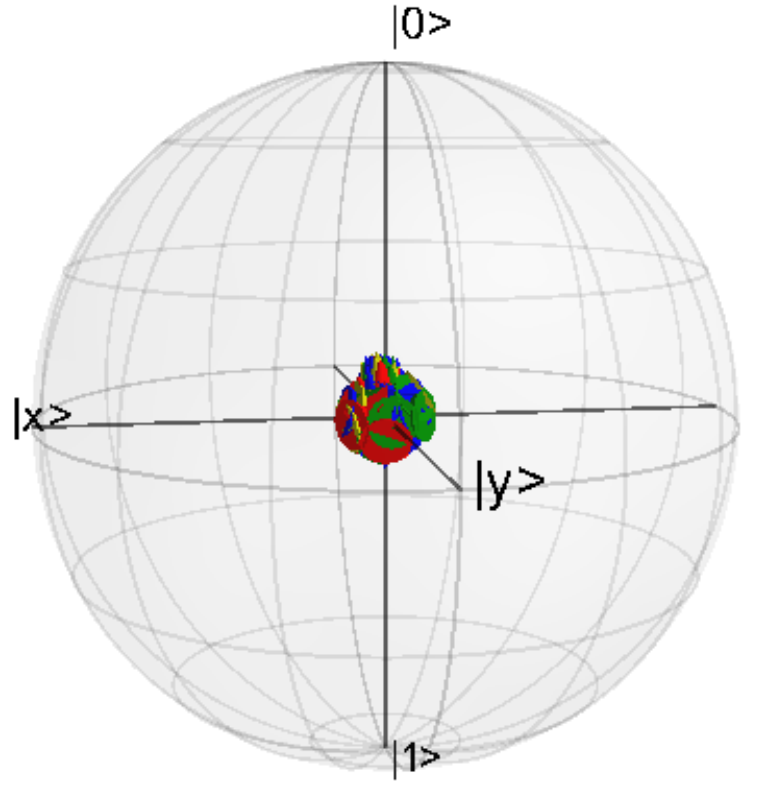}
    \includegraphics[width=0.13\textwidth]{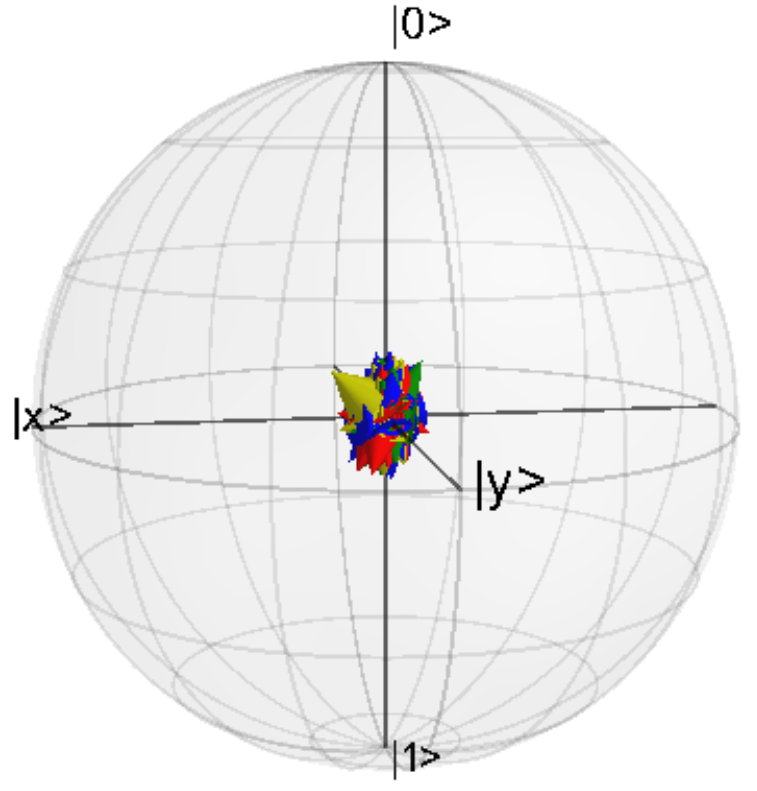}
    \includegraphics[width=0.13\textwidth]{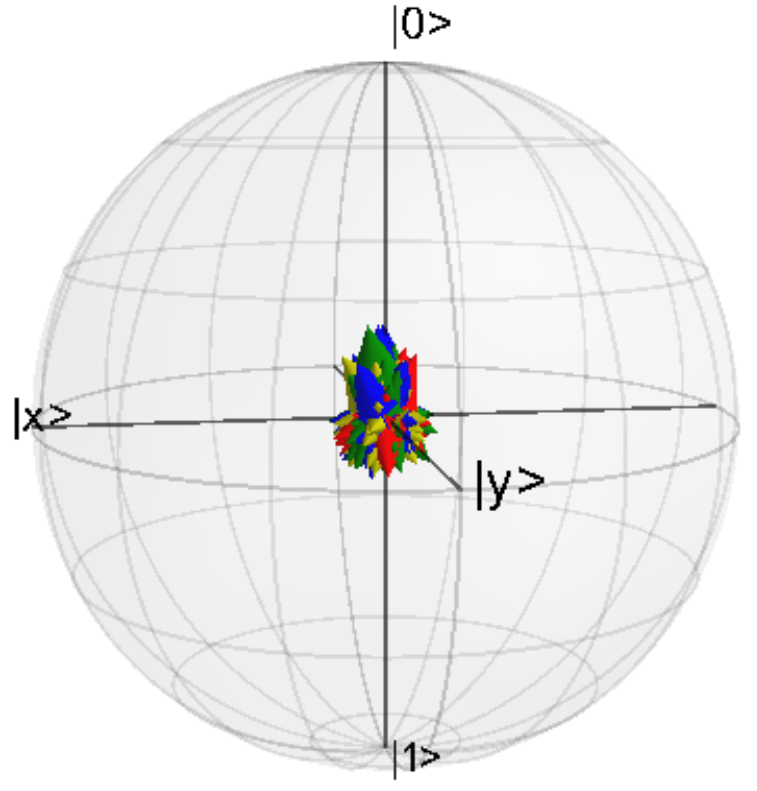}
    \includegraphics[width=0.13\textwidth]{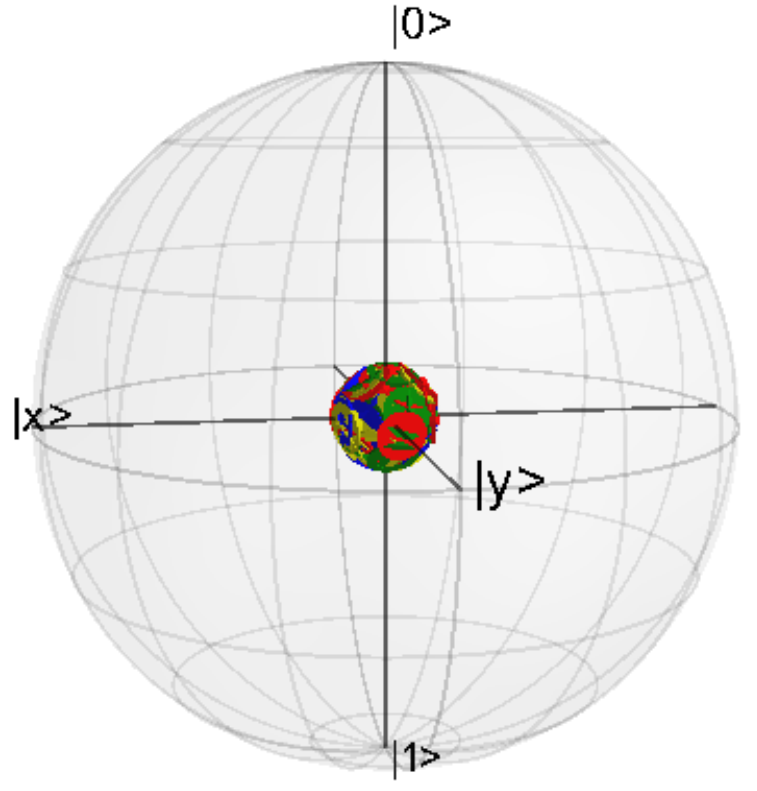}
    \includegraphics[width=0.13\textwidth]{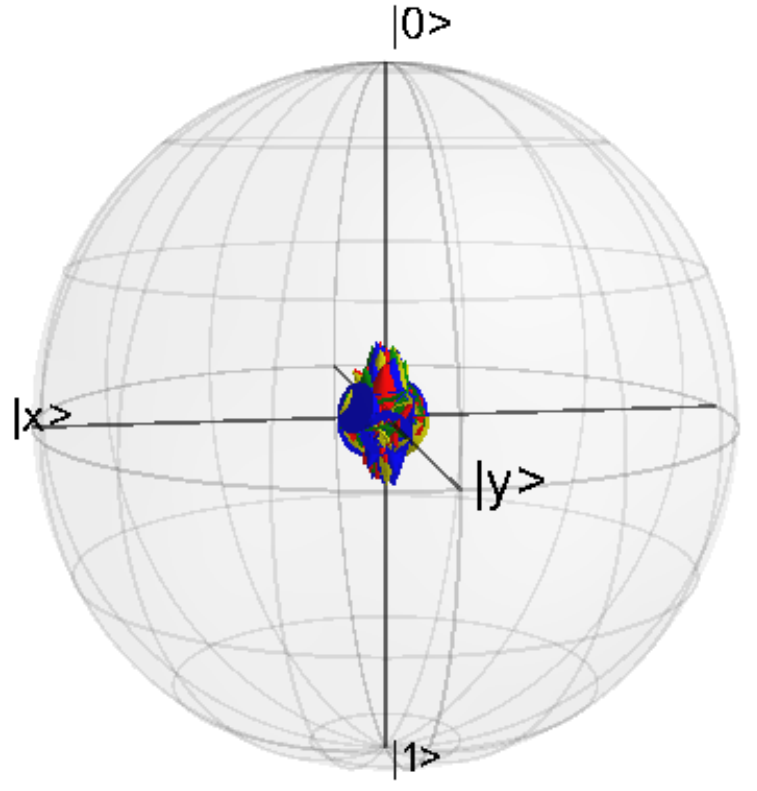}
    \includegraphics[width=0.13\textwidth]{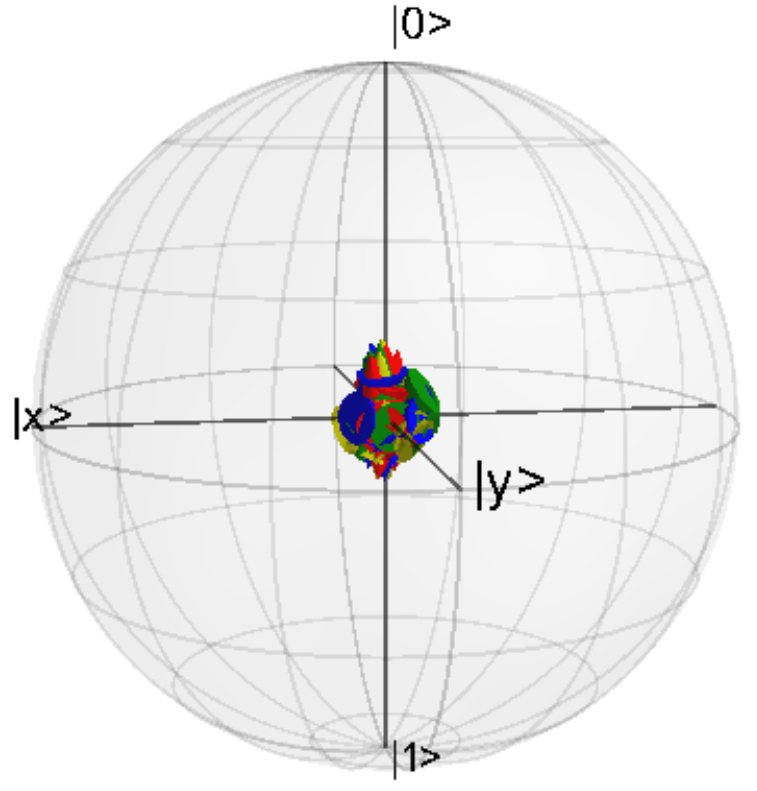}
    \includegraphics[width=0.13\textwidth]{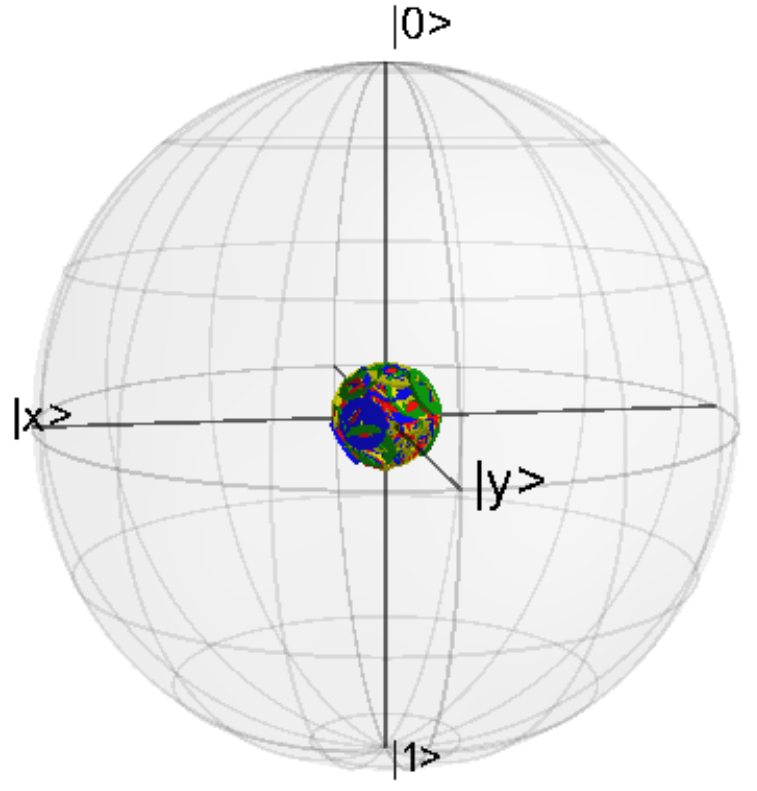}
    \includegraphics[width=0.13\textwidth]{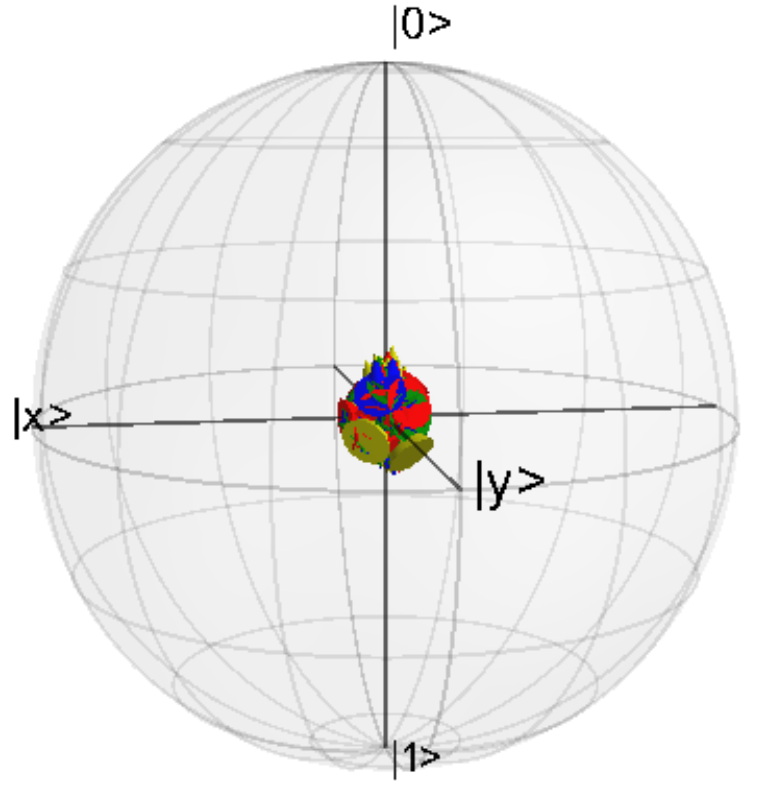}
    \includegraphics[width=0.13\textwidth]{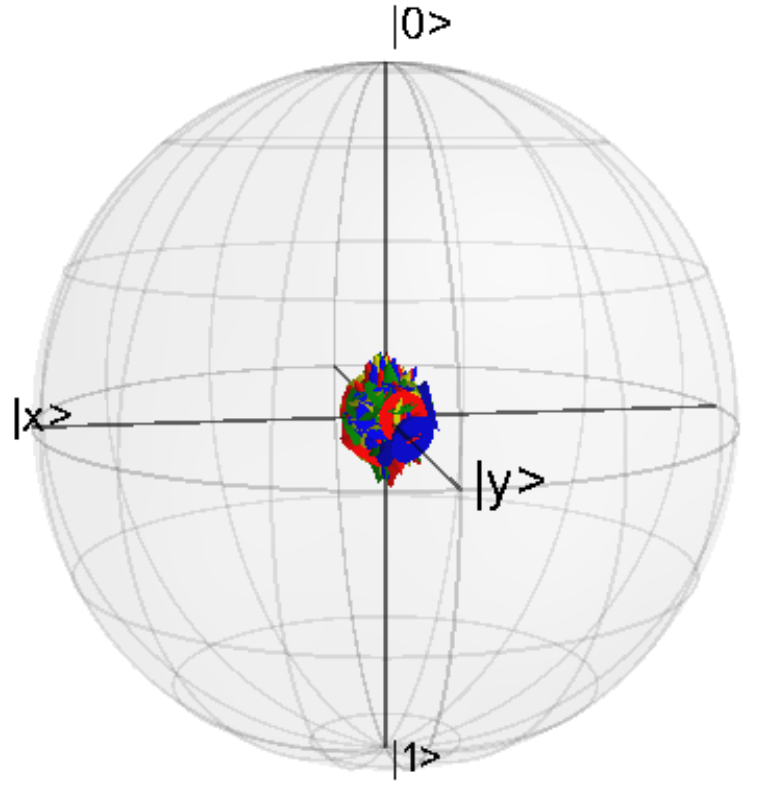}
    \includegraphics[width=0.13\textwidth]{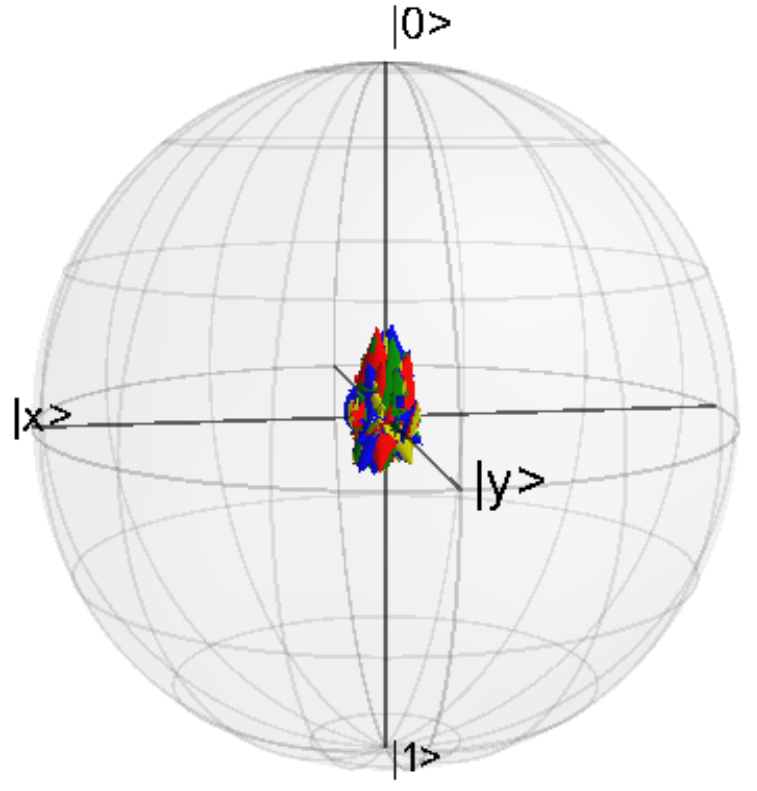}
    \includegraphics[width=0.13\textwidth]{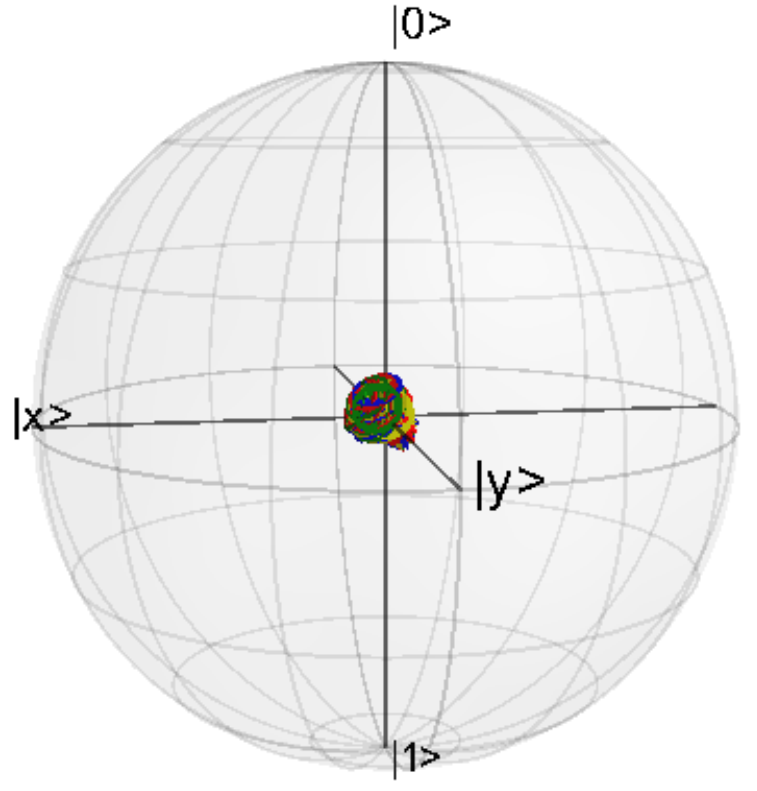}
    \includegraphics[width=0.13\textwidth]{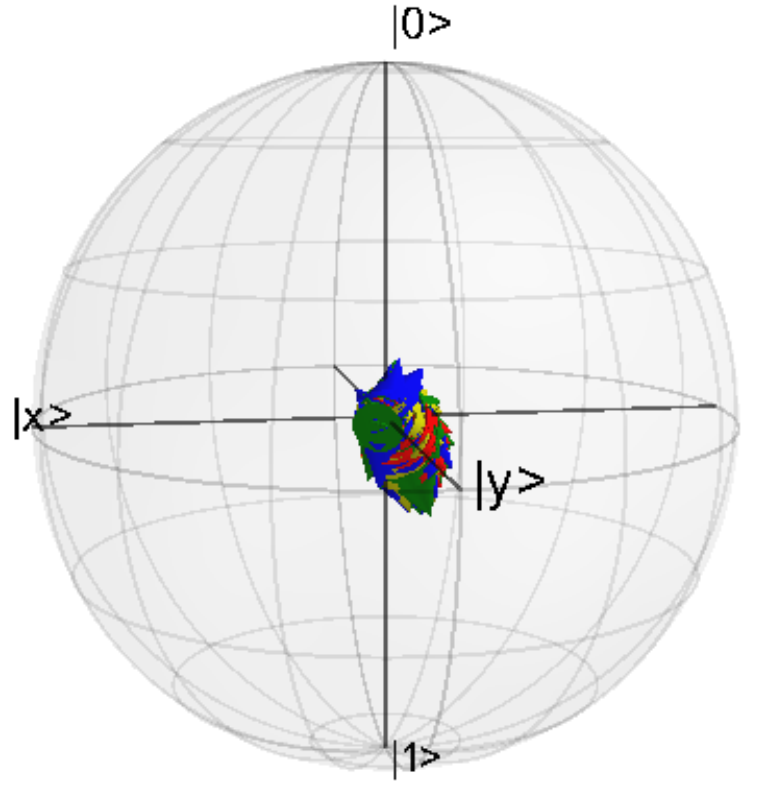}
    \includegraphics[width=0.13\textwidth]{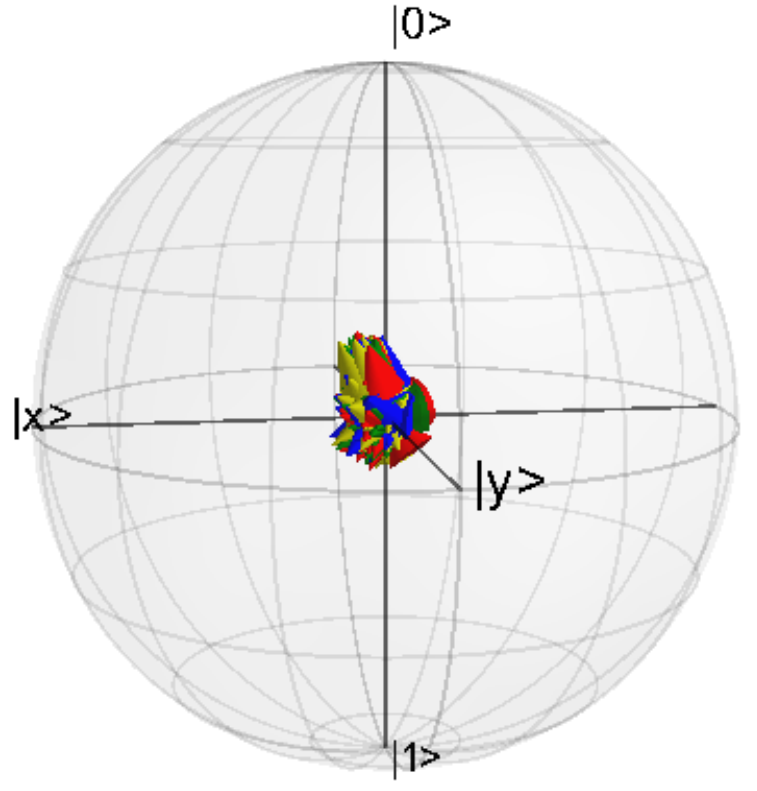}
    \includegraphics[width=0.13\textwidth]{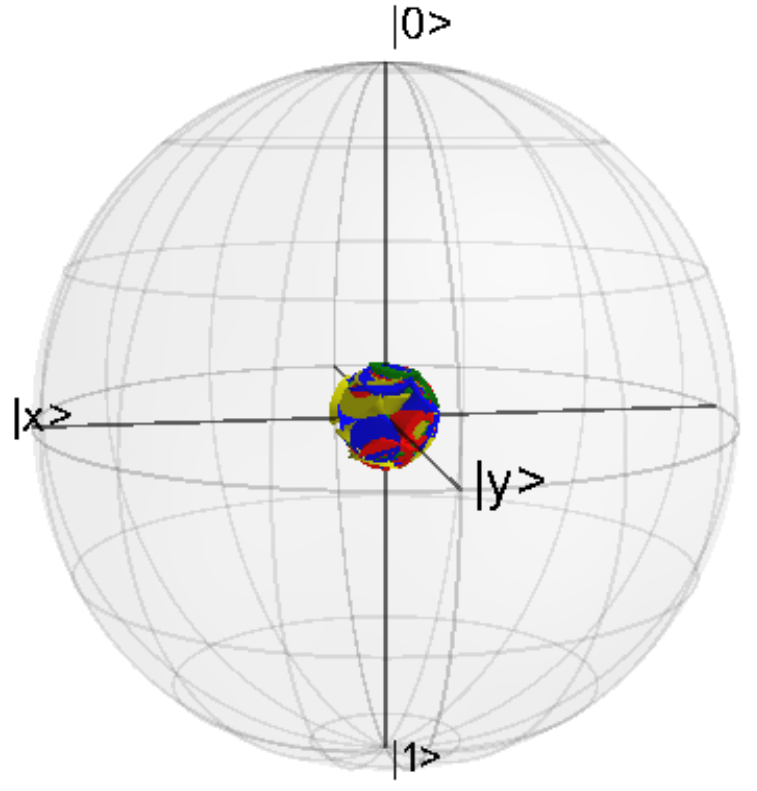}
    \includegraphics[width=0.13\textwidth]{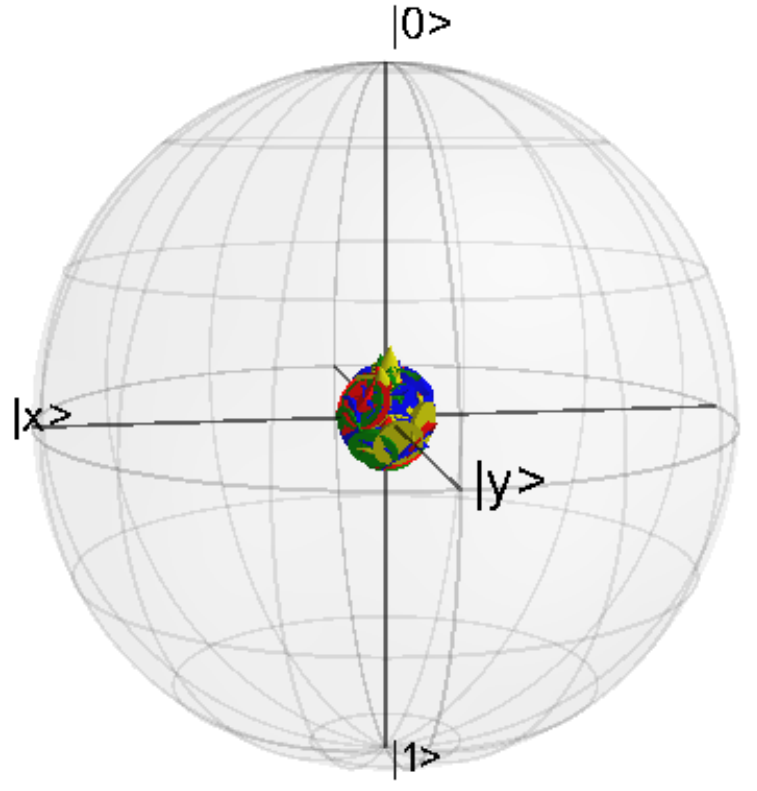}
    \includegraphics[width=0.13\textwidth]{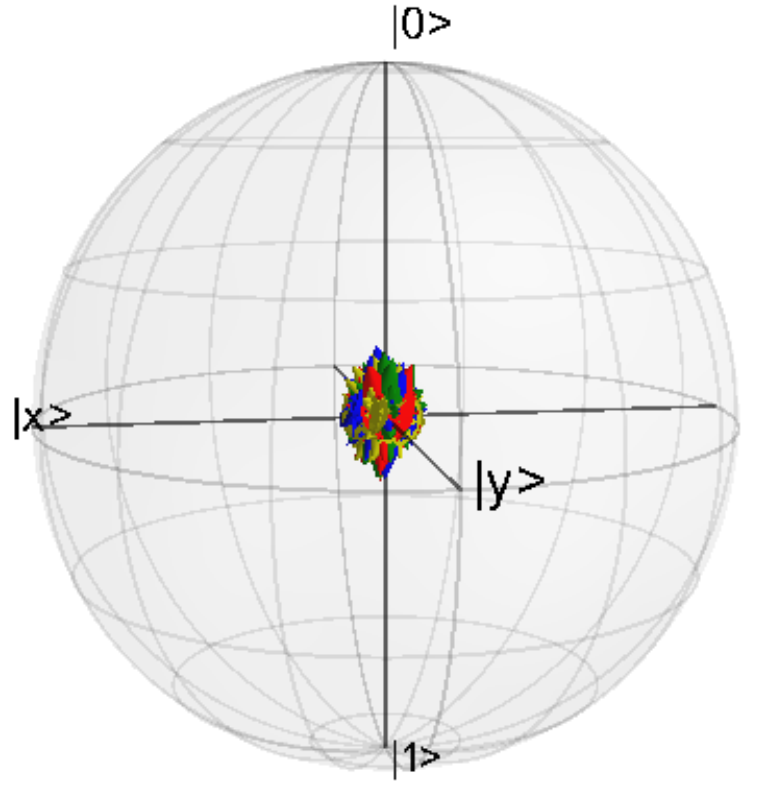}
    \includegraphics[width=0.13\textwidth]{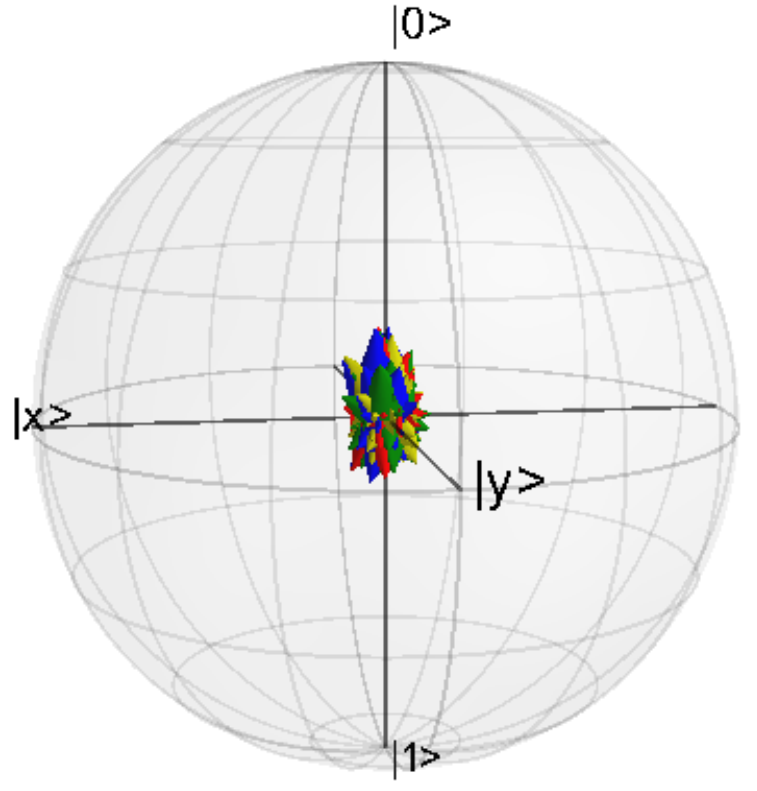}
    \includegraphics[width=0.13\textwidth]{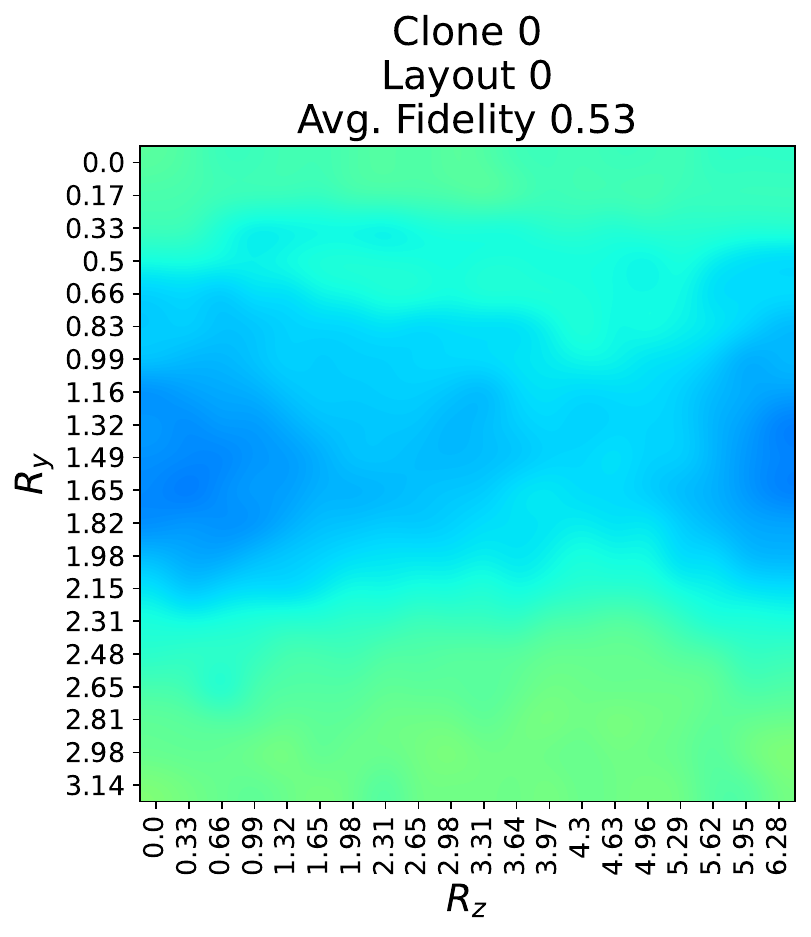}
    \includegraphics[width=0.13\textwidth]{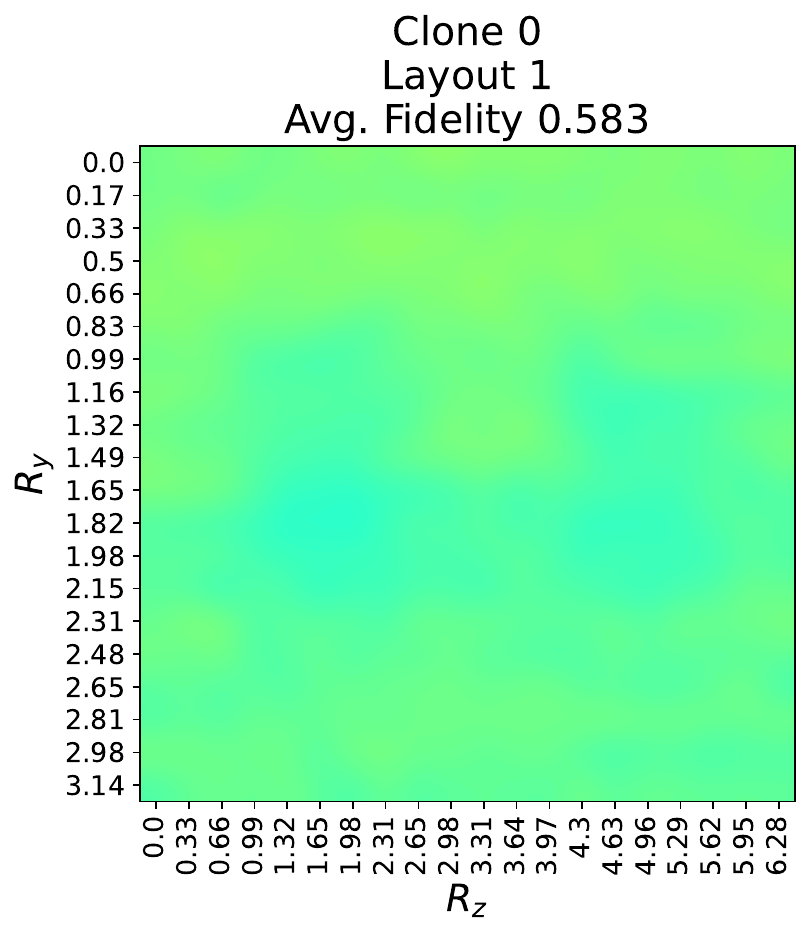}
    \includegraphics[width=0.13\textwidth]{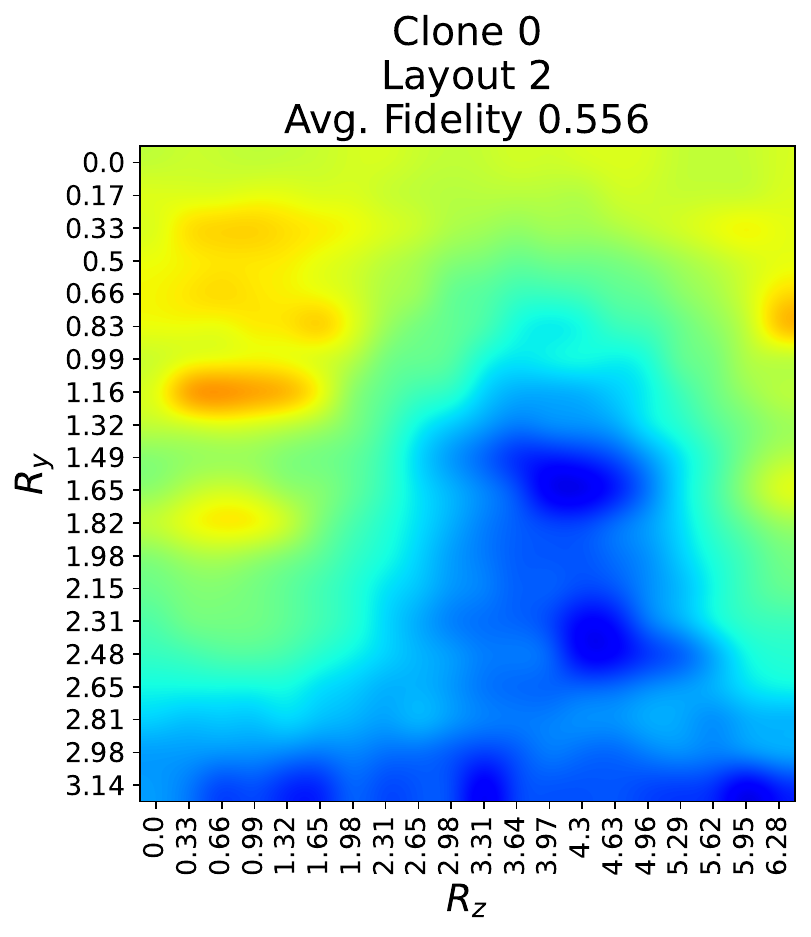}
    \includegraphics[width=0.13\textwidth]{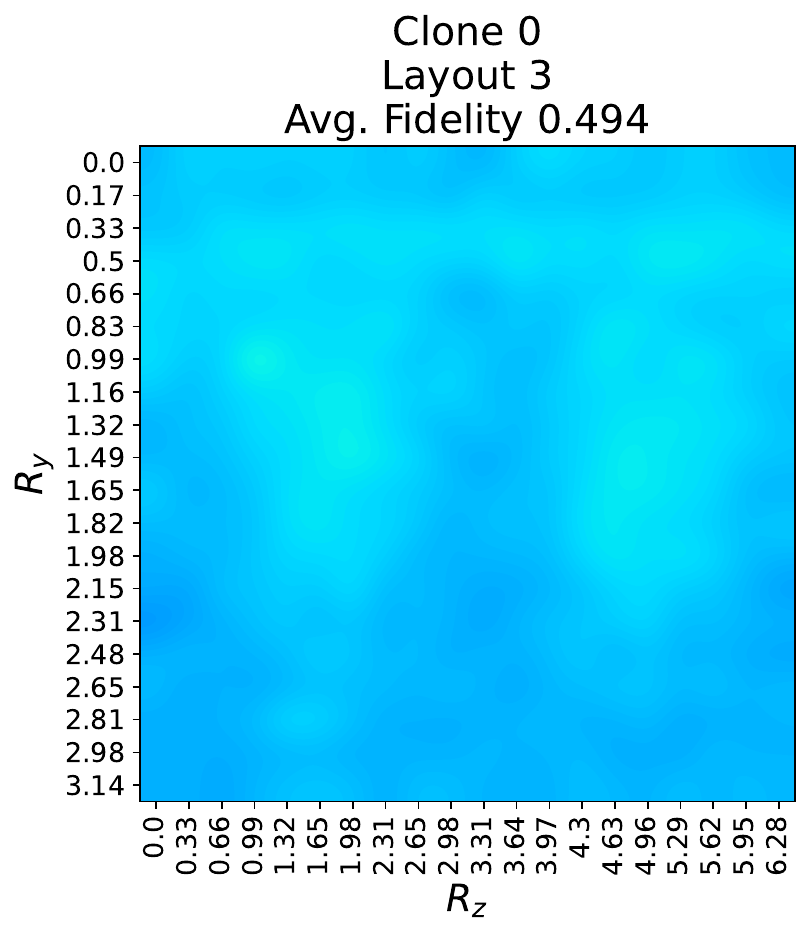}
    \includegraphics[width=0.13\textwidth]{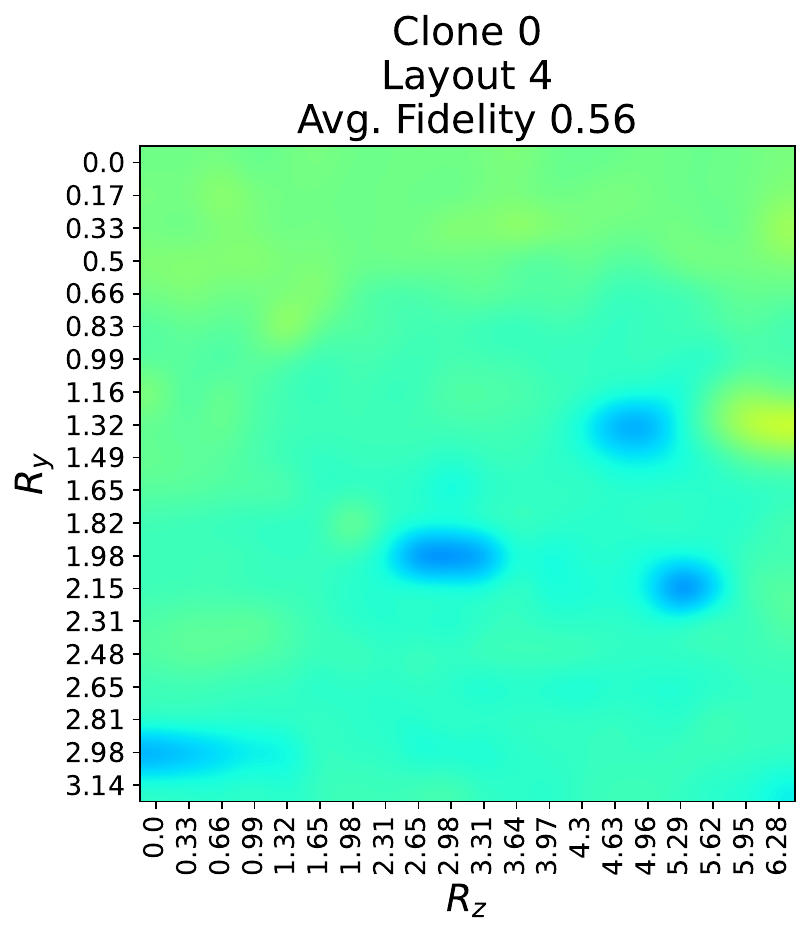}
    \includegraphics[width=0.13\textwidth]{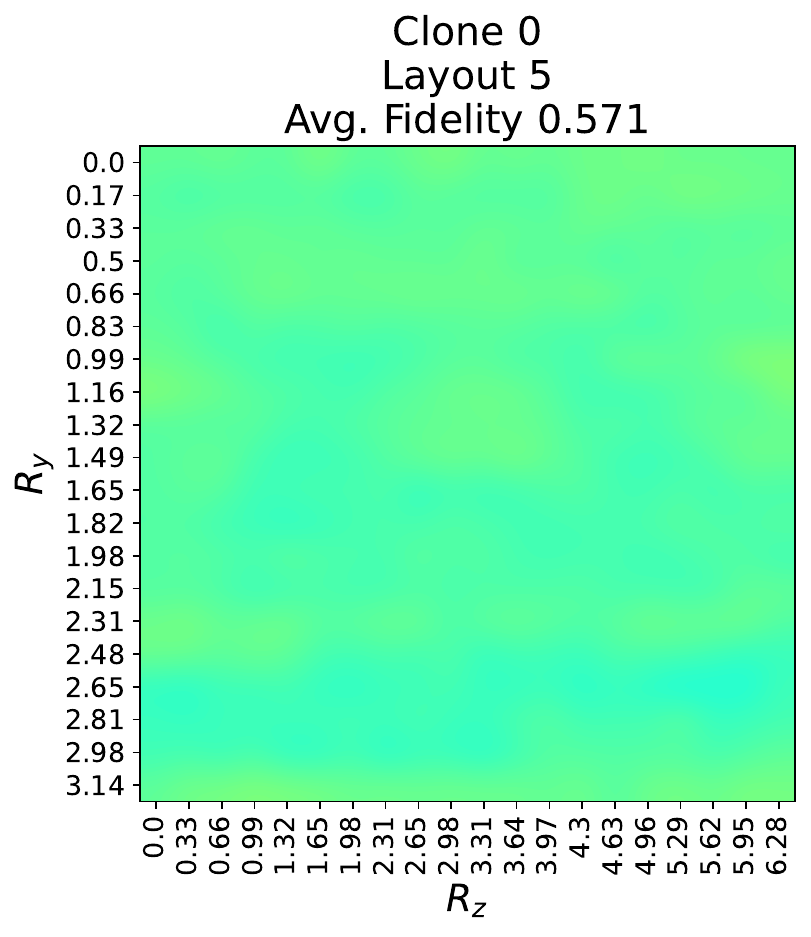}
    \includegraphics[width=0.13\textwidth]{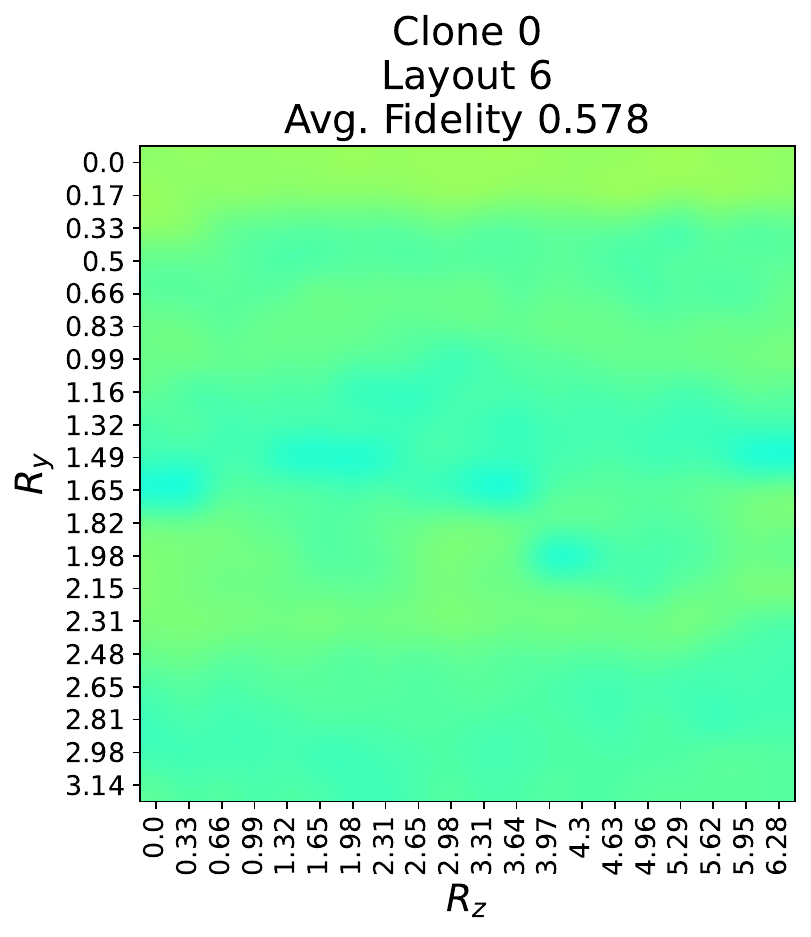}
    \includegraphics[width=0.13\textwidth]{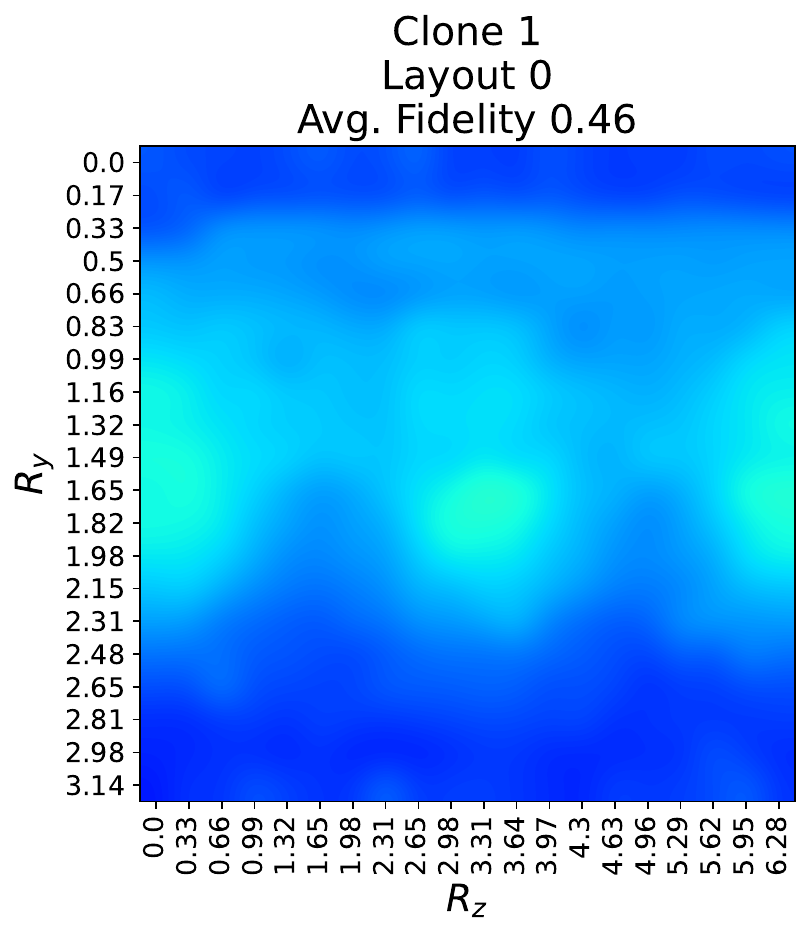}
    \includegraphics[width=0.13\textwidth]{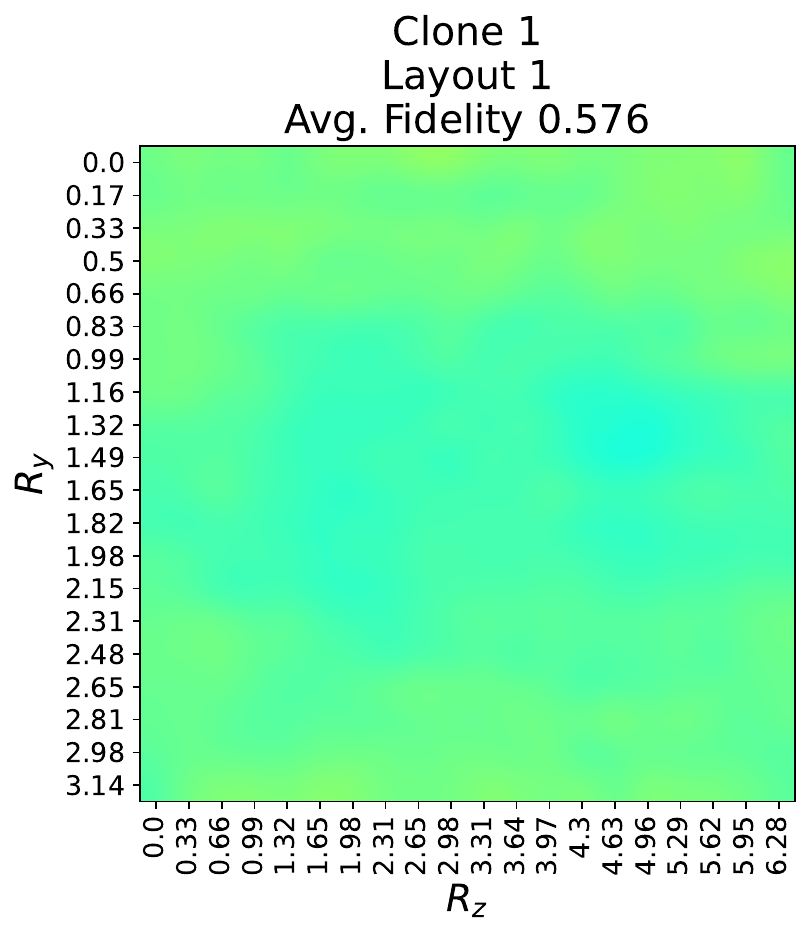}
    \includegraphics[width=0.13\textwidth]{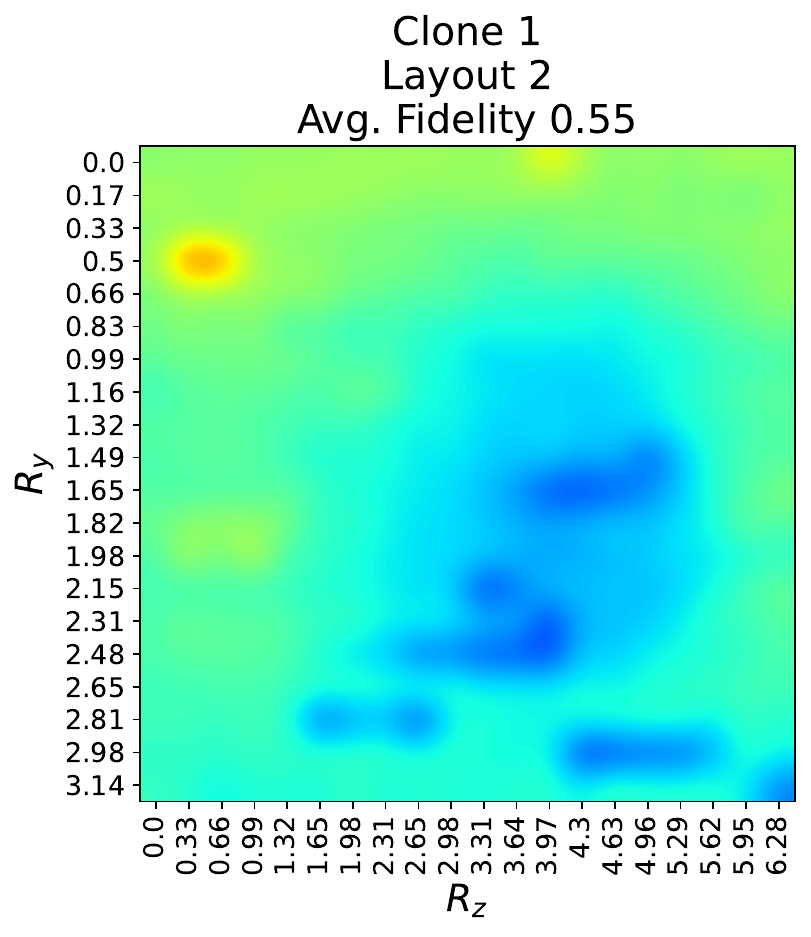}
    \includegraphics[width=0.13\textwidth]{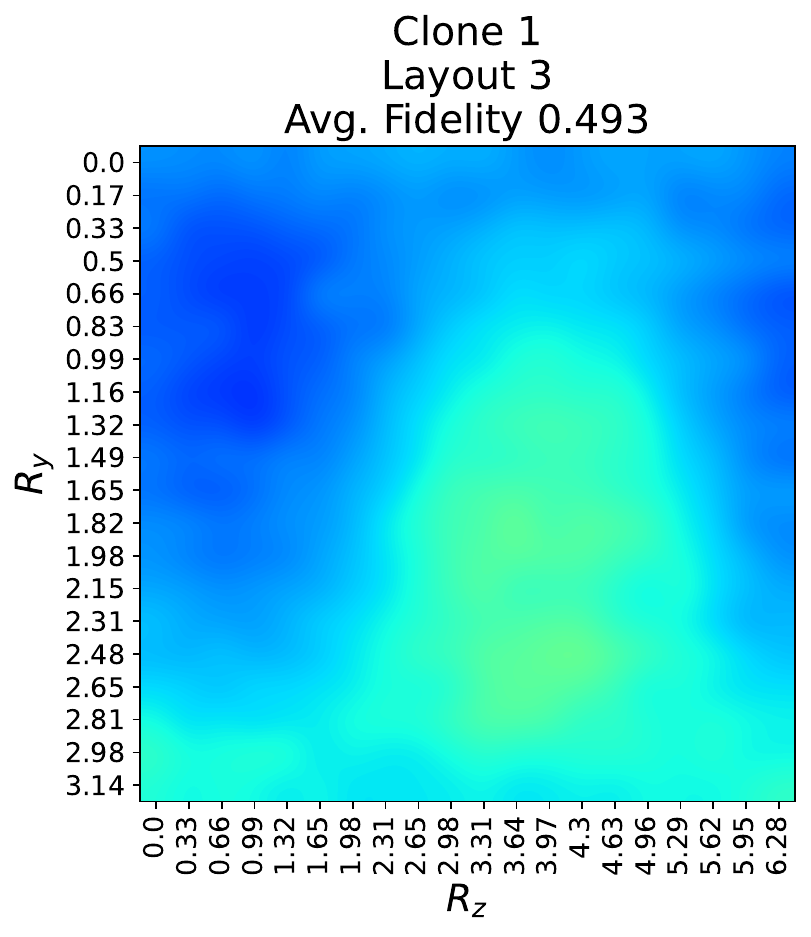}
    \includegraphics[width=0.13\textwidth]{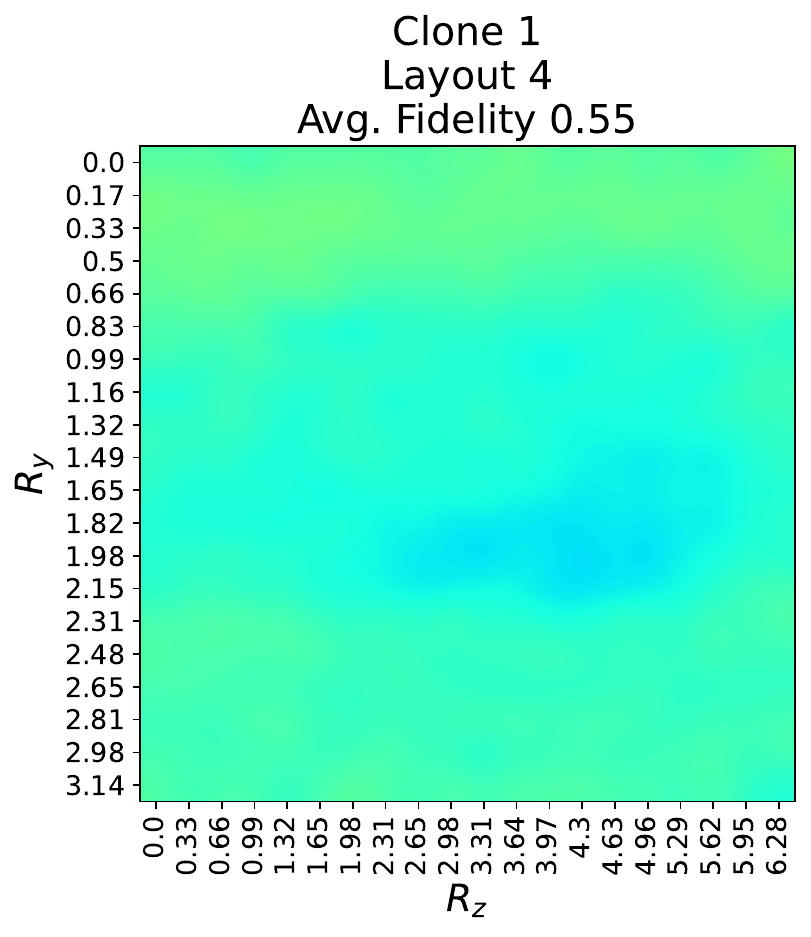}
    \includegraphics[width=0.13\textwidth]{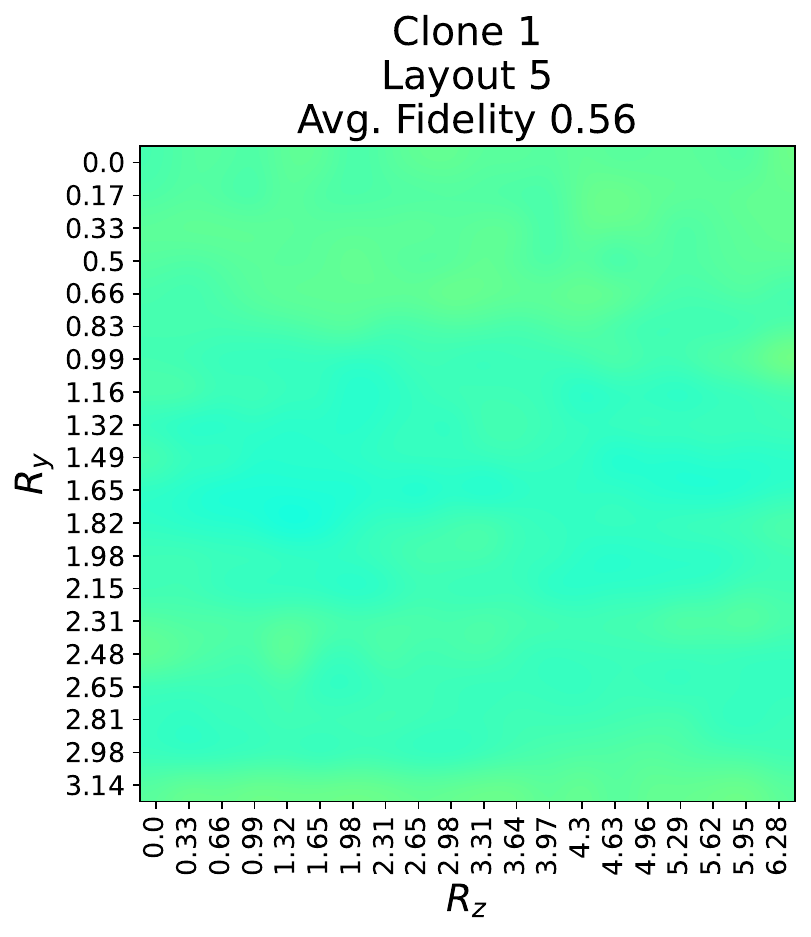}
    \includegraphics[width=0.13\textwidth]{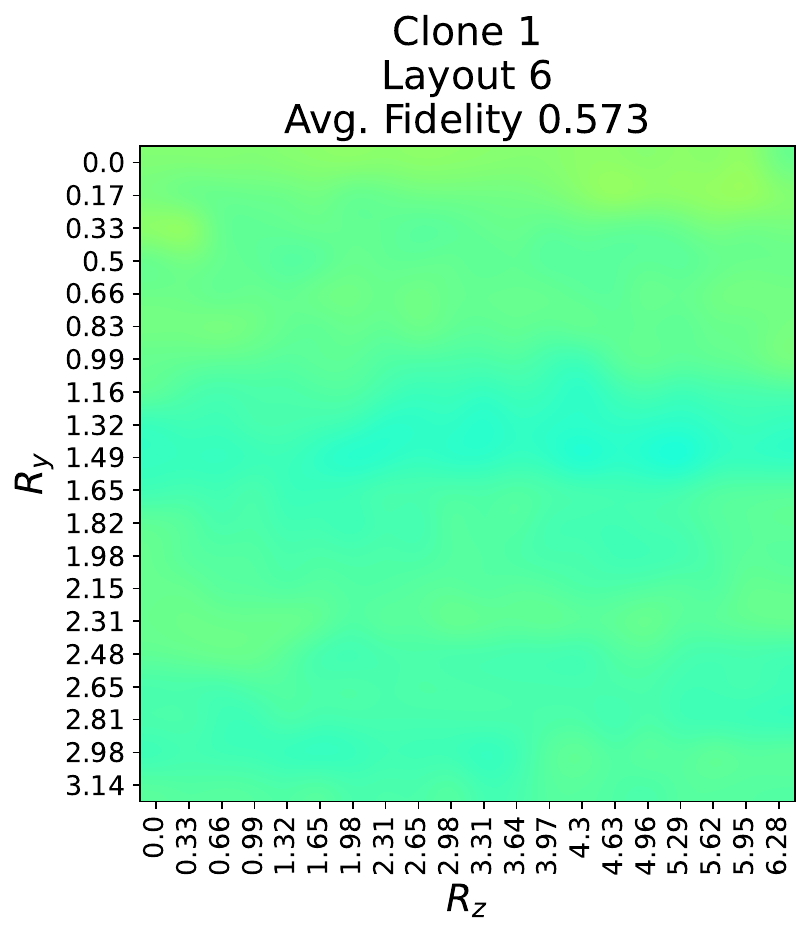}
    \includegraphics[width=0.13\textwidth]{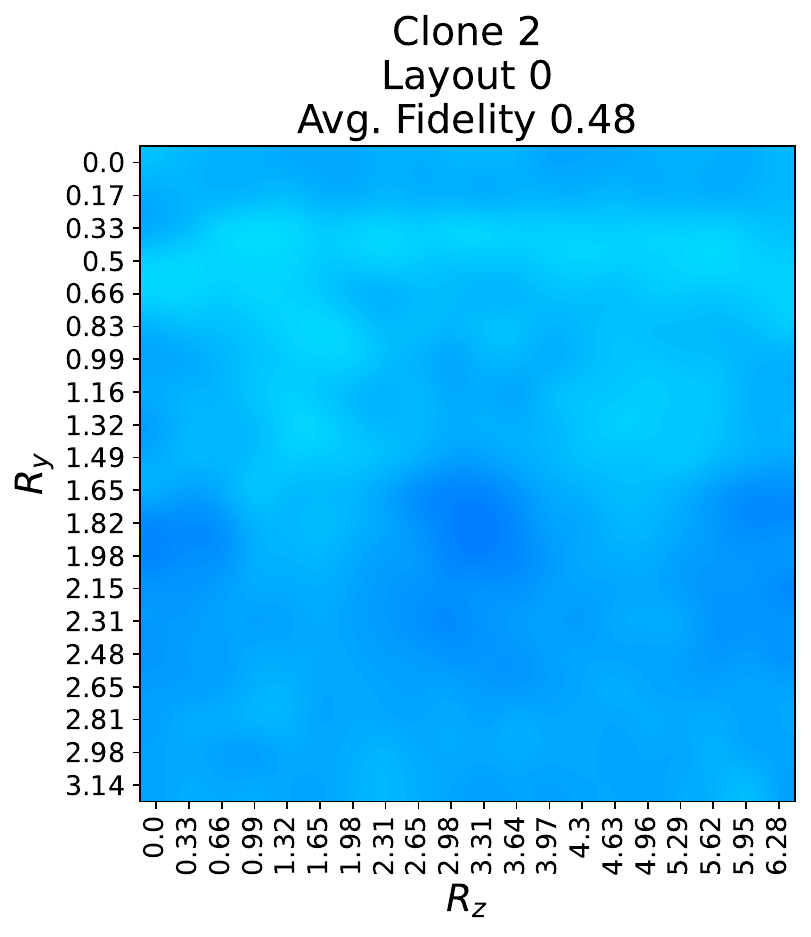}
    \includegraphics[width=0.13\textwidth]{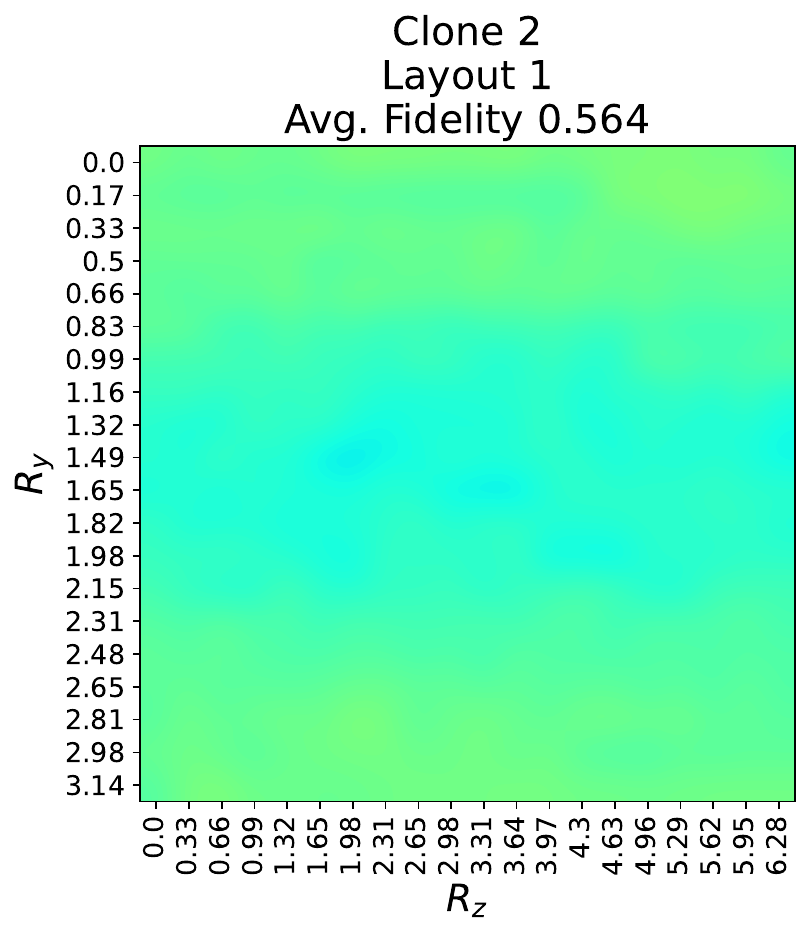}
    \includegraphics[width=0.13\textwidth]{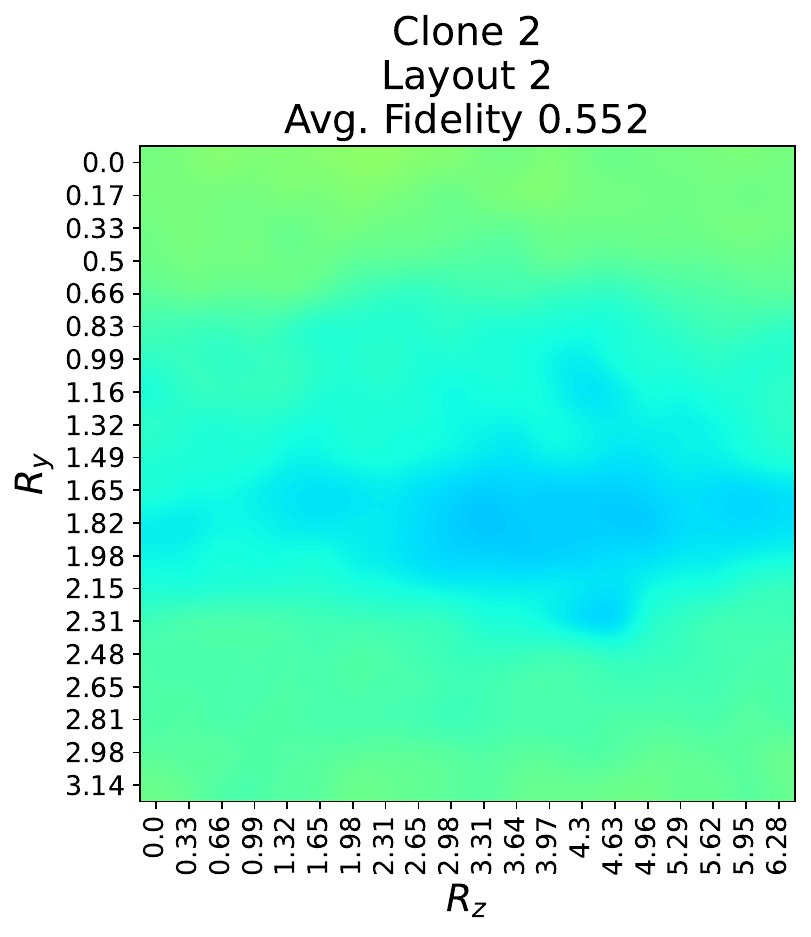}
    \includegraphics[width=0.13\textwidth]{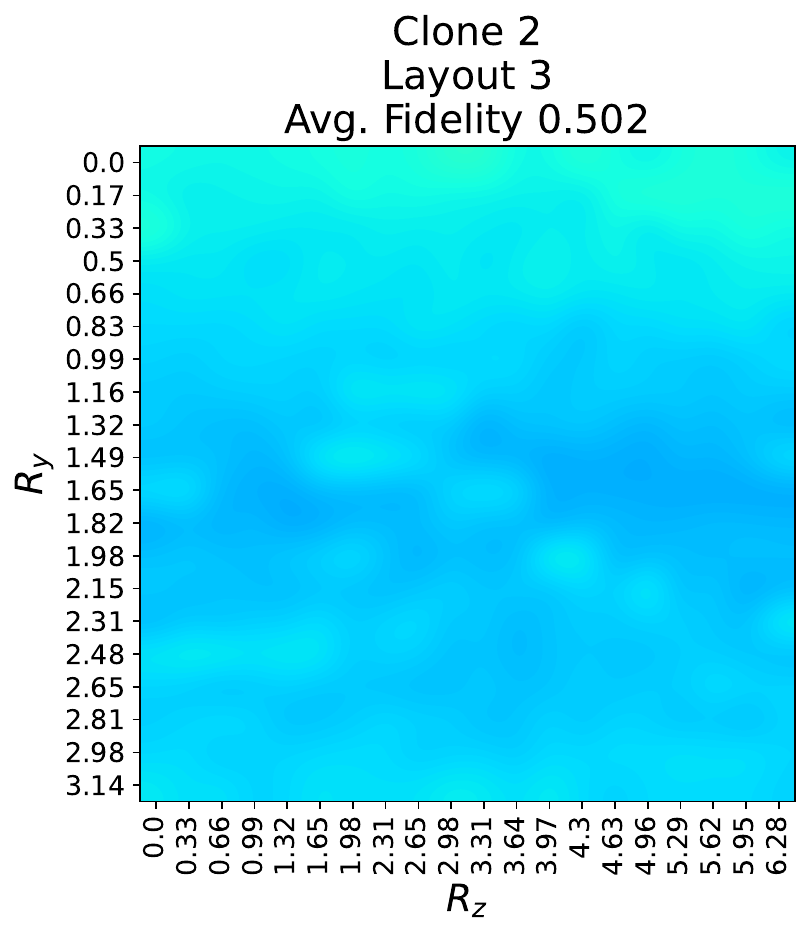}
    \includegraphics[width=0.13\textwidth]{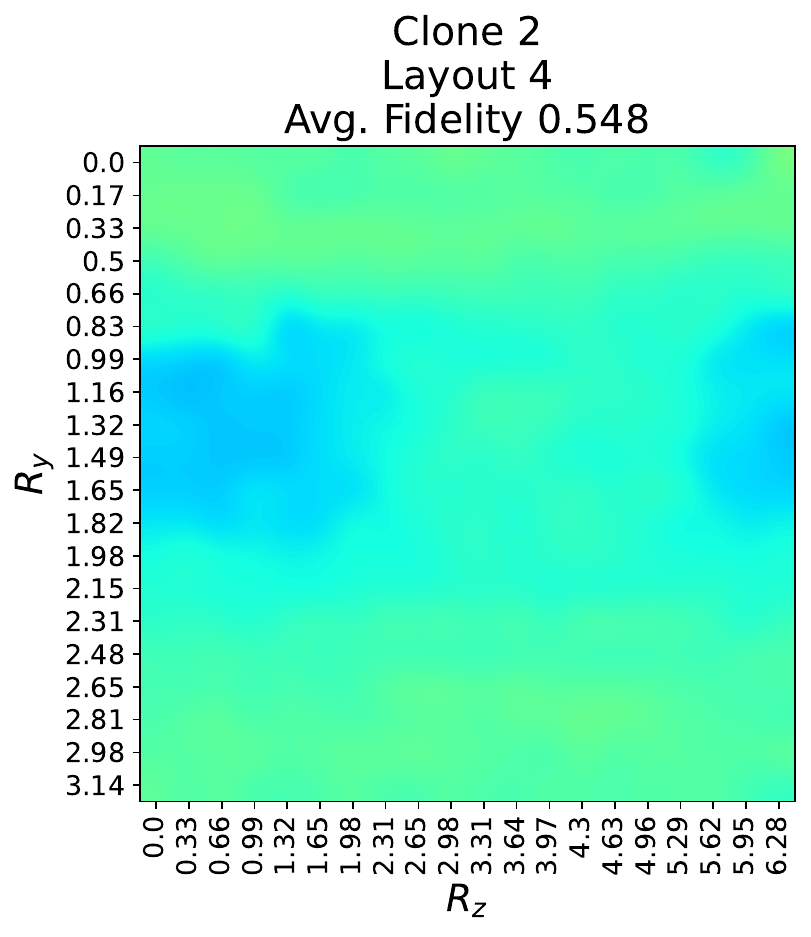}
    \includegraphics[width=0.13\textwidth]{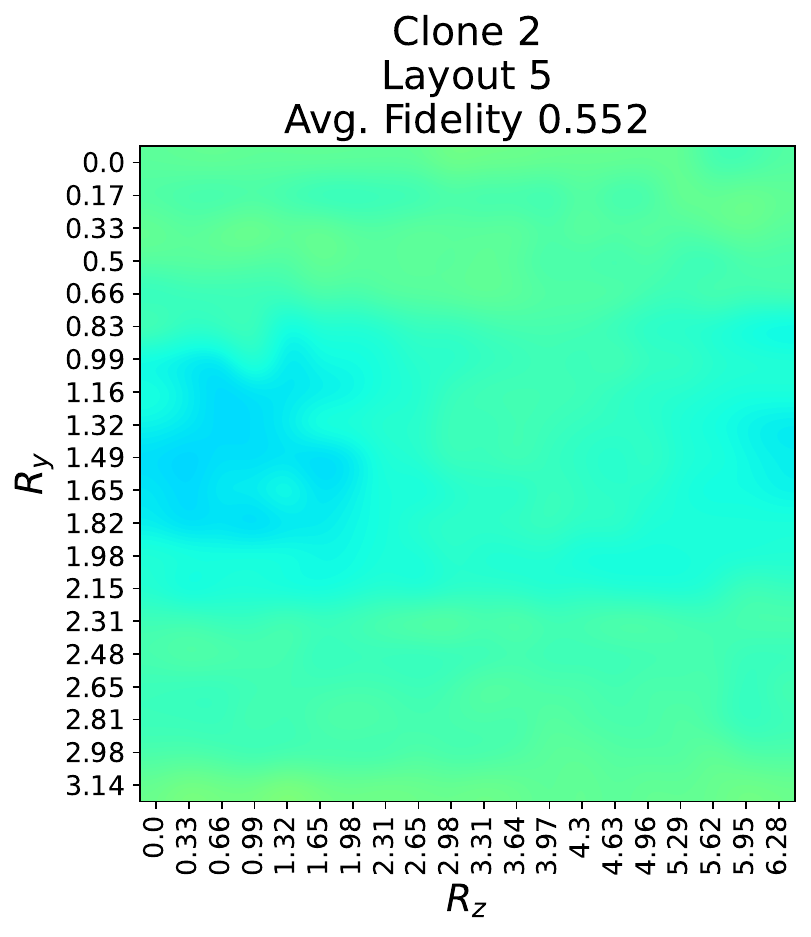}
    \includegraphics[width=0.13\textwidth]{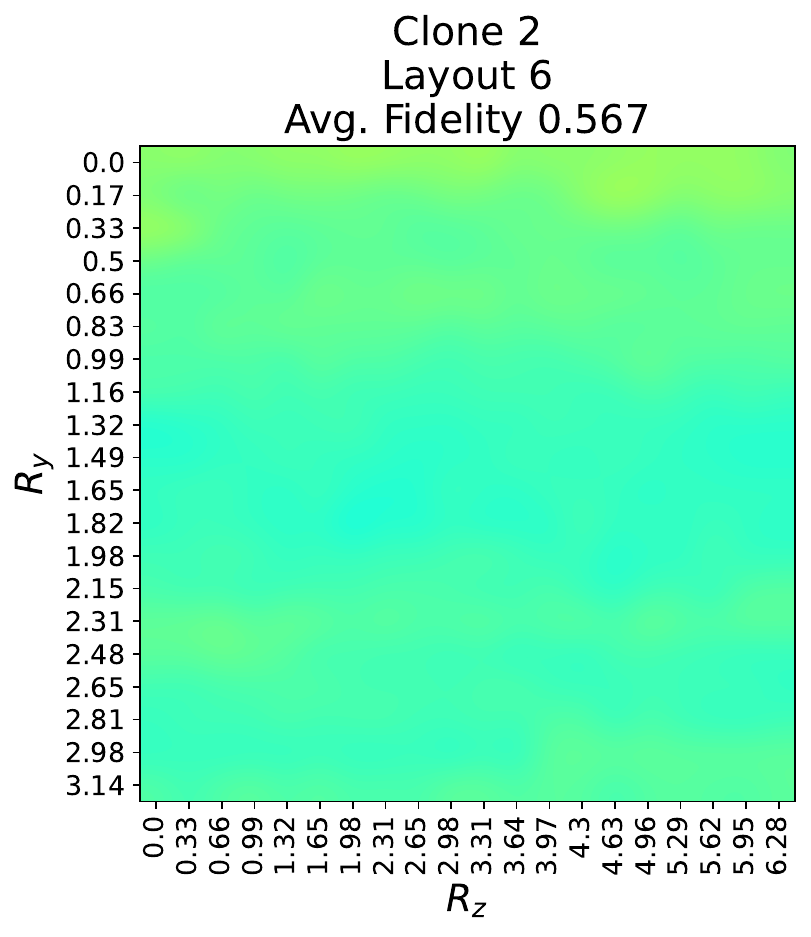}
    \includegraphics[width=0.13\textwidth]{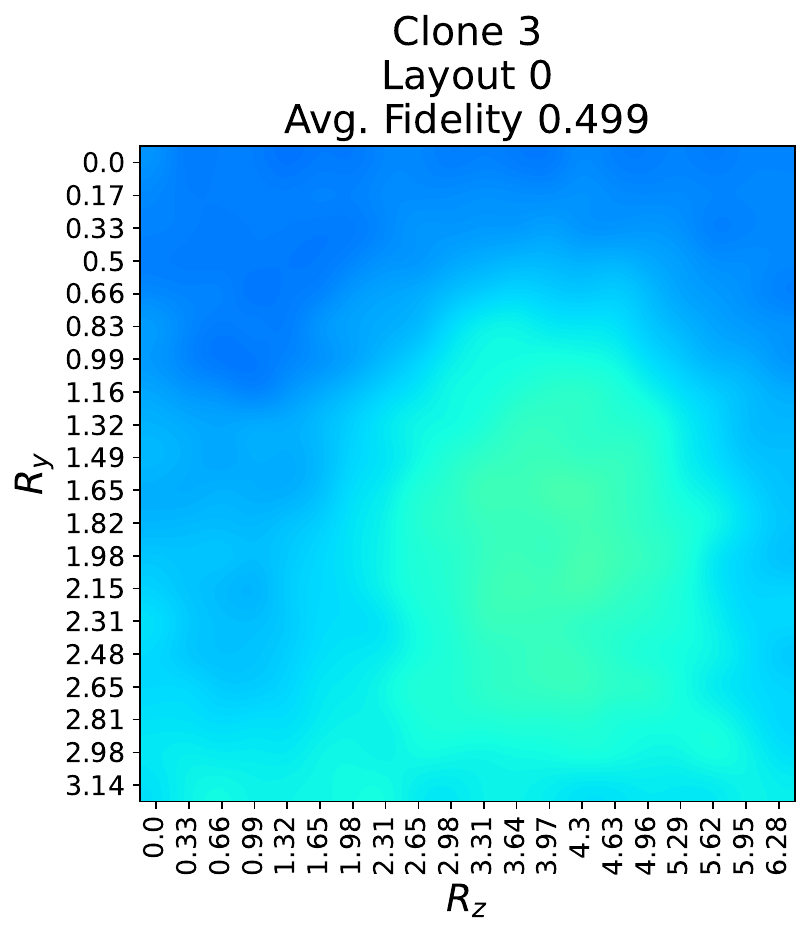}
    \includegraphics[width=0.13\textwidth]{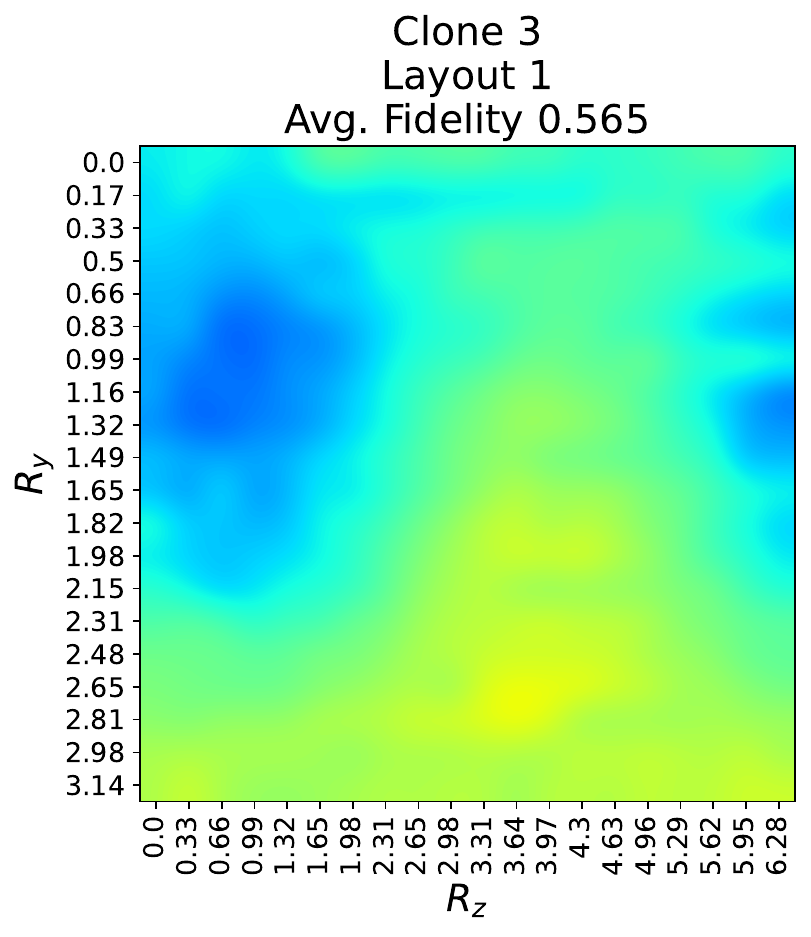}
    \includegraphics[width=0.13\textwidth]{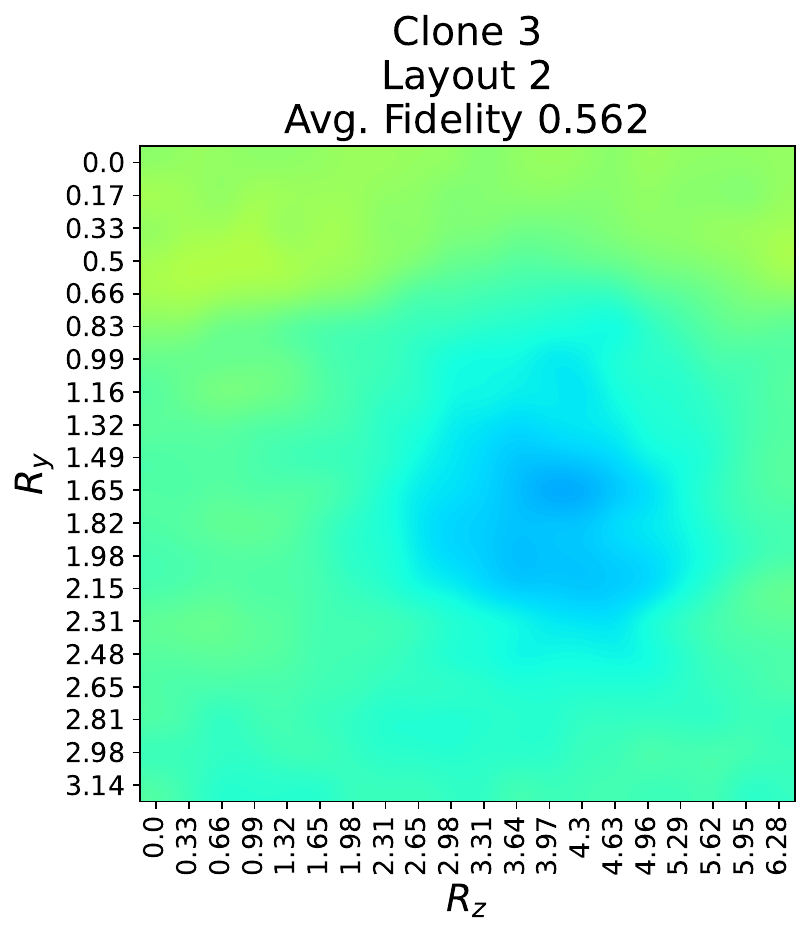}
    \includegraphics[width=0.13\textwidth]{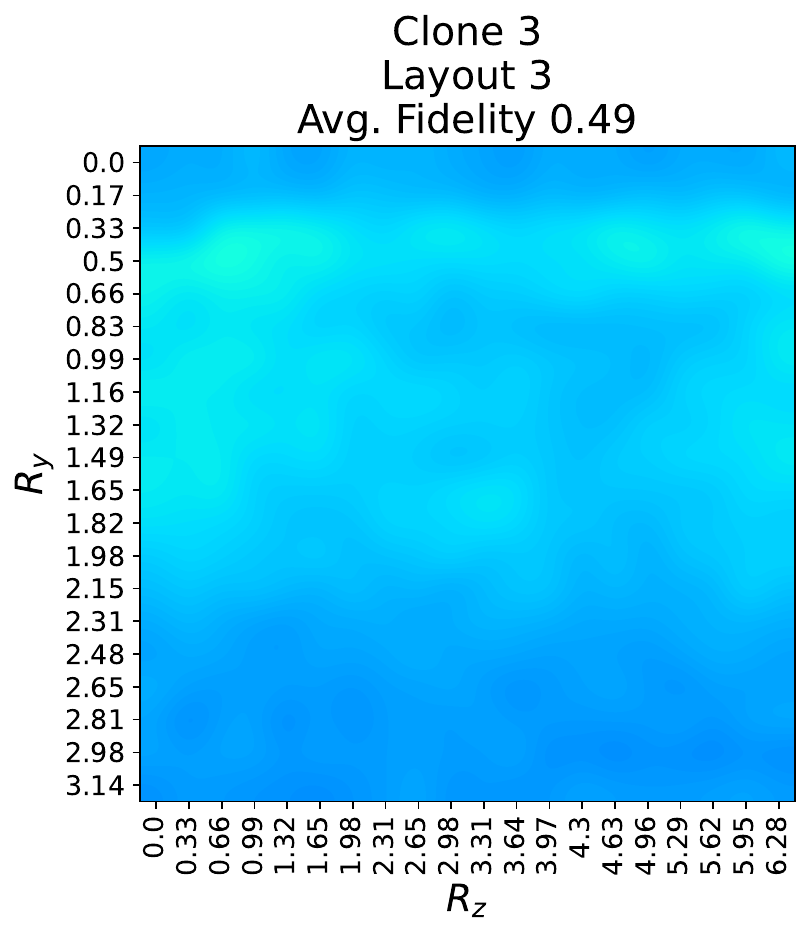}
    \includegraphics[width=0.13\textwidth]{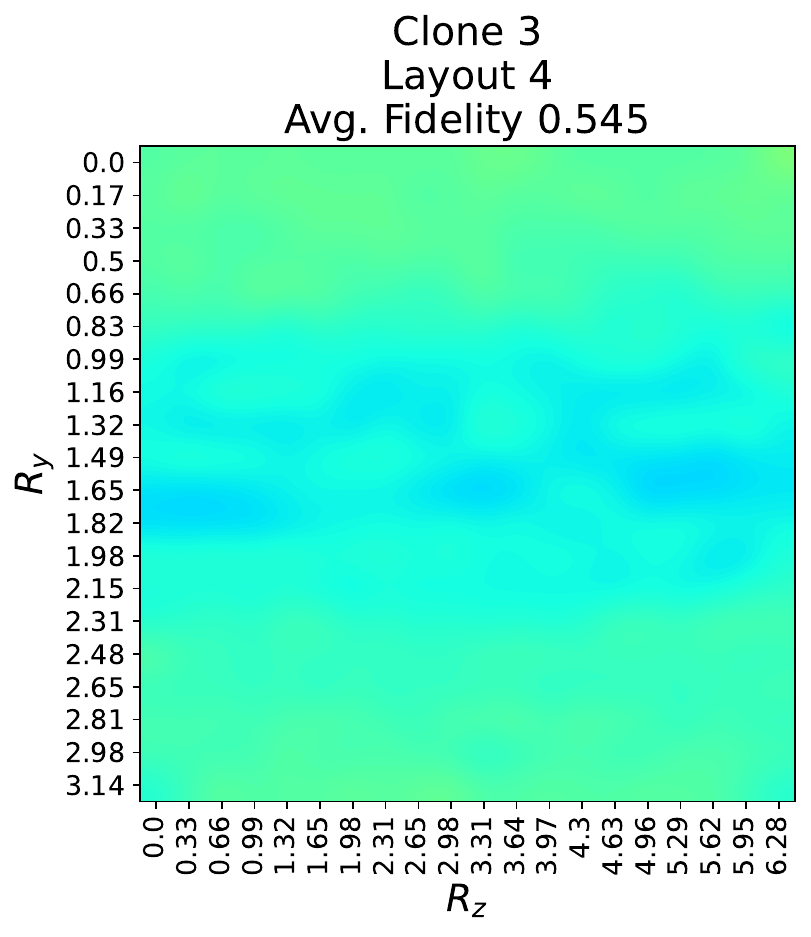}
    \includegraphics[width=0.13\textwidth]{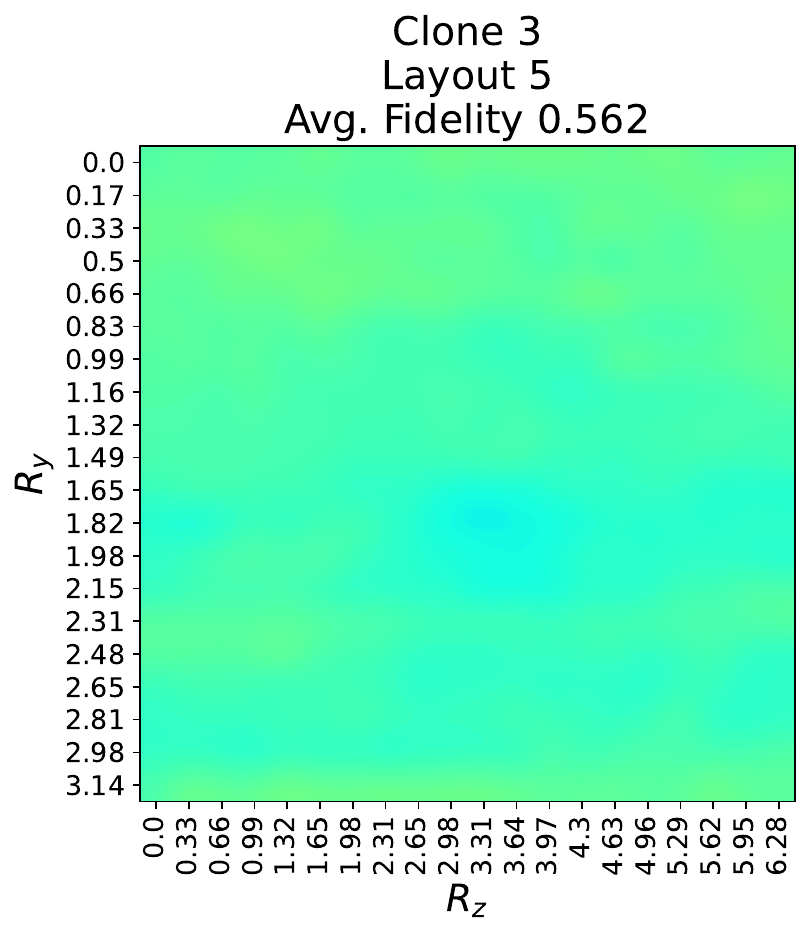}
    \includegraphics[width=0.13\textwidth]{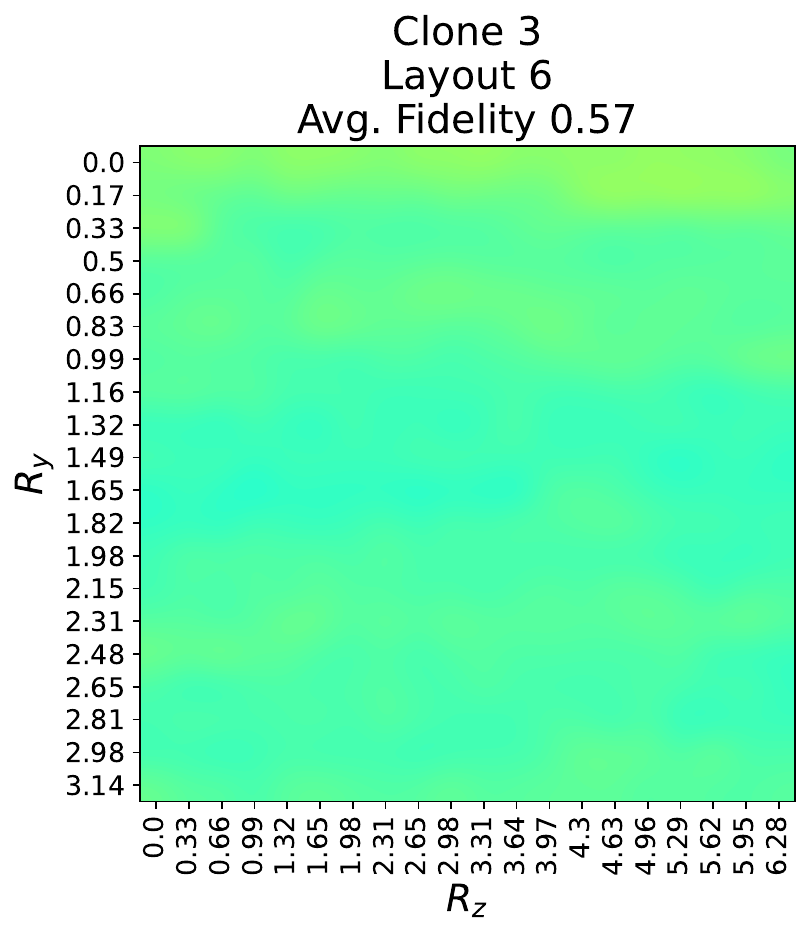}
    \includegraphics[width=0.43\textwidth]{figures/colorbar.pdf}\\
    \caption{Bloch sphere vector representations of the computed density matrices (top $4$ rows) and single qubit clone fidelity heatmaps (bottom $4$ rows) of the single  qubit clone fidelity heatmaps of the single qubit clones for $M=4$ (necessarily with ancilla qubits), executed with dynamical decoupling. Each column corresponds to the $7$ different compiled hardware layouts. Each row corresponds to the $4$ different single qubit clones. Data from \texttt{ibm\_hanoi}.  }
    \label{fig:fidelity_heatmaps_M4_ibm_hanoi_DD}
\end{figure*}

\begin{figure*}[th!]
    \centering
    \includegraphics[width=0.13\textwidth]{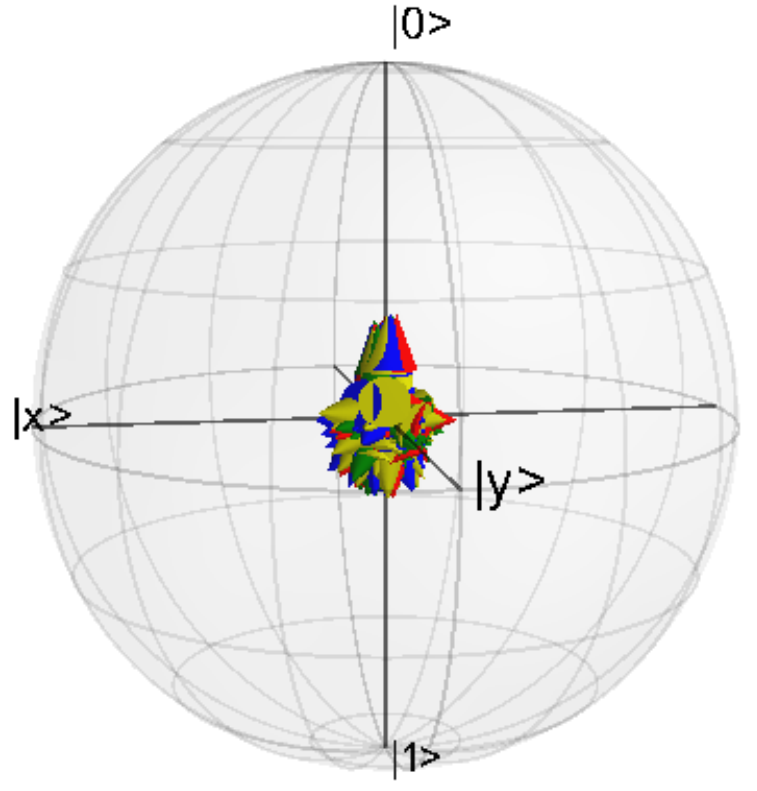}
    \includegraphics[width=0.13\textwidth]{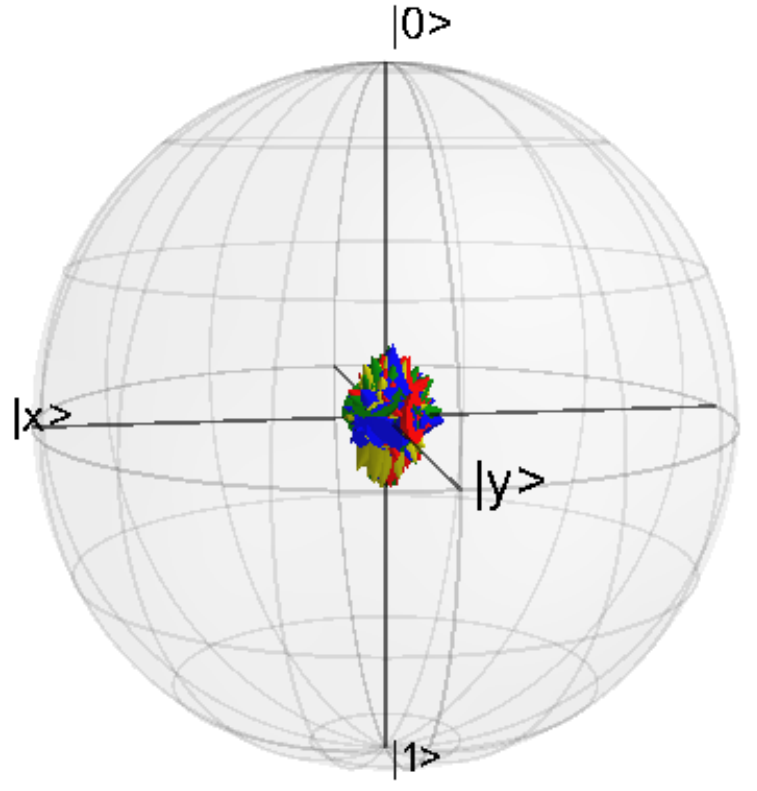}
    \includegraphics[width=0.13\textwidth]{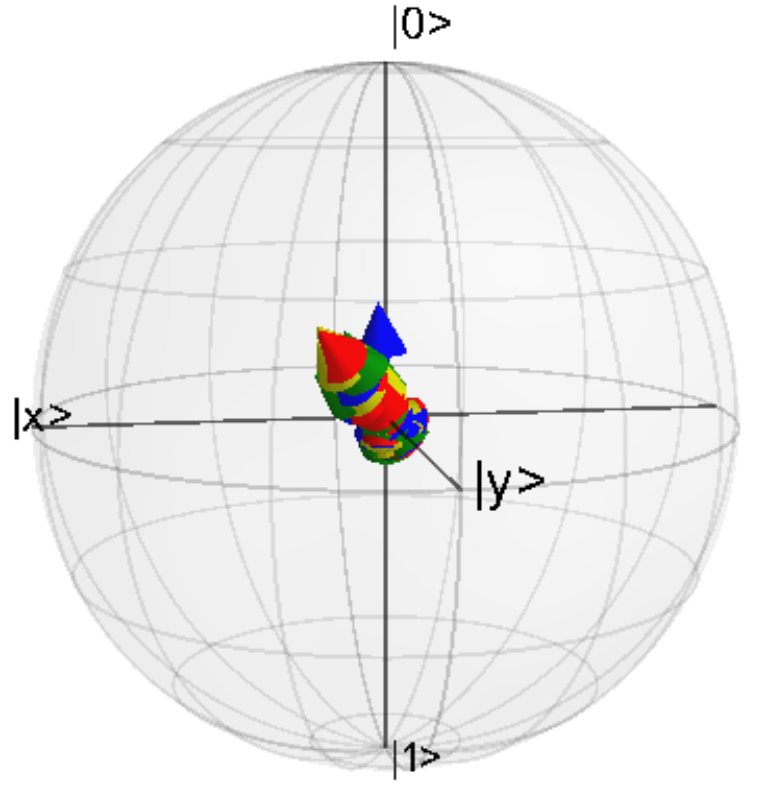}
    \includegraphics[width=0.13\textwidth]{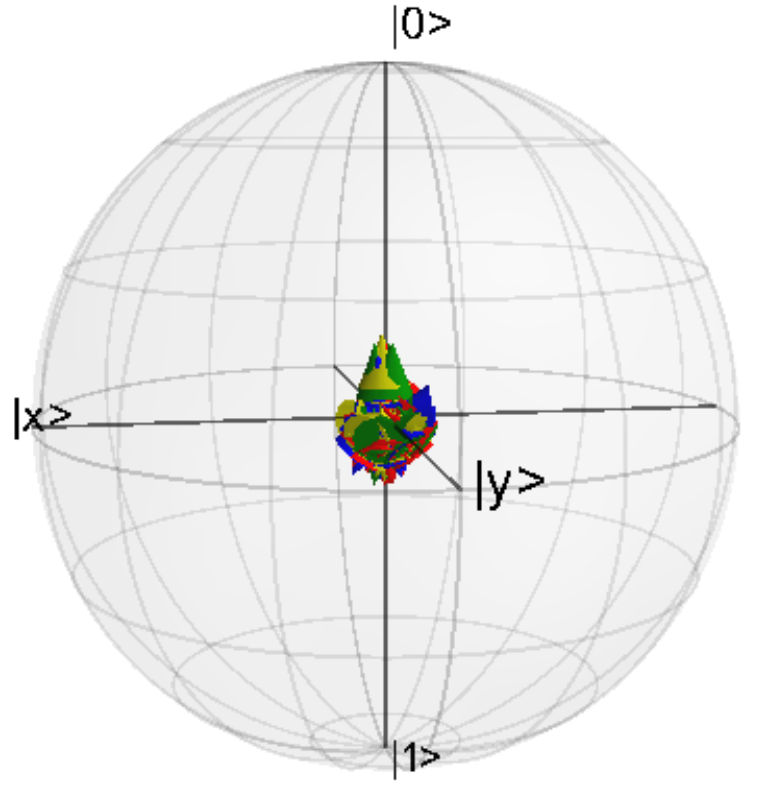}
    \includegraphics[width=0.13\textwidth]{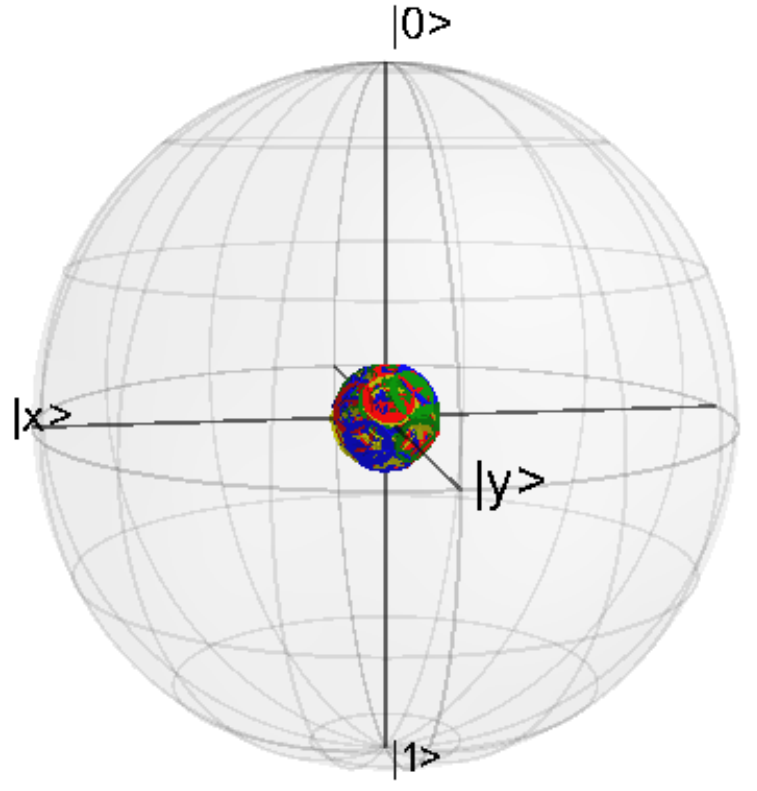}
    \includegraphics[width=0.13\textwidth]{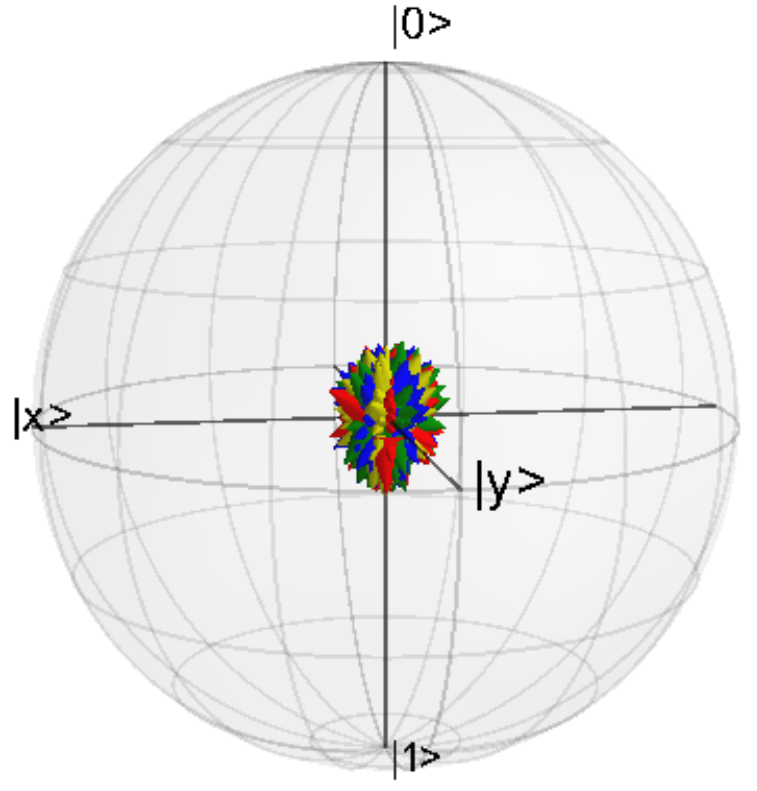}
    \includegraphics[width=0.13\textwidth]{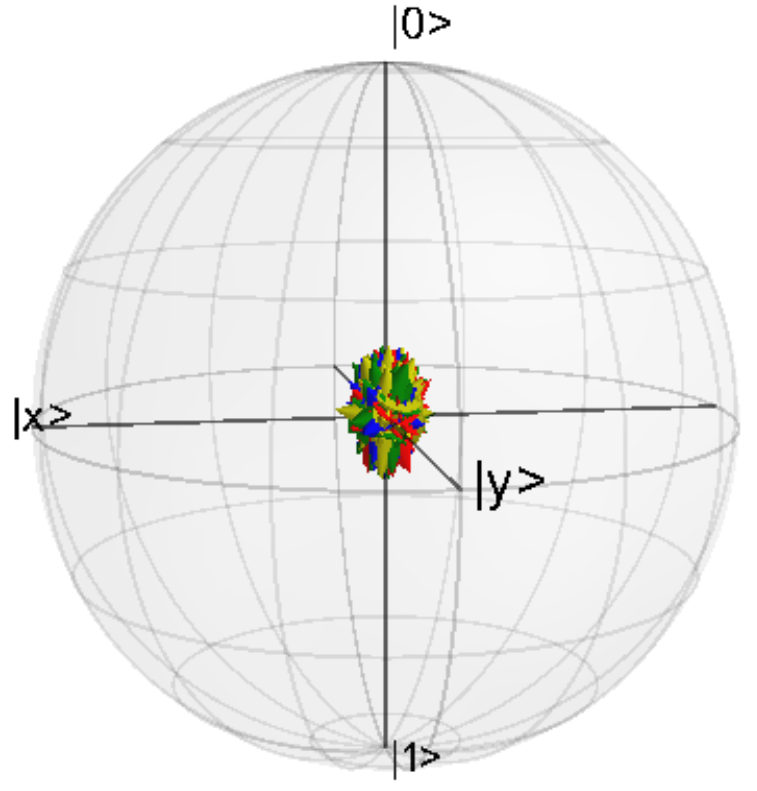}
    \includegraphics[width=0.13\textwidth]{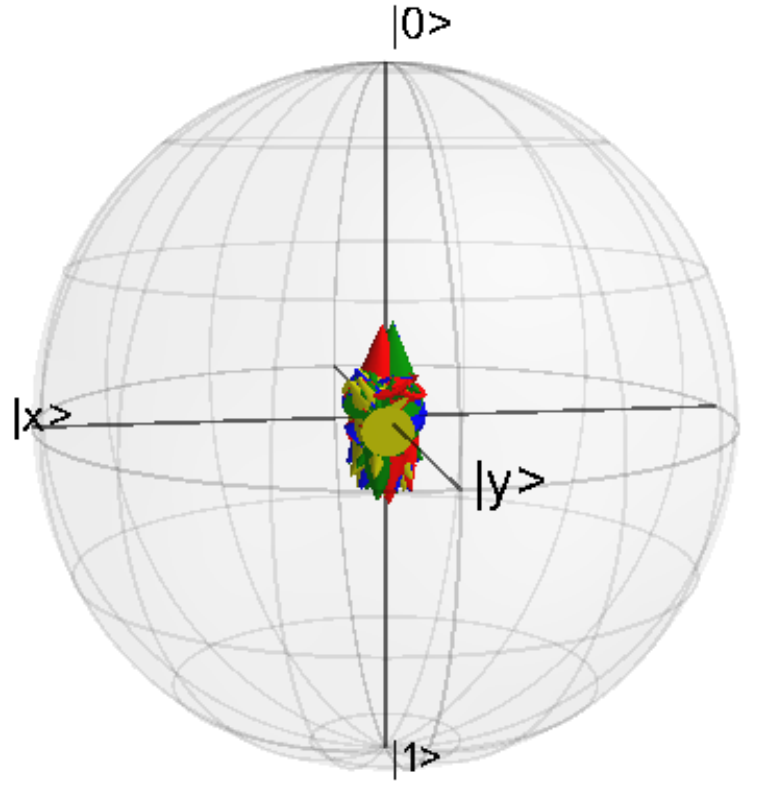}
    \includegraphics[width=0.13\textwidth]{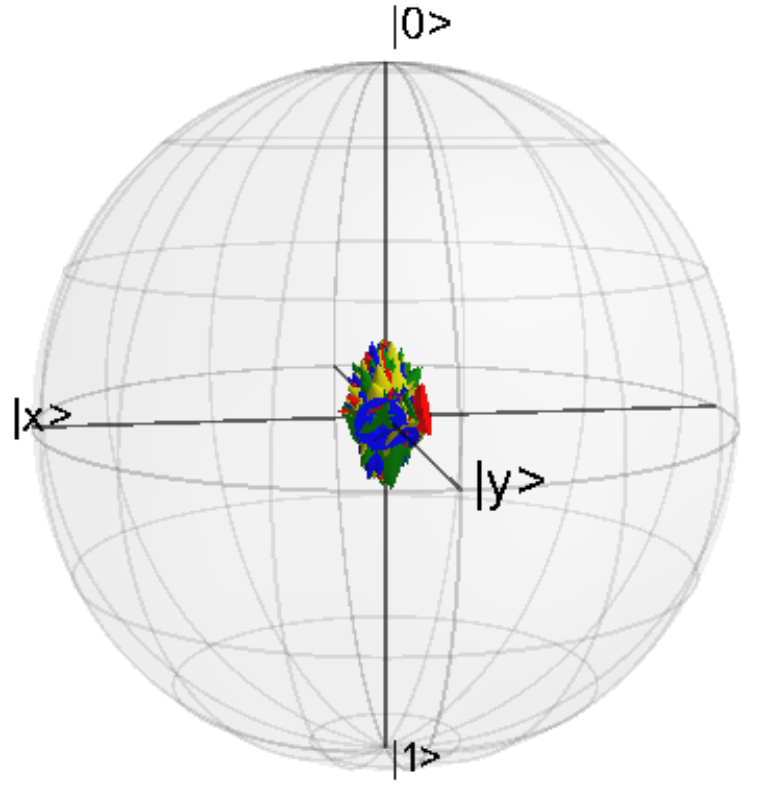}
    \includegraphics[width=0.13\textwidth]{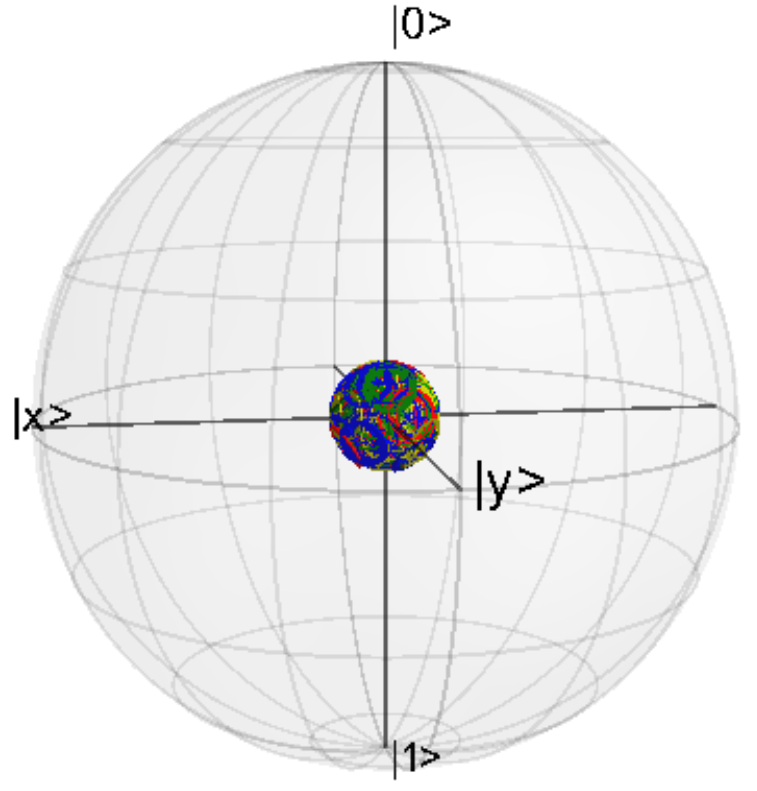}
    \includegraphics[width=0.13\textwidth]{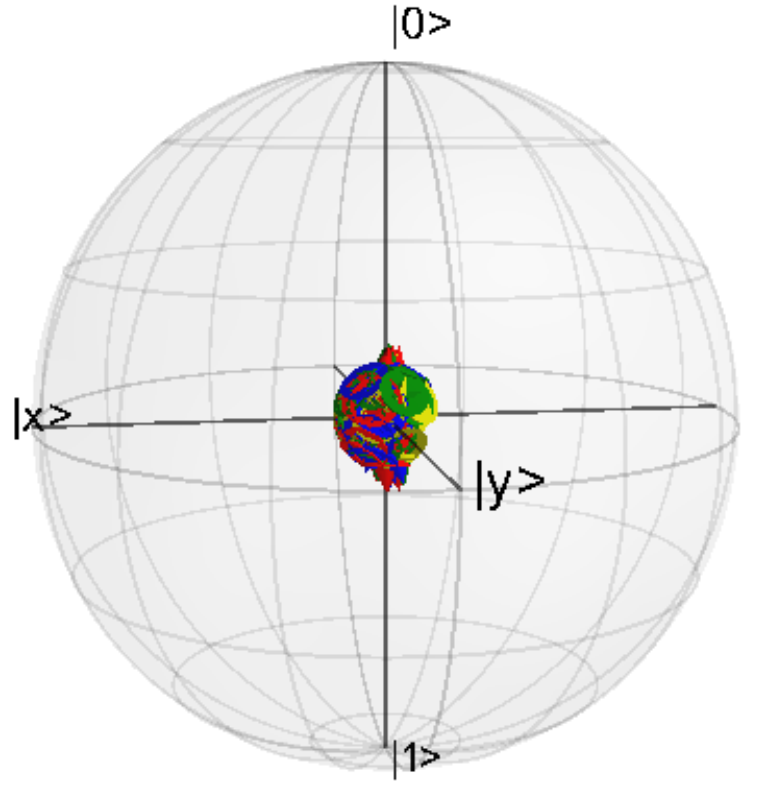}
    \includegraphics[width=0.13\textwidth]{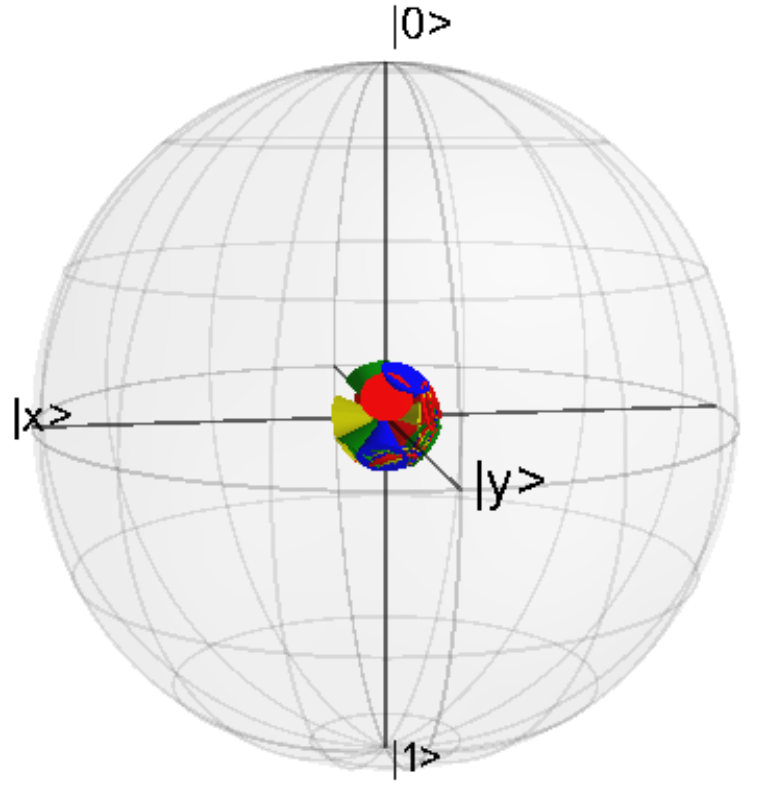}
    \includegraphics[width=0.13\textwidth]{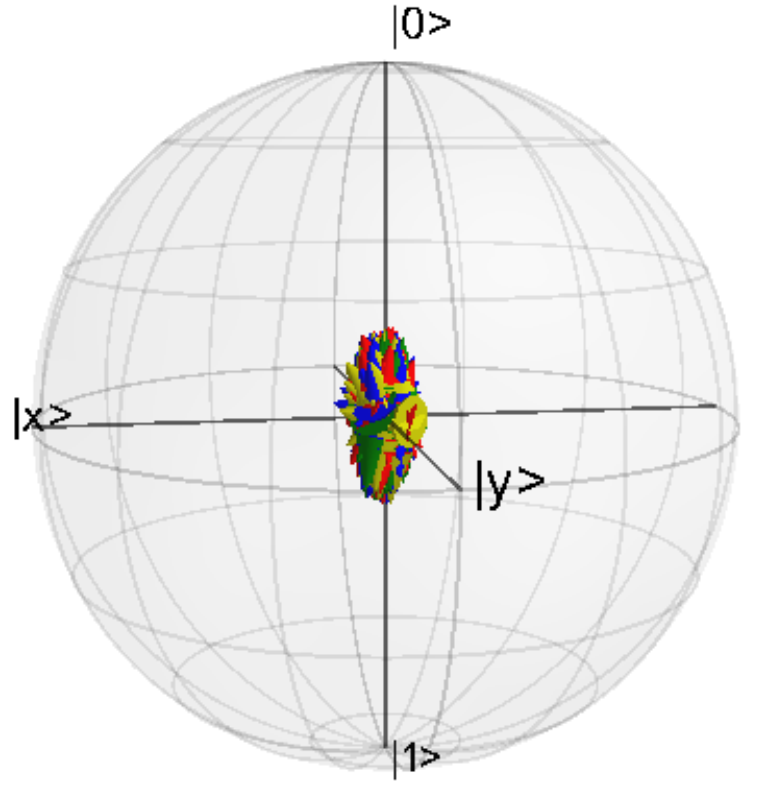}
    \includegraphics[width=0.13\textwidth]{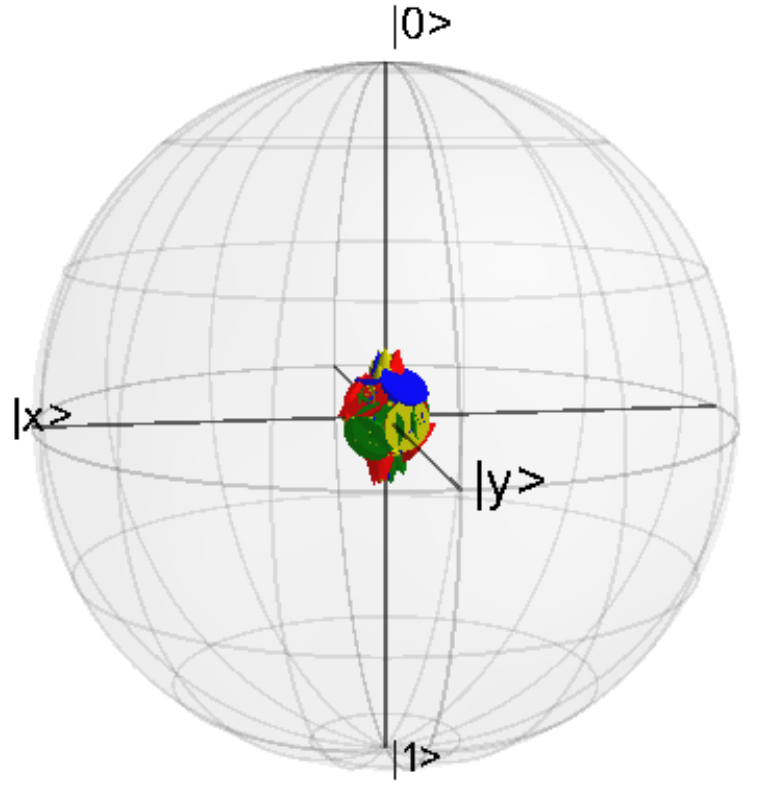}
    \includegraphics[width=0.13\textwidth]{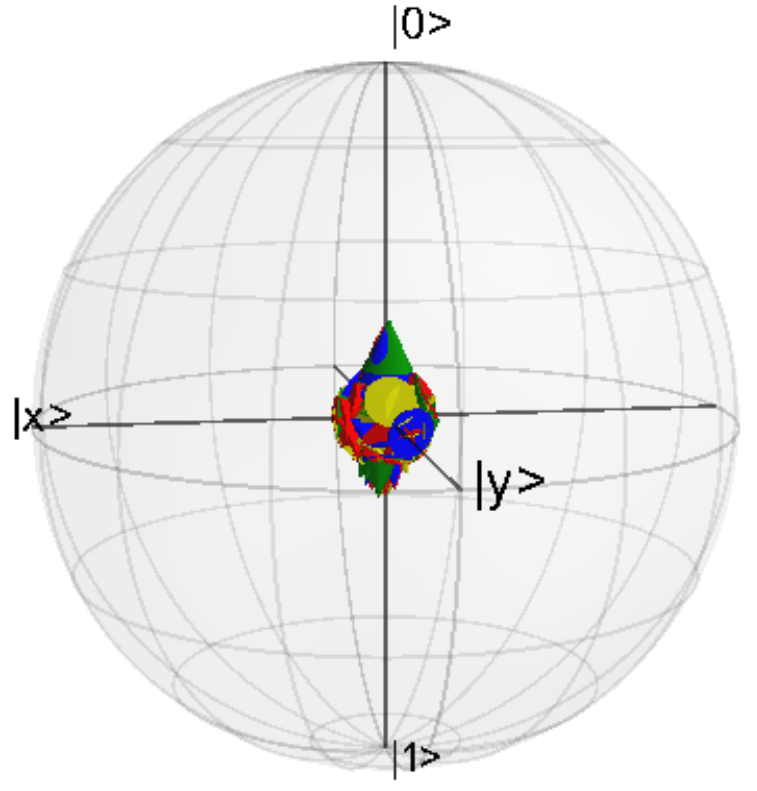}
    \includegraphics[width=0.13\textwidth]{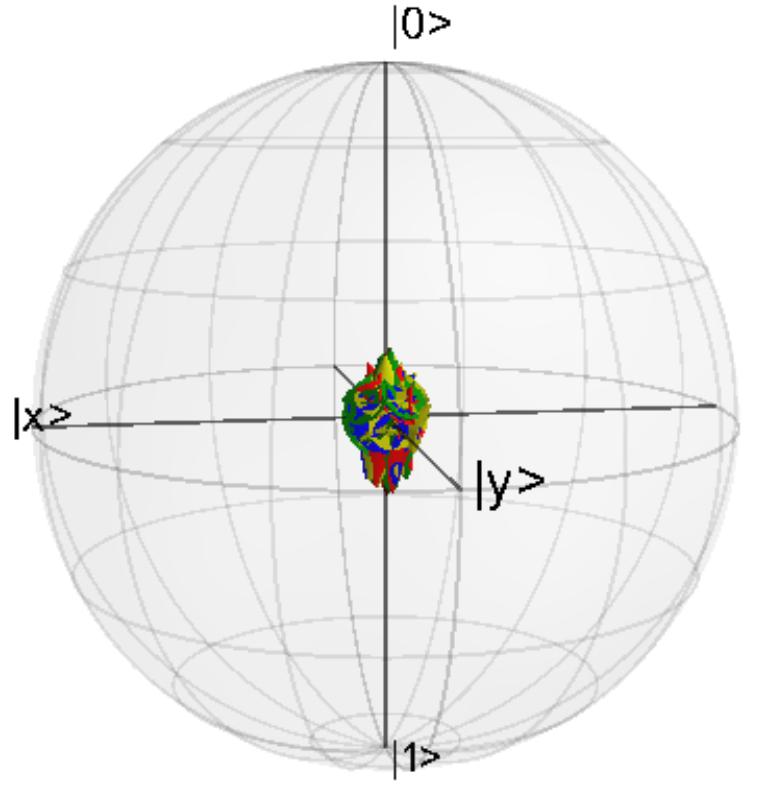}
    \includegraphics[width=0.13\textwidth]{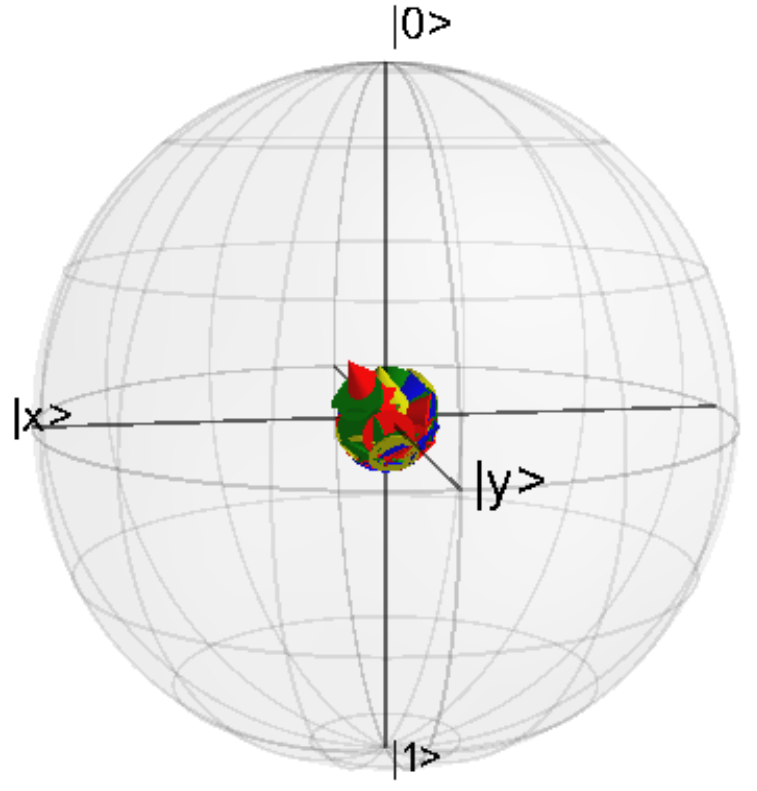}
    \includegraphics[width=0.13\textwidth]{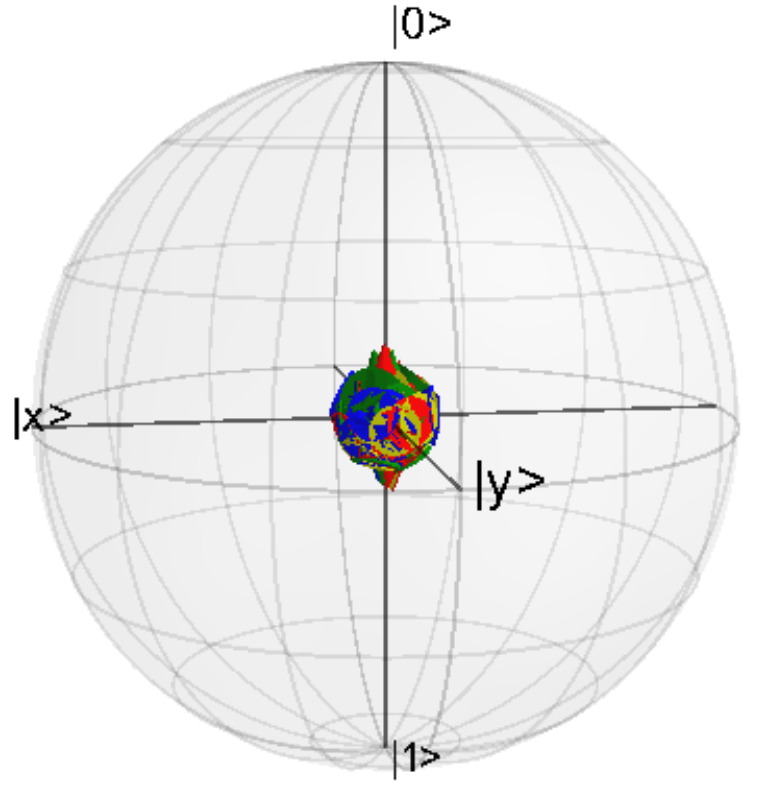}
    \includegraphics[width=0.13\textwidth]{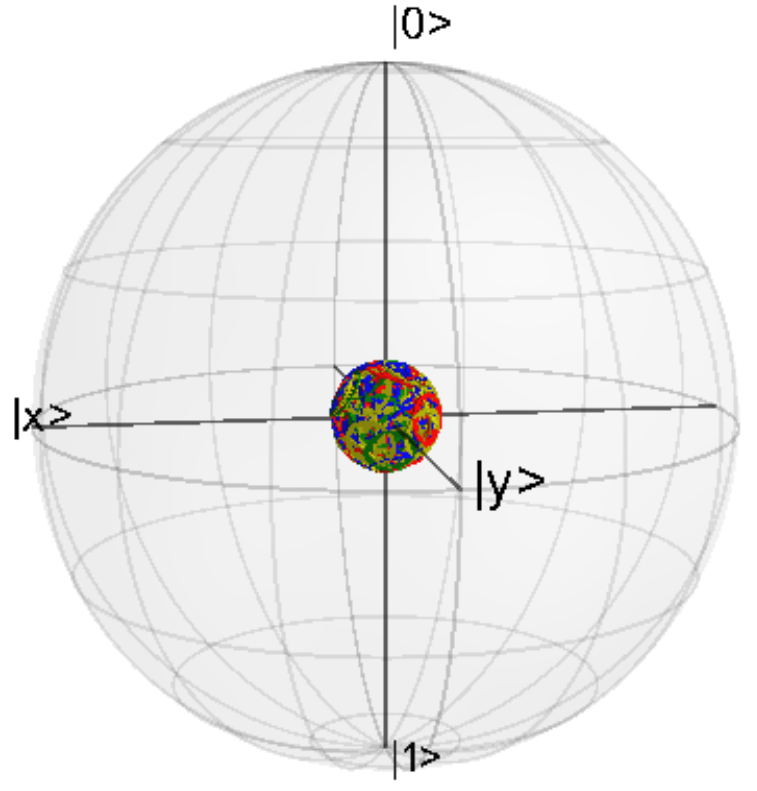}
    \includegraphics[width=0.13\textwidth]{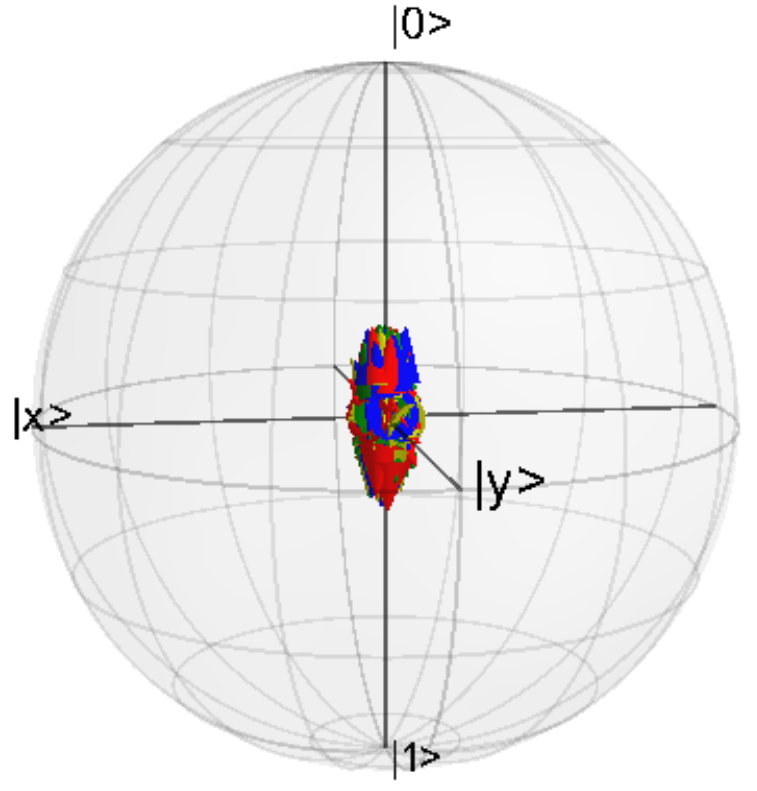}
    \includegraphics[width=0.13\textwidth]{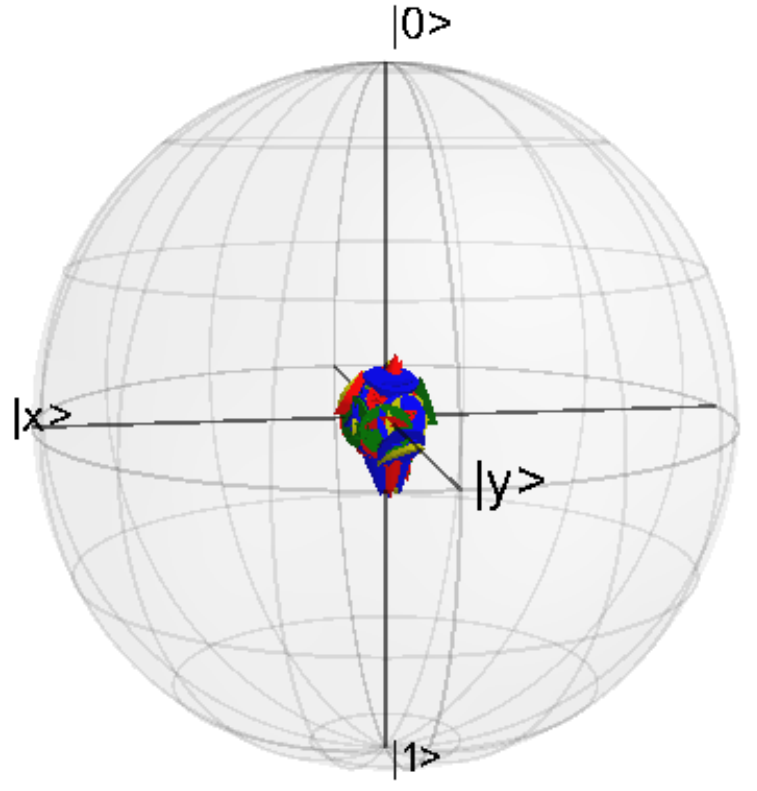}
    \includegraphics[width=0.13\textwidth]{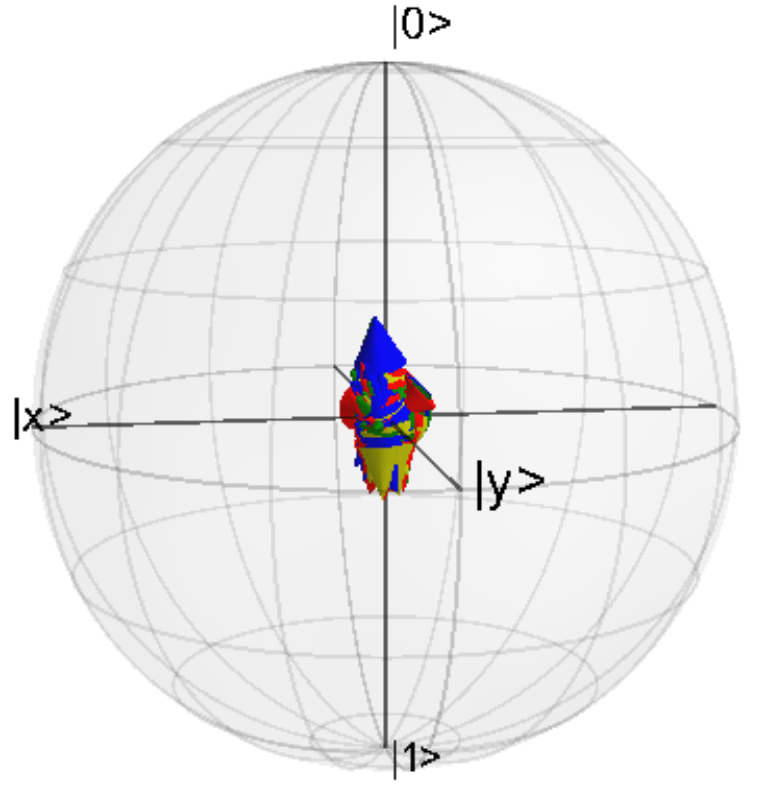}
    \includegraphics[width=0.13\textwidth]{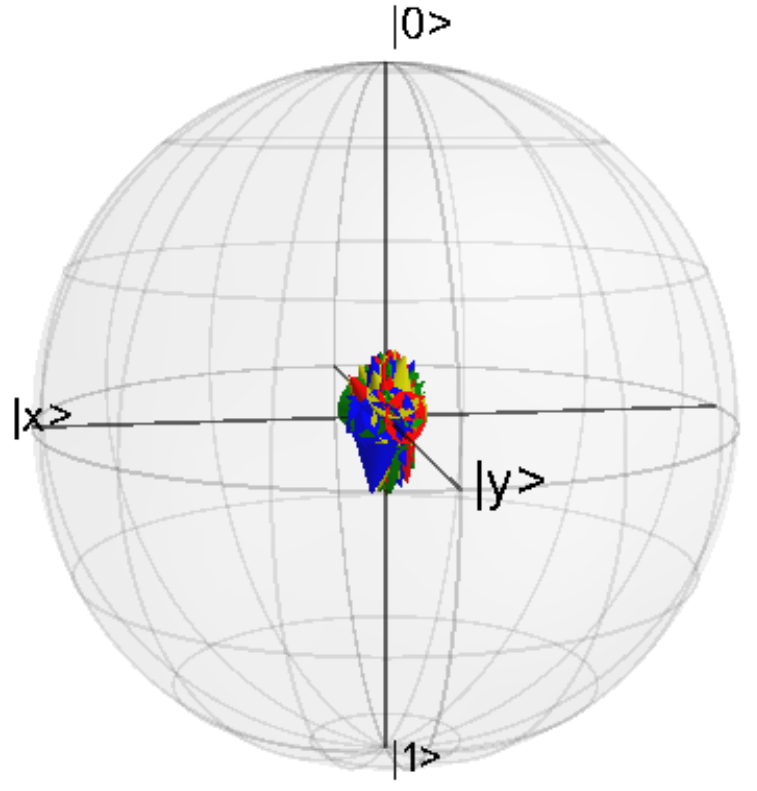}
    \includegraphics[width=0.13\textwidth]{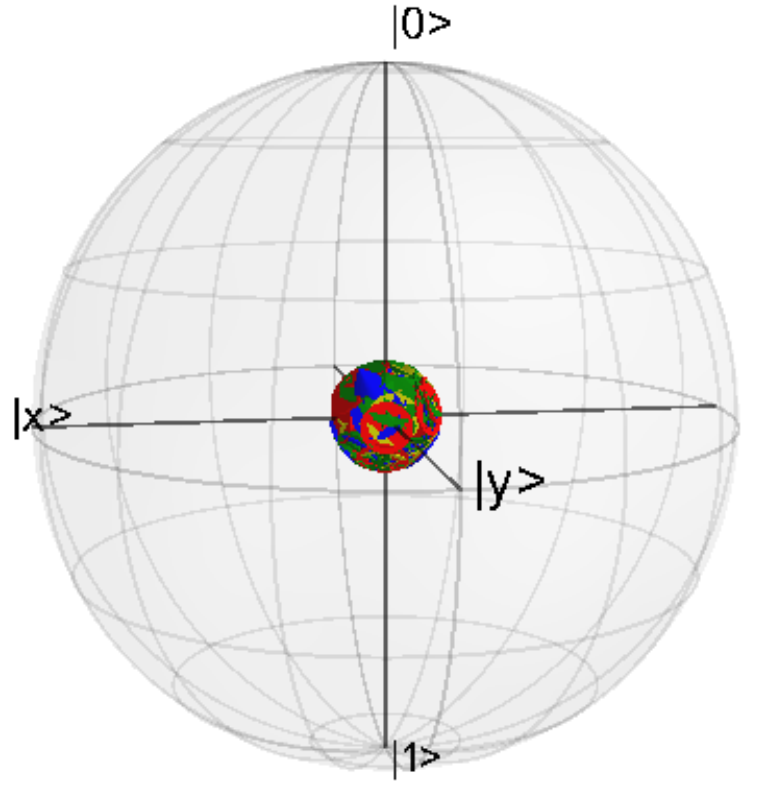}
    \includegraphics[width=0.13\textwidth]{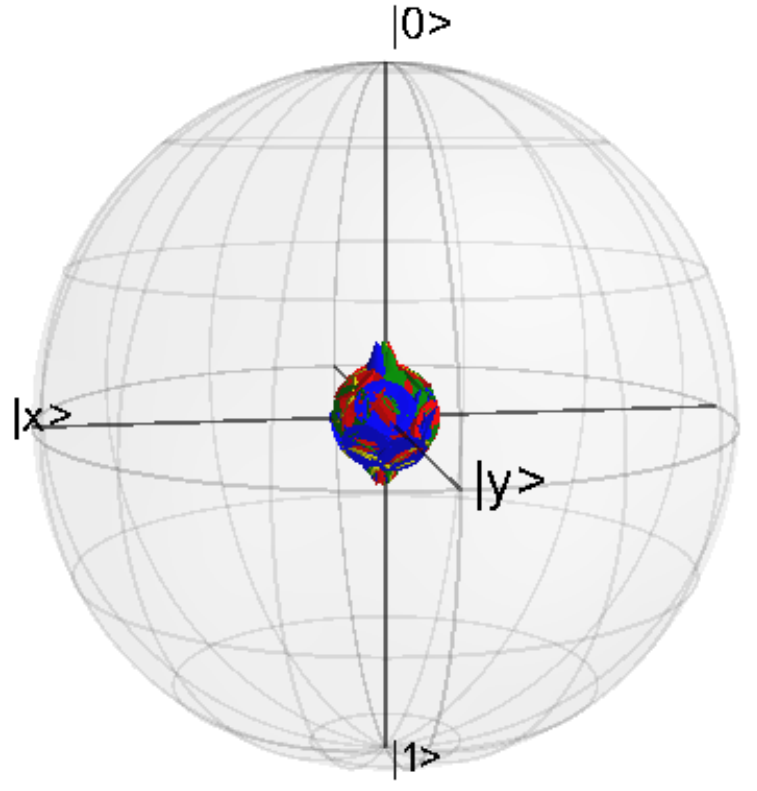}
    \includegraphics[width=0.13\textwidth]{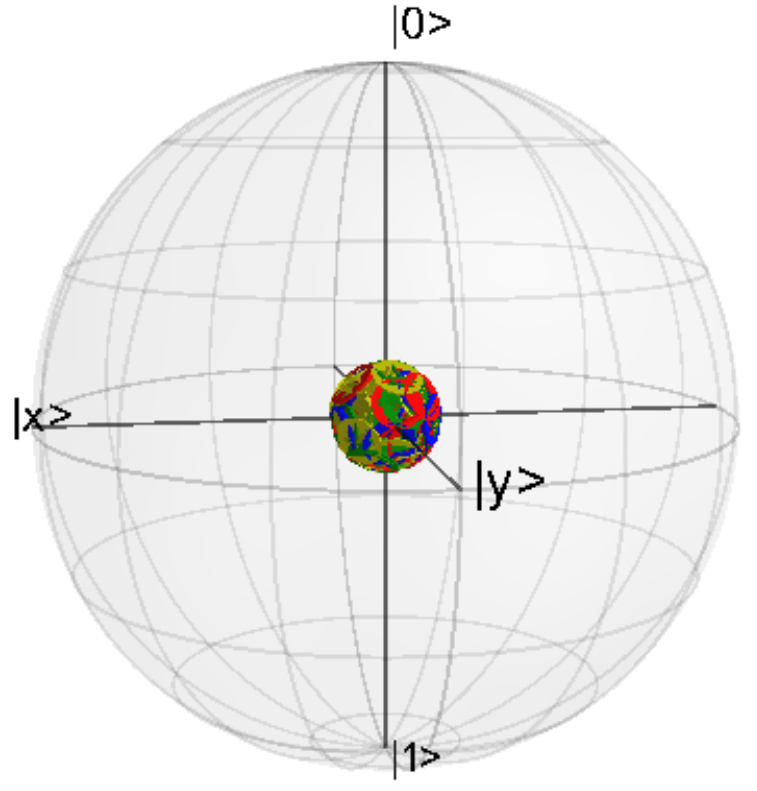}
    \includegraphics[width=0.13\textwidth]{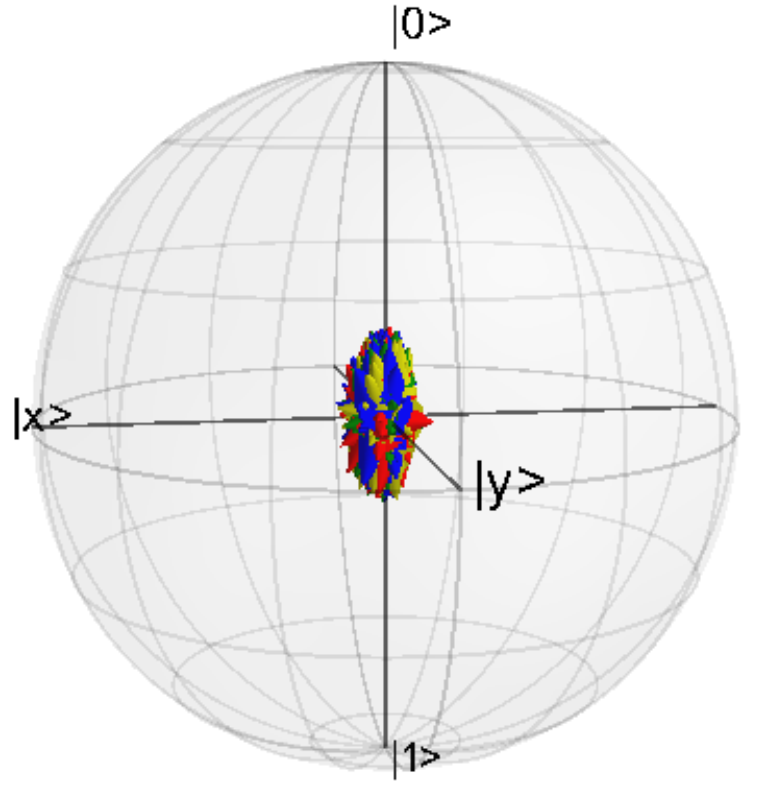}
    \includegraphics[width=0.13\textwidth]{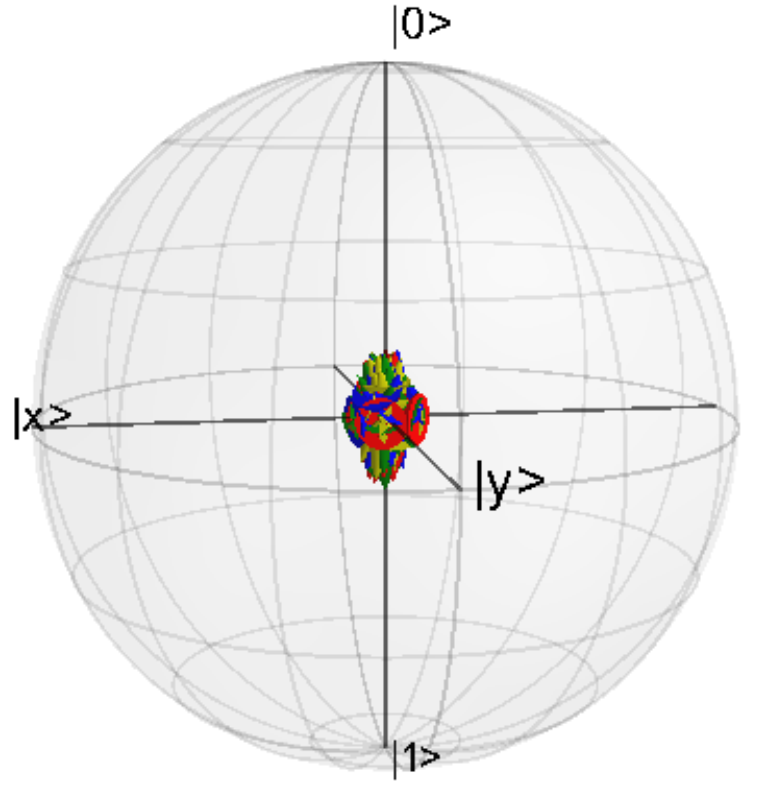}
    \includegraphics[width=0.13\textwidth]{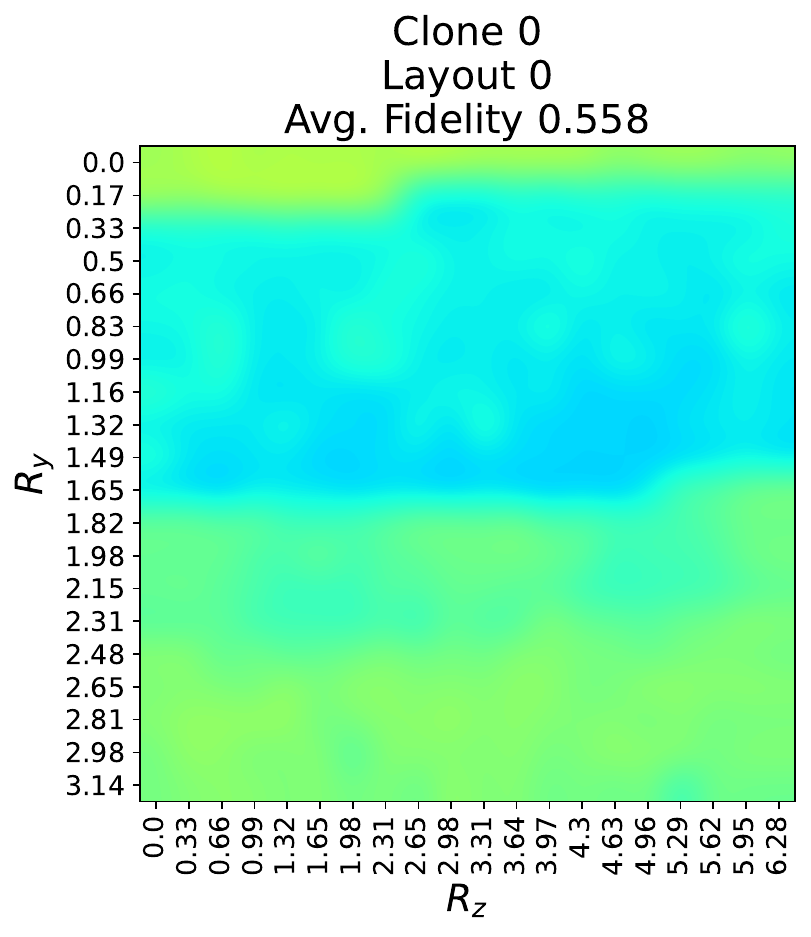}
    \includegraphics[width=0.13\textwidth]{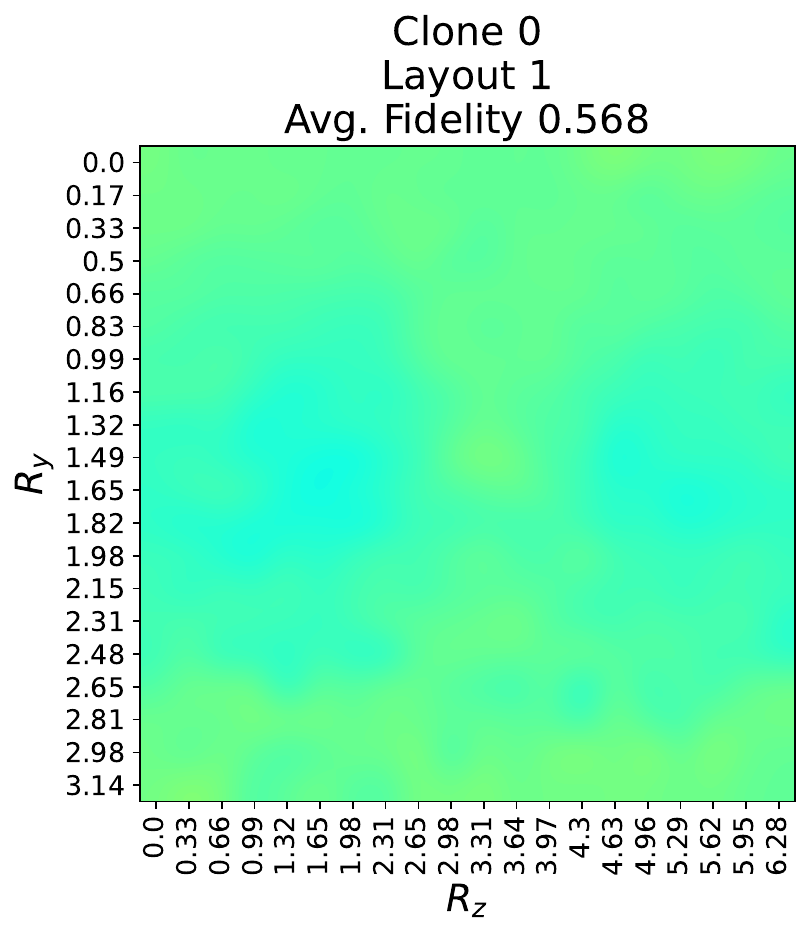}
    \includegraphics[width=0.13\textwidth]{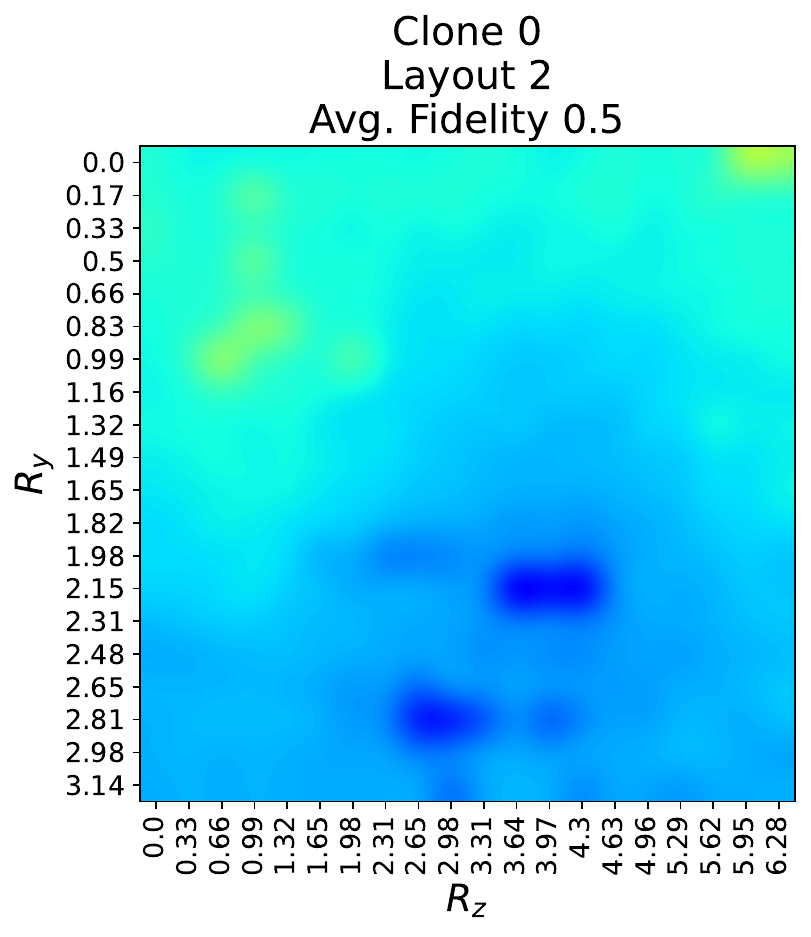}
    \includegraphics[width=0.13\textwidth]{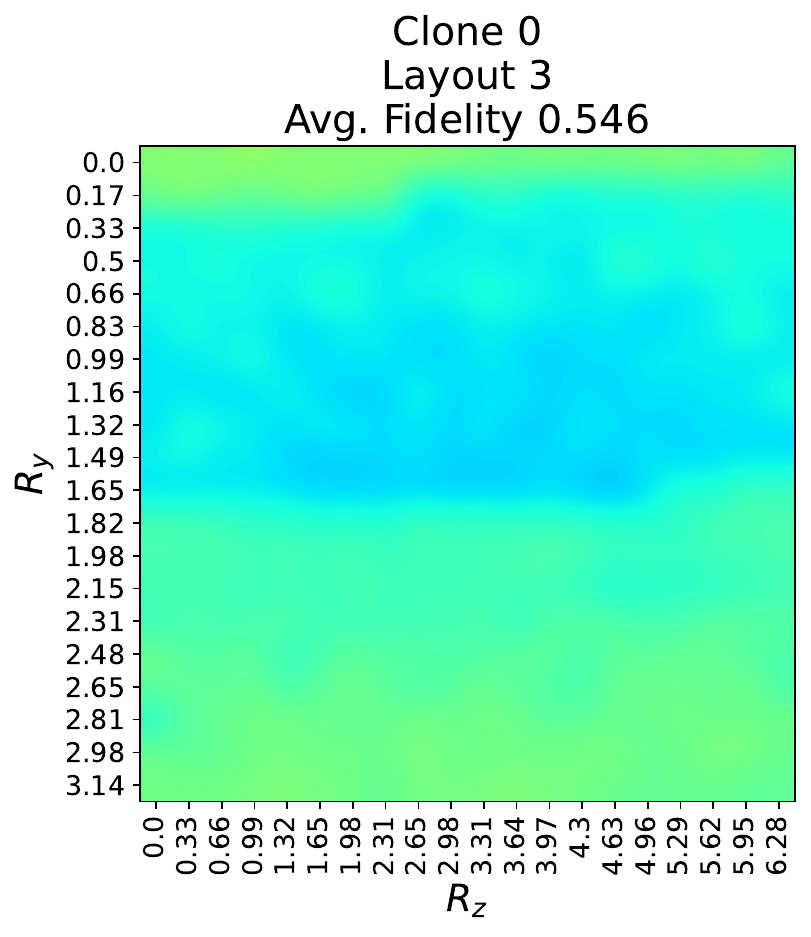}
    \includegraphics[width=0.13\textwidth]{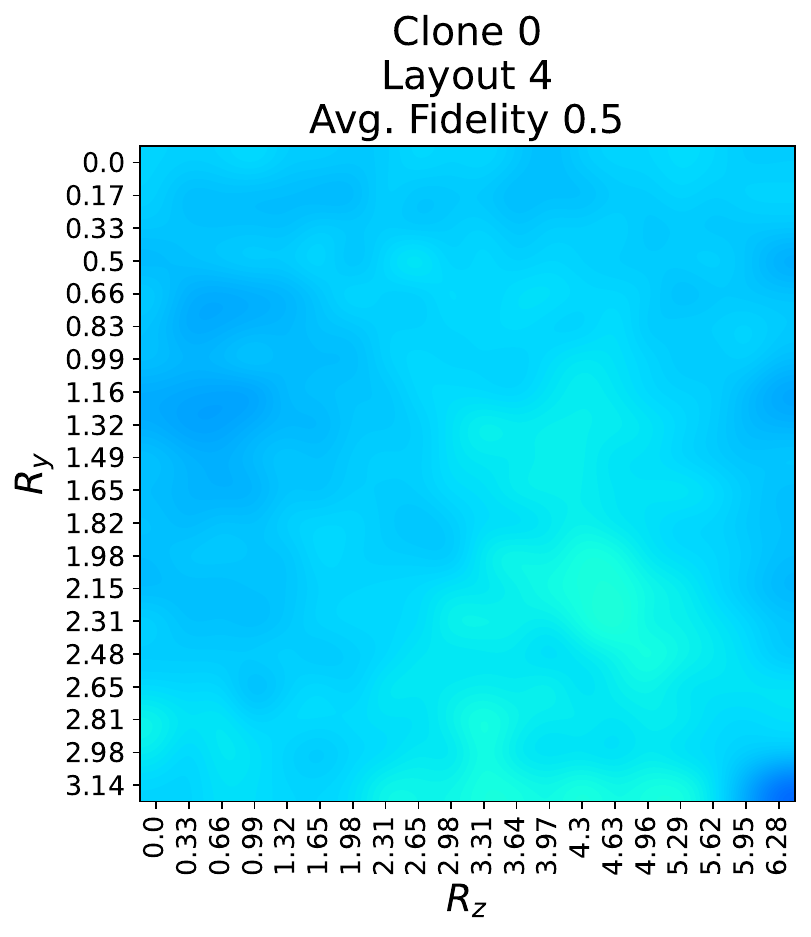}
    \includegraphics[width=0.13\textwidth]{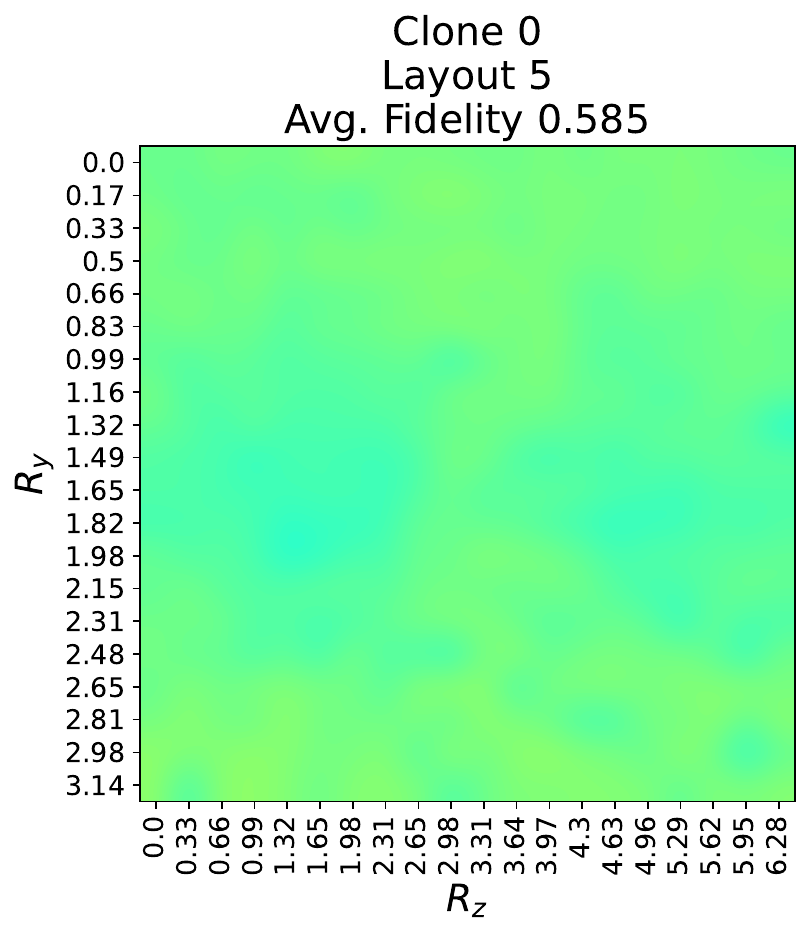}
    \includegraphics[width=0.13\textwidth]{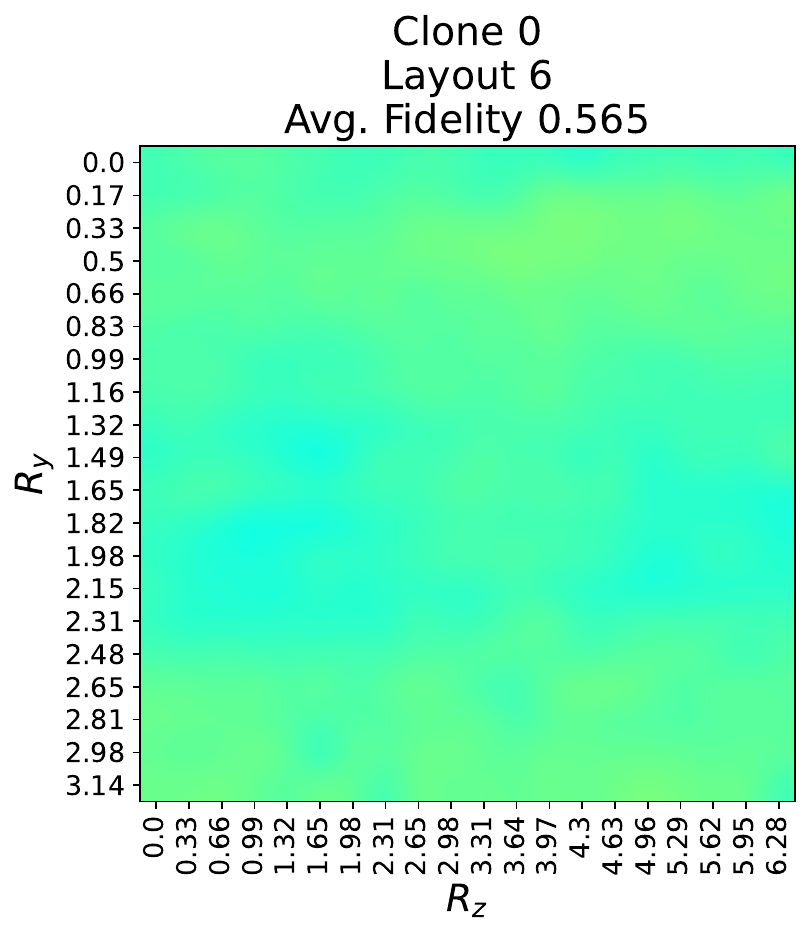}
    \includegraphics[width=0.13\textwidth]{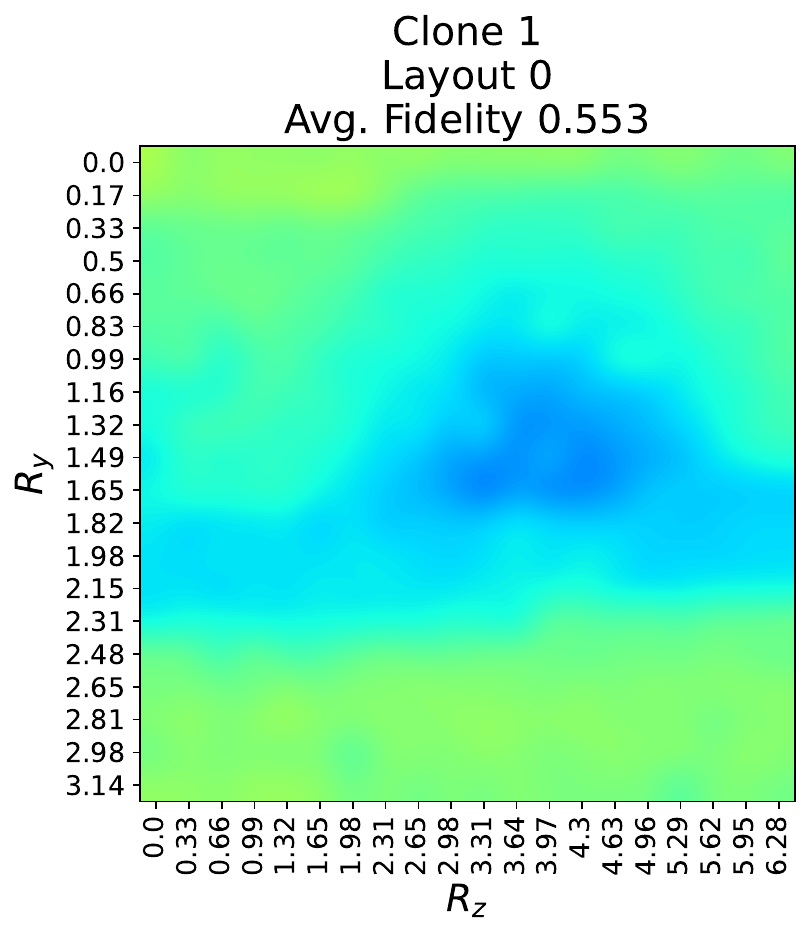}
    \includegraphics[width=0.13\textwidth]{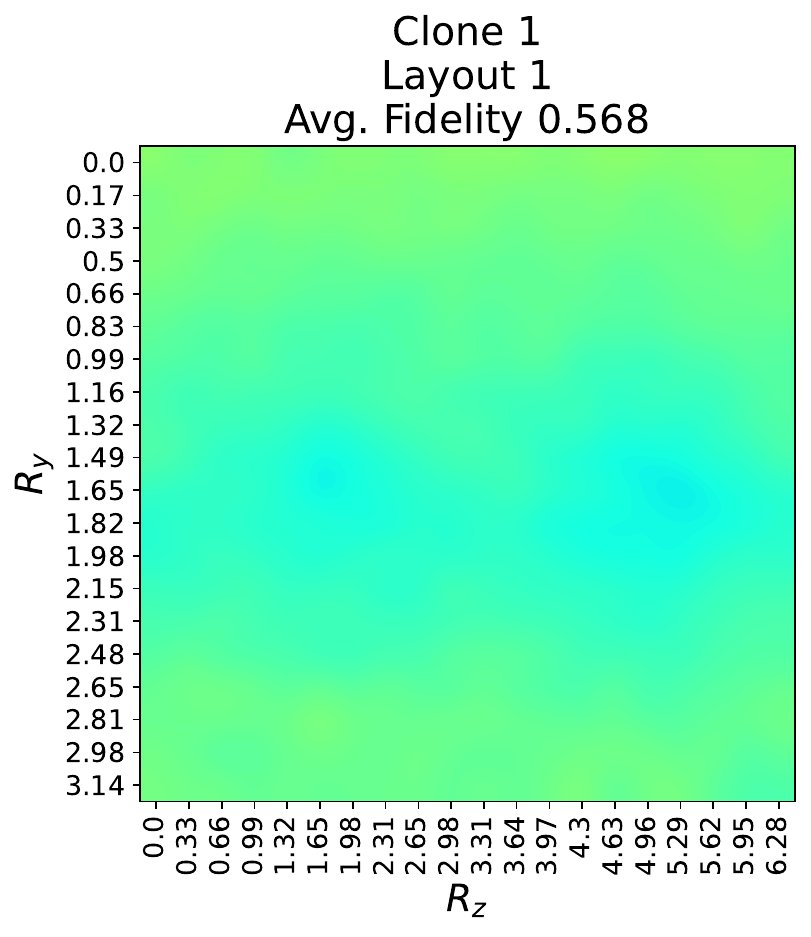}
    \includegraphics[width=0.13\textwidth]{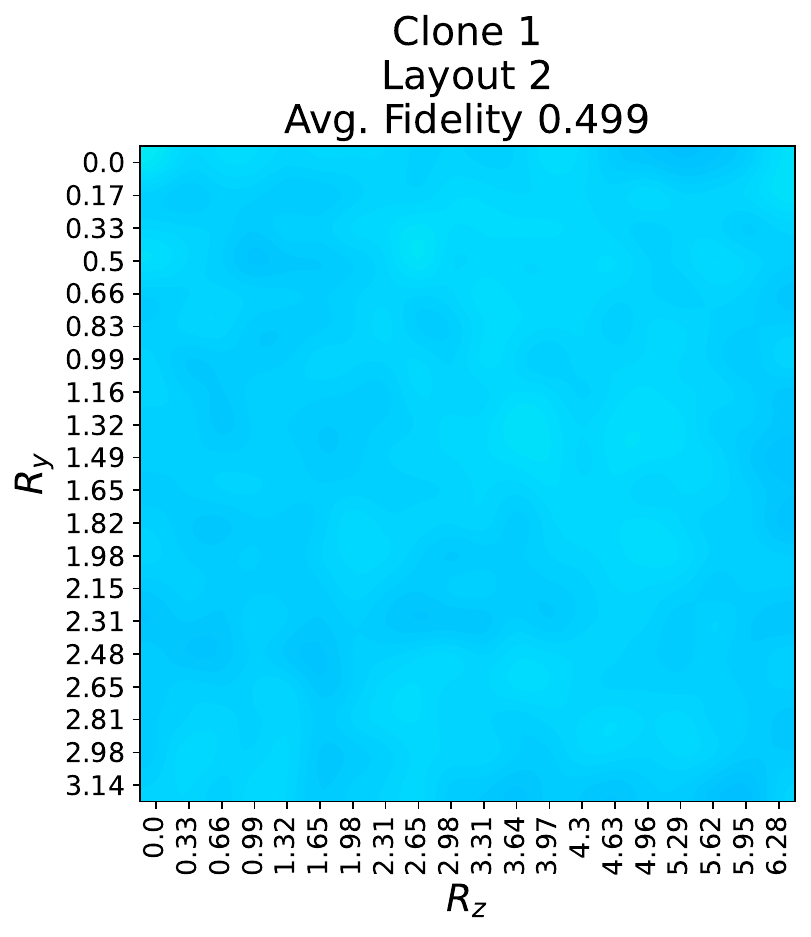}
    \includegraphics[width=0.13\textwidth]{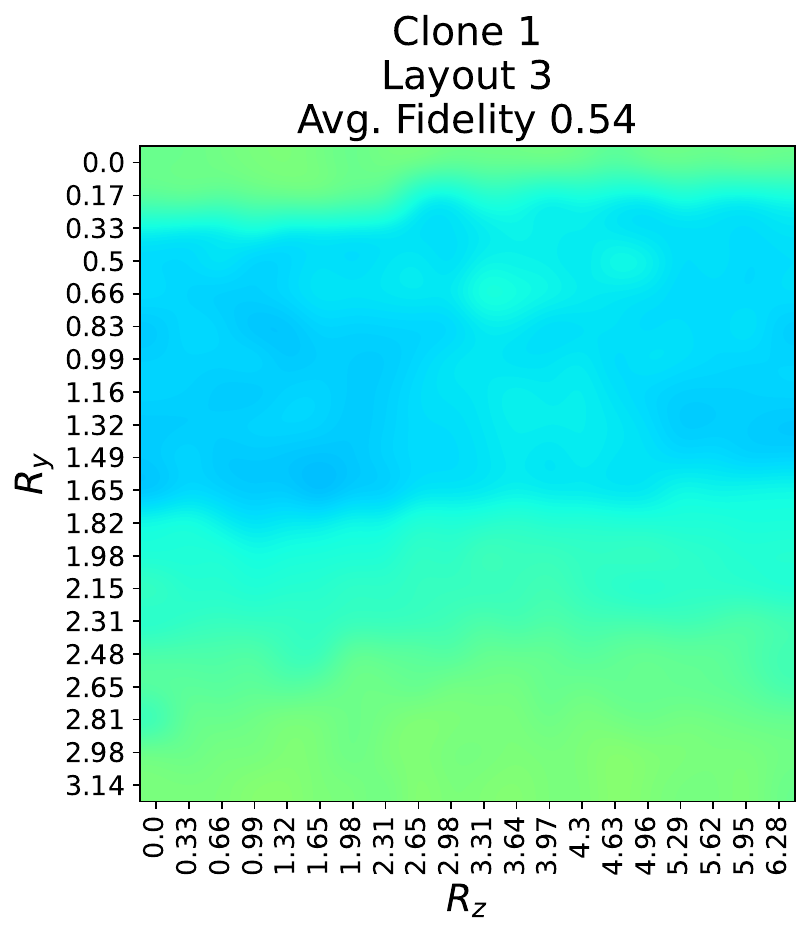}
    \includegraphics[width=0.13\textwidth]{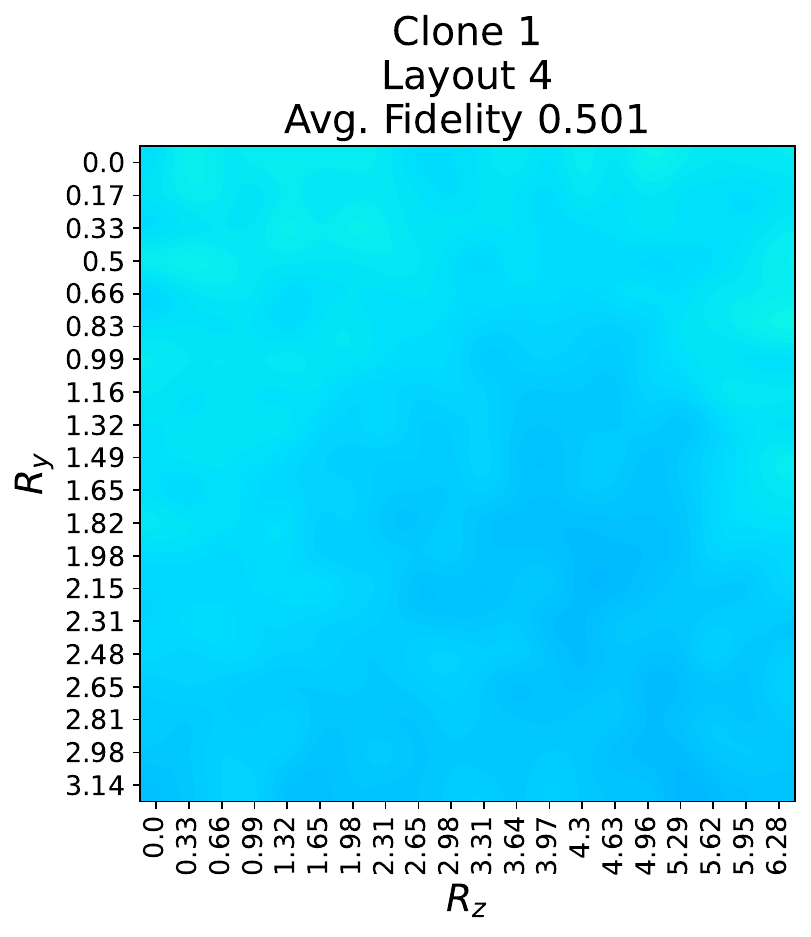}
    \includegraphics[width=0.13\textwidth]{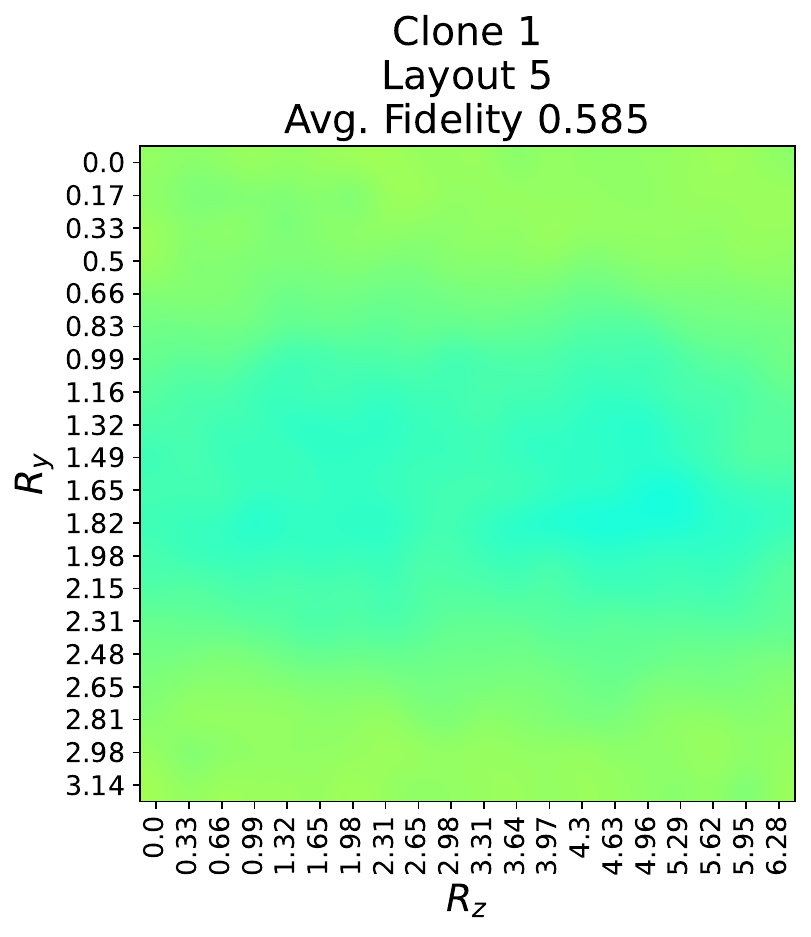}
    \includegraphics[width=0.13\textwidth]{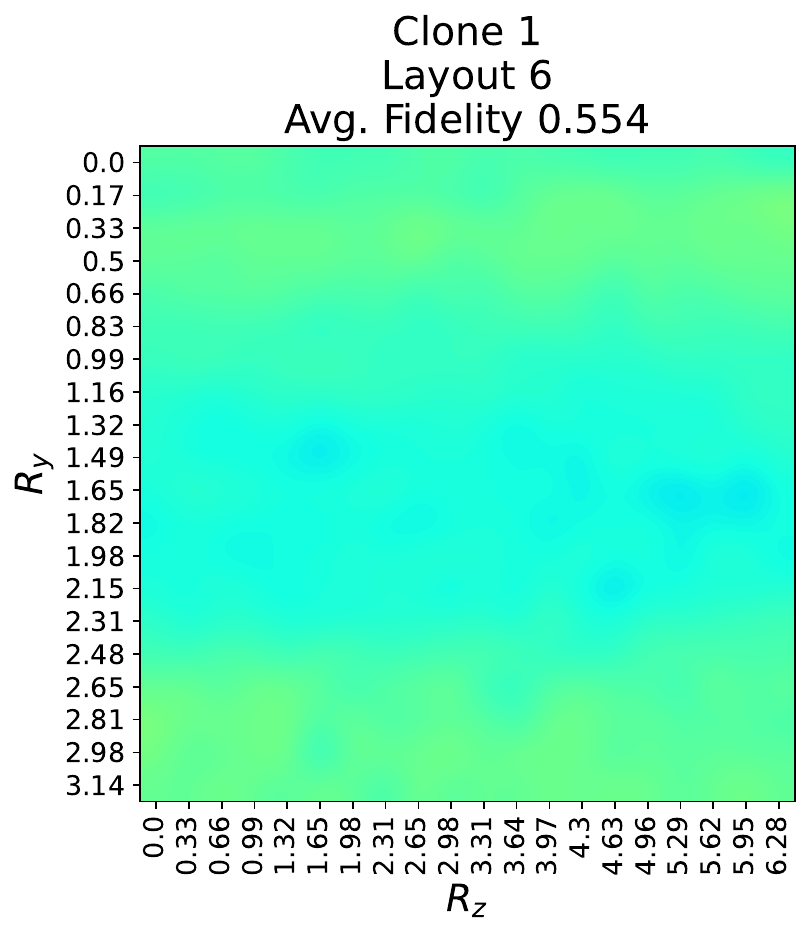}
    \includegraphics[width=0.13\textwidth]{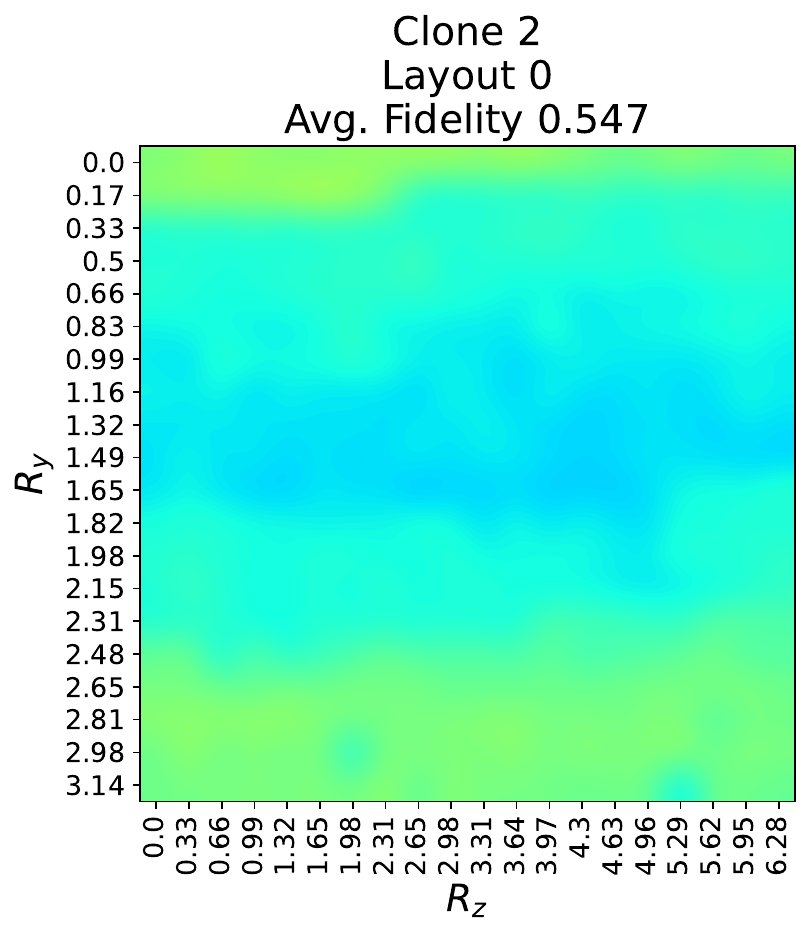}
    \includegraphics[width=0.13\textwidth]{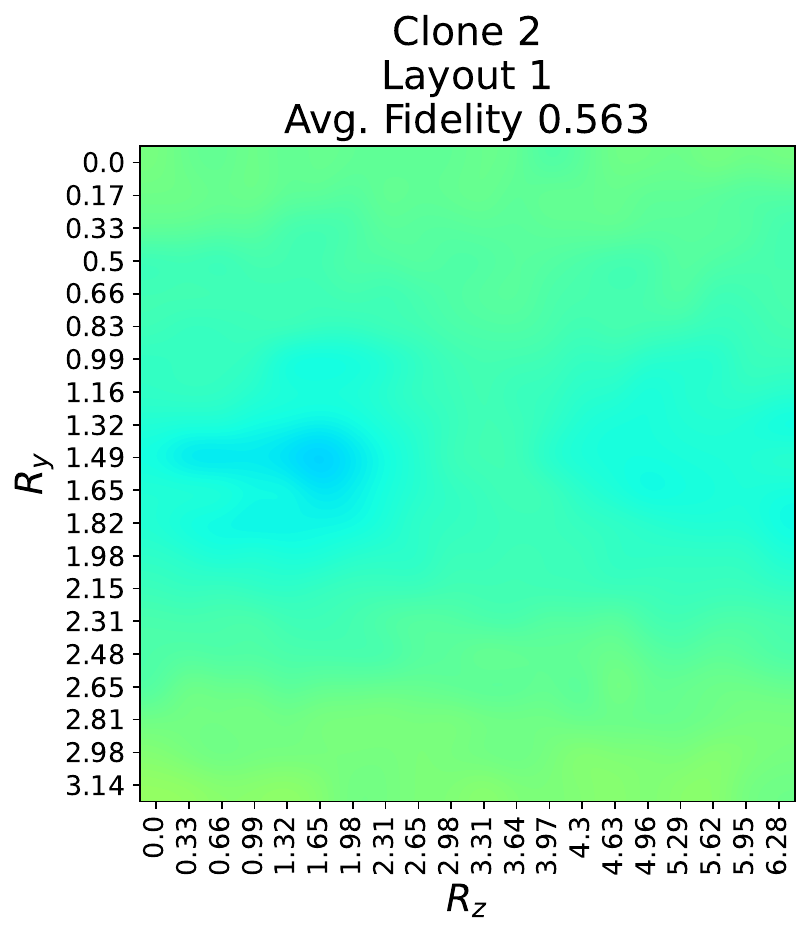}
    \includegraphics[width=0.13\textwidth]{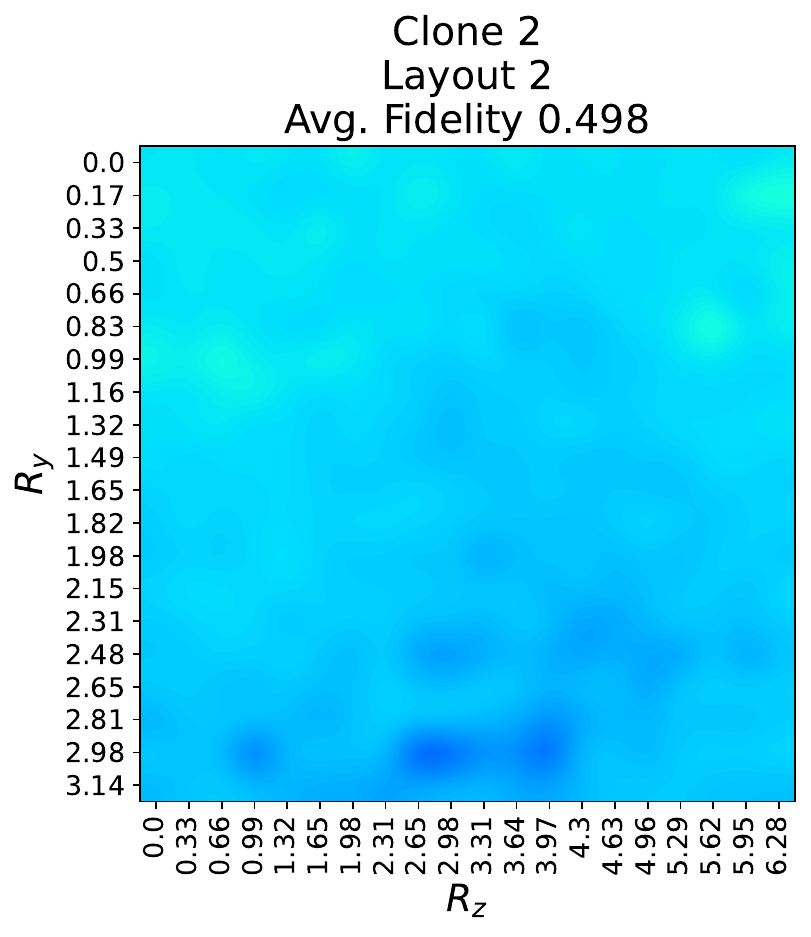}
    \includegraphics[width=0.13\textwidth]{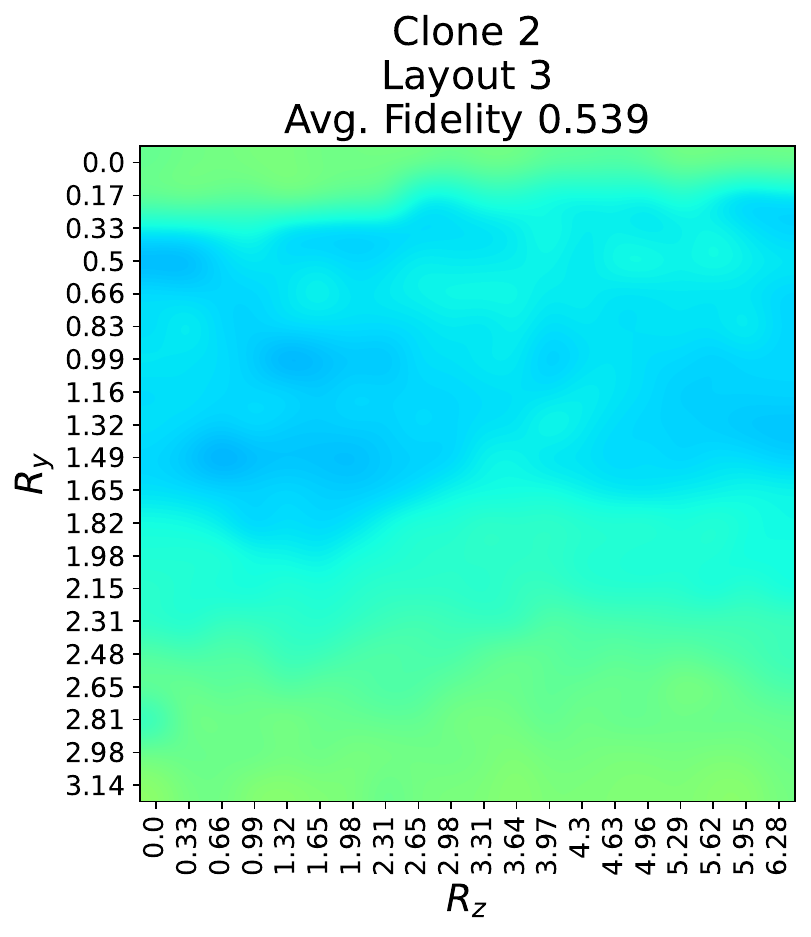}
    \includegraphics[width=0.13\textwidth]{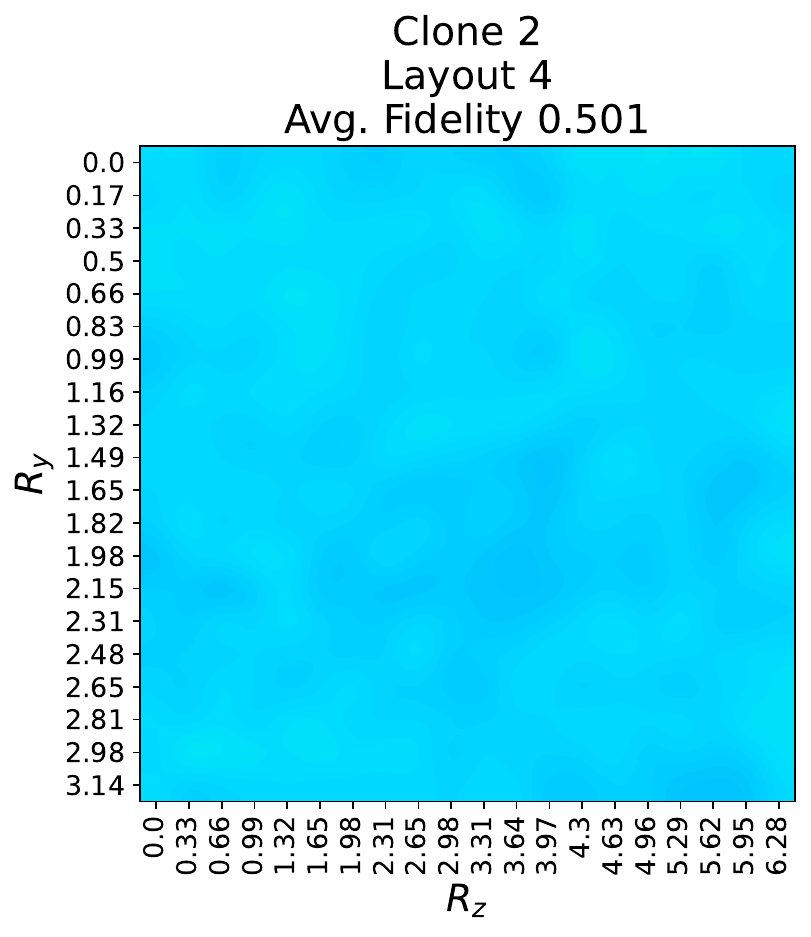}
    \includegraphics[width=0.13\textwidth]{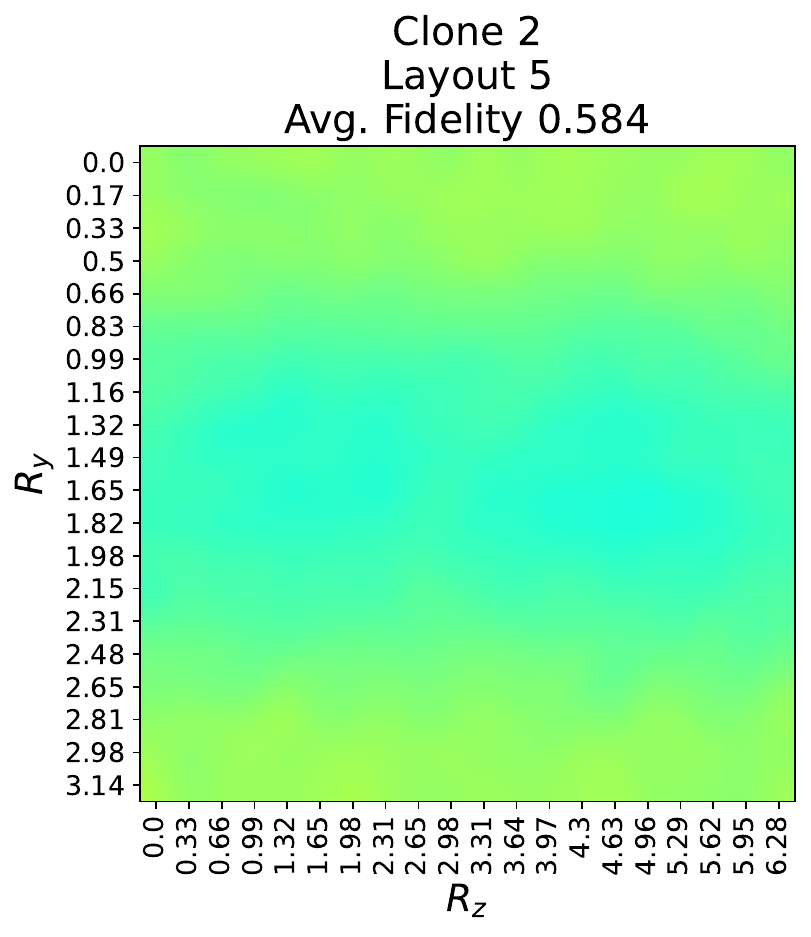}
    \includegraphics[width=0.13\textwidth]{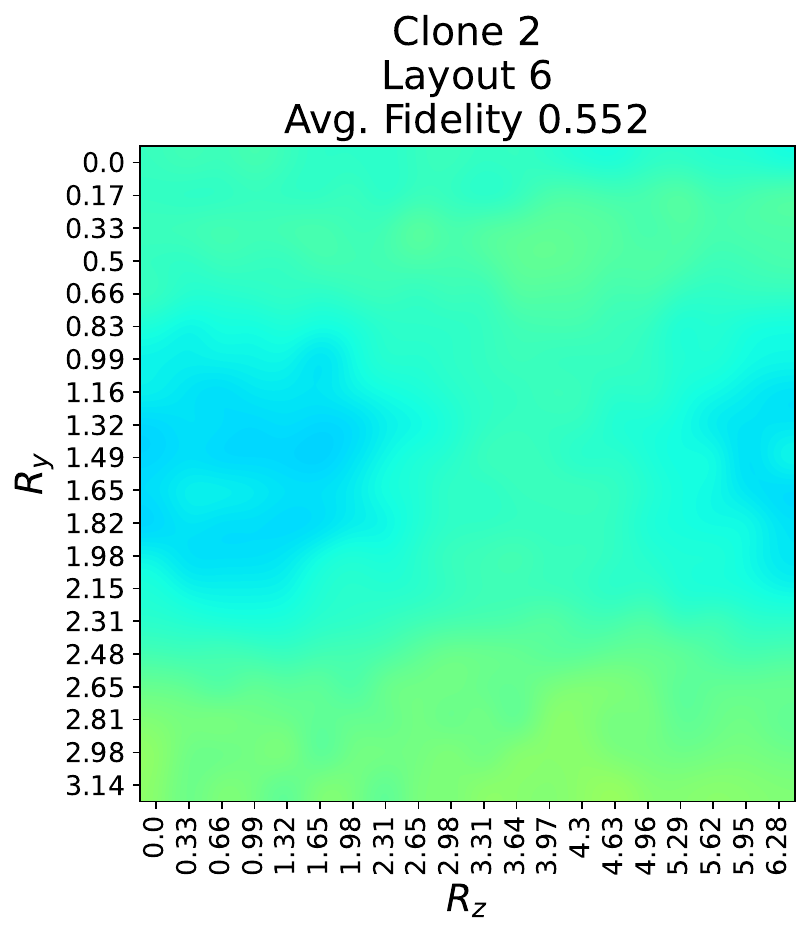}
    \includegraphics[width=0.13\textwidth]{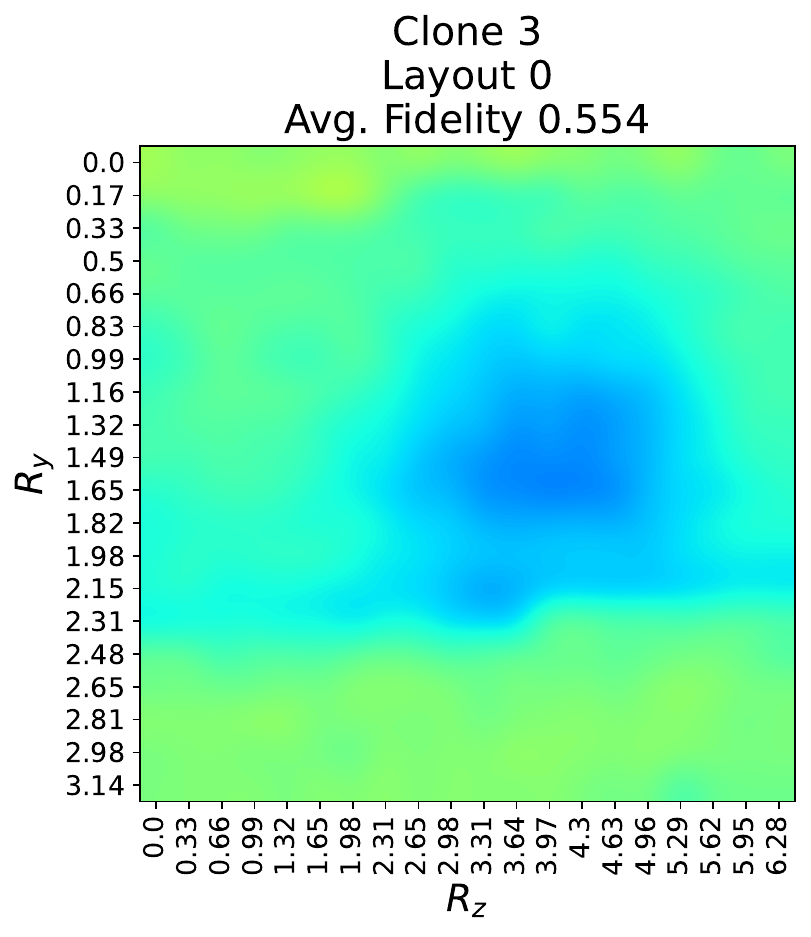}
    \includegraphics[width=0.13\textwidth]{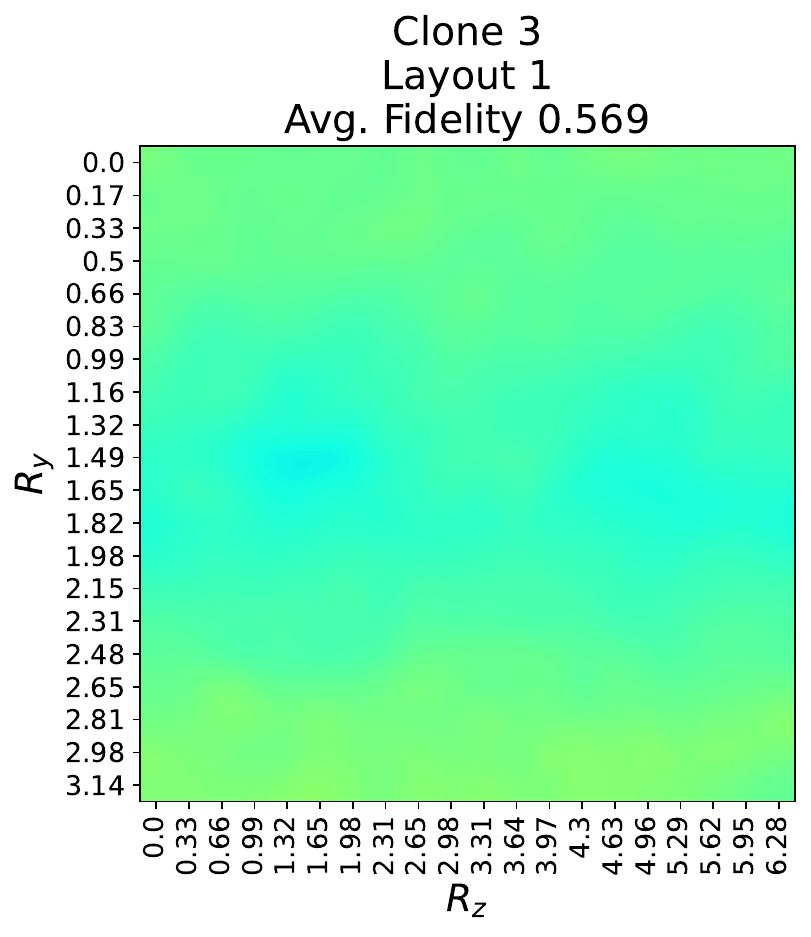}
    \includegraphics[width=0.13\textwidth]{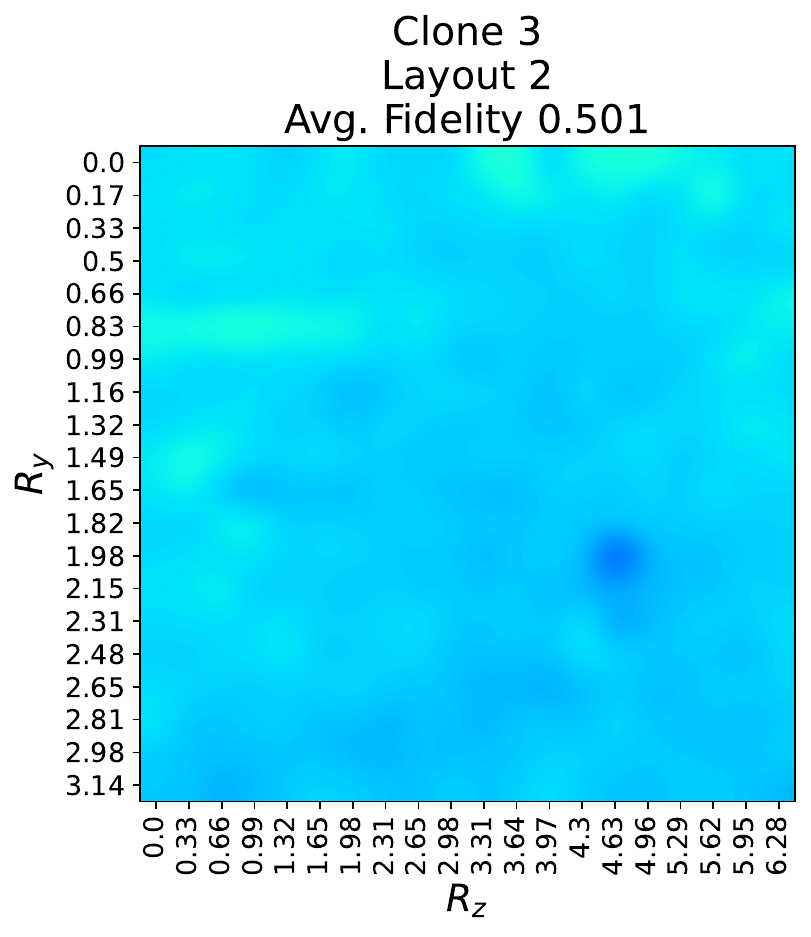}
    \includegraphics[width=0.13\textwidth]{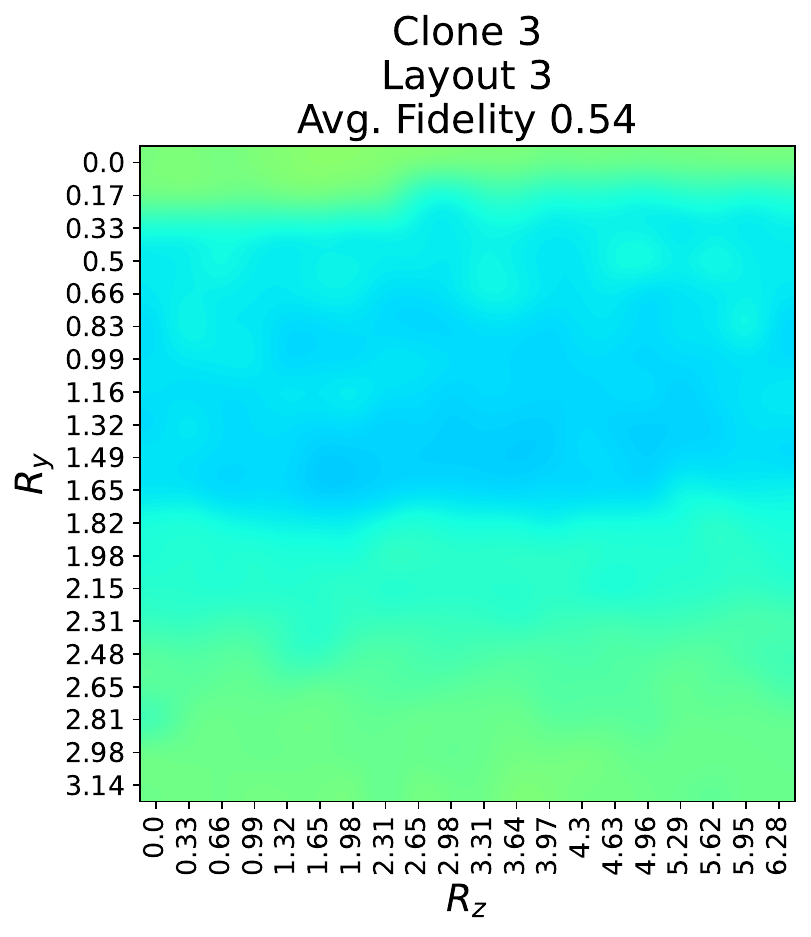}
    \includegraphics[width=0.13\textwidth]{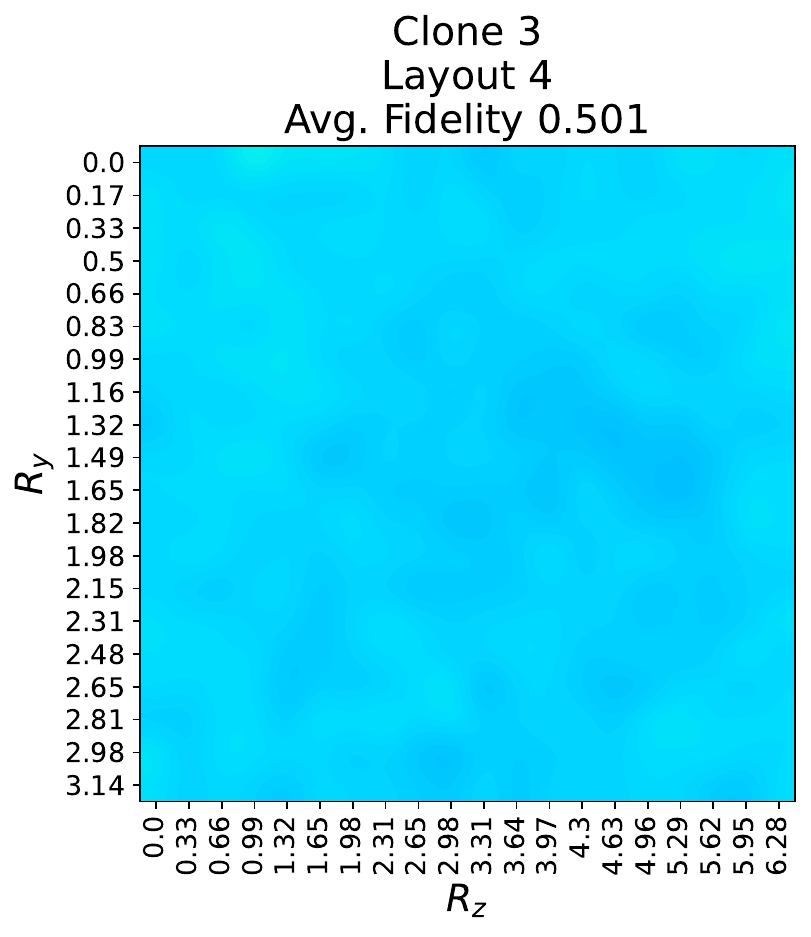}
    \includegraphics[width=0.13\textwidth]{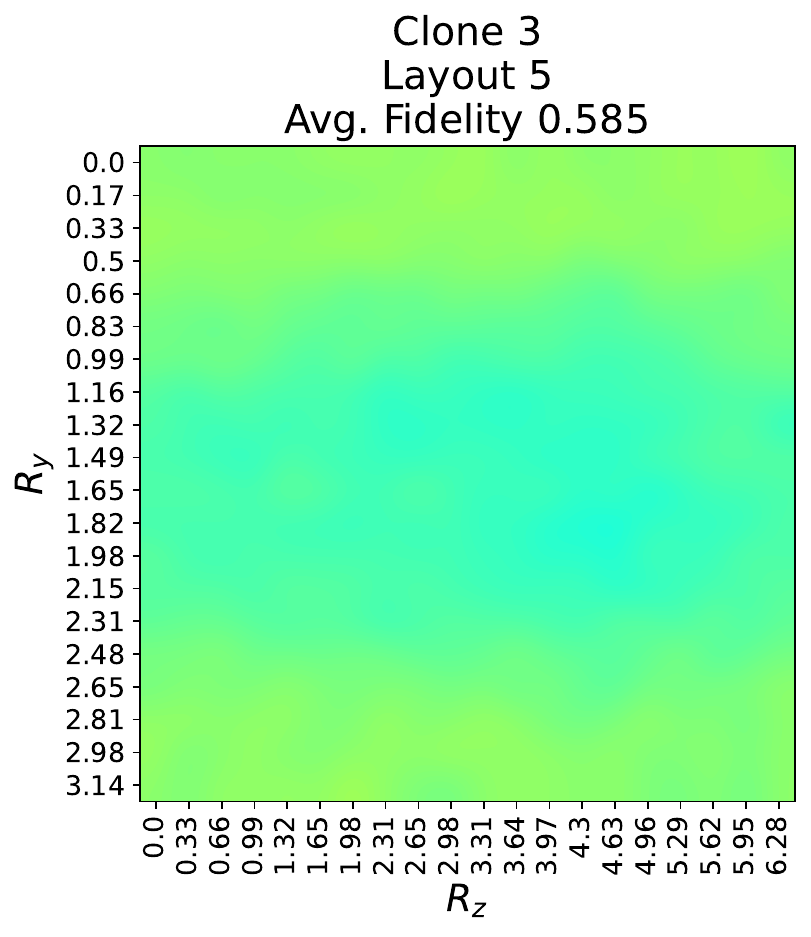}
    \includegraphics[width=0.13\textwidth]{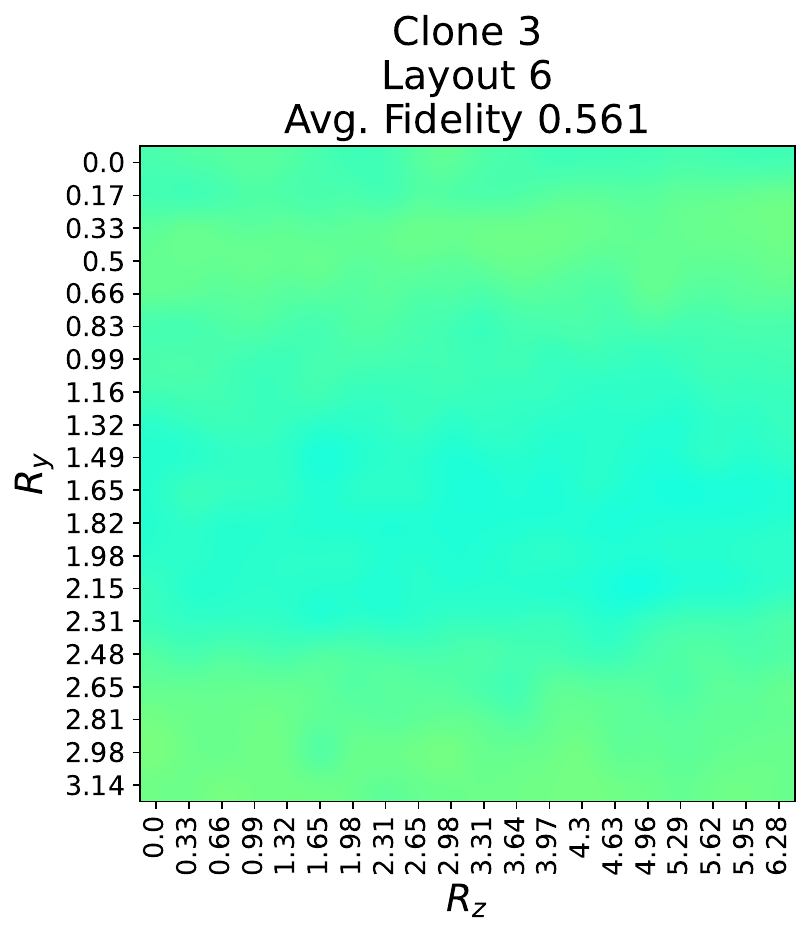}
    \includegraphics[width=0.43\textwidth]{figures/colorbar.pdf}\\
    \caption{Bloch sphere vector representations of the computed density matrices (top $4$ rows) and single qubit clone fidelity heatmaps (bottom $4$ rows) of the single qubit clone fidelity heatmaps of the single qubit clones for $M=4$ (necessarily with ancilla qubits), executed with dynamical decoupling. Each column corresponds to the $7$ different compiled hardware layouts. Each row corresponds to the $4$ different single qubit clones. Data from \texttt{ibm\_auckland}.  }
    \label{fig:fidelity_heatmaps_M4_ibm_auckland_DD}
\end{figure*}

Figure~\ref{fig:fidelity_heatmaps_M4_ibm_hanoi_DD} shows complete single qubit clone results for $M=4$ quantum telecloning circuits, necessarily using ancilla qubits, with dynamical decoupling sequences and run on \texttt{ibm\_hanoi}. Figure~\ref{fig:fidelity_heatmaps_M4_ibm_auckland_DD} shows the fidelity results for those same experimental settings ($M=4$, with dynamical decoupling), but executed on \texttt{ibm\_auckland}. The mean fidelity for the best used hardware layout for the experiments shown in Figures~\ref{fig:fidelity_heatmaps_M2_no_ancilla_ALAP_DD_ibm_auckland}, \ref{fig:fidelity_heatmaps_M2_no_ancilla_ALAP_DD_ibmq_mumbai}, \ref{fig:fidelity_heatmaps_M2_ibmq_kolkata_with_ancilla}, \ref{fig:fidelity_heatmaps_M2_ibm_cairo_with_ancilla}, \ref{fig:fidelity_heatmaps_M3_ibm_hanoi_no_ancilla_DD}, \ref{fig:fidelity_heatmaps_M3_ibm_algiers_no_ancilla_DD}, \ref{fig:fidelity_heatmaps_M4_ibm_hanoi_DD}, and \ref{fig:fidelity_heatmaps_M4_ibm_auckland_DD} are summarized, along with all of the other hardware experiments that were performed, in Table~\ref{table:summary_clone_fidelity}.

The theoretical properties of these quantum telecloning circuits are that the generated clones are \emph{optimal}, meaning they adhere to the maximum clone fidelity of eq.~\eqref{eq:theoretical-fidelity}, they are \emph{symmetric}, meaning each of the clones that are produced is indistinguishable from each other, and that they are \emph{universal} meaning that the clone fidelity is independent of the state that is cloned. We can examine the experimental results to see to what degree these theoretical properties are retained by the experiments. Across all of the fidelity heatmap plots, it can be seen that there is clear state dependence due to the noise characteristics of the quantum computers, and the clone fidelity does vary across the generated clones. This is not unexpected because of the highly variable noise sources on these devices. In particular in the clone fidelity heatmaps, there are visible trends that appear somewhat random and out of place, including, for instance, horizontal or vertical bands of high or low fidelity. The underlying cause of this instability is due to the circuits being executed on the devices at potentially far apart from each other in time -- up to several months. This occurs in particular because of total QPU and job time, queue times, backend down times, and transient job errors that need to be subsequently fixed. Current superconducting quantum computers have a reasonably high variability over time in terms of gate fidelity and qubit characteristics, which has been studied on a number of different devices~\cite{Pelofske_2022, dasgupta2021stability, e24020244, 9259941, dasgupta2023reliability, Pelofske_2023}, which then leads to potentially variable results for experiments that span a large amount of time. Examples of this variability due to data collection spanning a large amount of time can be seen in the fidelity heatmaps shown here (for example in Figure~\ref{fig:fidelity_heatmaps_M3_ibm_hanoi_no_ancilla_DD}).

Note that the ordering of the sub-figures in Figures~\ref{fig:fidelity_heatmaps_M2_no_ancilla_ALAP_DD_ibm_auckland}, \ref{fig:fidelity_heatmaps_M2_no_ancilla_ALAP_DD_ibmq_mumbai}, \ref{fig:fidelity_heatmaps_M2_ibmq_kolkata_with_ancilla}, \ref{fig:fidelity_heatmaps_M2_ibm_cairo_with_ancilla}, \ref{fig:fidelity_heatmaps_M3_ibm_hanoi_no_ancilla_DD}, \ref{fig:fidelity_heatmaps_M3_ibm_algiers_no_ancilla_DD}, \ref{fig:fidelity_heatmaps_M4_ibm_hanoi_DD}, and \ref{fig:fidelity_heatmaps_M4_ibm_auckland_DD} is the same between the Bloch sphere vector representations and fidelity heatmap representations of the data. This allows similarities between the fidelity heatmaps and the Bloch vectors to be seen for each individual clone qubit.

\section{Discussion and Conclusion}
\label{section:discussion_and_conclusion}

We have demonstrated the largest and most comprehensive analysis of quantum cloning, specifically the quantum cloning variant of quantum telecloning, that has been implemented to date with respect to the size of the experimental parameters studied, the largest size of the quantum telecloning circuit (of $M=10$), and the number of quantum computers used. This demonstration was performed using the real time classical conditional operations available on IBM Quantum devices, known as \emph{dynamic circuits}. Dynamic circuits allow the quantum telecloning protocol to be implemented natively on the hardware, as opposed to using post selection or deferred measurement techniques~\cite{telecloning_circuits_1}. We found that the clone fidelity sharply drops off to essentially noise for $M$ greater than $5$. We also found that the X-X digital dynamical decoupling sequences always improved the clone fidelity computation when it was applied to the dynamic quantum telecloning circuits. However, there exist many other dynamical decoupling schemes~\cite{9872062, sym15010062, ezzell2023dynamical, PhysRevApplied.18.024068, dd_crosstalk} that could more effectively suppress errors on idle qubits than the relatively simple pair of Pauli X gates scheme we tested, especially if tailored for the noise profile of these superconducting qubit processors. There are also other low-overhead protocols that could suppress errors in future telecloning experiments on noisy quantum computers such as randomized compiling (also known as Pauli twirling)~\cite{PhysRevA.94.052325, randomized_benchmark_2011, randomized_benchmark_2008}. 

As quantum processor hardware continues to scale, both with respect to number of qubits and lower error rates, these quantum telecloning circuit algorithms will allow ever larger quantum cloning systems to be implemented and empirically probed for purpose of testing the fundamental properties of quantum mechanics. Specifically, probing to what degree quantum cloning can be approximately performed is of fundamental interest for quantum information processing. 

These experiments serve as a benchmark of current superconducting qubit processor capabilities with respect to mid-circuit measurement and real time classical feedback control mechanisms. This is a very important feature to be evaluated, since it is a critical ingredient in quantum error correction~\cite{gupta2023encoding, govia2022randomized, Livingston_2022, Sundaresan_2023, Chen_2022}. 

The cloned qubit Bloch sphere vector representations show clearly that the NISQ computations are better when the qubit being cloned is closer to the states $\ket{1}$ and $\ket{0}$, and states at the equator of the Bloch sphere result in clones of lower fidelity. This points to a property of the superconducting qubits on the IBM Quantum computers which is interesting to observe, and is also reflected in the calibrated T1 and T2 coherence times of the devices. Similar single qubit fidelity patterns on IBM Quantum superconducting transmon qubits have been observed in previous studies~\cite{suau2022vector, telecloning_circuits_1}. 

Given that quantum cloning machines, such as quantum telecloning, can now be instantiated on quantum computers, there is an opportunity for quantum computers to serve as a quantum networking protocol test bench, as well as a simple way to understand the capabilities of current quantum computers~\cite{zhukov2019quantum}. Future quantum networking may need to utilize quantum cloning in some form, and quantum computers can serve as a mechanism to test those protocols on-chip so as to evaluate their effectiveness for use in real world quantum networks. There are not very many use cases that have been proposed for using quantum cloning in communication networks. We encourage research into use cases of quantum cloning in communication networks and information processing in general, since this is an important fundamental limit on copying quantum information. There exist several proposals for ways that quantum cloning could be used to improve aspects of quantum-classical information transmission, for example ref.~\cite{PhysRevLett.95.090504} suggests an application of quantum cloning that can be used to enhance the transmission fidelity over a noiseless lossy quantum channel. Ref.~\cite{PhysRevA.62.022301} gives two examples where quantum cloning could be used to improve quantum computation for certain tasks. As quantum error correction and quantum error suppression techniques continue to improve, and as hardware continues to scale in both fidelity and number of qubits, larger quantum cloning machines can be implemented to test these types of protocols beyond what could be verified classically, and therefore the circuit descriptions for the various types of quantum cloning will need to be developed. There are several open algorithmic questions in regards to quantum telecloning, and more generally quantum cloning specifically for quantum computation:

\begin{enumerate}[noitemsep]
    \item There does not yet exist a complete circuit model description for any type of quantum cloning, with the exception of quantum telecloning. Specific types of quantum cloning circuit models that would be interesting to develop include a universal quantum cloning machine, asymmetric quantum cloning, probabilistic quantum cloning, and qudit (d-dimensional quantum system) cloning. 
    \item It is not known whether it is possible to construct a $1 \rightarrow M$ quantum telecloning circuit that does not use ancilla qubits for $M \geq 4$. 
    \item We believe that the $1 \rightarrow 2$ and $1 \rightarrow 3$ telecloning circuits without ancilla qubits that are used in this study are highly optimized, but there could exist further optimizations to reduce the gate count or gate depth of these circuits. 
    \item A majority of the quantum cloning experiments that have been performed to date were for cloning a single qubit to multiple clones (e.g. $1 \rightarrow M$). However, it is certainly an interesting question of how to generally clone $N \rightarrow M$, where for instance the $N$ qubits are entangled, and what those circuit descriptions are. This specific topic of cloning states which are entangled is has largely not been studied~\cite{PhysRevLett.125.210502}. 
    \item The reverse of quantum telecloning, remote information concentration~\cite{PhysRevLett.86.352, PhysRevA.68.024303, PhysRevA.84.042310}, also has not been implemented in a circuit model description. 
    \item Determining if the optimized Dicke state preparation circuits in refs.~\cite{B_rtschi_2019, 9951196, 9774323} could be used for preparing other types of quantum cloning circuits besides quantum telecloning. 
\end{enumerate}

\section{Acknowledgments}
\label{section:acknowledgments}
This work was supported by the U.S. Department of Energy through the Los Alamos National Laboratory. Los Alamos National Laboratory is operated by Triad National Security, LLC, for the National Nuclear Security Administration of U.S. Department of Energy (Contract No. 89233218CNA000001). This work was supported by the NNSA's Advanced Simulation and Computing Beyond Moore's Law Program at Los Alamos National Laboratory. This research used resources provided by the Los Alamos National Laboratory Institutional Computing Program, which is supported by the U.S. Department of Energy National Nuclear Security Administration under Contract No. 89233218CNA000001.
We acknowledge the use of IBM Quantum services for this work. The views expressed are those of the authors, and do not reflect the official policy or position of IBM or the IBM Quantum team. E.P. thanks the folks at the 2023 IBM Quantum Internal Developer forum for helpful discussions on dynamic circuits and dynamical decoupling in Qiskit. This work has been assigned the LANL report number LA-UR-23-29397.


\appendix



\section{Detailed Quantum Telecloning Circuit Renderings}
\label{section:appendix_compiled_circuits}

\begin{figure*}[h!]
    \centering
    \hspace*{-0.55in}
    \includegraphics[width=1.05\textwidth]{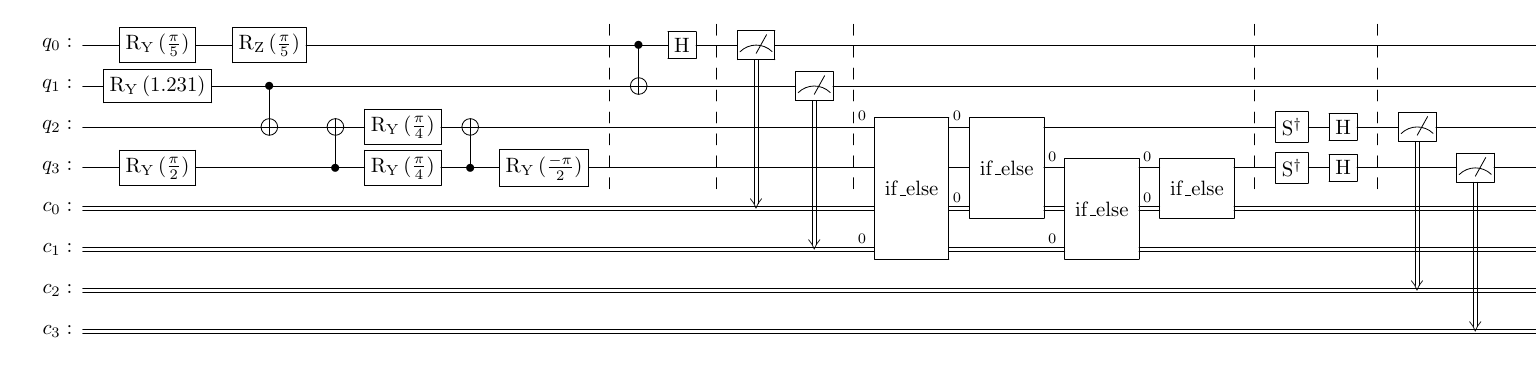}
    \hspace*{-0.55in}
    \includegraphics[width=1.05\textwidth]{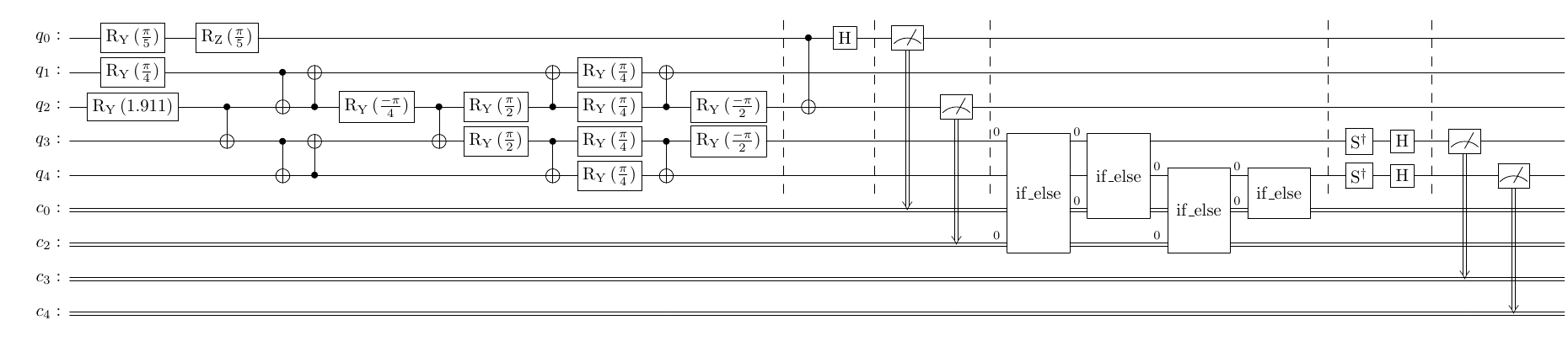}
    \caption{$1 \rightarrow 2$ quantum telecloning circuits with no ancilla qubits (top diagram) and with $1$ ancilla qubit (bottom diagram). Compiled to a Linear Nearest Neighbors (LNN) hardware graph, specifically targeting subsets of a heavy-hex graph (see Figure~\ref{fig:qubit_layouts}). }
    \label{fig:M2_detailed_circuits}
\end{figure*}

\begin{figure*}[h!]
    \centering
    \hspace*{-0.55in}
    \includegraphics[width=1.05\textwidth]{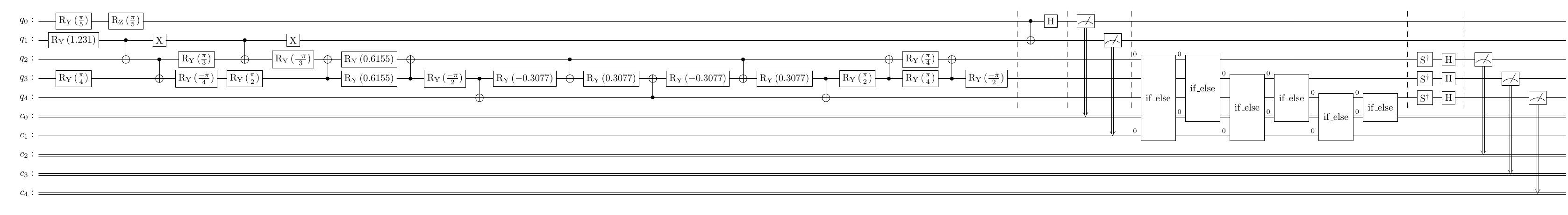}
    \hspace*{-0.55in}
    \includegraphics[width=1.05\textwidth]{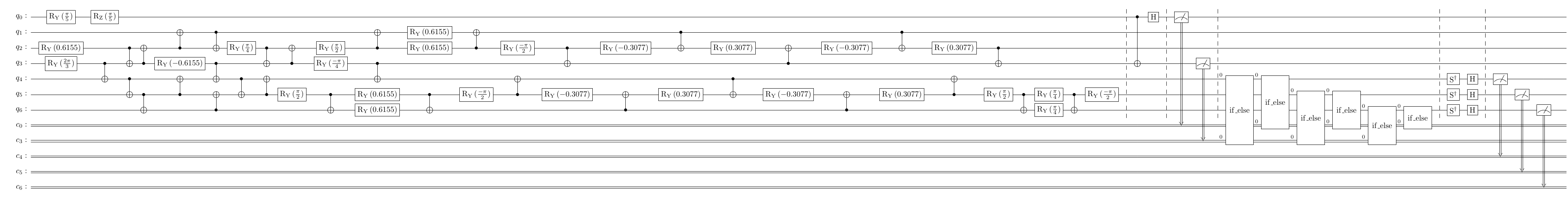}
    \caption{$1 \rightarrow 3$ quantum telecloning circuits with no ancilla qubits (top diagram) and with $2$ ancilla qubits (bottom diagram). Compiled to a (LNN) hardware graph, specifically targeting subsets of a heavy-hex graph (see Figure~\ref{fig:qubit_layouts}). }
    \label{fig:M3_detailed_circuits}
\end{figure*}

\begin{figure*}[th!]
    \centering
    \hspace*{-0.55in}
    \includegraphics[width=1.05\textwidth]{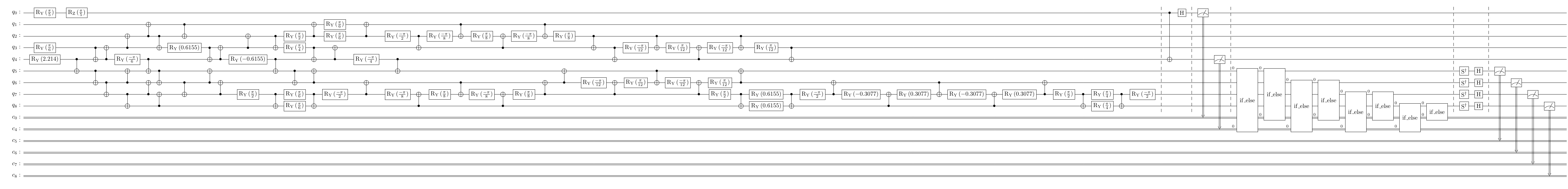}
    \caption{$1 \rightarrow 4$ quantum telecloning circuit with $3$ ancilla qubits. Compiled to a (LNN) hardware graph, specifically targeting subsets of a heavy-hex graph (see Figure~\ref{fig:qubit_layouts}). }
    \label{fig:M4_detailed_circuit}
\end{figure*}

The detailed quantum telecloning circuits shown in Figures~\ref{fig:M2_detailed_circuits},~\ref{fig:M3_detailed_circuits}, and~\ref{fig:M4_detailed_circuit} are all divided into $6$ distinct segments using barriers, representing different stages of the quantum telecloning protocol and the parallel single qubit state tomography. The first stage prepares the telecloning state, which is comprised of Dicke state unitaries (and possibly SCS unitaries); this is the computationally intensive stage and is step 2 in Algorithm~\ref{algorithm:telecloning}. The first stage is also where the message qubit is introduced (in this case, generated by parameterized $R_y$ and $R_z$ single qubit rotations), which is step 1 in Algorithm~\ref{algorithm:telecloning}. The second stage prepares a Bell state, using a CNOT and a Hadamard gate, between the port qubit and the message qubit. The third stage measures the state of the port qubit and message qubit, and stores the two classical bits into two classical registers. Combined, the second and third stage comprise the Bell measurement, which is step 3 in Algorithm~\ref{algorithm:telecloning}. The fourth stage uses a classical co-processor to execute classical channel conditional operations in order to optionally apply Pauli Z and or X gates (this is step 4 in Algorithm~\ref{algorithm:telecloning}), represented is if-else control blocks. The if-else control blocks visually do not show what gates are (potentially) applied, but it is X and then Z Pauli gates on each will-be clone qubit. The fifth stage applies single qubit gates to the clone qubits so as to put each of the clones into a Pauli X, Y, or Z basis - in this case all of the circuit figures in this section put the clone qubits into the Pauli Y basis. In the sixth stage, all of the clone qubits are measured and the results are stored in classical registers. Note that in the figures with ancilla qubits, the ancilla qubits are discarded and their states are not measured. Experimentally, when these circuits are executed on the quantum computers each circuit is executed with the three different Pauli basis and multiple circuit measurements (specifically $10,000$ samples per Pauli basis) to reconstruct the density matrices of the clone qubits that were prepared on the device. 

The quantum telecloning circuits in Figures~\ref{fig:M2_detailed_circuits},~\ref{fig:M3_detailed_circuits}, and~\ref{fig:M4_detailed_circuit} that have ancilla qubits all use the circuit optimization introduced in~\cite{telecloning_circuits_2}, reducing the total number of gate operations that are used. Figure~\ref{fig:M4_detailed_circuit} uses ancilla qubits because there is no known telecloning circuit that does not require ancilla qubits for $M \geq 4$ (for single qubit cloning). 

In Figures~\ref{fig:M2_detailed_circuits},~\ref{fig:M3_detailed_circuits}, and~\ref{fig:M4_detailed_circuit} the message qubit is prepared with arbitrarily chosen $R_y = \frac{\pi}{5}$ and $R_z = \frac{\pi}{5}$ for demonstration purposes.

\section{IBM Quantum Computer Circuit Execution Details}
\label{section:appendix_job_execution_metadata}

The total compute time utilized in this study was quite intensive. A total of $8,400$ dynamic circuits were executed for each entry in Table~\ref{table:summary_clone_fidelity}. There are $47$ entries in Table~\ref{table:summary_clone_fidelity}, meaning that $394,800$ dynamic circuits were executed in the entirety of the data shown in this study. This is equal to a total of $3,948,000,000$ individual circuit measurements. All of these circuits were run between December 2022 through February 2024. The total Quantum Processing Unit (QPU) time (this is the time taken on the device to run the circuits, not including queue times, networking communication time, etc) to obtain these measurements was $5,493,107$ seconds ($1,526$ hours). The cumulative queue wait time, across all jobs, was $24,731,058,876$ seconds ($784$ years), with a minimum queue wait time of $10.6$ seconds, a maximum queue wait time of $2,239,317$ seconds ($26$ days), and a mean job queue wait time of $65,426$ seconds ($18$ hours) with a standard deviation of $172,050$. These queue and compute times do not include jobs that were re-executed because of being in an error status. 

The scale of this study meant that numerous jobs encountered a variety of server-side or job related errors. In these cases, the circuits would need to be re-executed (sometimes multiple times) so that the complete set of data could be obtained. In all cases, the errors that were encountered were stochastic - there was typically no clear cause of the errors, and the same identical job would not deterministically cause the same error. The types of errors changed based on the backend used and also changed over time due to the software on the devices being updated (along with Qiskit version releases). This is a short, and likely incomplete, list of the error messages encountered when executing these quantum telecloning circuits:

\begin{enumerate}[noitemsep]
    \item \texttt{Internal Error}. This was the most frequently encountered error. 
    \item \texttt{Error preprocessing job.}
    \item \texttt{Error queueing job.}
    \item \texttt{Unknown error.}
    \item \texttt{Timed out waiting for job to finish.}
    \item \texttt{JSONDecodeError: Expecting value: line 1 column 1 (char 0)}
    \item \texttt{Internal Error while executing OpenQASM 3 circuit.}
    \item \texttt{not able to get the queue name for \newline ibmq\_mumbai}
    \item \texttt{BackendPropertyError: `Unable to process backend properties.'}
    \item ``" (an empty string)
\end{enumerate}

There were also two types of errors that did not explicitly cause the jobs to be in an error state, but were nevertheless unrecoverable errors that required re-executing the job:

\begin{enumerate}[noitemsep]
    \item Jobs that were in a \texttt{Cancelled} state, likely due to their QPU time usage exceeding an expected pre-defined limit. 
    \item Jobs that were in a completed, error free state, but when querying the job results in Qiskit, the result object datastructure was not in the correct format, and a JSON decoding error was raised: \texttt{json.decoder.JSONDecodeError} in Python 3
\end{enumerate}



\clearpage

\setlength\bibitemsep{0pt}
\printbibliography

%% file: references.bib
@inproceedings{telecloning_circuits_1,
	doi = {10.1109/qce53715.2022.00083},
  
	url = {https://doi.org/10.1109%2Fqce53715.2022.00083},
  
	year = 2022,
	month = {sep},
  
	publisher = {{IEEE}
},
  
	author = {Elijah Pelofske and Andreas Bartschi and Bryan Garcia and Boris Kiefer and Stephan Eidenbenz},
  
	title = {{Quantum Telecloning on NISQ Computers}},
  
	booktitle = {2022 {IEEE} International Conference on Quantum Computing and Engineering ({QCE})}
}

@INPROCEEDINGS{telecloning_circuits_2,

  author={Pelofske, Elijah and Bärtschi, Andreas and Eidenbenz, Stephan},

  booktitle={2022 IEEE International Conference on Rebooting Computing (ICRC)}, 

  title={Optimized Telecloning Circuits: Theory and Practice of Nine NISQ Clones}, 

  year={2022},

  volume={},

  number={},

  pages={51-56},

  doi={10.1109/ICRC57508.2022.00009}}

@article{Cross_2022,
	doi = {10.1145/3505636},
  
	url = {https://doi.org/10.1145%2F3505636},
  
	year = 2022,
	month = {sep},
  
	publisher = {Association for Computing Machinery ({ACM})},
  
	volume = {3},
  
	number = {3},
  
	pages = {1--50},
  
	author = {Andrew Cross and Ali Javadi-Abhari and Thomas Alexander and Niel De Beaudrap and Lev S. Bishop and Steven Heidel and Colm A. Ryan and Prasahnt Sivarajah and John Smolin and Jay M. Gambetta and Blake R. Johnson},
  
	title = {{OpenQASM}~3: A Broader and Deeper Quantum Assembly Language},
  
	journal = {{ACM} Transactions on Quantum Computing}
}

@misc{https://doi.org/10.48550/arxiv.1707.03429,
  doi = {10.48550/ARXIV.1707.03429},
  
  url = {https://arxiv.org/abs/1707.03429},
  
  author = {Cross, Andrew W. and Bishop, Lev S. and Smolin, John A. and Gambetta, Jay M.},
  
  title = {Open Quantum Assembly Language},
  
  publisher = {arXiv},
  
  year = {2017},
  
  copyright = {arXiv.org perpetual, non-exclusive license}
}

@article{russo2021toqito,
  title={toqito--Theory of quantum information toolkit: A Python package for studying quantum information},
  author={Russo, Vincent},
  journal={Journal of Open Source Software},
  volume={6},
  number={61},
  pages={3082},
  year={2021}
}

@misc{Qiskit,
    author = {{Qiskit contributors}},
    title = {Qiskit: An open-source framework for quantum computing},
    year = {2023},
    doi = {10.5281/zenodo.2573505}
}

@misc{thomas_a_caswell_2021_5194481,
	title	= {matplotlib/matplotlib},
	version	= {v3.4.3},
	doi	= {10.5281/zenodo.5194481},
	author	= {Thomas A Caswell and others},
}

@Article{Hunter:2007,
	Author	= {Hunter, J. D.},
	Title	= {{Matplotlib: A 2D graphics environment}},
	Journal	= {Computing in Science \& Engineering},
	Volume	= {9},
	Number	= {3},
	Pages	= {90--95},
	doi	= {10.1109/MCSE.2007.55},
	year	= {2007},
}

@article{JOHANSSON20131234,
title = {QuTiP 2: A Python framework for the dynamics of open quantum systems},
journal = {Computer Physics Communications},
volume = {184},
number = {4},
pages = {1234-1240},
year = {2013},
issn = {0010-4655},
doi = {https://doi.org/10.1016/j.cpc.2012.11.019},
url = {https://www.sciencedirect.com/science/article/pii/S0010465512003955},
author = {J.R. Johansson and P.D. Nation and Franco Nori}
}

@article{JOHANSSON20121760,
title = {QuTiP: An open-source Python framework for the dynamics of open quantum systems},
journal = {Computer Physics Communications},
volume = {183},
number = {8},
pages = {1760-1772},
year = {2012},
issn = {0010-4655},
doi = {https://doi.org/10.1016/j.cpc.2012.02.021},
url = {https://www.sciencedirect.com/science/article/pii/S0010465512000835},
author = {J.R. Johansson and P.D. Nation and Franco Nori}
}

@article{ramachandran2011mayavi,
  title={{Mayavi: 3D Visualization of Scientific Data}},
  author={Ramachandran, P. and Varoquaux, G.},
  journal={Computing in Science \& Engineering},
  volume={13},
  number={2},
  pages={40--51},
  issn={1521-9615},
  year={2011},
  publisher={IEEE}
}

@article{dieks1982communication,
  title={Communication by EPR devices},
  author={Dieks, DGBJ},
  journal={Physics Letters A},
  volume={92},
  number={6},
  pages={271--272},
  year={1982},
  publisher={Elsevier},
  doi={10.1016/0375-9601(82)90084-6}
}

@article{wootters1982single,
  title={A single quantum cannot be cloned},
  author={Wootters, William K and Zurek, Wojciech H},
  journal={Nature},
  volume={299},
  pages={802--803},
  year={1982},
  publisher={Springer},
  doi={10.1038/299802a0}
}

@article{Bu_ek_1996,
	doi = {10.1103/physreva.54.1844},
  
	url = {https://doi.org/10.1103%2Fphysreva.54.1844},
  
	year = 1996,
	month = {sep},
  
	publisher = {American Physical Society ({APS})},
  
	volume = {54},
  
	number = {3},
  
	pages = {1844--1852},
  
	author = {V. Bu{\v{z}
}ek and M. Hillery},
  
	title = {Quantum copying: Beyond the no-cloning theorem},
  
	journal = {Physical Review A}
}

@article{Fan_2001,
	doi = {10.1103/physreva.65.012304},
  
	url = {https://doi.org/10.1103%2Fphysreva.65.012304},
  
	year = 2001,
	month = {dec},
  
	publisher = {American Physical Society ({APS})},
  
	volume = {65},
  
	number = {1},
  
	author = {Heng Fan and Keiji Matsumoto and Xiang-Bin Wang and Miki Wadati},
  
	title = {Quantum cloning machines for equatorial qubits},
  
	journal = {Physical Review A}
}

@article{jozsa1994fidelity,
  title={Fidelity for mixed quantum states},
  author={Jozsa, Richard},
  journal={Journal of modern optics},
  volume={41},
  number={12},
  pages={2315--2323},
  year={1994},
  publisher={Taylor \& Francis},
  doi={10.1080/09500349414552171}
}

@article{Cerf_1997,
	doi = {10.1103/physreva.56.1721},
  
	url = {https://doi.org/10.1103%2Fphysreva.56.1721},
  
	year = 1997,
	month = {sep},
  
	publisher = {American Physical Society ({APS})},
  
	volume = {56},
  
	number = {3},
  
	pages = {1721--1732},
  
	author = {Nicolas J. Cerf and Richard Cleve},
  
	title = {Information-theoretic interpretation of quantum error-correcting codes},
  
	journal = {Physical Review A}
}

@article{Bennett_2014,
	doi = {10.1016/j.tcs.2014.05.025},
  
	url = {https://doi.org/10.1016%2Fj.tcs.2014.05.025},
  
	year = 2014,
	month = {dec},
  
	publisher = {Elsevier {BV}
},
  
	volume = {560},
  
	pages = {7--11},
  
	author = {Charles H. Bennett and Gilles Brassard},
  
	title = {Quantum cryptography: Public key distribution and coin tossing},
  
	journal = {Theoretical Computer Science}
}

@article{PhysRevA.72.032301,
  title = {Security of two quantum cryptography protocols using the same four qubit states},
  author = {Branciard, Cyril and Gisin, Nicolas and Kraus, Barbara and Scarani, Valerio},
  journal = {Phys. Rev. A},
  volume = {72},
  issue = {3},
  pages = {032301},
  numpages = {18},
  year = {2005},
  month = {Sep},
  publisher = {American Physical Society},
  doi = {10.1103/PhysRevA.72.032301},
  url = {https://link.aps.org/doi/10.1103/PhysRevA.72.032301}
}

@article{PhysRevLett.67.661,
  title = {Quantum cryptography based on Bell's theorem},
  author = {Ekert, Artur K.},
  journal = {Phys. Rev. Lett.},
  volume = {67},
  issue = {6},
  pages = {661--663},
  numpages = {0},
  year = {1991},
  month = {Aug},
  publisher = {American Physical Society},
  doi = {10.1103/PhysRevLett.67.661},
  url = {https://link.aps.org/doi/10.1103/PhysRevLett.67.661}
}

@article{PhysRevA.73.012337,
  title = {Performance of two quantum-key-distribution protocols},
  author = {Fung, Chi-Hang Fred and Tamaki, Kiyoshi and Lo, Hoi-Kwong},
  journal = {Phys. Rev. A},
  volume = {73},
  issue = {1},
  pages = {012337},
  numpages = {19},
  year = {2006},
  month = {Jan},
  publisher = {American Physical Society},
  doi = {10.1103/PhysRevA.73.012337},
  url = {https://link.aps.org/doi/10.1103/PhysRevA.73.012337}
}

@article{Bru__2000,
	doi = {10.1103/physreva.62.012302},
  
	url = {https://doi.org/10.1103%2Fphysreva.62.012302},
  
	year = 2000,
	month = {jun},
  
	publisher = {American Physical Society ({APS})},
  
	volume = {62},
  
	number = {1},
  
	author = {Dagmar Bru{\ss} and Mirko Cinchetti and G. Mauro D'Ariano and Chiara Macchiavello},
  
	title = {Phase-covariant quantum cloning},
  
	journal = {Physical Review A}
}

@article{PhysRevA.75.012317,
  title = {Experimental realization of $\mathbf{1}\mathbf{\ensuremath{\rightarrow}}\mathbf{2}$ asymmetric phase-covariant quantum cloning},
  author = {Chen, Hongwei and Zhou, Xianyi and Suter, Dieter and Du, Jiangfeng},
  journal = {Phys. Rev. A},
  volume = {75},
  issue = {1},
  pages = {012317},
  numpages = {5},
  year = {2007},
  month = {Jan},
  publisher = {American Physical Society},
  doi = {10.1103/PhysRevA.75.012317},
  url = {https://link.aps.org/doi/10.1103/PhysRevA.75.012317}
}

@misc{gottesman2000introduction,
      title={An Introduction to Quantum Error Correction}, 
      author={Daniel Gottesman},
      year={2000},
      eprint={quant-ph/0004072},
      archivePrefix={arXiv},
      primaryClass={quant-ph}
}

@article{bouchard2017high,
  title={High-dimensional quantum cloning and applications to quantum hacking},
  author={Bouchard, Fr{\'e}d{\'e}ric and Fickler, Robert and Boyd, Robert W and Karimi, Ebrahim},
  journal={Science advances},
  volume={3},
  number={2},
  pages={e1601915},
  year={2017},
  publisher={American Association for the Advancement of Science},
  doi={10.1126/sciadv.1601915}
}

@article{roffe2019quantum,
  title={Quantum error correction: an introductory guide},
  author={Roffe, Joschka},
  journal={Contemporary Physics},
  volume={60},
  number={3},
  pages={226--245},
  year={2019},
  publisher={Taylor \& Francis},
  doi={10.1080/00107514.2019.1667078}
}

@article{RevModPhys.77.1225,
  title = {Quantum cloning},
  author = {Scarani, Valerio and Iblisdir, Sofyan and Gisin, Nicolas and Ac\'{\i}n, Antonio},
  journal = {Rev. Mod. Phys.},
  volume = {77},
  issue = {4},
  pages = {1225--1256},
  numpages = {0},
  year = {2005},
  month = {Nov},
  publisher = {American Physical Society},
  doi = {10.1103/RevModPhys.77.1225},
  url = {https://link.aps.org/doi/10.1103/RevModPhys.77.1225}
}

@article{PhysRevLett.79.2153,
  title = {Optimal Quantum Cloning Machines},
  author = {Gisin, N. and Massar, S.},
  journal = {Phys. Rev. Lett.},
  volume = {79},
  issue = {11},
  pages = {2153--2156},
  numpages = {0},
  year = {1997},
  month = {Sep},
  publisher = {American Physical Society},
  doi = {10.1103/PhysRevLett.79.2153},
  url = {https://link.aps.org/doi/10.1103/PhysRevLett.79.2153}
}

@article{fan2014quantum,
  title={Quantum cloning machines and the applications},
  author={Fan, Heng and Wang, Yi-Nan and Jing, Li and Yue, Jie-Dong and Shi, Han-Duo and Zhang, Yong-Liang and Mu, Liang-Zhu},
  journal={Physics Reports},
  volume={544},
  number={3},
  pages={241--322},
  year={2014},
  publisher={Elsevier},
  doi={10.1016/j.physrep.2014.06.004}
}

@article{PhysRevA.59.156,
  title = {Quantum telecloning and multiparticle entanglement},
  author = {Murao, M. and Jonathan, D. and Plenio, M. B. and Vedral, V.},
  journal = {Phys. Rev. A},
  volume = {59},
  issue = {1},
  pages = {156--161},
  numpages = {0},
  year = {1999},
  month = {Jan},
  publisher = {American Physical Society},
  doi = {10.1103/PhysRevA.59.156},
  url = {https://link.aps.org/doi/10.1103/PhysRevA.59.156}
}

@article{PhysRevLett.106.180404,
  title = {Experimental Demonstration of Probabilistic Quantum Cloning},
  author = {Chen, Hongwei and Lu, Dawei and Chong, Bo and Qin, Gan and Zhou, Xianyi and Peng, Xinhua and Du, Jiangfeng},
  journal = {Phys. Rev. Lett.},
  volume = {106},
  issue = {18},
  pages = {180404},
  numpages = {4},
  year = {2011},
  month = {May},
  publisher = {American Physical Society},
  doi = {10.1103/PhysRevLett.106.180404},
  url = {https://link.aps.org/doi/10.1103/PhysRevLett.106.180404}
}

@article{Cacciapuoti_2020,
	doi = {10.1109/mnet.001.1900092},
  
	url = {https://doi.org/10.1109%2Fmnet.001.1900092},
  
	year = 2020,
	month = {jan},
  
	publisher = {Institute of Electrical and Electronics Engineers ({IEEE})},
  
	volume = {34},
  
	number = {1},
  
	pages = {137--143},
  
	author = {Angela Sara Cacciapuoti and Marcello Caleffi and Francesco Tafuri and Francesco Saverio Cataliotti and Stefano Gherardini and Giuseppe Bianchi},
  
	title = {Quantum Internet: Networking Challenges in Distributed Quantum Computing},
  
	journal = {{IEEE} Network}
}

@article{PhysRevLett.86.352,
  title = {Remote Information Concentration Using a Bound Entangled State},
  author = {Murao, Mio and Vedral, Vlatko},
  journal = {Phys. Rev. Lett.},
  volume = {86},
  issue = {2},
  pages = {352--355},
  numpages = {0},
  year = {2001},
  month = {Jan},
  publisher = {American Physical Society},
  doi = {10.1103/PhysRevLett.86.352},
  url = {https://link.aps.org/doi/10.1103/PhysRevLett.86.352}
}

@article{PhysRevA.68.024303,
  title = {Remote information concentration by a Greenberger-Horne-Zeilinger state and by a bound entangled state},
  author = {Yu, Yafei and Feng, Jian and Zhan, Mingsheng},
  journal = {Phys. Rev. A},
  volume = {68},
  issue = {2},
  pages = {024303},
  numpages = {3},
  year = {2003},
  month = {Aug},
  publisher = {American Physical Society},
  doi = {10.1103/PhysRevA.68.024303},
  url = {https://link.aps.org/doi/10.1103/PhysRevA.68.024303}
}

@article{PhysRevA.84.042310,
  title = {Remote information concentration and multipartite entanglement in multilevel systems},
  author = {Wang, Xin-Wen and Zhang, Deng-Yu and Yang, Guo-Jian and Tang, Shi-Qing and Xie, Li-Jun},
  journal = {Phys. Rev. A},
  volume = {84},
  issue = {4},
  pages = {042310},
  numpages = {8},
  year = {2011},
  month = {Oct},
  publisher = {American Physical Society},
  doi = {10.1103/PhysRevA.84.042310},
  url = {https://link.aps.org/doi/10.1103/PhysRevA.84.042310}
}

@article{PhysRevA.56.3446,
	doi = {10.1103/physreva.56.3446},
  
	url = {https://doi.org/10.1103%2Fphysreva.56.3446},
  
	year = 1997,
	month = {nov},
  
	publisher = {American Physical Society ({APS})},
  
	volume = {56},
  
	number = {5},
  
	pages = {3446--3452},
  
	author = {V. Bu{\v{z}}ek and S. L. Braunstein and M. Hillery and D. Bru{\ss}},
  
	title = {Quantum copying: A network},
  
	journal = {Physical Review A}
}

@article{PhysRevA.58.4377,
  title = {Optimal copying of one quantum bit},
  author = {Niu, Chi-Sheng and Griffiths, Robert B.},
  journal = {Phys. Rev. A},
  volume = {58},
  issue = {6},
  pages = {4377--4393},
  numpages = {0},
  year = {1998},
  month = {Dec},
  publisher = {American Physical Society},
  doi = {10.1103/PhysRevA.58.4377},
  url = {https://link.aps.org/doi/10.1103/PhysRevA.58.4377}
}

@misc{PhysRevLett.81.5003,
      title={Universal Optimal Cloning of Arbitrary Quantum States: From Qubits to Quantum Registers}, 
      author={Vladimir Buzek and Mark Hillery},
      year={1998},
      eprint={quant-ph/9801009},
      archivePrefix={arXiv},
      primaryClass={quant-ph},
        journal = {Phys. Rev. Lett.},
  volume = {81},
  issue = {22},
  pages = {5003--5006},
  numpages = {0},
  year = {1998},
  month = {Nov},
  publisher = {American Physical Society},
  doi = {10.1103/PhysRevLett.81.5003},
  url = {https://link.aps.org/doi/10.1103/PhysRevLett.81.5003}
}

@article{PhysRevA.57.2368,
  title = {Optimal universal and state-dependent quantum cloning},
  author = {Bru\ss{}, Dagmar and DiVincenzo, David P. and Ekert, Artur and Fuchs, Christopher A. and Macchiavello, Chiara and Smolin, John A.},
  journal = {Phys. Rev. A},
  volume = {57},
  issue = {4},
  pages = {2368--2378},
  numpages = {0},
  year = {1998},
  month = {Apr},
  publisher = {American Physical Society},
  doi = {10.1103/PhysRevA.57.2368},
  url = {https://link.aps.org/doi/10.1103/PhysRevA.57.2368}
}

@article{PhysRevLett.81.2598,
  title = {Optimal Universal Quantum Cloning and State Estimation},
  author = {Bruss, Dagmar and Ekert, Artur and Macchiavello, Chiara},
  journal = {Phys. Rev. Lett.},
  volume = {81},
  issue = {12},
  pages = {2598--2601},
  numpages = {0},
  year = {1998},
  month = {Sep},
  publisher = {American Physical Society},
  doi = {10.1103/PhysRevLett.81.2598},
  url = {https://link.aps.org/doi/10.1103/PhysRevLett.81.2598}
}

@article{cerf2006optical,
  title={Optical quantum cloning},
  author={Cerf, Nicolas J and Fiurasek, Jaromir},
  journal={Progress in Optics},
  volume={49},
  pages={455},
  year={2006}
}

@article{PhysRevLett.86.4942,
	doi = {10.1103/physrevlett.86.4942},
  
	url = {https://doi.org/10.1103%2Fphysrevlett.86.4942},
  
	year = 2001,
	month = {may},
  
	publisher = {American Physical Society ({APS})},
  
	volume = {86},
  
	number = {21},
  
	pages = {4942--4945},
  
	author = {Jarom{\'{\i}
}r Fiur{\'{a}}{\v{s}}ek},
  
	title = {Optical Implementation of Continuous-Variable Quantum Cloning Machines},
  
	journal = {Physical Review Letters}
}

@article{cerf2000asymmetric,
  title={Asymmetric quantum cloning in any dimension},
  author={Cerf, Nicolas J},
  journal={Journal of modern optics},
  volume={47},
  number={2-3},
  pages={187--209},
  year={2000},
  publisher={Taylor \& Francis}
}

@article{gisin1998quantum,
  title={Quantum cloning without signaling},
  author={Gisin, Nicolas},
  journal={Physics Letters A},
  volume={242},
  number={1-2},
  pages={1--3},
  year={1998},
  publisher={Elsevier},
  doi={10.1016/S0375-9601(98)00170-4}
}

@article{hardy1999no,
  title={No signalling and probabilistic quantum cloning},
  author={Hardy, Lucien and Song, David D},
  journal={Physics Letters A},
  volume={259},
  number={5},
  pages={331--333},
  year={1999},
  publisher={Elsevier},
  doi={10.1016/S0375-9601(99)00448-X}
}

@article{zhang2006cavity,
  title={Cavity-QED scheme to implement the optimal symmetric approximate quantum telecloning},
  author={Zhang, Wen-Hai and Ye, Liu},
  journal={Physics Letters A},
  volume={354},
  number={5-6},
  pages={344--352},
  year={2006},
  publisher={Elsevier},
  doi={10.1016/j.physleta.2006.01.101}
}

@article{PhysRevLett.110.173601,
  title = {Experimental Eavesdropping Based on Optimal Quantum Cloning},
  author = {Bartkiewicz, Karol and Lemr, Karel and \ifmmode \check{C}\else \v{C}\fi{}ernoch, Anton\'{\i}n and Soubusta, Jan and Miranowicz, Adam},
  journal = {Phys. Rev. Lett.},
  volume = {110},
  issue = {17},
  pages = {173601},
  numpages = {5},
  year = {2013},
  month = {Apr},
  publisher = {American Physical Society},
  doi = {10.1103/PhysRevLett.110.173601},
  url = {https://link.aps.org/doi/10.1103/PhysRevLett.110.173601}
}

@article{PhysRevA.67.022317,
  title = {Phase-covariant quantum cloning of qudits},
  author = {Fan, Heng and Imai, Hiroshi and Matsumoto, Keiji and Wang, Xiang-Bin},
  journal = {Phys. Rev. A},
  volume = {67},
  issue = {2},
  pages = {022317},
  numpages = {5},
  year = {2003},
  month = {Feb},
  publisher = {American Physical Society},
  doi = {10.1103/PhysRevA.67.022317},
  url = {https://link.aps.org/doi/10.1103/PhysRevA.67.022317}
}

@article{PhysRevA.72.042328,
  title = {Multipartite asymmetric quantum cloning},
  author = {Iblisdir, S. and Ac\'{\i}n, A. and Cerf, N. J. and Filip, R. and Fiur\'a\ifmmode \check{s}\else \v{s}\fi{}ek, J. and Gisin, N.},
  journal = {Phys. Rev. A},
  volume = {72},
  issue = {4},
  pages = {042328},
  numpages = {4},
  year = {2005},
  month = {Oct},
  publisher = {American Physical Society},
  doi = {10.1103/PhysRevA.72.042328},
  url = {https://link.aps.org/doi/10.1103/PhysRevA.72.042328}
}

@article{PhysRevA.62.042302,
  title = {Probabilistic quantum cloning via Greenberger-Horne-Zeilinger states},
  author = {Zhang, Chuan-Wei and Li, Chuan-Feng and Wang, Zi-Yang and Guo, Guang-Can},
  journal = {Phys. Rev. A},
  volume = {62},
  issue = {4},
  pages = {042302},
  numpages = {8},
  year = {2000},
  month = {Sep},
  publisher = {American Physical Society},
  doi = {10.1103/PhysRevA.62.042302},
  url = {https://link.aps.org/doi/10.1103/PhysRevA.62.042302}
}

@article{PhysRevLett.126.060503,
  title = {All-Optical Optimal $N$-to-$M$ Quantum Cloning of Coherent States},
  author = {Liu, Shengshuai and Lou, Yanbo and Chen, Yingxuan and Jing, Jietai},
  journal = {Phys. Rev. Lett.},
  volume = {126},
  issue = {6},
  pages = {060503},
  numpages = {6},
  year = {2021},
  month = {Feb},
  publisher = {American Physical Society},
  doi = {10.1103/PhysRevLett.126.060503},
  url = {https://link.aps.org/doi/10.1103/PhysRevLett.126.060503}
}

@article{PhysRevLett.95.090504,
  title = {Separating the Classical and Quantum Information via Quantum Cloning},
  author = {Ricci, M. and Sciarrino, F. and Cerf, N. J. and Filip, R. and Fiur\'a\ifmmode \check{s}\else \v{s}\fi{}ek, J. and De Martini, F.},
  journal = {Phys. Rev. Lett.},
  volume = {95},
  issue = {9},
  pages = {090504},
  numpages = {4},
  year = {2005},
  month = {Aug},
  publisher = {American Physical Society},
  doi = {10.1103/PhysRevLett.95.090504},
  url = {https://link.aps.org/doi/10.1103/PhysRevLett.95.090504}
}

@article{cerf2002cloning,
  title={Cloning a qutrit},
  author={Cerf, Nicolas and Durt, Thomas and Gisin, Nicolas},
  journal={Journal of modern optics},
  volume={49},
  number={8},
  pages={1355--1373},
  year={2002},
  publisher={Taylor \& Francis},
  doi={10.1080/09500340110109043}
}

@article{PhysRevA.69.032313,
  title = {Security of quantum key distributions with entangled qudits},
  author = {Durt, Thomas and Kaszlikowski, Dagomir and Chen, Jing-Ling and Kwek, L. C.},
  journal = {Phys. Rev. A},
  volume = {69},
  issue = {3},
  pages = {032313},
  numpages = {11},
  year = {2004},
  month = {Mar},
  publisher = {American Physical Society},
  doi = {10.1103/PhysRevA.69.032313},
  url = {https://link.aps.org/doi/10.1103/PhysRevA.69.032313}
}

@article{PhysRevA.71.042327,
  title = {Economical phase-covariant cloning of qudits},
  author = {Buscemi, Francesco and D'Ariano, Giacomo Mauro and Macchiavello, Chiara},
  journal = {Phys. Rev. A},
  volume = {71},
  issue = {4},
  pages = {042327},
  numpages = {7},
  year = {2005},
  month = {Apr},
  publisher = {American Physical Society},
  doi = {10.1103/PhysRevA.71.042327},
  url = {https://link.aps.org/doi/10.1103/PhysRevA.71.042327}
}

@article{Preskill2018quantumcomputingin,
  doi = {10.22331/q-2018-08-06-79},
  url = {https://doi.org/10.22331/q-2018-08-06-79},
  title = {Quantum {C}omputing in the {NISQ} era and beyond},
  author = {Preskill, John},
  journal = {{Quantum}},
  issn = {2521-327X},
  publisher = {{Verein zur F{\"{o}}rderung des Open Access Publizierens in den Quantenwissenschaften}},
  volume = {2},
  pages = {79},
  month = aug,
  year = {2018}
}

@ARTICLE{9774323,

  author={Aktar, Shamminuj and Bärtschi, Andreas and Badawy, Abdel-Hameed A. and Eidenbenz, Stephan},

  journal={IEEE Transactions on Quantum Engineering}, 

  title={A Divide-and-Conquer Approach to Dicke State Preparation}, 

  year={2022},

  volume={3},

  number={},

  pages={1-16},

  doi={10.1109/TQE.2022.3174547}}

@INPROCEEDINGS{9951196,

  author={Bärtschi, Andreas and Eidenbenz, Stephan},

  booktitle={2022 IEEE International Conference on Quantum Computing and Engineering (QCE)}, 

  title={Short-Depth Circuits for Dicke State Preparation}, 

  year={2022},

  volume={},

  number={},

  pages={87-96},

  doi={10.1109/QCE53715.2022.00027}}

@incollection{B_rtschi_2019,
	doi = {10.1007/978-3-030-25027-0_9},
  
	url = {https://doi.org/10.1007%2F978-3-030-25027-0_9},
  
	year = 2019,
	publisher = {Springer International Publishing},
  
	pages = {126--139},
  
	author = {Andreas Bärtschi and Stephan Eidenbenz},
  
	title = {Deterministic Preparation of Dicke States},
  
	booktitle = {Fundamentals of Computation Theory}
}

@article{PhysRevLett.87.247901,
  title = {Telecloning of Continuous Quantum Variables},
  author = {van Loock, P. and Braunstein, Samuel L.},
  journal = {Phys. Rev. Lett.},
  volume = {87},
  issue = {24},
  pages = {247901},
  numpages = {4},
  year = {2001},
  month = {Nov},
  publisher = {American Physical Society},
  doi = {10.1103/PhysRevLett.87.247901},
  url = {https://link.aps.org/doi/10.1103/PhysRevLett.87.247901}
}

@article{PhysRevA.72.032320,
  title = {Security bound of two-basis quantum-key-distribution protocols using qudits},
  author = {Nikolopoulos, Georgios M. and Alber, Gernot},
  journal = {Phys. Rev. A},
  volume = {72},
  issue = {3},
  pages = {032320},
  numpages = {10},
  year = {2005},
  month = {Sep},
  publisher = {American Physical Society},
  doi = {10.1103/PhysRevA.72.032320},
  url = {https://link.aps.org/doi/10.1103/PhysRevA.72.032320}
}

@article{PhysRevA.84.034302,
  title = {Unified universal quantum cloning machine and fidelities},
  author = {Wang, Yi-Nan and Shi, Han-Duo and Xiong, Zhao-Xi and Jing, Li and Ren, Xi-Jun and Mu, Liang-Zhu and Fan, Heng},
  journal = {Phys. Rev. A},
  volume = {84},
  issue = {3},
  pages = {034302},
  numpages = {4},
  year = {2011},
  month = {Sep},
  publisher = {American Physical Society},
  doi = {10.1103/PhysRevA.84.034302},
  url = {https://link.aps.org/doi/10.1103/PhysRevA.84.034302}
}

@article{cozzolino2019high,
  title={High-dimensional quantum communication: benefits, progress, and future challenges},
  author={Cozzolino, Daniele and Da Lio, Beatrice and Bacco, Davide and Oxenl{\o}we, Leif Katsuo},
  journal={Advanced Quantum Technologies},
  volume={2},
  number={12},
  pages={1900038},
  year={2019},
  publisher={Wiley Online Library},
  doi={10.1002/qute.201900038}
}

@article{10.5555/2535680.2535689,
author = {Kay, Alastair and Ramanathan, Ravishankar and Kaszlikowshi, Dagomir},
title = {Optimal Asymmetric Quantum Cloning for Quantum Information and Computation},
year = {2013},
issue_date = {September 2013},
publisher = {Rinton Press, Incorporated},
address = {Paramus, NJ},
volume = {13},
number = {9–10},
issn = {1533-7146},
journal = {Quantum Info. Comput.},
month = {sep},
pages = {880–900},
numpages = {21}
}

@article{PhysRevA.79.064306,
  title = {Probabilistic ancilla-free phase-covariant telecloning of qudits with the optimal fidelity},
  author = {Wang, Xin-Wen and Yang, Guo-Jian},
  journal = {Phys. Rev. A},
  volume = {79},
  issue = {6},
  pages = {064306},
  numpages = {4},
  year = {2009},
  month = {Jun},
  publisher = {American Physical Society},
  doi = {10.1103/PhysRevA.79.064306},
  url = {https://link.aps.org/doi/10.1103/PhysRevA.79.064306}
}

@article{PhysRevA.72.032331,
  title = {Broadcasting of entanglement at a distance using linear optics and telecloning of entanglement},
  author = {Ghiu, Iulia and Karlsson, Anders},
  journal = {Phys. Rev. A},
  volume = {72},
  issue = {3},
  pages = {032331},
  numpages = {10},
  year = {2005},
  month = {Sep},
  publisher = {American Physical Society},
  doi = {10.1103/PhysRevA.72.032331},
  url = {https://link.aps.org/doi/10.1103/PhysRevA.72.032331}
}

@article{PhysRevA.67.012323,
  title = {Asymmetric quantum telecloning of d-level systems and broadcasting of entanglement to different locations using the ``many-to-many'' communication protocol},
  author = {Ghiu, Iulia},
  journal = {Phys. Rev. A},
  volume = {67},
  issue = {1},
  pages = {012323},
  numpages = {12},
  year = {2003},
  month = {Jan},
  publisher = {American Physical Society},
  doi = {10.1103/PhysRevA.67.012323},
  url = {https://link.aps.org/doi/10.1103/PhysRevA.67.012323}
}

@article{srikara2020continuous,
  title={Continuous variable B92 quantum key distribution protocol using single photon added and subtracted coherent states},
  author={Srikara, S and Thapliyal, Kishore and Pathak, Anirban},
  journal={Quantum Information Processing},
  volume={19},
  pages={1--16},
  year={2020},
  publisher={Springer},
  doi={10.1007/s11128-020-02872-6}
}

@article{lu2019experimental,
  title={Experimental quantum network coding},
  author={Lu, He and Li, Zheng-Da and Yin, Xu-Fei and Zhang, Rui and Fang, Xiao-Xu and Li, Li and Liu, Nai-Le and Xu, Feihu and Chen, Yu-Ao and Pan, Jian-Wei},
  journal={npj Quantum Information},
  volume={5},
  number={1},
  pages={89},
  year={2019},
  publisher={Nature Publishing Group UK London},
  doi={10.1038/s41534-019-0207-2}
}

@article{PhysRevLett.125.210502,
  title = {Cloning of Quantum Entanglement},
  author = {Peng, Li-Chao and Wu, Dian and Zhong, Han-Sen and Luo, Yi-Han and Li, Yuan and Hu, Yi and Jiang, Xiao and Chen, Ming-Cheng and Li, Li and Liu, Nai-Le and Nemoto, Kae and Munro, William J. and Sanders, Barry C. and Lu, Chao-Yang and Pan, Jian-Wei},
  journal = {Phys. Rev. Lett.},
  volume = {125},
  issue = {21},
  pages = {210502},
  numpages = {6},
  year = {2020},
  month = {Nov},
  publisher = {American Physical Society},
  doi = {10.1103/PhysRevLett.125.210502},
  url = {https://link.aps.org/doi/10.1103/PhysRevLett.125.210502}
}

@article{zhukov2019quantum,
  title={Quantum communication protocols as a benchmark for programmable quantum computers},
  author={Zhukov, AA and Kiktenko, Evgeniy O and Elistratov, AA and Pogosov, WV and Lozovik, Yu E},
  journal={Quantum Information Processing},
  volume={18},
  number={1},
  pages={31},
  year={2019},
  publisher={Springer},
  doi={10.1007/s11128-018-2144-y}
}

@misc{gupta2023encoding,
      title={Encoding a magic state with beyond break-even fidelity}, 
      author={Riddhi S. Gupta and Neereja Sundaresan and Thomas Alexander and Christopher J. Wood and Seth T. Merkel and Michael B. Healy and Marius Hillenbrand and Tomas Jochym-O'Connor and James R. Wootton and Theodore J. Yoder and Andrew W. Cross and Maika Takita and Benjamin J. Brown},
      year={2023},
      eprint={2305.13581},
      archivePrefix={arXiv},
      primaryClass={quant-ph}
}

@article{yang2021experimental,
  title={Experimental demonstration of entanglement-enabled universal quantum cloning in a circuit},
  author={Yang, Zhen-Biao and Han, Pei-Rong and Huang, Xin-Jie and Ning, Wen and Li, Hekang and Xu, Kai and Zheng, Dongning and Fan, Heng and Zheng, Shi-Biao},
  journal={npj Quantum Information},
  volume={7},
  number={1},
  pages={44},
  year={2021},
  publisher={Nature Publishing Group UK London},
  doi={10.1038/s41534-021-00375-5}
}

@article{McKay_2017,
	doi = {10.1103/physreva.96.022330},
  
	url = {https://doi.org/10.1103%2Fphysreva.96.022330},
  
	year = 2017,
	month = {aug},
  
	publisher = {American Physical Society ({APS})},
  
	volume = {96},
  
	number = {2},
  
	author = {David C. McKay and Christopher J. Wood and Sarah Sheldon and Jerry M. Chow and Jay M. Gambetta},
  
	title = {Efficient Z-Gates for Quantum Computing},
  
	journal = {Physical Review A}
}

@article{PhysRevX.10.011022,
  title = {Topological and Subsystem Codes on Low-Degree Graphs with Flag Qubits},
  author = {Chamberland, Christopher and Zhu, Guanyu and Yoder, Theodore J. and Hertzberg, Jared B. and Cross, Andrew W.},
  journal = {Phys. Rev. X},
  volume = {10},
  issue = {1},
  pages = {011022},
  numpages = {19},
  year = {2020},
  month = {Jan},
  publisher = {American Physical Society},
  doi = {10.1103/PhysRevX.10.011022},
  url = {https://link.aps.org/doi/10.1103/PhysRevX.10.011022}
}

@article{PhysRevLett.108.070502,
  title = {Efficient Method for Computing the Maximum-Likelihood Quantum State from Measurements with Additive Gaussian Noise},
  author = {Smolin, John A. and Gambetta, Jay M. and Smith, Graeme},
  journal = {Phys. Rev. Lett.},
  volume = {108},
  issue = {7},
  pages = {070502},
  numpages = {4},
  year = {2012},
  month = {Feb},
  publisher = {American Physical Society},
  doi = {10.1103/PhysRevLett.108.070502},
  url = {https://link.aps.org/doi/10.1103/PhysRevLett.108.070502}
}

@article{Vidal_2002,
	doi = {10.1103/physreva.65.032314},
  
	url = {https://doi.org/10.1103%2Fphysreva.65.032314},
  
	year = 2002,
	month = {feb},
  
	publisher = {American Physical Society ({APS})},
  
	volume = {65},
  
	number = {3},
  
	author = {G. Vidal and R. F. Werner},
  
	title = {Computable measure of entanglement},
  
	journal = {Physical Review A}
}

@article{PhysRevA.58.883,
  title = {Volume of the set of separable states},
  author={{\.Z}yczkowski, Karol and Horodecki, Pawe{\l} and Sanpera, Anna and Lewenstein, Maciej},  journal = {Phys. Rev. A},
  volume = {58},
  issue = {2},
  pages = {883--892},
  numpages = {0},
  year = {1998},
  month = {Aug},
  publisher = {American Physical Society},
  doi = {10.1103/PhysRevA.58.883},
  url = {https://link.aps.org/doi/10.1103/PhysRevA.58.883}
}

@misc{eisert2006entanglement,
      title={Entanglement in quantum information theory}, 
      author={J. Eisert},
      year={2006},
      eprint={quant-ph/0610253},
      archivePrefix={arXiv},
      primaryClass={quant-ph}
}

@article{Hill_1997,
	doi = {10.1103/physrevlett.78.5022},
  
	url = {https://doi.org/10.1103%2Fphysrevlett.78.5022},
  
	year = 1997,
	month = {jun},
  
	publisher = {American Physical Society ({APS})},
  
	volume = {78},
  
	number = {26},
  
	pages = {5022--5025},
  
	author = {Scott Hill and William K. Wootters},
  
	title = {Entanglement of a Pair of Quantum Bits},
  
	journal = {Physical Review Letters}
}

@article{PhysRevLett.80.2245,
  title = {Entanglement of Formation of an Arbitrary State of Two Qubits},
  author = {Wootters, William K.},
  journal = {Phys. Rev. Lett.},
  volume = {80},
  issue = {10},
  pages = {2245--2248},
  numpages = {0},
  year = {1998},
  month = {Mar},
  publisher = {American Physical Society},
  doi = {10.1103/PhysRevLett.80.2245},
  url = {https://link.aps.org/doi/10.1103/PhysRevLett.80.2245}
}

@article{Livingston_2022,
	doi = {10.1038/s41467-022-29906-0},
  
	url = {https://doi.org/10.1038%2Fs41467-022-29906-0},
  
	year = 2022,
	month = {apr},
  
	publisher = {Springer Science and Business Media {LLC}
},
  
	volume = {13},
  
	number = {1},
  
	author = {William P. Livingston and Machiel S. Blok and Emmanuel Flurin and Justin Dressel and Andrew N. Jordan and Irfan Siddiqi},
  
	title = {Experimental demonstration of continuous quantum error correction},
  
	journal = {Nature Communications}
}

@misc{govia2022randomized,
      title={A randomized benchmarking suite for mid-circuit measurements}, 
      author={L. C. G. Govia and P. Jurcevic and S. T. Merkel and D. C. McKay},
      year={2022},
      eprint={2207.04836},
      archivePrefix={arXiv},
      primaryClass={quant-ph}
}

@article{Sundaresan_2023,
	doi = {10.1038/s41467-023-38247-5},
  
	url = {https://doi.org/10.1038%2Fs41467-023-38247-5},
  
	year = 2023,
	month = {may},
  
	publisher = {Springer Science and Business Media {LLC}
},
  
	volume = {14},
  
	number = {1},
  
	author = {Neereja Sundaresan and Theodore J. Yoder and Youngseok Kim and Muyuan Li and Edward H. Chen and Grace Harper and Ted Thorbeck and Andrew W. Cross and Antonio D. C{\'{o}}rcoles and Maika Takita},
  
	title = {Demonstrating multi-round subsystem quantum error correction using matching and maximum likelihood decoders},
  
	journal = {Nature Communications}
}

@article{Chen_2022,
	doi = {10.1103/physrevlett.128.110504},
  
	url = {https://doi.org/10.1103%2Fphysrevlett.128.110504},
  
	year = 2022,
	month = {mar},
  
	publisher = {American Physical Society ({APS})},
  
	volume = {128},
  
	number = {11},
  
	author = {Edward H. Chen and Theodore J. Yoder and Youngseok Kim and Neereja Sundaresan and Srikanth Srinivasan and Muyuan Li and Antonio D. C{\'{o}
}rcoles and Andrew W. Cross and Maika Takita},
  
	title = {Calibrated Decoders for Experimental Quantum Error Correction},
  
	journal = {Physical Review Letters}
}

@article{PhysRevA.62.062302,
  title = {Approximate quantum cloning and the impossibility of superluminal information transfer},
  author = {Bruss, D. and D'Ariano, G. M. and Macchiavello, C. and Sacchi, M. F.},
  journal = {Phys. Rev. A},
  volume = {62},
  issue = {6},
  pages = {062302},
  numpages = {4},
  year = {2000},
  month = {Nov},
  publisher = {American Physical Society},
  doi = {10.1103/PhysRevA.62.062302},
  url = {https://link.aps.org/doi/10.1103/PhysRevA.62.062302}
}

@article{PhysRevLett.88.187901,
  title = {Approximate Quantum Cloning with Nuclear Magnetic Resonance},
  author = {Cummins, Holly K. and Jones, Claire and Furze, Alistair and Soffe, Nicholas F. and Mosca, Michele and Peach, Josephine M. and Jones, Jonathan A.},
  journal = {Phys. Rev. Lett.},
  volume = {88},
  issue = {18},
  pages = {187901},
  numpages = {4},
  year = {2002},
  month = {Apr},
  publisher = {American Physical Society},
  doi = {10.1103/PhysRevLett.88.187901},
  url = {https://link.aps.org/doi/10.1103/PhysRevLett.88.187901}
}

@article{PhysRevA.105.042604,
  title = {Progress toward practical quantum cryptanalysis by variational quantum cloning},
  author = {Coyle, Brian and Doosti, Mina and Kashefi, Elham and Kumar, Niraj},
  journal = {Phys. Rev. A},
  volume = {105},
  issue = {4},
  pages = {042604},
  numpages = {35},
  year = {2022},
  month = {Apr},
  publisher = {American Physical Society},
  doi = {10.1103/PhysRevA.105.042604},
  url = {https://link.aps.org/doi/10.1103/PhysRevA.105.042604}
}

@misc{gupta10achieving,
  title={Achieving improved fidelity for quantum cloning on the IBM quantum computer},
  author={Gupta, Udit and Behera, Bikash K and Panigrahi, Prasanta K},
  publisher={DOI},
  doi={10.13140/RG.2.2.18174.13128}
}

@article{PhysRevLett.92.047901,
  title = {Teleportation Scheme Implementing the Universal Optimal Quantum Cloning Machine and the Universal NOT Gate},
  author = {Ricci, M. and Sciarrino, F. and Sias, C. and De Martini, F.},
  journal = {Phys. Rev. Lett.},
  volume = {92},
  issue = {4},
  pages = {047901},
  numpages = {4},
  year = {2004},
  month = {Jan},
  publisher = {American Physical Society},
  doi = {10.1103/PhysRevLett.92.047901},
  url = {https://link.aps.org/doi/10.1103/PhysRevLett.92.047901}
}

@article{haw2016surpassing,
  title={Surpassing the no-cloning limit with a heralded hybrid linear amplifier for coherent states},
  author={Haw, Jing Yan and Zhao, Jie and Dias, Josephine and Assad, Syed M and Bradshaw, Mark and Blandino, R{\'e}mi and Symul, Thomas and Ralph, Timothy C and Lam, Ping Koy},
  journal={Nature communications},
  volume={7},
  number={1},
  pages={13222},
  year={2016},
  publisher={Nature Publishing Group UK London},
  doi={10.1038/ncomms13222}
}

@article{dan2010universal,
  title={Universal Quantum Cloning Machine in Circuit Quantum Electrodynamics},
  author={Dan-Dan, Lv and Hong, Lu and Ya-Fei, Yu and Xun-Li, Feng and Zhi-Ming, Zhang},
  journal={Chinese Physics Letters},
  volume={27},
  number={2},
  pages={020302},
  year={2010},
  publisher={IOP Publishing},
  doi={10.1088/0256-307X/27/2/020302}
}

@article{Maruyama_2003,
	doi = {10.1103/physreva.67.032303},
  
	url = {https://doi.org/10.1103%2Fphysreva.67.032303},
  
	year = 2003,
	month = {mar},
  
	publisher = {American Physical Society ({APS})},
  
	volume = {67},
  
	number = {3},
  
	author = {K. Maruyama and P. L. Knight},
  
	title = {Upper bounds for the number of quantum clones under decoherence},
  
	journal = {Physical Review A}
}

@article{PhysRevLett.82.2417,
  title = {Dynamical Decoupling of Open Quantum Systems},
  author = {Viola, Lorenza and Knill, Emanuel and Lloyd, Seth},
  journal = {Phys. Rev. Lett.},
  volume = {82},
  issue = {12},
  pages = {2417--2421},
  numpages = {0},
  year = {1999},
  month = {Mar},
  publisher = {American Physical Society},
  doi = {10.1103/PhysRevLett.82.2417},
  url = {https://link.aps.org/doi/10.1103/PhysRevLett.82.2417}
}

@article{Pokharel_2023,
	doi = {10.1103/physrevlett.130.210602},
  
	url = {https://doi.org/10.1103%2Fphysrevlett.130.210602},
  
	year = 2023,
	month = {may},
  
	publisher = {American Physical Society ({APS})},
  
	volume = {130},
  
	number = {21},
  
	author = {Bibek Pokharel and Daniel A. Lidar},
  
	title = {Demonstration of Algorithmic Quantum Speedup},
  
	journal = {Physical Review Letters}
}

@article{PhysRevLett.121.220502,
  title = {Demonstration of Fidelity Improvement Using Dynamical Decoupling with Superconducting Qubits},
  author = {Pokharel, Bibek and Anand, Namit and Fortman, Benjamin and Lidar, Daniel A.},
  journal = {Phys. Rev. Lett.},
  volume = {121},
  issue = {22},
  pages = {220502},
  numpages = {6},
  year = {2018},
  month = {Nov},
  publisher = {American Physical Society},
  doi = {10.1103/PhysRevLett.121.220502},
  url = {https://link.aps.org/doi/10.1103/PhysRevLett.121.220502}
}

@misc{ezzell2023dynamical,
      title={Dynamical decoupling for superconducting qubits: a performance survey}, 
      author={Nic Ezzell and Bibek Pokharel and Lina Tewala and Gregory Quiroz and Daniel A. Lidar},
      year={2023},
      eprint={2207.03670},
      archivePrefix={arXiv},
      primaryClass={quant-ph}
}

@ARTICLE{9872062,

  author={Niu, Siyuan and Todri-Sanial, Aida},

  journal={IEEE Transactions on Quantum Engineering}, 

  title={Effects of Dynamical Decoupling and Pulse-Level Optimizations on IBM Quantum Computers}, 

  year={2022},

  volume={3},

  number={},

  pages={1-10},

  doi={10.1109/TQE.2022.3203153}}

@article{Souza_2012,
	doi = {10.1098/rsta.2011.0355},
  
	url = {https://doi.org/10.1098%2Frsta.2011.0355},
  
	year = 2012,
	month = {oct},
  
	publisher = {The Royal Society},
  
	volume = {370},
  
	number = {1976},
  
	pages = {4748--4769},
  
	author = {Alexandre M. Souza and Gonzalo A. {\'{A}
}lvarez and Dieter Suter},
  
	title = {Robust dynamical decoupling},
  
	journal = {Philosophical Transactions of the Royal Society A: Mathematical, Physical and Engineering Sciences}
}

@misc{suau2022vector,
      title={Vector Field Visualization of Single-Qubit State Tomography}, 
      author={Adrien Suau and Marc Vuffray and Andrey Y. Lokhov and Lukasz Cincio and Carleton Coffrin},
      year={2022},
      eprint={2205.02483},
      archivePrefix={arXiv},
      primaryClass={quant-ph}
}

@article{Pelofske_2022,
	doi = {10.1109/tqe.2022.3184764},
  
	url = {https://doi.org/10.1109%2Ftqe.2022.3184764},
  
	year = 2022,
	publisher = {Institute of Electrical and Electronics Engineers ({IEEE})},
  
	volume = {3},
  
	pages = {1--19},
  
	author = {Elijah Pelofske and Andreas Bärtschi and Stephan Eidenbenz},
  
	title = {{Quantum Volume in Practice: What Users Can Expect From {NISQ} Devices}},
  
	journal = {{IEEE} Transactions on Quantum Engineering}
}

@misc{dasgupta2021stability,
      title={Stability of noisy quantum computing devices}, 
      author={Samudra Dasgupta and Travis S. Humble},
      year={2021},
      eprint={2105.09472},
      archivePrefix={arXiv},
      primaryClass={quant-ph}
}

@Article{e24020244,
AUTHOR = {Dasgupta, Samudra and Humble, Travis S.},
TITLE = {Characterizing the Reproducibility of Noisy Quantum Circuits},
JOURNAL = {Entropy},
VOLUME = {24},
YEAR = {2022},
NUMBER = {2},
ARTICLE-NUMBER = {244},
URL = {https://www.mdpi.com/1099-4300/24/2/244},
PubMedID = {35205538},
ISSN = {1099-4300},
DOI = {10.3390/e24020244}
}

@INPROCEEDINGS{9259941,

  author={Dasgupta, Samudra and Humble, Travis S.},

  booktitle={2020 IEEE International Conference on Quantum Computing and Engineering (QCE)}, 

  title={Characterizing the Stability of NISQ Devices}, 

  year={2020},

  volume={},

  number={},

  pages={419-429},

  doi={10.1109/QCE49297.2020.00059}}

@misc{dasgupta2023reliability,
      title={Reliability of Noisy Quantum Computing Devices}, 
      author={Samudra Dasgupta and Travis S. Humble},
      year={2023},
      eprint={2307.06833},
      archivePrefix={arXiv},
      primaryClass={quant-ph}
}

@article{Pelofske_2023,
	doi = {10.1088/2058-9565/accbe6},
  
	url = {https://doi.org/10.1088%2F2058-9565%2Faccbe6},
  
	year = 2023,
	month = {apr},
  
	publisher = {{IOP} Publishing},
  
	volume = {8},
  
	number = {3},
  
	pages = {035005},
  
	author = {Elijah Pelofske and Georg Hahn and Hristo N Djidjev},
  
	title = {Noise dynamics of quantum annealers: estimating the effective noise using idle qubits},
  
	journal = {Quantum Science and Technology}
}

@article{PhysRevA.66.042304,
  title = {Relative error of state-dependent cloning},
  author = {Rastegin, A. E.},
  journal = {Phys. Rev. A},
  volume = {66},
  issue = {4},
  pages = {042304},
  numpages = {6},
  year = {2002},
  month = {Oct},
  publisher = {American Physical Society},
  doi = {10.1103/PhysRevA.66.042304},
  url = {https://link.aps.org/doi/10.1103/PhysRevA.66.042304}
}

@article{adhikari2007hybrid,
  title={Hybrid quantum cloning machine},
  author={Adhikari, Satyabrata and Pati, Arun K and Chakrabarty, Indranil and Choudhury, Binayak S},
  journal={Quantum Information Processing},
  volume={6},
  number={4},
  pages={197--219},
  year={2007},
  publisher={Springer},
  doi={10.1007/s11128-007-0053-6}
}

@article{PhysRevA.62.022301,
  title = {Cloning and quantum computation},
  author = {Galv\~ao, Ernesto F. and Hardy, Lucien},
  journal = {Phys. Rev. A},
  volume = {62},
  issue = {2},
  pages = {022301},
  numpages = {5},
  year = {2000},
  month = {Jul},
  publisher = {American Physical Society},
  doi = {10.1103/PhysRevA.62.022301},
  url = {https://link.aps.org/doi/10.1103/PhysRevA.62.022301}
}

@article{PhysRevA.58.3484,
  title = {Quantum cloning in $d$ dimensions},
  author = {Zanardi, Paolo},
  journal = {Phys. Rev. A},
  volume = {58},
  issue = {5},
  pages = {3484--3490},
  numpages = {0},
  year = {1998},
  month = {Nov},
  publisher = {American Physical Society},
  doi = {10.1103/PhysRevA.58.3484},
  url = {https://link.aps.org/doi/10.1103/PhysRevA.58.3484}
}

@misc{bäumer2023efficient,
      title={Efficient Long-Range Entanglement using Dynamic Circuits}, 
      author={Elisa Bäumer and Vinay Tripathi and Derek S. Wang and Patrick Rall and Edward H. Chen and Swarnadeep Majumder and Alireza Seif and Zlatko K. Minev},
      year={2023},
      eprint={2308.13065},
      archivePrefix={arXiv},
      primaryClass={quant-ph}
}

@article{Adami_2015,
   title={Black holes are almost optimal quantum cloners},
   volume={48},
   ISSN={1751-8121},
   url={http://dx.doi.org/10.1088/1751-8113/48/23/23FT01},
   DOI={10.1088/1751-8113/48/23/23ft01},
   number={23},
   journal={Journal of Physics A: Mathematical and Theoretical},
   publisher={IOP Publishing},
   author={Adami, Christoph and Steeg, Greg Ver},
   year={2015},
   month=may, pages={23FT01} }

@Article{sym15010062,
AUTHOR = {Mena López, Arturo and Wu, Lian-Ao},
TITLE = {Protectability of IBMQ Qubits by Dynamical Decoupling Technique},
JOURNAL = {Symmetry},
VOLUME = {15},
YEAR = {2023},
NUMBER = {1},
ARTICLE-NUMBER = {62},
URL = {https://www.mdpi.com/2073-8994/15/1/62},
ISSN = {2073-8994},
DOI = {10.3390/sym15010062}
}

@article{PhysRevApplied.18.024068,
  title = {Suppression of Crosstalk in Superconducting Qubits Using Dynamical Decoupling},
  author = {Tripathi, Vinay and Chen, Huo and Khezri, Mostafa and Yip, Ka-Wa and Levenson-Falk, E.M. and Lidar, Daniel A.},
  journal = {Phys. Rev. Appl.},
  volume = {18},
  issue = {2},
  pages = {024068},
  numpages = {24},
  year = {2022},
  month = {Aug},
  publisher = {American Physical Society},
  doi = {10.1103/PhysRevApplied.18.024068},
  url = {https://link.aps.org/doi/10.1103/PhysRevApplied.18.024068}
}

@article{Zanardi1999,
  doi = {10.1016/S0375-9601(99)00365-5},
  title={Symmetrizing evolutions},
  author={Zanardi, P.},
  journal={Physics Letters A},
  volume={258},
  number={2--3},
  pages={77--82},
  year={1999},
  publisher={Elsevier}
}

@misc{dd_crosstalk,
  title = {{Quantum Crosstalk Robust Quantum Control}},
  author = {Zhou, Zeyuan and Sitler, Ryan and Oda, Yasuo and Schultz, Kevin and Quiroz, Gregory},
  journal = {Physical Review Letters},
  volume = {131},
  number = {21},
  pages = {210802},
  numpages = {7},
  year = {2023},
  month = {Nov},
  publisher = {American Physical Society},
  doi = {10.1103/PhysRevLett.131.210802},
  url = {https://link.aps.org/doi/10.1103/PhysRevLett.131.210802},
}

@article{randomized_benchmark_2008,
	title = {Randomized benchmarking of quantum gates},
    year = {2008},
	volume = {77},
	url = {https://link.aps.org/doi/10.1103/PhysRevA.77.012307},
	doi = {10.1103/PhysRevA.77.012307},
	number = {1},
	journal = {Physical Review A},
	shortjournal = {Phys. Rev. A},
	author = {Knill, E. and Leibfried, D. and Reichle, R. and Britton, J. and Blakestad, R. B. and Jost, J. D. and Langer, C. and Ozeri, R. and Seidelin, S. and Wineland, D. J.},
	urldate = {2023-11-15},
	date = {2008-01-08},
	note = {Publisher: American Physical Society},
}

@article{randomized_benchmark_2011,
	title = {{Scalable and Robust Randomized Benchmarking of Quantum Processes}},
	volume = {106},
    year = {2011},
	url = {https://link.aps.org/doi/10.1103/PhysRevLett.106.180504},
	doi = {10.1103/PhysRevLett.106.180504},
	pages = {180504},
	number = {18},
	journal = {Physical Review Letters},
	shortjournal = {Phys. Rev. Lett.},
	author = {Magesan, Easwar and Gambetta, J. M. and Emerson, Joseph},
	urldate = {2023-11-15},
	date = {2011-05-06},
	note = {Publisher: American Physical Society},
}

@article{PhysRevA.94.052325,
  title = {Noise tailoring for scalable quantum computation via randomized compiling},
  author = {Wallman, Joel J. and Emerson, Joseph},
  journal = {Phys. Rev. A},
  volume = {94},
  issue = {5},
  pages = {052325},
  numpages = {9},
  year = {2016},
  month = {Nov},
  publisher = {American Physical Society},
  doi = {10.1103/PhysRevA.94.052325},
  url = {https://link.aps.org/doi/10.1103/PhysRevA.94.052325}
}

@article{PhysRevLett.107.080502,
  title = {{Simple All-Microwave Entangling Gate for Fixed-Frequency Superconducting Qubits}},
  author = {Chow, Jerry M. and C\'orcoles, A. D. and Gambetta, Jay M. and Rigetti, Chad and Johnson, B. R. and Smolin, John A. and Rozen, J. R. and Keefe, George A. and Rothwell, Mary B. and Ketchen, Mark B. and Steffen, M.},
  journal = {Phys. Rev. Lett.},
  volume = {107},
  issue = {8},
  pages = {080502},
  numpages = {5},
  year = {2011},
  month = {Aug},
  publisher = {American Physical Society},
  doi = {10.1103/PhysRevLett.107.080502},
  url = {https://link.aps.org/doi/10.1103/PhysRevLett.107.080502},
      eprint={1106.0553},
      archivePrefix={arXiv},
}

@article{C_rcoles_2021,
   title={Exploiting Dynamic Quantum Circuits in a Quantum Algorithm with Superconducting Qubits},
   volume={127},
   ISSN={1079-7114},
   url={http://dx.doi.org/10.1103/PhysRevLett.127.100501},
   DOI={10.1103/physrevlett.127.100501},
   number={10},
   journal={Physical Review Letters},
   publisher={American Physical Society (APS)},
   author={Córcoles, A. D. and Takita, Maika and Inoue, Ken and Lekuch, Scott and Minev, Zlatko K. and Chow, Jerry M. and Gambetta, Jay M.},
   year={2021},
   month=aug }
